\documentclass[12pt]{extreport}
\renewcommand{\baselinestretch}{1.5}
\usepackage{cite}
\usepackage[dvips]{graphicx}
\usepackage{graphicx}
\usepackage{uathes}
\usepackage{amsfonts,eucal,bm}
\usepackage{a4wide}
\usepackage{amssymb}
\usepackage{amsmath}
\begin{document}
\hoffset=-29.0pt \voffset=-4.0cm \textwidth=16.0cm \textheight=24.0cm
% Create the title page
%\vskip 4 true cm
\pagestyle{empty}
$$ $$
\vskip 1 true cm
\centerline{{\normalsize {PARTICLE PRODUCTION IN MATTER}}} 
\centerline{{\normalsize {AT EXTREME CONDITIONS}}}
\bigskip
 \centerline{{\normalsize {by}}}
\centerline{{\normalsize {Inga Vladimirovna Kuznetsova}}}
\vskip 1.5 true cm
%\centerline{{\normalsize {____}}}
\centerline{\rule{5cm}{0.02cm}}
\centerline{{\normalsize{Copyright
$\copyright$ Inga Vladimirovna Kuznetsova 2009}}}
\vskip 2 true cm

\begin{center}
{\normalsize {The Dissertation Submitted to the Faculty of \\
THE DEPARTMENT OF PHYSICS}}
\end{center}
\begin{center}
{\normalsize {in Partial Fulfillments of the Requirements
\\ For the Degree of}}
\end{center}
\begin{center}
{\normalsize  {DOCTOR OF PHILOSOPHY }}
\end{center}
\begin{center}
{\normalsize {In the Graduate College \\
UNIVERSITY OF ARIZONA}}
\end{center}

\vskip 2.5 true cm

\centerline{2009} \vfill

%Insert correct approval form here AND PAGE NUMBER!!!
\pagestyle{empty}

\topmargin 1 cm\oddsidemargin = 1.5cm\evensidemargin = 1.5cm
\begin{figure}
\centering
 \includegraphics[width=16.0cm, height=25.0cm]{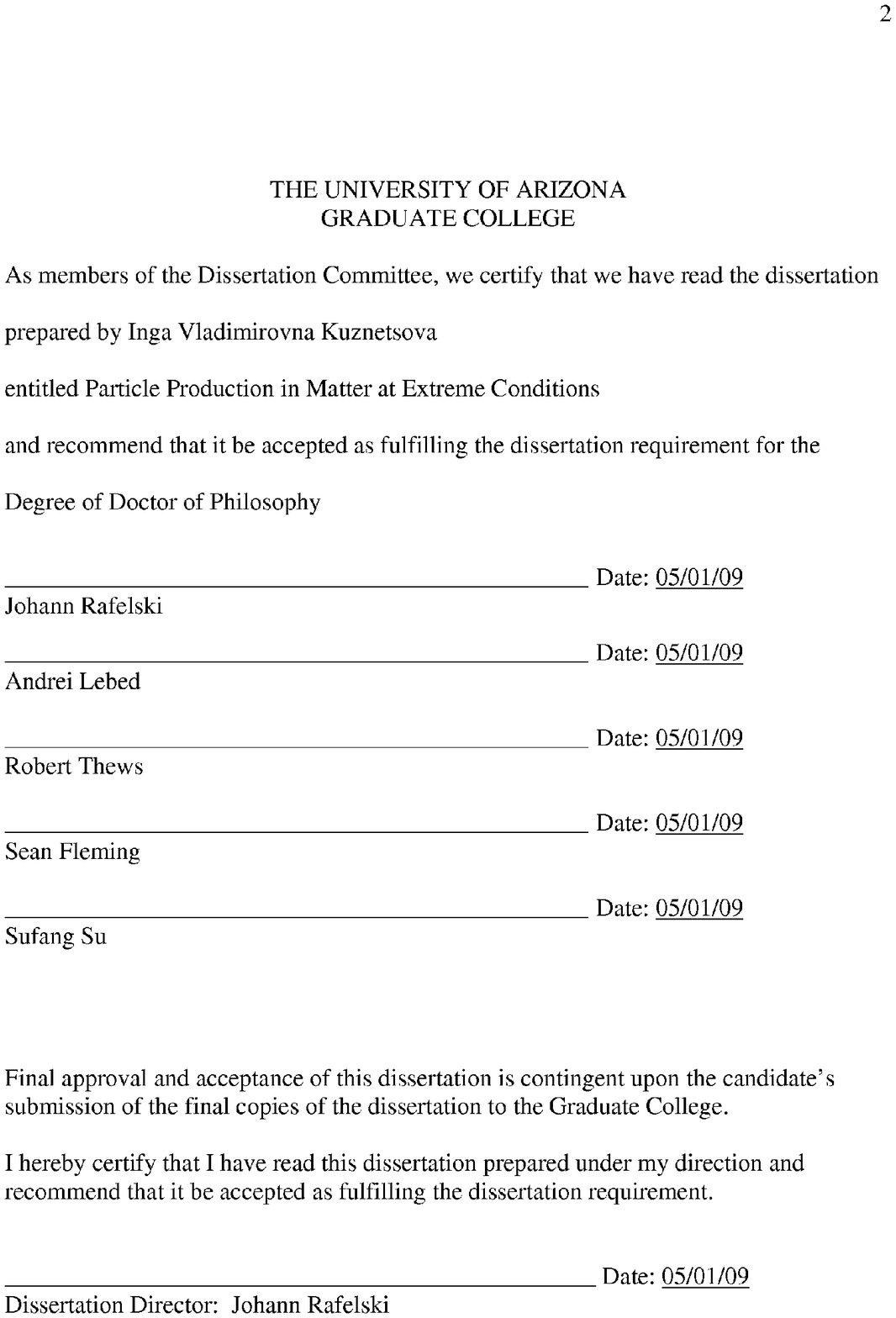}
\end{figure}

 \newpage
 
\topmargin 5.5cm\oddsidemargin = 1.5cm\evensidemargin = 1.5cm
%\newpage 
%\hoffset=-29.0pt \voffset=-2.0cm \textwidth=16.0cm \textheight=24.0cm

\begin{center}
STATEMENT BY AUTHOR
\end{center}

This dissertation has been submitted in partial fulfillments of requirements for an advanced degree 
at The University of Arizona and is deposited in the University Library to be made available for borrowers
under rules of the library.
\\

Brief quotations from this dissertation are allowable without special permission, provided that accurate
acknowledgment of source is made. Requests for permission to extended quotation from or reproduction of this manuscript in whole or 
in part may be granted by the copyright holder.
\vskip 1 true cm
$\,\,\,\,\,\,\,\,\,\,\,\,\,\,\,\,\,\,\,\,\,\,\,\,\,\,\,\,\,\,\,\,\,\,\,\,\,\,\,\,\,\,\,\,\,\,\,\,\,\,\,\,\,\,\,\,\,\,\,\,\,\,\,\,\,\,\,\,\,\,\,\,\,\,\,\,\,\,$ SIGNED: \underline{Inga Vladimirovna Kuznetsova}
%
%\input{statbyauth}
%\statementbyauthor
%\input{staut}

% Insert the approval form
%\addcontentsline{toc}{chapter}{FINAL EXAMINING COMMITTEE APPROVAL FORM}
%Insert correct approval form here AND PAGE NUMBER!!!
%\begin{figure}[h]
%\centering
%  \psfig{figure=page2.ps,width=15cm}
%\end{figure}

%\newpage
%newpage
%\input{page2}
% Include the ``Statement by Author''
%% Include the ``Abstract''
\hoffset=-29.0pt \voffset=-6.0cm \textwidth=16.0cm \textheight=24.0cm
\acknowledgements

"{\it When you really want something, all the universe conspires in helping you to achieve it }"
Paulo Coelho "Alchemist"\\

I would like to thank all people who directly or indirectly contributed to this work and to my decision to go to graduate school. 

First I would like to thank my advisor Johann Rafelski for guiding me through four long years. 
I got a lot of inspiration and support from him during these years.
His approaches to research and presentations definitely had much impact on me. 
Thank you for your patience, optimism, intuition and for opportunities you gave me to travel and present my research.

The basis of what I know comes from my professors in Moscow Institute of Physics and Technology.
I'm grateful to my advisor there Dr. Vasily Beskin and also my advisor in University of Illinois at Chicago Alexander Fridman for great research experience and support to go to the graduate school in the United State. 

I would like also to thank the organizers  of Strangeness in Quark Matter 2006,    ELI Workshop and School on "Fundamental Physics with Ultra-high Fields", Summer Nuclear Physics School 2007, 24th workshop on Nuclear Dynamics, APS Division of Nuclear 
Physics  for travel support.

My research in University of Arizona was supported by a grant from: the U.S. Department of Energy  DE-FG02-04ER4131.

I would like to thank all hard working experimentalists who worked to provide the experimental results used in my work.

I'm very thankful to my husband, Ivan, for spiritual and financial support, understanding and all his help during these years. This work would be never finished without him. I also thank him and my son Danny for all fun and patience during these years.  

I thank my mom Lidia Kuznetsova that she gave me the opportunity to study physics and mathematics. I always had a much of support from her, despite I think it was difficult for her that I left home.
 
I want also to say thank you to all my friends for support of my spirit. Especially Svetlana, Anna G.,
Anna P., Nick, Kamilla, Olga, Igor  who had directly helped me during study and transfer to grad school in the United States.

I am grateful to the members of my dissertation committee Dr. Robert Thews, Dr. Andrei Lebed,
Dr. Sean Fleming and Dr. Shufang Su for serving my committee and reviewing this work. My grateful acknowledgments go to our graduate study director Dr. Keith Dienes and all University of Arizona Department of Physics for support, great atmosphere at the department and excellent
classes.
% Include the ``Dedication''
%\input{dedication}
%\newpage
\begin{dedication}
%\bf {DEDICATION}
%\end{center}
%\begin{center}
Dedicated to my loving family Danny, Ivan and mom.
\end{dedication}

% This is needed in order to prevent overwriting of dropped header

% Create a ``Table of Contents''

% 
\pagestyle{headings}
\renewcommand{\baselinestretch}{1.2}
\tableofcontents
\textheight=22.0cm
% This is needed in order to prevent overwriting of dropped header

\listoffigures

% Create a ``List of Tables''
\listoftables

%\input{acknowledgements}
% Include the various chapters
\begin{abstract}

We study particle production and its density evolution and equilibration in hot
dense medium, such as hadronic gas after quark gluon plasma
hadronization and relativistic electron positron photon plasma. For
this study we use kinetic momentum integrated equations for
particles density evolution with Lorentz invariant reaction rates.
We extend these equations, used before for two-to-two particles
reactions ($1+2\leftrightarrow 3+4$), to the case of two-to-one and
backward reactions ($1+2 \leftrightarrow 3$). One type of hot dense
medium, which we study, is hadronic gas produced at quark gluon plasma hadronization 
in heavy ions collisions in SPS, RHIC and LHC experiments. We study hadron
production at quark gluon plasma hadronization and their evolution
in thermal hadronic gas phase. We consider non-equilibrium
hadronization model, for which the yields of the light quark hadrons are
defined by entropy conservation. Yields of hadrons containing heavier
(strange, charm, bottom) quarks are mainly controlled by flavor conservation.
We predict yields of charm and bottom hadrons within this
non-equilibrium statistical hadronization model. Then we use this
non-equilibrium hadronization as the initial condition in the study of hadronic
kinetic phase. During this time period some hadronic resonances can be produced in
lighter hadrons fusion. This reaction is opposite to resonance
decay. Production of resonances is dominant over decay if there is non-equilibrium
excess of decay products. Within this model we explain apparently contradictory experimental
results reported in RHIC experiments: 
$\Sigma(1385)$ yield is enhanced while $\Lambda(1520)$ yield is
suppressed compared to the statistical hadronization model expectation obtained without
kinetic phase.  We also predict $\Delta(1232)$ enhancement. The second type of plasma medium we consider is the relativistic
electron positron photon plasma (EP$^3$) drop. This plasma is
expected to be produced in decay of supercritical field created in ultrashort laser pulse. We
study at what conditions this plasma drop is opaque for photons and
therefore may reach thermal and chemical equilibrium. Further we consider muon and pion production
in this plasma also as a diagnostic tool. Such heavy particles can be diagnostic tool to study the properties
of EP$^3$ plasma, similar to the role taken by heavy hadrons production in heavy ions collisions. 
Finally all these theoretical developments can be applied to begin a study of particles evolution in early universe in temperatures domain
from QGP hadronization (160 MeV) to nucleosynthesis (0.1 MeV). The first results on pion equilibration are presented here.

\end{abstract}
% Include the ``Acknowledgements''
% Create a ``List of Figures''

%%%%%%%%%%%%%%%%%%%%%%%%%%%%%%%%%%%%%%%%%%%%%%%%%%%%%%%%%%%%%%%%%%%%%%%%%%%%%%%%%%%%%%%%
\chapter{PARTICLE PRODUCTION IN MATTER AT EXTREME CONDITIONS}\label{inthist}
%%%%%%%%%%%%%%%%%%%%%%%%%%%%%%%%%%%%%%%%%%%%%%%%%%%%%%%%%%%%%%%%%%%%%%%%%%%%

%%%%%%%%%%%%%%%%%%%%%%%%%%%%%%%%%
\section{OUTLINE}
%%%%%%%%%%%%%%%%%%%%%%%%%%%%%%%%%

In this dissertation I present particle production and equilibration in different types of plasma medium.
In this chapter~\ref{inthist} I overview the challenges and earlier developments, related to quark gluon plasma (QGP) in relativistic heavy ions collisions and electron-positron-photon plasma production in strong laser field that provide motivation for our research.  In this chapter I discuss the hadronization process, in which quark gluon plasma breaks (freeze-out) into hadrons.
The hadronization conditions have dominant influence on almost all final hadrons yields, even though these yields change during kinetic phase. For electron-positron-photon plasma we consider possibilities of its creation in strong laser field in future experiments. 

In chapter~\ref{sthad1} I discuss the statistical hadronization model (SHM) in greater depth and show how to estimate numerically $\gamma_q$ from entropy conservation an $\gamma_{i}$ $i=s,\,c,\,b$ from corresponding flavor conservation strangeness  conservation. In chapter~\ref{hfhad} we considered heavy flavor (charm, bottom) hadron production within statistical hadronization model. The new feature
compared to the others studies is that we assume entropy and strangeness conservation during hadronization, accounting  in this way
for higher light quark and strangeness content in QGP. We study how this model improvement influences the yields of heavy 
flavor hadrons.  We studied in depth how the (relative) yields of strange and non-strange
charmed mesons vary with strangeness content. 

As the result of high strangeness environment we find also a relative suppression of the
multi-heavy hadrons, except when they contain strangeness. The degree of this suppression depends on both, strangeness and light quarks content. When phase space occupancy of
light and strange quark is relatively high the probability for charm quarks to make hadrons with strange quarks increases and probability to find the second charm quark
among light and strange quarks decreases.
These results have been published in~\cite{Kuznetsova:2006bh, Kuznetsova:2006hx}.

In chapter~\ref{dpeq} I derive equations for Lorentz invariant rates and particle density evolution for decay reaction 1-to-2 particles and backward reaction 2-to-1 (particle fusion). In this approach we connected the particle decay time in vacuum with kinetically modified particle decay time in medium, and with relaxation time for the backward reaction for the resonance production in two particles fusion. We calculated the relaxation times in medium for reactions: $\rho\leftrightarrow \pi+\pi$; $\Sigma(1385)\leftrightarrow \Lambda+\pi$; $\pi^0\leftrightarrow \gamma+\gamma$. We are going to publish these results in~\cite{KuznKodRafl:2008}.
 
In chapter~\ref{respr} I apply equations derived in chapter~\ref{dpeq} to baryon resonance densities evolution in thermal hadron gas after quark
gluon plasma hadronization. The goal is to explain ratios $\Sigma(1385)/\Lambda^0$ and $\Lambda(1520)/\Lambda^0$ reported by RHIC-experiments
and also to predict $\Delta(1232)/N$ ratio. In this chapter I also take into account non-equilibrium
condition at hadronization, defined by entropy conservation, used also in the model presented in chapter \ref{hfhad}.
I find that a significant additional yields of $\Delta$(1232), $\Sigma$(1385)  can be produced by the back-reaction of the over-abundance of the decay products of resonances: $\pi+N \rightarrow\Delta(1231)$, $\pi+\Lambda \rightarrow \Sigma(1385)$. A more complex situation arises for a relatively narrow
resonance such as $\Lambda$ (1520), which can be also seen as a stable state, which is depopulated to increase the heavier resonance yield ($\Lambda(1520)+\pi\rightarrow\Sigma^*$). I find that a suppression of the yield of such resonances, as compared to statistical hadronization model, is possible. The pattern of deviation of hadron resonance yields from expectations based on statistical
hadronization model is another characteristic signature for a fast hadronization of entropy rich QGP. The total yield of the ground state baryons used in analysis of data (such as N, $\Lambda$) is not affected. The results are in agreement with yields of these resonances reported by RHIC experiments.
This part of thesis is published in references~\cite{Kuznetsova:2008zr} and~\cite{Kuznetsova:2008hb}.

In chapter~\ref{eeg} I consider $e^+e^-\gamma$ plasma. I investigate the size and temperature limits of thermally and chemically equilibrated  plasma drops, created by sub-optical wavelength high energy light-laser pulses. The plasma to become equilibrated must be opaque to electron and photon interactions. Opaqueness condition is determined by comparing plasma size with the free electron and photon paths, which are calculated using thermal Lorentz invariant reaction rates for pair production and Compton scattering. These results are in preparation~\cite{Inga2}. 

In this chapter I also study
heavy particles (pion, muon) production in this plasma 
at a temperature $T\ll m_\mu,\,m_\pi $. I argue that the observation of pions and muons
can be a diagnostic tool in the study of the initial properties of such a plasma formed by means of
strong  laser  fields. Conversely, properties of  muons and pions in thermal environment become
accessible to precise experimental study. 

In this chapter \ref{earlyun} I consider the pion equilibration in early universe. This chapter is part of  reference~\cite{KuznKodRafl:2008}.
In chapter~\ref{sumconc} I present summary of results from each chapter, and conclusions.

{\bf List of publications related to dissertation, including works in preparation:}\\

1.) Chemical Equilibration Involving Decaying Particles at Finite Temperature\\
\underline{Inga Kuznetsova}, Takeshi Kodama,  and Johann Rafelski; (in preparation)\\

{\it We study kinetic master equations for chemical reactions involving the formation and the natural decay of
particles in a thermal bath. We consider decay channel into two particles and the inverse process, the
fusion of two thermal particles into one. We derive  chemical equilibrium
condition for the particle density. We evaluate the thermal invariant rate
using as input the free space (vacuum) decay time.
A particularly interesting application of our
formalism   is the  $\pi ^{0}\leftrightarrow \gamma +\gamma$ evolution
in the early Universe.}\\

\centerline{\rule{5cm}{0.03cm}}

2.) Equilibration size limit of e-, e+, gamma plasma, accessible to high energy light pulse.\\
\underline{Inga Kuznetsova}, Johann Rafelski; (in preparation)\\

{\it We investigate the size and temperature limits of thermally and chemically equilibrated $e^+e^-\gamma$ plasma drops, created by sub-optical wavelength laser pulses. For the plasma to be equilibrated it must be opaque to electrons and photons interactions. Opaque condition is determined by comparing plasma size with the free electron and photon paths, which are calculated using thermal Lorentz invariant reaction rates for pair production and Compton scattering} \\

\centerline{\rule{5cm}{0.03cm}}

3.) Charmed hadrons from strangeness-rich QGP.\\
\underline{Inga Kuznetsova} and Johann Rafelski. May 2006. 6pp.\\ 
Contributed to International Conference on Strangeness in Quark Matter (SQM2006), Los Angeles, California, 26-31 Mar 2006. \\
J.Phys.G32:S499-S504,2006;
e-Print: hep-ph/0605307\\

4.) Heavy flavor hadrons in statistical hadronization of strangeness-rich QGP.\\
\underline{Inga Kuznetsova} and Johann Rafelski . Jun 2006. 18pp. \\
Eur.Phys.J.C51:113-133,2007;
 hep-ph/0607203\\
 
5.) Heavy Ion Collisions at the LHC - Last Call for Predictions.
N. Armesto, \underline{I. Kuznetsova} (ed.)  et al. Nov 2007. 185pp.\\ 
Presented at Workshop on Heavy Ion Collisions at the LHC: Last Call for Predictions, Geneva, Switzerland, 14 May - 8 Jun 2007. \\
Published in J.Phys.G35:054001,2008. 
e-Print: arXiv:0711.0974 [hep-ph]\\

6.) Non-Equilibrium Heavy Flavored Hadron Yields from Chemical Equilibrium Strangeness-Rich QGP.
\underline{Inga Kuznetsova}, Johann Rafelski Jan 2008. 6pp. \\
Presented at International Conference on Strangeness in Quark Matter (SQM 2007), Levoca, Slovakia, 24-29 Jun 2007. \\
J.Phys.G35:044043,2008;
 arXiv:0801.0788 [hep-ph]\\

{\it In above 4 papers we study   $b$, $c$ quark hadronization from QGP.  We obtain
the yields of charm and bottom flavored hadrons within the statistical
hadronization model. The important novel feature  of this study
is that we take into
account the high strangeness and entropy content of QGP, conserving
strangeness and entropy yields at hadronization.}\\

\centerline {\rule{5cm}{0.03cm}}

7.) Pion and muon production in e-, e+, gamma plasma.\\
\underline{Inga Kuznetsova}, Dietrich Habs, Johann Rafelski. Mar 2008. 14pp. \\
Phys.Rev.D78:014027,2008. 
arXiv:0803.1588 [hep-ph]\\

{\it We study production and equilibration of  pions and muons in
relativistic electron-positron-photon plasma  at
a temperature $T\ll m_\mu,\,m_\pi $. We argue that the observation of pions and muons
can be a diagnostic tool in the study of the initial properties of such a plasma formed by means of
strong  laser  fields. Conversely, properties of  muons and pions in thermal environment become
accessible to precise experimental study.}\\

\centerline{\rule{5cm}{0.03cm}}

8.) Enhanced Production of Delta and Sigma(1385) Resonances.\\
\underline{Inga Kuznetsova}, Johann Rafelski (Arizona U. and Munich U. and Munich, Tech. U.) . Apr 2008. 6pp. \\
Phys.Lett.B668:105-110,2008;
arXiv:0804.3352 [nucl-th]\\

{\it Yields of    $\Delta(1230)$, $\Sigma(1385)$  resonances produced in
heavy ion collisions are studied within   the
framework of a kinetic master equation.   The  time evolution
is driven by the  process $\Delta \leftrightarrow N \pi $, $\Sigma(1385) \leftrightarrow \Lambda \pi $.  We obtain
  resonance yield  both below and  above chemical equilibrium, depending on initial
hadronization condition and separation of kinetic and chemical freeze-out.}\\

\centerline{\rule{5cm}{0.03cm}}

9.) Resonance Production in Heavy Ion Collisions: Suppression of  $\Lambda$ (1520) and Enhancement of  $\Sigma$(1385)\\
\underline{Inga Kuznetsova}, Johann Rafelski, Nov 2008, (Arizona U. and Munich U), 13 pp.\\
Phys.\ Rev.\  C {\bf 79}, 014903 (2009)
arXiv:0811.1409 [nucl-th]\\

10.) Resonances Do Not Equilibrate.\\
\underline{I. Kuznetsova}, J. Letessier, J. Rafelski\\
in Fourth Workshop on Particle Correlations and Femtoscopy (WPCF2008), Krakow, September 11-14, 2008,  arXiv:0902.2550v1 [nucl-th]\\

{\it In last two works we investigate the yield of $\Lambda(1520)$ resonance in heavy ion
collisions within the framework of a kinetic master equation without
the assumption of chemical equilibrium. We show that reactions such
as $\Lambda(1520)+\pi \leftrightarrow \Sigma^*$ can favor $\Sigma^*$
production, thereby decreasing the $\Lambda(1520)$ yield.  Within
the same approach we thus find
a yield  enhancement  for  $\Sigma(1385)$ and a yield suppression for $\Lambda(1520)$.}

In this dissertation we consider two types of extreme matter condition: hadronic matter in relativistic heavy ions collisions and 
$e^+e^-\gamma$ plasma created in strong laser field. 

\section{Relativistic Heavy Ions Collisions}\label{rhic}

\subsection{Overview}

The quarks differ by quantum number called flavor. Six flavors are known. $u$ and $d$ are light quarks, $s$ is strange, $c$ is charm and $b$ is bottom quark. These quarks are arranged into doublets:
$$\left(\begin{array}{c} u \\ d   \end{array}\right),\quad \left(\begin{array}{c} c \\ s  \end{array}\right), \quad \left(\begin{array}{c} t \\ b   \end{array}\right).$$
The quarks on the top of doublets ($u$, $c$, $t$) have charge +2/3, the quarks on the bottom of doublets ($d$, $s$, $b$) have charge -1/3. 
Quarks $u$ and $d$ are the lightest. It is known that $m_u \approx 0.5 m_d$, both are in the range 2-8 MeV, and their average mass  
$\bar{m}=(m_u+m_d)/2 \approx 2.5-5 MeV$. All stable matter is made of only $u$ and $d$ quarks.
$m_s$ is between 75-125 MeV, $m_c$ is about 1.3 GeV, $m_b \approx$ 4.2 GeV and $m_t\approx 174$ GeV.  

In nature quarks are always confined in hadrons. This is one of the reasons that mass of quarks is not well defined. Even if a lot of energy is applied to separate $q\bar q$ pair, at some point of separation it becomes energetically preferred to create one more confined $q\bar q$ pair. Therefore in the result we have two confined quark-antiquark pairs.

In relativistic heavy ions collisions hot and dense fireball of nuclear matter is
created. We believe that at very high energy the deconfinement of
partons is expected and results to new phase of matter, quark gluon
plasma (QGP). The temperature of the fireball drops with expansion. At
critical hadronization temperature the hadrons are formed again.
Observing final yields of hadrons we try to learn about physics of QGP formed in heavy ions collisions. The difference in hadrons yields
between pp collisions and heavy nuclear collisions can justify the
existence of QGP state.~\cite{Letessier:2002gp}

In part of this dissertation we develop models that explain (or predict) yields of some baryon resonances and heavy flavor hadrons, reported in heavy ions collision experiments.
We use experimental data reported in Relativistic Hadronic Collider (RHIC) experiments and predict yield of heavy flavor hadrons at Large Hadronic Collider (LHC) energies.
RHIC  is built in Brookhaven National Laboratory to create and search the new phase of matter (QGP), colliding Au  
ions with energy in center of mass frame ranging from $\sqrt s = 20$ to 200 {\it A}GeV. 
The much higher energies are expected at the Large Hadronic Collider (LHC), which began to operate recently at CERN. 
Here we will also refer to Super Photon Synchrotron (SPS) accelerator with fixed target results at CERN, where energies of accelerated
Pb ions up to 158 {\it A}GeV were achieved.

In our approach we assume that
the following evolution stages are present in heavy-ions
collisions:
\begin{enumerate}
\item
Primary partons collide, practically all heavy $c,\, b$ quarks are
produced;
\item
A thermalized parton state within $\tau=\tau_{th}\simeq
0.25-1 \, \rm{fm/c}$ is formed.
By the end of this stage nearly all entropy is
produced.
\item
The subsequent chemical equilibration:
diverse thermal particle production reactions
occur, allowing first the approach to chemical equilibrium
by gluons $g$ and light non-strange quarks $q=u,d$.
\item
The strangeness chemical equilibration within $\tau\sim 5\,\rm{fm/c}$).
\item
Chemical freeze-out (hadronization) near $\tau\sim 10$ fm/c), when quarks become confined in hadrons.
\item
Kinetic phase, where hadrons can scatter and be regenerated.
\item
Kinetic freeze-out.
\end{enumerate} 
%%%%%%%%%%%%%%%%%%%%%%%%%%%%%%%%%%%%%%%%%%%%%%%%%%%%%%%%%%%%%%%%%%%%%%%%%%%%%%%%
\subsection{Statistical hadronization model} \label{sthad}
%%%%%%%%%%%%%%%%%%%%%%%%%%%%%%%%%%%%%%%%%%%%%%%%%%%%%%%%%%%%%%%%%%%%%%%%%%%%%%%%

The transition from quark gluon plasma to hadrons can be described by statistical hadronization model.
SHM arises from the  Fermi multi-particle production
model~\cite{Fermi:1950jd}. In Fermi model it was micro-canonical ensemble the flavor and energy are conserved exactly. It was developed further by Landau~\cite{Landau:1953gs} and Hagedorn~\cite{Hagedorn:1965st}. In~\cite{Hagedorn:1965st} the infinite number of hadrons with increasing mass results to exponential mass spectrum, which diverges at critical temperature.
This was the first evidence toward the phase transition from confined hadrons to QGP. 
There was a transition towards the finite size of hadrons, composed of quarks, and phase transition 
between QGP and hadronic gas in Hagedorn and Rafelski work~\cite{Hagedorn:1980kb, Hagedorn:1980cv}. 

The transition from micro-canonical to
canonical, and grand-canonical ensembles, where averaged flavor is conserved, simplifies the
computational effort considerably~\cite{Hagedorn:1984hz}.
This important step does not in our context introduce
the hadron phase, although before the understanding of QGP this of
course was the reaction picture: a highly compressed hadron gas
matter evaporates particles. Today, it is the highly compressed hot quark-gluon matter that evaporates
particles.

Yields of most hadrons are described by statistical hadronization model without
kinetic phase. For resonances with strong decay there are the deviations of their yields ratios reported in experiment from predicted by statistical hadronization only. This makes necessary to
introduce the kinetic phase, where yields of hadrons can be changed by
reactions between hadrons.

The SHM is related to assumption that system also described by hydrodynamic model. This includes assumption that for strong interactions relaxation times are small enough that prehadronic matter (QGP) is in thermal equilibrium and the system can be treated as a relativistic fluid. 

In a hydrodynamic description with flow in the longitudinal and transverse directions the current
of particle with mass $m$ is
\begin{equation}
j^{\mu} = \frac{1}{(2\pi)^3}\int d^{4}p{2\delta({\bf p}^2-m^2)\theta{(p_0)}p^{\mu}_{s}}f({\bf p}) =  \frac{1}{(2\pi)^3}\int \frac{d^{3}p}{E}p^{\mu}f({\bf p, \gamma}); \label{cur}
\end{equation}
${\bf p}$ is 4-vector momentum of particle:
\begin{equation}
p^{\mu}=\left(E, \vec{p}\right) \label{4vp}
\end{equation}

In this equation we use Lorentz invariant distribution functions for bosons and fermions:
\begin{equation}
f(p, \gamma, \lambda)=\frac{1}{\gamma^{-1}\lambda^{-1}e^{u\cdot p/T}\mp 1};\label{distrhd}
\end{equation}
%\begin{equation}
%f(p)= {1\over \gamma^{-1}\lambda^{-1} e^{E/T}\pm 1} \to \gamma\lambda e^{-E/T}, \label{distrhd}
%\end{equation}
where the Fermi `$(+)$' and Bose `$(-)$' distributions are indicated.
$g$ is the degeneracy factor, $T$ is the temperature, 
$u^{\mu}$ is the frame four-vector, which in the observer rest frame of is
\begin{equation}
u^{\mu}=\left(1,\vec{0}\right).\label{4v}
\end{equation}

Then number of particles within the element of 3-dimensional freeze-out surface in the Minkovski space-time $d\sigma_{\mu}=(d^3x, \vec{0})$ is defined by
\begin{equation}
dN(\sigma) = j^{\mu}d\sigma_{\mu}
\end{equation}
The distribution of particles at this surface is given by the Cooper-Fryer formula~\cite{Cooper:1974mv}:
\begin{equation}
E\frac{dN}{d^3p} = \frac{g}{2\pi^3}\int_{\sigma}f(x,p)p^{\mu}d\sigma_{\mu}.  \label{cffor}
\end{equation}
In Boltzmann limit, which is good approximation for considered temperatures for most hadrons, we can omit `1' in determinator of distribution function:
\begin{equation} 
f(p,x) \to \gamma\lambda e^{-u\cdot p/T}. \label{fbol}
\end{equation}

The important parameters
of the SHM, which control the relative yields
of particles, are the particle specific fugacity factor ${\lambda}$ and
space occupancy factor ${\gamma}$. The fugacity is related to chemical
potential ${\mu} = T{\ln{\lambda}}$. The occupancy ${\gamma}$ is, nearly,
the ratio of produced   particles to the number of particle
expected in chemical equilibrium ($\gamma=1$ is chemical equilibrium).

The fugacity ${\lambda}$ is associated with a
conserved quantum number, such as  net-baryon number, net-strangeness, heavy flavor.
Thus antiparticles
have inverse value of  ${\lambda}$, and  ${\lambda}$  evolution during
the reaction process is related to the
changes in   densities due to dynamics such as expansion.
Here we always consider particle anti-particle symmetry or $\lambda = 1$. This condition is 
almost satisfied for $RHIC$ and $LHC$ energies. 
 ${\gamma}$  is the same for particles and antiparticles.  Its value
changes as a function of time even if the system does not expand, it describes buildup of
the particular particle species.
For this reason ${\gamma}$ is changes rapidly during the reaction,
while ${\lambda}$ is more constant. Thus it is ${\gamma}$ which carries
the information about the time history of the reaction and the precise
condition of particle production referred to as chemical freeze-out.

These distribution functions with phase space occupancy $\gamma$  can describe quark in QGP and
also hadrons multiplicity after hadronization. For phase space occupancy for hadron is a product
of $\gamma$s for each quark in this hadron. For example for kaon $K$ $\gamma_K = \gamma_q\gamma_s$,
for pion $\gamma_{\pi}=\gamma_q^2$, etc. 

Then we have, as expected for Boltzmann limit from Eq.(\ref{cffor}) and (\ref{fbol}):
\begin{equation}
E\frac{dN}{d^3p} = \frac{g}{2\pi^3}\lambda Ee^{-E/T}V
\end{equation}

Integrating this equation over momentum, the number of particles of type `$i$' with mass $m_i$ per unit of rapidity is
in our approach given by:
\begin{equation}
N_i={\gamma_i}n_i^{\rm eq}V.  \label{dist}
\end{equation}
Here $dV$ is system the volume of a fireball,
and $n_i^{\rm eq}$ is a Boltzmann particle density in chemical equilibrium:
\begin{equation}\label{BolzDis}
n_i^{\rm eq}=g_i\int\frac{d^3p}{(2\pi)^3}\lambda_i\exp(-\sqrt{p^2+m_i^2}/T)=\lambda_i\frac{T^3}{2\pi^2}g_iW(m_i/T), 
\end{equation}
  and
\begin{equation}\label{Wdistapr}
W(x)=x^2K_2(x)\rightarrow 2\  {\rm for}\ x\rightarrow 0.
\end{equation}
 Both, $m_ic^2\to m_i$, and $kT\to T$, are measured in
energy units when $\hbar,c,k\to 1$.

In non-relativistic Boltzmann limit Eq.(\ref{BolzDis}) can be expand as:
\begin{equation}
%
%{dN_i\over dV}  \gamma_i
%
n_i^{\rm eq}= \frac{g_iT^3}{2\pi^2}  \lambda_i\sqrt\frac{{\pi}m_i^3}{2T^3}{\exp}(-m_i/T)\left(1+\frac{15T}{8m_i}+
\frac{105}{128}\left(\frac{T}{m_i}\right)^2\ldots\right).
\label{dist1}
\end{equation}
Often  we can use the first term alone for heavy flavor hadrons, since $T/m<<1$,
however the asymptotic series in Eq.\,(\ref{dist1}) converges slowly (if at all) 
and one should proceed with caution.

We use occupancy factors $\gamma^\mathrm{Q}_i$ and
$\gamma^\mathrm{H}_i$ for QGP and hadronic gas phase respectively, tracking
every quark flavor ($i$ = q, s, b, c) .
We assume that in the QGP phase
the light quarks and gluons are adjusting fast to the
ambient conditions, and thus are in chemical equilibrium with
$\gamma^{\mathrm{Q}}_{q,G}\to 1$.
For heavy, and strange flavor, the value of
$\gamma^{\mathrm{Q}}_i$ at hadronization condition is  given  by the number of
particles present, generated by prior kinetic processes, see Eq.\,(\ref{dist}).

The  yields of different quark flavors
originate in different physical processes, such as production in
initial collisions for $c,b,s$, and for $s$ also production in
thermal plasma processes.
In general we thus cannot expect  that   $\gamma^{\mathrm{Q}}_{c,b}$ will be near unity
at hadronization. However, the thermal strangeness production process  $GG\to s\bar s$
can nearly chemically equilibrate strangeness flavor in plasma
formed at RHIC and/or LHC~\cite{Letessier:2006wn},
and we will always consider, among other cases the limit
$\gamma^{\mathrm{Q}}_s\to 1$ prior to
hadronization.

Differing from  other recent studies which assume that
the hadron yields after hadronization
are in chemical equilibrium at list in respect of light quarks or light and strange quarks together ($\gamma_{q(s)}=1$)~\cite{Becattini:2005hb, Andronic:2003zv}, to evaluate yields of final state hadrons we enforce
conservation of entropy (determine $\gamma_q$), and the flavor $s,c,b$ quark pair number (determine corresponded $\gamma_i$)
during phase transition or transformation. 

The faster the transition, the less likely is that there is significant
change in strange quark pair yield. Similarly, any entropy
production is minimized when the entropy rich QGP breakup
into the entropy poor HG occurs  rapidly. The entropy conservation
constraint fixes the final light quark yield. We assume a fast
transition between QGP and HG phases, such that all hadron yields
are at the same physical conditions as in QGP breakup.

Assuming that in the hadronization process the number of
$b$, $c$, $s$ quark pairs does not change,
the three unknown $\gamma^\mathrm{H}_s$,
$\gamma^\mathrm{H}_c$,
$\gamma^\mathrm{H}_b$ can be determine from their
values in the QGP phase, $\gamma^\mathrm{Q}_s$,
$\gamma^\mathrm{Q}_c$,
$\gamma^\mathrm{Q}_b$ (or $N^Q_i$) and the three flavor conservation
equations,
\begin{equation} \label{flcons}
N^\mathrm{H}_i=N^\mathrm{Q}_i=dN_i, \quad i=s,c,b.
\end{equation}
In order to conserve entropy:
\begin{equation}
{S^\mathrm{H}}=S^\mathrm{Q} = S, \label{Scons}
\end{equation} 
a value $\gamma^\mathrm{H}_q > 1$ is required for $T > 180$ MeV when in
the QGP phase $\gamma^\mathrm{Q}_{q,G} = 1$. $\gamma^\mathrm{H}_q > 1$ is needed to compensate 
entropy lost because of decreasing number of degrees of freedom in hadronic gas and increasing mass of hadrons.
This implies that yields of hadrons with light quark content
are, in general, not in chemical equilibrium, unless there is some extraordinary circumstance
allowing a prolonged period of time  in which hadron reactions can occur after hadronization. Chemical
non-equilibrium thus will influence the yields of  heavy
flavored particles in final state as we shall discuss in this work.

As noted at the beginning of this section,
the use of the hadron phase space (denoted by H above)
does not imply the presence of a real physical `hadron matter' phase: the  SHM
particle yields will be attained  solely on the basis of availability of this  phase
space as noted at the beginning of this section.  Another way to argue this is to
imagine a pot of quark matter with hadrons evaporating . Which kind
of hadron emerges and at which momentum is entirely determined by the access to
the phase space, and there are only free-streaming particles in the final state.

Thinking in these terms, one can imagine that especially
for heavy quark hadrons some particles are
pre-formed in the deconfined plasma, and thus the heavy hadron yields
may be based on a value of temperature which is higher than the global
value expected  for other hadrons.
For this reason we will study in this work  a range $140<T<260$ MeV
and also consider sensitivity to this type of two-temperature chemical
freeze-out of certain heavy hadron yield ratios.

We use this non-equilibrium hadronization model to predict heavy flavor hadrons yields for
RHIC and LHC conditions.

Here we also use grand canonical hadronization model for heavy flavor hadronization.
The grand-canonical hadronization condition is conservation of symmetry of
strange and antistrange hadrons in reactions:
\begin{equation}
<n_i>-<n_{\bar i}>=0.
\end{equation}
When number of flavor is small this condition must be replaced by sharper condition:
\begin{equation}
<n_i-n_{\bar i}>=0.
\end{equation}
This more exact condition results to canonical suppression of given flavor yield. There is the large effect when
the quarks multiplicity for this flavor is about unity and effect increase for multiflavor hadrons. 
This shows that the smaller multiplicity of given flavor(s) quarks  is, it is more difficult for them  
to find each other to bind to hadron. Moreover here for hadrons with one heavy quark we normalize yield by quark multiplicity,
and canonical suppression is the same for both yields and canceled. 

For LHC even multiplicity of c quarks per rapidity is large $dN_c/dy \approx 10$ and for hidden charm ($c\bar{c}$) mesons we almost do not have 
canonical suppression. There is small canonical effect for hidden charm mesons at RIHC, which we did not consider.
This effect may be large for $B_c$, $b\bar b$ mesons. 

%%%%%%%%%%%%%%%%%%%%%%%%%%%%%%%%%%%%%%%%%%%%%%%%%%%
\subsection{Strangeness enhancement}\label{sten}
%%%%%%%%%%%%%%%%%%%%%%%%%%%%%%%%%%%%%%%%%%%%%%%%%%%%%
Strangeness enhancement in relativistic heavy ions collisions, compared to pp collisions, was proposed to be an indication of presence of deconfined state by Rafelski and Muler~\cite{Rafelski:1982pu, Rafelski:1982ii, Rafelski:1983hg}.
In~\cite{Rafelski:1982pu, Koch:1986ud} it was shown that
strangeness is mostly produced in QGP phase in reactions
\begin{eqnarray}
&&q+\bar{q} \rightarrow s + \bar{s}, \label{qqss}\\
&&g+g \rightarrow s + \bar{s}. \label{ggss}
\end{eqnarray}
Strangeness production in the reaction (\ref{ggss}) is dominant in qgp. 
Strangeness can also be produced in hadronic gas in reactions
\begin{eqnarray}
&&\pi+ N \rightarrow K+Y; \label{sthg1}\\
&&\pi+\pi \rightarrow K+K,\label{sthg2}
\end{eqnarray}
where $Y$ is strange baryon (hyperon).
However the strangeness production rate in thermal QGP the strangeness production rate
is an order of magnitude larger than in hadronic gas~\cite{Rafelski:1982pu, Koch:1986ud}.

Therefore  strangeness is mostly produced in QGP phase. Before hadronization in QGP strangeness multiplicity  is near equilibrium. Then in our model during fast hadronization
number of strange quarks does not have time to change. As the result after hadronization the multiplicity of strange hadrons is higher  than for models which just put $\gamma^{HG}_s = 1$ the same as in QGP. The explanation is that in hadronic gas strange hadrons are more massive than free strange quark in QGP. Therefore distribution function with the same $\gamma_s$ multiplicity of strangeness may be smaller than in QGP. This effect depends on hadronization temperature. In order to conserve strangeness the $\gamma^{HG}_s > 1$ may be needed. We will add quantative explanation in
chapter~\ref{strSec}.

Strangeness content is defined by strangeness to entropy ratio
$s/S$, which can be measured. $s/S$ increases with energy of
collision increase. The explanation is that the hot state where the
energy of gluons exceeds the threshold for strangeness production
lives longer.  For RHIC at central collisions $s/S \approx 0.03$, for LHC it is expected
to be about $0.04$.

$s/S$ increases for higher energy of collision.
Possible explanation for higher energies the hot state with temperature above threshold for strangeness production lives longer. 
$s/S$ increases with centrality of collision or number of participants $N_{part}$~\cite{Letessier:2005kc}

Also experimentally strangeness content can be found by measuring ratios of yields of hadrons with different strangeness content,
for example $K/\pi$ or $\Lambda/p$ which are $\propto \gamma_s/\gamma_q$. $s/S$ increases with $\gamma_s/\gamma_q$ grow. The numerical results for connection between ratio $s/S$ and $\gamma_s/\gamma_q$
are shown in chapter~\ref{hfhad}

\begin{figure}[htb]
\begin{center}
\includegraphics[width=9.6cm,height=8.5cm]{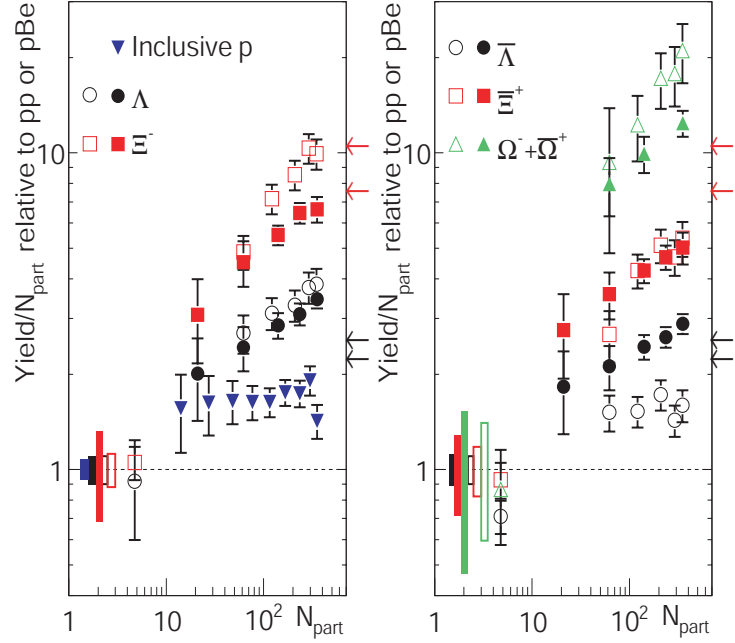}
\caption{ (color online) {$E(i)$} as a function of $N_{part}$ for
$\Lambda$, $\bar{\Lambda}$,  $\Xi^+$, $\Xi^-$, $\Omega^+$, $\Omega^-$
and inclusive p. Boxes at
unity show statistical and systematical uncertainties combined in
the p$+$p (p$+$Be) data. Error bars on the data points represent those
from the heavy-ions. The solid markers are for Au+Au at $\sqrt s$=200
GeV and the open symbols for Pb+Pb at $\sqrt s$=17.3
GeV. The arrows on the right axes mark the
predictions from a GC (grand canonical) formalism model when varying {\it T} from 165 MeV
({\it $E(\Xi^-)$}=10.7, $E(\Lambda)$=2.6) to 170 MeV ({$E(\Xi^-)$}=7.5,
{\it E($\Lambda$)}=2.2)~\cite{Cleymans:1990mn}. The red arrows indicate the predictions for $\Xi$ and
the black arrows those for $\Lambda$. } \label{Fig:Enhancement}
\end{center}
\end{figure}

In figure~\ref{Fig:Enhancement} we show experimental results for ground states of strange baryons
$\Lambda$, $\Xi$,  $\Omega$ yields, normalized to the yields in pp or pBe collisions $E(i)$ as a function of $N_{pat}$ for Au+Au and Pb+Pb and pp collisions.
For each species, $i$, the yield enhancement, $E(i)$, above
that expected from $N_{part}$ scaling was calculated using:
\begin{equation}
E(i) =\frac{ Yield^{AA}(i)}{\langle N_{part}^{AA}\rangle}\frac{\langle N_{part}^{NN}\rangle}{
Yield^{NN}(i)}
\end{equation}
The inclusive proton data illustrate the effects for
non-strange baryons. The figure is taken from~\cite{Abelev:2007xp}.

The difference between particles and antiparticles is due chemical potential, which is higher for lower energy collisions. 
We see increases in the yield of strangeness containing hadrons with $N_{part}$ as predicted in~\cite{Letessier:2005kc}.
For $\Xi$ yield ($\propto \gamma_s^2$) the increase is larger than for $\Lambda$ ($\propto \gamma_s$) and the largest effect
is for $\Omega$ yield ($\propto \gamma_s^3$).

%%%%%%%%%%%%%%%%%%%%%%%%%%%%%%%%%%%%%%%%%%% 
\subsection{$J/\Psi$ suppression}
%%%%%%%%%%%%%%%%%%%%%%%%%%%%%%%%%%%%%%%%%

The observation of  heavy hadrons containing more than
one heavy quark, for example charmonium can give information about deconfined QGP phase (Matsui and Satz 1986). 
In deconfined phase charm and anticharm quarks from different nucleon-nucleon collisions diffuse, meet and can produce charmonium. The probability of charmonium production and therefore the enhancement of observed yield is expected as compared to single nucleon-nucleon collisions without deconfined phase
~\cite{Thews:2001hy, Schroedter:2000ek, Becattini:2005hb}. Similar enhancement is possible for $B_c$ mesons.

However in SPS (NA50 experiment) and RHIC experiments otherwise the
$J/\Psi$ suppression is reported~\cite{Abreu:2000zb, Atomssa:2007zz}. 
This  suppression  can not be explained by standard nuclear absorption model.
In NA50 experiment $J/\Psi$ yield is reconstructed through observation of decay $J/\Psi \rightarrow \mu\mu$ only.
Then nuclear absorption model with in medium mass modification was proposed. Small shift of $J/\Psi$ mass makes decay
$J/\Psi \rightarrow D\bar D$ also possible. However in medium mass modification has not been observed. 

Our present work suggests that it is important to account for the binding of
heavy flavor with strangeness, an effect which depletes the eligible
supply of heavy flavor quarks which could form )  $J/\Psi (c\bar c)$ and Bc($b\bar c,\bar b
c$)~\cite{Kuznetsova:2006hx}.

The strangeness enhancement and $J/\Psi$ suppression are considered
as the two most profound evidence of QGP state. There are also
indications that  after hadrons form there is a kinetic phase where
hadrons can rescatter. Heavier hadrons do not decay only but they
can also be recreated by lighter particles. After kinetic freeze out
hadrons decay only and their yields are reconstructed by observation
of their decay products.

%%%%%%%%%%%%%%%%%%%%%%%%%%%%%%%%%%%%%%%%%%%%%%%%%%%%%%%%%%%%%%%%%%%%%%%%%%%%%%%%%%%%%%%%%%%%%%%
\section{Kinetic theory particles production and equilibration in QGP and thermal hadronic gas.}\label{partpr}
%%%%%%%%%%%%%%%%%%%%%%%%%%%%%%%%%%%%%%%%%%%%%%%%%%%%%%%%%%%%%%%%%%%%%%%%%%%%%%%%%%%%%%%%%%%%%%%
\subsection{Boltzman equation}
Relativistic Boltzmann equation is
\begin{equation}
\left(\frac{1}{m}p^{\mu}\frac{\partial}{\partial x^{\mu}}+F^{\mu}\frac{\partial}{\partial p^{\mu}}\right)f_i(x,p)=
=\sum_{q}\left(\eta_{i}^{q}-\chi_{i}^{q}f_{i}(x,p)\right)  ,\nonumber\label{bol}
\end{equation}
where $f_{i}(p,x)$ are particles
distribution functions, the index
$i$ denotes the type of particle, $p^{\mu}$ is its 4-momentum, $F^{\mu}$ is external force and $\eta_{i}^{q}$
and $\chi_{i}^{q}$ are the emission and the absorption coefficients for the
production of a particle of type \textquotedblleft$i$" via the physical
process labeled by $q$. The collision term presentation, right hand side is taken from~\cite{Aksenov:2009dy}. 
Here we will consider cases with $F^{\mu}=0$.
For two-to-two particles reaction:
\begin{equation}
1+2 \leftrightarrow 3  +4 \label{1234}
\end{equation}%
the collision integral for absorption of particle '3' can be written (similar to~\cite{Aksenov:2009dy}):
\begin{equation}
\chi^{^{\mathrm{q}}}f_{3}=\int d\mathbf{p}_4d\mathbf{p_1}%
d\mathbf{p}_{2}W_{\mathbf{p}_{3},\mathbf{p}_{4};\mathbf{p_1}%
,\mathbf{p_2}}f_{4}(\mathbf{p_4},x)f_{3}(\mathbf{p_3},x),
\label{colint}%
\end{equation}
where $W_{\mathbf{p}_{3},\mathbf{p}_{4};\mathbf{p_1},\mathbf{p_2}}$ is transition function,
connected to transition probability $w_{\mathbf{p}_{3},\mathbf{p}_{4};\mathbf{p_1},\mathbf{p_2}}V=W_{\mathbf{p}_{3},\mathbf{p}_{4};\mathbf{p_1},\mathbf{p_2}}$.
Similar for particle '3' emission coefficient:
\begin{equation}
\eta_i{^{\mathrm{q}}}=\int d\mathbf{p}_4d\mathbf{p_1}%
d\mathbf{p}_{2}W_{\mathbf{p_1},\mathbf{p_2};\mathbf{p}_{3},\mathbf{p}_{4}}f_{1}(\mathbf{p_4},x)f_{2}(\mathbf{p_3},x),
\label{colint1}%
\end{equation}
Here we took $d\mathbf{p}=d^4{\bf p}\delta({{\bf p^2_i}-m_i^2})\theta(p^0_i)=d^3p/(2E)$.
If we assume that in reaction (\ref{1234}) particles obey Fermi or Bose statistic
the probability of transition $34 \rightarrow 12$ is
\begin{equation}
w_{\mathbf{p}_{3},\mathbf{p}_{4};\mathbf{p_1},\mathbf{p_2}}=\frac{1}{(2\pi)^2}\langle p_{3}p_{4}\left|M\right|p_{1}p_2\rangle^{2}
\delta^{4}\left(p_{1}+p_{2}-p_{3}-p_4\right)
\left(1 \pm f(x, p_1)\right)\left(1 \pm f(x, p_2)\right).
\end{equation}
Similar we can write a probability for particle '3' emission:
\begin{equation}
w_{\mathbf{p_1},\mathbf{p_2};\mathbf{p}_{3},\mathbf{p}_{4}}=\frac{1}{(2\pi)^2}\langle p_{1}p_2 \left|M\right|p_{3}p_{4}\rangle^{2}
\delta^{4}\left(p_{1}+p_{2}-p_{3}-p_4\right)
\left(1 \pm f(x, p_3)\right)\left(1 \pm f(x, p_4)\right).
\end{equation}
The factors $\left(1 \pm f(x, p)\right)$ are Bose enhancement or Fermi suppression factors, which shows that we can have more than
one or only one particle in given final state.  
The transition probability is proportional to the matrix element $\langle p_{1}p_2\left|M\right|p_{3}p_{4}\rangle^{2}$, which we
took as known from others studies.

\subsection{Equation for particle density evolution and reaction rates}

In this dissertation we investigate resonances evolution during kinetic phase (or thermal hadronic gas).
The kinetic equations, we use, are similar to those used before for strangeness production and equilibration in QGP~\cite{Koch:1984tz, Rafelski:1982pu, Koch:1986ud, Matsui:1985eu, Altherr:1993fd}. These equations can be obtained by integration of Boltzman equation~\ref{bol} over momentum of studied particle $i$. 
We can do it if we know momentum dependence of distribution function. It is possible in case if thermal equilibrium establishes faster by scattering reactions than chemical equilibrium of particle $i$. Then to study particle '3' yield evolution in reaction \ref{1234} we can assume thermal distribution for all particles: 
\begin{equation}
f_{b, f}(\Upsilon, p)=\frac{1}{\Upsilon^{-1}e^{u\cdot p/T}\mp 1};\label{bf}
\end{equation}
Here historically we changed $\lambda\gamma$ to $\Upsilon$, which we call fugacity.

We assume that reaction (\ref{1234}) does not change momentum distribution, the $\Upsilon$ changes. Temperature can change with volume expansion.
In our case from entropy conservation $T^3V \approx$ constant.

For example, strangeness is produced mostly in thermal gluons fusion reaction~(\ref{ggss}). Backward reaction also becomes important near equilibrium point. These reactions studied before in approach considered here  are two-to-two particles reactions~\ref{1234}.

In example of strangeness production these reactions are bosons fusion into fermion-antifermion pair and backward:
\begin{equation}
b+b \leftrightarrow f+\bar{f}. \label{bbff}
\end{equation}
The others possible reactions, which we will consider here are
\begin{eqnarray}
&&f+\bar{f}\leftrightarrow f^{\prime}+\bar{f^{\prime}}; \label{ffff}\\
&&b + f (\bar{f}) \leftrightarrow b + f(\bar{f}). \label{bfbf}
\end{eqnarray}

Under assumptions described above we integrate Boltzmann equation \ref{bol} for particle 3(or 4) over momentum phase space 
$d{\bf p_{i}} = d^4p_{i}\delta({{\bf p^2_i}-m_i^2})\theta(p^0_i)$ we obtain
the evolution equation for current of produced particle 3 or 4  in reactions (\ref{1234}):
\begin{equation}
j^{\mu}_{;\mu}=\frac{dW_{12\rightarrow 3\bar 4}}{dVdt}
-\frac{dW_{34 \rightarrow 12}}{dVdt}.\label{popeq1}
\end{equation}
The current of produced particle $j^{\mu}$  is defined by Eq.(\ref{cur}).
The covariant derivative is
\begin{equation}
j^{\mu}_{;\mu} =
\frac{1}{\sqrt{-g}}\partial_{\mu}(\sqrt{-g}j^{\mu}).
\end{equation}
 
The Lorentz invariant particle production and annihilation rates are
\begin{eqnarray}
&&\frac{dW_{12 \rightarrow 34}}{dVdt} = \frac{\Upsilon_1\Upsilon_2}{1+I}\frac{g_1g_2}{(2\pi)^{12}}
\int\frac{d^{3}p_{1}}{2E_{1}}f_{b,f}(p_1)\int\frac{d^{3}p_{2}}{2E_{2}}f_{b,f}(p_2)
\int\frac{d^{3}p_{3}}{2E_{3}}\int\frac{d^{3}p_{3}}{2E_{4}}\left(2\pi\right)^{4}\times \nonumber\\
&&\sum_{spin}\left|
\langle p_{1}p_2\left| M\right|p_{3}p_{4}\rangle\right|^{2}\delta^{4}\left(p_{1}+p_{2}-p_{3}-p_4\right)
\left(1 \pm f_{b,f}(\Upsilon_{3}, p_3)\right)\left(1 \pm f_{b,f}(\Upsilon_{4}, p_4)\right).
\label{pp4}\\
&&\frac{dW_{34 \rightarrow 12}}{dVdt} = \frac{\Upsilon_3\Upsilon_4}{2}\frac{g_3g_4}{(2\pi)^{12}}
\int\frac{d^{3}p_{3}}{2E_{3}}f_{b,f}(p_3)\int\frac{d^{3}p_{4}}{2E_{4}}f_{b,f}(p_4)
\int\frac{d^{3}p_{3}}{2E_{3}}\int\frac{d^{3}p_{3}}{2E_{4}}\left(2\pi\right)^{4}\times \nonumber\\
&&\sum_{spin}\left|
\langle p_{3}p_{4}\left|M\right|p_{1}p_2\rangle\right|^{2}\delta^{4}\left(p_{1}+p_{2}-p_{3}-p_4\right)
\left(1 \pm f_{b,f}(\Upsilon_{1}, p_1)\right)\left(1 \pm f_{b,f}(\Upsilon_{2}, p_2)\right).
\label{ff}
\end{eqnarray}
Using the relation
\begin{equation} \label{FBrel}
 1\mp f_{f,b} =\Upsilon_i^{-1}e^{u \cdot p_{i}/T}f_{b,f} ,
\end{equation}
and time reversal invariance of matrix element $\left|\langle p_{3}p_{4}\left|M\right|p_{1}p_2\rangle\right|^2=\left|\langle p_{1}p_{2}\left|M\right|p_{3}p_4\rangle\right|^2$,
we obtain equation connecting fermions pair production and annihilation rates:
\begin{equation}
\frac{1}{\Upsilon_1\Upsilon_2}\frac{dW_{12 \rightarrow 34}}{dVdt} =\frac{1}{\Upsilon_3\Upsilon_4} \frac{dW_{34 \rightarrow 12}}{dVdt}=R_{12 \leftrightarrow 34} \label{db}
\end{equation}
From this equation we see that fermions production and annihilation rates are equal and reactions are in equilibrium when 
$\Upsilon_b = \Upsilon_f = 1$. Local equilibrium is also possible when $\Upsilon_b = \Upsilon_f \ne 1$. However in the system with many reactions
it may be impossible to satisfy equilibrium conditions for all reactions when $\Upsilon_i \ne 1$. Also in QGP the gluons are
likely at their highest possible density with $\Upsilon_g = 1$.

Using Eq.~(\ref{db}), we can rewrite Eq.~(\ref{popeq1}) as
\begin{equation}
 j^{\mu}_{;\mu}= (\Upsilon_1\Upsilon_2-\Upsilon_3\Upsilon_4)R_{12 \leftrightarrow 34}.
\end{equation}

For homogeneous expansion this equation can be written as
\begin{equation}
\frac{1}{V}\frac{dN_{3(4)}}{dt} = (\Upsilon_1\Upsilon_2-\Upsilon_3\Upsilon_4)R_{12 \leftrightarrow 34}.
\end{equation}
We can rewrite this equation as equation for $\Upsilon_{3(4)}$:
\begin{equation}
\frac{dn_{3(4)}}{d\Upsilon_{3(4)}}\frac{d\Upsilon_{3(4)}}{dt} + \frac{d(n_{3(4)})}{dT}{\dot T} + n_{3(4)}\frac{\dot{V}}{V} =(\Upsilon_1\Upsilon_2-\Upsilon_3\Upsilon_4)R_{12 \leftrightarrow 34}.\label{Upsgen}
\end{equation}

We introduce reaction relaxation time 
\begin{equation}
\tau_{12 \leftrightarrow 34} = \frac{dn_{3(4)}/d{\Upsilon_{3(4)}}}{2\sqrt{(\Upsilon_1\Upsilon_2)}R_{12 \leftrightarrow 34}}.\label{taurel}
\end{equation}
This time is on the order of magnitude of time needed to reach equilibrium condition for fugacities:
\begin{equation}
\Upsilon^{eq}_1\Upsilon_2^{eq} = \Upsilon_3^{eq}\Upsilon_4^{eq}.
\end{equation}
In simpler example $V=$const, $T=$const, $\Upsilon_1 = \Upsilon_2=\Upsilon$ and $\Upsilon_3 = \Upsilon_4=\Upsilon^{\prime}$ Eq. (\ref{Upsgen}) is
\begin{equation}
\frac{1}{\Upsilon}\frac{d\Upsilon^{\prime}}{dt} = \left(1-\frac{\Upsilon^{\prime\,2}}{\Upsilon^2}\right)\frac{1}{2\tau_{12 \leftrightarrow 34}},
\end{equation}
which  has for $\Upsilon_{\prime}(t=0)=0$ the simple analytical solution~\cite{Rafelski:1982pu}:
  \begin{equation}
\Upsilon^{\prime}=\Upsilon\tanh (t/2\tau_{12 \leftrightarrow 34}).
\end{equation}
For $t\to \infty$, near to chemical equilibrium,  $\Upsilon^{\prime}/\Upsilon \to 1-e^{-t/\tau_{12 \leftrightarrow 34}}$, while
 for $t\to 0$, at the onset of particle production with small $\Upsilon^{\prime}$ we have
$\Upsilon^{\prime} ={t/(2\tau^{\prime})}$. Hence,  near to chemical equilibrium  it
is appropriate to use factor $2$ in definition of relaxation time Eq.(\ref{taurel}).

For Boltzmann distribution, for example for strange particles production in hadronic gas (reactions~(\ref{sthg1}), (\ref{sthg2})), we may use reaction cross section in center mass frame $\sigma$ to estimate production rates in two body processes  
and the relation. Then Lorentz invariant reaction rate is~\cite{Koch:1984tz, Letessier:2002gp}:
\begin{equation}
R_{12 \rightarrow 34} = \left\langle\sigma v_{rel} \right\rangle n_1n_2,
\end{equation}
where $v_{rel}$ is relative velocity of particle 1 in respect of particles 2.
The cross section is connected to matrix element by
\begin{equation}
\sigma v_{12}E_1E_2 = \int d^4p_3\delta(p_3^2-m_3)\theta(p_3^0)d^4p_4\delta(p_4^2-m_4)\theta(p_4^0)\delta^4(p_1+p_2-p_3-p_4)\langle p_{1}p_2\left|M\right|p_{3}p_{4}\rangle^{2}
\end{equation}
The rate can be evaluated, using 
\begin{equation}
v_{12}E_1E_2=\lambda_2(s) = (s-(m_{1}+m_2)^2)(s-(m_1-m_2)^2),
\end{equation}
as~\cite{Koch:1984tz, Letessier:2002gp}, where $\sqrt{s}=(E_1+E_2)$ is total energy of interacting particles 1 and 2 in the center mass frame.
$m_1$ and $m_2$ are masses of  the initial interacting particles
\begin{equation} \label{rcrsec}
{R_{12 \rightarrow 34}} = \frac{g_1g_2}{32\pi^4}\frac{T}{1+I}
\int_{s_{th}}^{\infty}ds\sigma(s)\frac{\lambda_2(s)}{\sqrt{s}}K_1(\sqrt{s}/T),
\end{equation}
(compared to reference~\cite{Letessier:2002gp} our definition is changed 
$R_{12\rightarrow 34} \rightarrow 
R_{12 \rightarrow 34}/(\Upsilon_1 \Upsilon_2)$)
where
 $g_1$, $g_2$ and $\Upsilon_1$, $\Upsilon_2$ are degeneracies and fugacities of  the initial interacting particles.

%From this equation we can derive equation for $\Upsilon$, which is (in case of spherical expansion)
%\begin{equation}
%\frac{dn_f}{d\Upsilon_}\frac{\Upsilon}
%\end{equation}

%%%%%%%%%%%%%%%%%%%%%%%%%%%%%%%%%%%%%%%%%%%%%%%%%%%%%%%%%%%%%%
\subsection{Resonance production during kinetic phase}
%%%%%%%%%%%%%%%%%%%%%%%%%%%%%%%%%%%%%%%%%%%%%%%%%%%%%%%%%%%%%%%
 
%%%%%%%%%%%%%%%%%%%%%%%%%%%%%%%%%%%%%%%%%%%%%%Fig 6
\begin{figure}
\centering
\includegraphics[width=7.5 cm, height=7.5 cm]{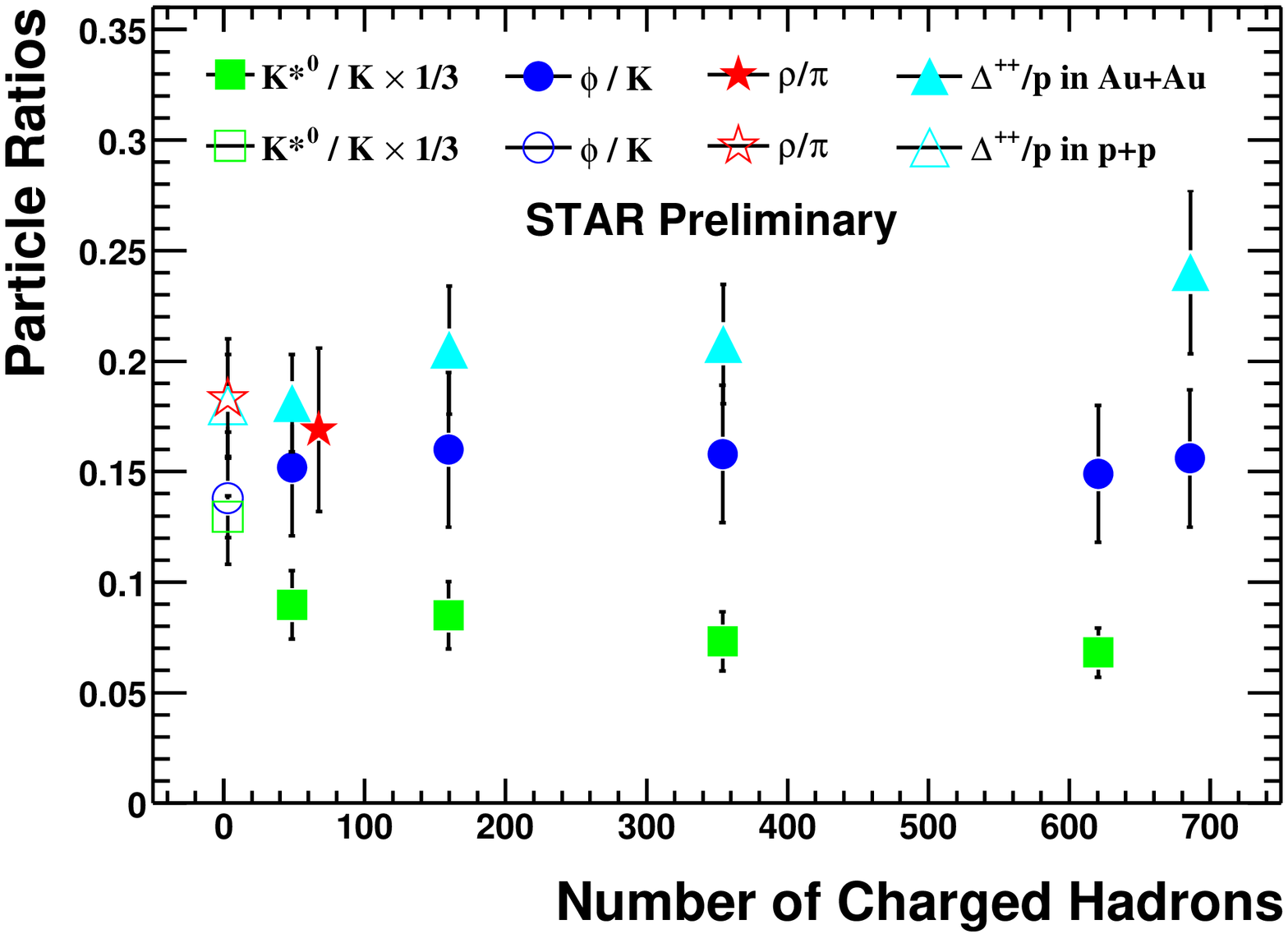}
\includegraphics[width=7.5 cm, height=7.5 cm]{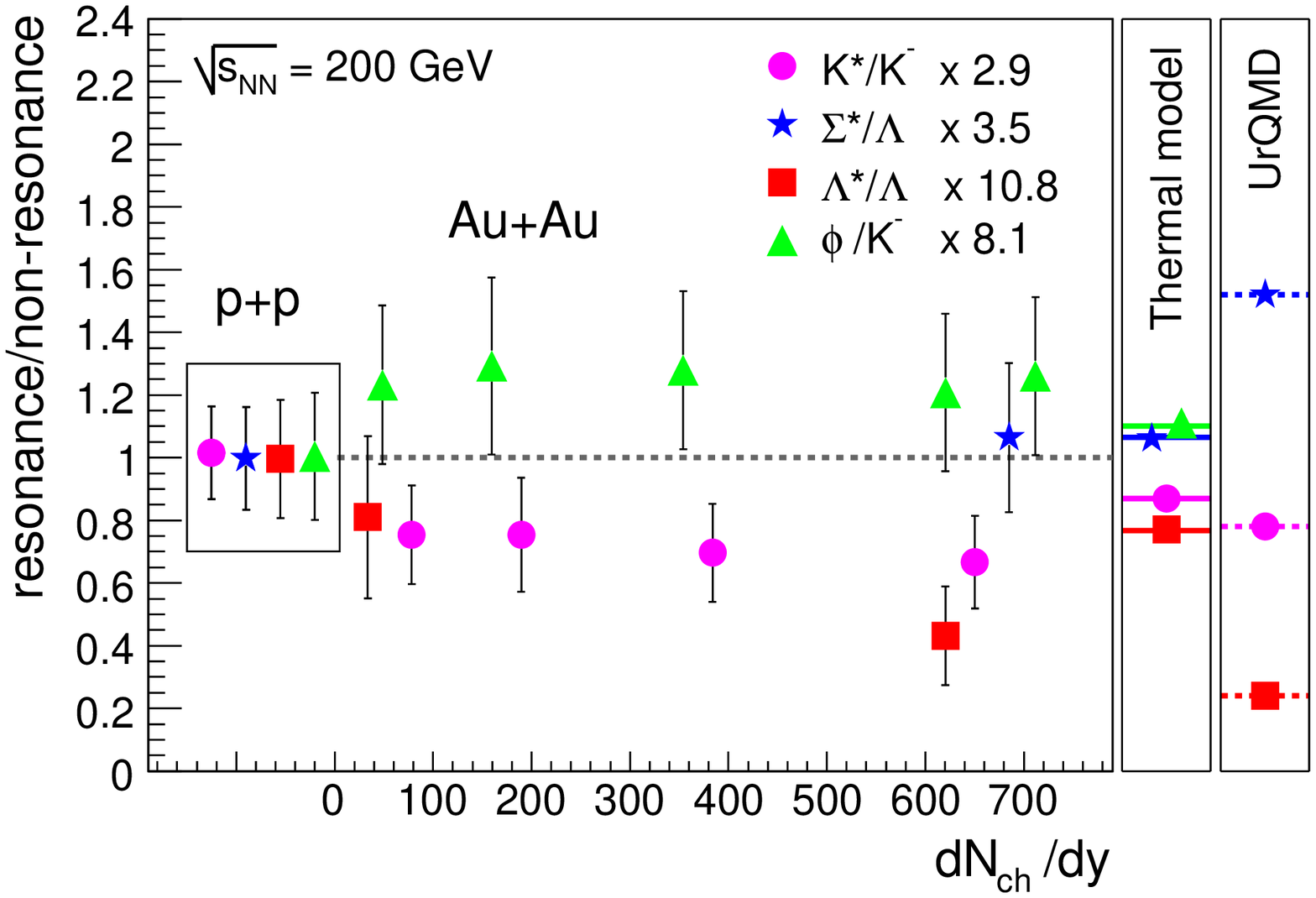}
\caption{\small{Left~\cite{Zhang:2004rj}: The $K^{*0}/K$, $\rho^0/\pi$, $\Delta^{++}/p$ and
$\phi/K$ ratios as a function of number of charged xhadrons in p+p
(open symbols) and various centralities in Au+Au (solid symbols)
collisions. Right~\cite{Abelev:2008yz}:Resonance to stable particle ratios for
$p+p$ and $Au+Au$ collisions. The ratios are normalized to unity in
$p+p$ and compared to thermal and UrQMD model predictions for
central $Au+Au$.}} \label{resrat}
\end{figure}
%%%%%%%%%%%%%%%%%%%%%%%%%%%%%%%%%%%%%%%%%%%%%%%%%%%%%%

Resonances are very short lived hadrons, baryons and mesons, with width (inverse lifespan) on the order of 1 - 100 MeV.
Because of their very short lifespan yields of resonances can not be observed directly. Their yields are reconstructed, using their decay invariant mass method. 

Some of resonance to similar non-resonance hadron yields ratios reported by RHIC and SPS experiments show deviations from those observed in pp
collisions and  from predicted by statistical hadronization model along. In figure~\ref{resrat} we show experimental resonance to non-resonance ratios as a functions of centrality (number of participants) in Au-Au collisions  reported in~\cite{Zhang:2004rj}(preliminary) on the left and in~\cite{Abelev:2008yz} on the right.

From figure~\ref{resrat} we see that some of resonance to non-resonance ratios shows noticeable dependence on collision centrality (enhancement or
suppression). The possible explanation is that reactions in kinetic phase of hadronic gas can influence. The idea that resonance can be regenerated
in kinetic phase was pointed out within ultra relativistic quantum molecular dynamics (UrQMD) model~\cite{Bleicher:2002dm}. 

In chapter 4 we present model, which explains the $\Sigma(1385)$ and $\Delta(1230)$ enhancement
and $\Lambda(1520)$ suppression compared to statistical hadronization without kinetic phase and pp collisions in good agreement with experimental results in figure~\label{resrat}. 
In our approach we assume thermally equilibrated hadronic gas in kinetic phase. We assume that hadronic gas temperature changes together with volume ($T^3V \approx const$ from entropy conservation). We consider only relevant 
\begin{equation} 
1+2 \leftrightarrow 3; \label{123}
\end{equation}
reactions, estimating resonance yield. Their rates in most cases are much faster then for 2-to-2 
particles reactions. Threshold energy for resonance production is smaller. It is possible that 2-to-2
particles reactions also have influence in some cases. Non-equilibrium can accelerate these reactions (rate is proportional to corresponding $\Upsilon$s). This question we leave for future research.
Then we use equations similar to equations from sections~\ref{partpr} extended to the case of reactions~(\ref{123}) for detailed balance and particles evolution. We will derive these equations in chapter~\ref{dpeq}.
The new feature in our detailed balance for reactions~\ref{123} is that we connect Lorentz invariant rate for reactions in both direction in thermal medium with resonance decay time in vacuum, which is known. This way we do not need to know cross section or matrix element for resonance production. 

%%%%%%%%%%%%%%%%%%%%%%%%%%%%%%%%%%%%%%%%%%%%%%%%%%%%
\subsection{Entropy in QGP fireball}\label{sseentr}
%%%%%%%%%%%%%%%%%%%%%%%%%%%%%%%%%%%%%%%%%%%%%%%%%%%%%
When heavy nuclear collide at high energy we expect that QGP is formed.
The QGP consist of quarks (fermions) and gluons (massless bosons). Number of degrees of freedom in this phase is 
\begin{equation}
g = g_g + \frac{7}{4}g_q; \label{qgpdeg}
\end{equation}
where gluon degeneracy $g_g = 2_s(N_c^2-1)=16$ ($2_s$ is spin degeneracy, $N_c=3$ is the number of colors) and
quark degeneracy $g_q = 2_sN_cn_f = 6n_f$, $n_f$ is number of light flavor ($m_f < T$). If semi-massive strange quarks are present 
$n_f = 2.5$. Factor $7/4$ in Eq.~(\ref{qgpdeg}) shows the presence of particles and antiparticles (factor 2) and the smaller fermion phase space compared to bosons, defined by exclusion principal. 

The  entropy content is seen in the  final state multiplicity of
particles produced after hadronization. More specifically, there is a relation
between entropy and particles multiplicities, once we note that the
entropy per particle in a gas is:
\begin{equation}
\frac{S_{\mathrm {B}}}{N}=3.61,\ \ \ \ \ \frac{S_\mathrm{cl}}{N}=4,\
\ \ \ \ \frac{S_\mathrm{F}}{N}=4.2,
\end{equation}
for massless Bose, classical (Boltzmann) and Fermi gases, respectively.
Effectively, for QGP with $u,s,d,G$ degrees of freedom,
$S^{\mathrm {Q}}/N^{\mathrm {Q}}\sim 4$ is applicable for large
range of masses. Thus:
\begin{equation}
{dS^{\mathrm{Q}}} \approx 4\,{dN^Q}. \label{ent}
\end{equation}
This in turn means that final state particle multiplicity provides
us with information about the primary entropy content generated in the
initial state of the QGP phase.

It is today generally believed that there is entropy conserving hydrodynamic
expansion of the QGP liquid. Entropy is conserved in the
fireball, and the conservation of entropy density $\sigma$ flow is expressed by:
\begin{equation}
\frac{\partial_\mu(\sigma u^{\mu})}{\partial{x^{\mu}}}=0,
\label{entrfl}
\end{equation}
where $u^{\mu}$ is local four  velocity vector.  A special case of
interest is   the Bj{\o}rken scenario~\cite{Bjorken:1982qr}.
In this scenario, we assume that\\
1. the energy of the colliding particles is so large that the flow of energy and matter after heavy ions collision remains unidirectional along the original collision axis;\\
2. the transverse extend of the system is so large that the existence of the edge of mater in a direction transverse to the collisional axis
is of a little relevance.\\

This scenario suggests that the natural variables for the dynamics of rapidly expanding in longitudinal direction flow are 
proper time $\tau (t,z)$ and rapidity $y(t,z)$:
\begin{equation}
\tau = (t^2 - z^2)^{1/2}, \quad y=\frac{1}{2}\ln\left(\frac{t+z}{t-z}\right),
\end{equation}  
where $z$ is coordinate in longitudinal direction.
In this case Eq.\,(\ref{entrfl}) can be solved exactly assuming as
initial condition scaling of the physical properties
as a function of rapidity. This implies that there is no preferred
frame of reference, a situation expected in very high energy collisions.
Even if highly idealized, this  simple reaction picture allows
a good estimate of many physical features. Of relevance here is that the exact
solution of hydrodynamics in  (1+1) dimensions implies
\begin{equation}
\frac{dS}{dy}=\rm{Const}.\label{entcon1}\,.
\end{equation}
 
Thus entropy $S$ is not only conserved globally in the hydrodynamic
expansion, but also per unit of rapidity. Though we have (1+3) expansion,
Eq.({\ref{entcon1}}) holds as long as there is, in rapidity, a
flat plateau of particles yields. Namely, each of the
domains of rapidity is equivalent,  excluding  the projectile-target
domains. However, at RHIC and LHC energies  these
are causally disconnected from the central rapidity bin, where
we study the evolution of heavy flavor. The entropy we observe in
the final hadron state has been  to a large extent   produced after the heavy
flavor is produced, during the initial parton thermalization phase, but before strangeness has been produced.
In order to model production of hadrons for different chemical freeze-out
scenarios of the same reaction, we need to relate the entropy content,
temperature and volume of the QGP domain.   We  consider for a
$u,d,G$-chemically equilibrated QGP, and allowing for partial chemical
equilibration of strangeness, the entropy content.

The entropy density $\sigma$ can be obtained from the equation
\begin{equation}
\sigma\equiv\frac{S}{V}=-\frac{1}{V}\frac{d{F_\mathrm{Q}}}{dT}, \label{entden}
\end{equation}
where the thermodynamic potential is:
\begin{equation}
F_\mathrm{Q}(T, \lambda_q, V)=-T\ln {Z(_\mathrm{Q}T, \lambda_q, V)}_\mathrm{Q}.
\label{thpot}
\end{equation}
Inside QGP the partition function is a product of partition
function of gluons $Z_g$, light quarks $Z_q$ and strange quarks  $Z_s$, hence:
\begin{equation}
\ln{Z}=\ln{Z_g}+\ln{Z_q}+\ln{Z_s};
\end{equation}
where for massless particles with $\lambda_q=1$
\begin{eqnarray}
\ln{Z_g}&=&\frac{g_g\pi^2}{90}VT^3,\\
\ln{Z_q}&=&\frac{7}{4}\frac{g_q\pi^2}{90}VT^3.
\end{eqnarray}
Here we take into account that the number of degrees of freedom of
quarks and gluons is influenced by  strongly  interactions, characterized
by strong coupling constant $\alpha_\mathrm{s}$:
\begin{eqnarray}
g_g &=& 2_s\,8_c\,\left(1-\frac{15}{4\pi}{\alpha_\mathrm{s}}+\ldots\right); \\
g_q&=&2_s\,3_c\,2_f\,\left(1-\frac{50}{21\pi}{\alpha_\mathrm{s}}+\ldots\right).
\end{eqnarray}

The case of strange quarks is somewhat more complicated, since we have
to consider the mass, the degree of chemical equilibration, and guess-estimate
the strength of QCD perturbative interactions. We have in Boltzmann approximation:
\begin{eqnarray}
\ln{Z_s}&=&2_{\rm p/a}\frac{g_s}{\pi^2} VT^3,\\
\label{gsk}
g_s &=& 2_s3_c\gamma_s^{\rm Q} \,0.5W(m_s/T)\left(1-k\frac{\alpha_s}{\pi}\right).
\end{eqnarray}
$W(x)=x^2K_2(x)$, where $K_2(x)$ is Bessel function of the 2nd order.
We allow both for strange and antistrange quarks,
factor $2_{\rm p/a}$ (which is for massless fermions $2\cdot 7/8=7/4$).
$k$ at this point is a temperature dependent parameter.
Even in the lowest order perturbation theory it
has not been evaluated for massive quarks at finite
temperature. We know that for massless quarks  $k\simeq 2$.
Considering expansion in $m/T$, for large masses
the correction reverses sign~\cite{Kapusta:1979fh},
which result supports the reduction in value of $k$
for $m\simeq T$. We will
use here the value $k=1$~\cite{Letessier:2006wn}.

The entropy density  following from Eq.\,(\ref{entden}) is:
\begin{equation}\label{entrcons2}
\sigma=\frac{4\pi^2}{90}(g_g + \frac{7}{4}g_q)T^3
      +\frac{4}{\pi^2}2_{\rm p/a}g_sT^3
      +\frac{\cal{A}}{T}.
\end{equation}
For strange quarks in the  second term in Eq.\,(\ref{entrcons2})
we set the entropy per strange quarks to 4 units.
In choosing $S_s/N_s=4$ irrespective of the effect of interaction
and mass value $m_s/T$ we are minimizing the influence of  unknown
QCD interaction effect.

The last term  in Eq.\,(\ref{entrcons2}) comes from differentiation of the
strong coupling constant $\alpha_\mathrm{s}$ in the partition function
with respect to  $T$, see Eq.({\ref{entden}}). Up to two loops in
the $\beta$-function of the renormalization group the correction
term is~\cite{Hamieh:2000fh}:
\begin{equation}
{\cal{A}}=(b_0{\alpha^2_s}+b_1{\alpha^3_s})\left[\frac{2\pi}{3}T^4+\frac{n_f5\pi}{18}T^4\right]
\end{equation}
with $n_f$ being the number of active fermions in the quark loop, $n_f\simeq 2.5$, and
\begin{eqnarray}
b_0 = \frac{1}{2\pi}\left(11-\frac{2}{3}n_f\right),\
b_1 = \frac{1}{4\pi^2}\left(51-\frac{19}{3}n_f\right).
\end{eqnarray}
For the strong coupling constant $\alpha_\mathrm{s}$ we use
\begin{equation}
\alpha_s(T) \simeq \frac{\alpha_s(T_c)}{1+{C}\ln(T/T_c)}, \ \ \
C=0.760\pm 0.002,
\end{equation}
where $T_{c}=0.16$ GeV~\cite{Letessier:2002gp}. This expression arises from running of $\alpha_s(\mu)$, the energy scale at $\mu=2\pi T$, and the value $\alpha_s(M_Z)=0.118$ ($M_z=$91.19 GeV).
A much more sophisticated study
of the entropy in the QGP phase is possible~\cite{Kapusta}, what we use
here is an effective model which agrees  with the
lattice data~\cite{Letessier:2003uj}.

%%%%%%%%%%%%%%%%%%%%%%%%%%%%%%%%%%%%%%%%%%%%%%%%%%%%%%%%%%%%%%%%%%%%%%%%%%%%%
\section{Electron-positron-photon plasma}
%%%%%%%%%%%%%%%%%%%%%%%%%%%%%%%%%%%%%%%%%%%%%%%%%%%%%%%%%%%%%%%%%%%%%%%%%%%%%%

\subsection{Electron-positron-photon generation and equilibration}

The other part of dissertation studies the possibility of $e^+e^-\gamma$ plasma production and equilibration in strong laser field and also heavy (muon, pion) particles production and equilibration
in this plasma. We also included here subsection about early universe. In the laboratory these heavy particles can be diagnostic tool for properties of $e^+e^-\gamma$ plasma, similar as in  case of heavy ions collisions. 

The electron-positron plasma can be found in many astrophysical objects as active galactic nuclear, pulsars, gamma ray bursts. 
Over last fifteen years the huge progress in laser intensity was achieved. The formation of the relativistic, electron-positron-photon $e^-, e^+, \gamma$  plasma  (EP$^3$, temperature $T$ in MeV range) in the laboratory using ultra-short pulse lasers is one of the current topics of interest and forthcoming experimental effort~\cite{TajMou,TajMouBoul}.   

One of questions we study if this electron-positron plasma can be in thermal and chemical equilibrium with photons. 
Here we show that to have this equilibrated opaque plasma it is necessary to focus pulse energy in small size 10-1 nm. This follows
from opaqueness condition. On the other hand in this case of small size less pulse energy is needed to create large electric field, close to breakdown Schwinger field $E_s$. 

The Schwinger field is a field necessary for virtual electron-positron pair to gain the energy $2m_ec^2$ during the time $\delta t$ defined by
Heisenberg uncertainty principal $\delta t=h/m_ec^2$. The energy gain length is $c\delta t = \lambda_b$, where $\lambda_b$ is de Broil wave length.
Then the Schwinger field is
\begin{equation}
E_s = \frac{2m_ec^2}{\lambda_b}=10^{16} {\mathrm V/cm}.  
\end{equation} 

The laser field is connected to laser intensity $I_l$ as  $E^2=Z_0I_l$, where $Z_0$ is vacuum impedance, $Z_0=377\,\Omega$. We find that
to create $E_s$ intensity $I_s=10^{30} W/cm^2$ is necessary~\cite{TajMou}.

In~\cite{Labun:2008re} the time scale necessary to convert laser field energy into $e^+e^-$ pairs energy was evaluated to be in oder of $1-10^{-2}$ fs for the corresponded  field $E \approx (0.4 \div 1.0)E_s$. Pulse duration has to correspond to this field to plasma conversion time. This time and field ranges may be enough to produce desirable high density $e^+e^-$ plasma.

\begin{figure}
\centering
\centerline{\includegraphics[width=10 cm, height=6 cm]{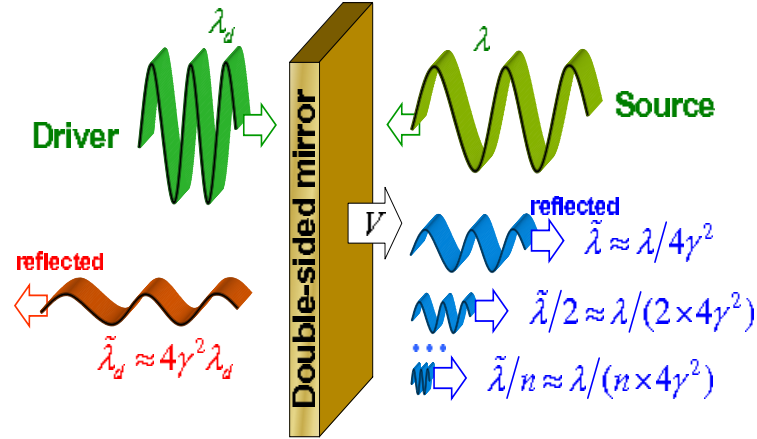}}
\caption{~\cite{Bulanov:2003zz} The scheme of the double-surface mirror.
The ultra-intense driver going from the left
accelerates the mirror in the radiation pressure dominant regime.
In its turn, the mirror reflects the 
intense source sent from the right.
\label{mirror}
}
\end{figure}

These physical conditions should become accessible in the foreseeable future upon the development 
of wavelength compression technology employing an  optical wavelength laser beam
reflected from a relativistic mirror, generated by a pulsed high intensity 
laser~\cite{Bulanov:2003zz}. In~\cite{Bulanov:2003zz} the thin plasma slap plays mirror role accelerating in the radiation 
pressure dominant regime. The flying mirror reflects counter-propagating radiation causing its frequency multiplication by squared Lorentz
factor of mirror because of double dopler effect. The scheme of radiation reflection from accelerated double-surface mirror is shown in figure~\ref{mirror}.

\subsection{Pion and muon production in $e^+e^-\gamma$ plasma}

We also study the production of heavy particles in $e^+e^-\gamma$ plasma. The purpose of this research is to use observation of heavy 
particles (pion, muon) yields as a tool for study of  properties, similar to what we do in the case of heavy ions collisions. Also it may be 
useful for study of the heavy particles production reactions itself. 

The $\pi^0$, $\pi^{\pm}$, $\mu^{\pm}$  can be produced in $e^+e^-\gamma$ plasma.
For $T<<m_{\pi_0}$ (starting from temperature about few MeV) the neutral pions are most effectively produced in two photons fusion: 
\begin{eqnarray} 
\gamma+\gamma \leftrightarrow \pi^0. \label{ggpi0}
\end{eqnarray}
$\pi^{\pm}$ can be produced in $\pi_0\pi_0$ charge exchange scattering:
\begin{equation}
\pi^0 + \pi^0 \rightarrow \pi^{+} + \pi^{-}, \label{pppp}
\end{equation}
as well as  in two photon, and  in electron-positron fusion processes
\begin{eqnarray}
\gamma+\gamma \rightarrow \pi^{+} + \pi^{-}, \label{ffpp} \\[0.2cm]
e^+ + e^- \rightarrow \pi^{+} + \pi^{-}.  \label{eepp}
\end{eqnarray}
We find  that for $\pi^{\pm}$ production, the last two processes are much
slower compared to the   first,  in case that $\pi_0$ density is near chemical equilibrium.
Similarly, the two photon fusion  to two $\pi^0$:
\begin{equation}
\gamma + \gamma \rightarrow \pi^0 + \pi^0, \label{ggpi0pi0}
\end{equation}
as expected, has rate much smaller than rate of one $\pi^0$ production. It is
a reaction of higher order in $\alpha$ and the energy is shared between two
final particles.

In the plasma under consideration, muons can be directly  produced in the reactions:
\begin{eqnarray}
\gamma + \gamma \rightarrow \mu^{+} + \mu^{-}, \label{ggmu}\\
e^{+} + e^{-} \rightarrow \mu^{+} + \mu^{-}. \label{eemu}
\end{eqnarray}

We will show in section~\ref{pimu} that already at temperatures $\geq 5$ MeV the large yield of pion and muon can be 
observed from $e^+e^-\gamma$ plasma.  

These pion and muon production reactions take place in early universe and $pi^0$ production does not freeze-out 
with universe expansion, as we will show here in section~\ref{earlyun}.  This results that density of $\pi^0$ is relatively large, comparable to
proton $p$ and neutron $n$ densities up to the temperatures of few MeV. $\pi^{\pm}$ and muon production freeze out about few MeV. However
up to this temperatures their density is also comparable to $n$ and $p$ densities.
\chapter{STATISTICAL HADRONIZATION AND ESTIMATION OF PHASE SPACE OCCUPANCY FACTORS}\label{sthad1}
\section{Introduction}

As we discussed in section~\ref{sthad} we assume fast hadronizaton. Then physical parameters of fireball as temperature,
volume, entropy, strangeness and heavier flavor multiplicities do not change during hadronization.

In this chapter
we introduce the notion of conservation of entropy  in section   \ref{entroSec}
and strangeness in section  \ref{strSec}, expected to be valid  in
the  fast hadronization process at LHC,
and discuss   how this impacts the SHM statistical parameters.
 We consider the entropy in
a system with evolving strangeness in subsection \ref{ssecdof} and
show that the number of active degrees of freedom in a QGP is nearly
constant. Another highlight is the discussion of sudden hadronization
of strangeness and the associated values of hadron phase space
parameters in subsection \ref{noneqSec}. Throughout this paper we will use explicitly and implicitly
the properties of QGP fireball and hadron phase space regarding entropy  and strangeness content
developed in these two sections \ref{entroSec}  and  \ref{strSec}.

%%%%%%%%%%%%%%%%%%%%%%%%%%%%%%%%%%%%%%%%%%%%%%%%%%%%%%%%%%%%%%%%%%%%%%%%
\section{Entropy conservation at hadronization}\label{entroSec}
%%%%%%%%%%%%%%%%%%%%%%%%%%%%%%%%%%%%%%%%%%%%%%%%%%%%%%%%%%%%%%%%%%%%%%%
\subsection{Number of degrees of freedom in QGP}\label{ssecdof}
%%%%%%%%%%%%%%%%%%%%%%%%%%%%%%%%%%%%%%%%%%%%%%%%%%%%%
The entropy density  following from Eq.\,(\ref{entden}) is:
\begin{equation}\label{entrcons}
\sigma=\frac{4\pi^2}{90}(g_g + \frac{7}{4}g_q)T^3
      +\frac{4}{\pi^2}2_{\rm p/a}g_sT^3
      +\frac{\cal{A}}{T}.
\end{equation}
For strange quarks in the  second term in Eq.\,(\ref{entrcons})
we set the entropy per strange quarks to 4 units.
In choosing $S_s/N_s=4$ irrespective of the effect of interaction
and mass value $m_s/T$ we are minimizing the influence of  unknown
QCD interaction effect.

The last term  in Eq.\,(\ref{entrcons}) comes from differentiation of the
strong coupling constant $\alpha_\mathrm{s}$ in the partition function
with respect to  $T$, see Eq.({\ref{entden}}). Up to two loops in
the $\beta$-function of the renormalization group the correction
term is~\cite{Hamieh:2000fh}:
\begin{equation}
{\cal{A}}=(b_0{\alpha^2_s}+b_1{\alpha^3_s})\left[\frac{2\pi}{3}T^4+\frac{n_f5\pi}{18}T^4\right]
\end{equation}
with $n_f$ being the number of active fermions in the quark loop, $n_f\simeq 2.5$, and
\begin{eqnarray}
b_0 = \frac{1}{2\pi}\left(11-\frac{2}{3}n_f\right),\
b_1 = \frac{1}{4\pi^2}\left(51-\frac{19}{3}n_f\right).
\end{eqnarray}
For the strong coupling constant $\alpha_\mathrm{s}$ we use
\begin{equation}
\alpha_s(T) \simeq \frac{\alpha_s(T_c)}{1+{C}\ln(T/T_c)}, \ \ \
C=0.760\pm 0.002,
\end{equation}
where $T_{c}=0.16\,$GeV. This expression arises from renormalization group running
of $\alpha_s(\mu)$, the energy scale at $\mu=2\pi T$, and the value $\alpha_s(M_Z)=0.118$.
A much more sophisticated study
of the entropy in the QGP phase is possible~\cite{Kapusta}, what we use
here is an effective model which agrees  with the
lattice data~\cite{Letessier:2003uj}.

Eq.\,(\ref{entrcons}) suggests that we  introduce an effective degeneracy of the QGP
based on the expression we use for entropy:
\begin{equation}\label{gS}
g_{\rm eff}^Q(T) = g_g(T) + \frac{7}{4}g_q(T)
+ 2g_s\frac{90}{\pi^4}+\frac{\cal{A}}{T^4}\frac{90}{4\pi^2}.
\end{equation}
Which allows us to write:
\begin{equation}\label{entrcons1}
\sigma=\frac{4\pi^2}{90}g_{\rm eff}^QT^3,
\end{equation}
and
\begin{equation}\label{SdVdy}
\frac{dS}{dy}=\frac{4\pi^2}{90}g_{\rm eff}^QT^3\frac{dV}{dy}\simeq\rm{Const}.
\end{equation}

\begin{figure}
\centering
\includegraphics[width=8cm]{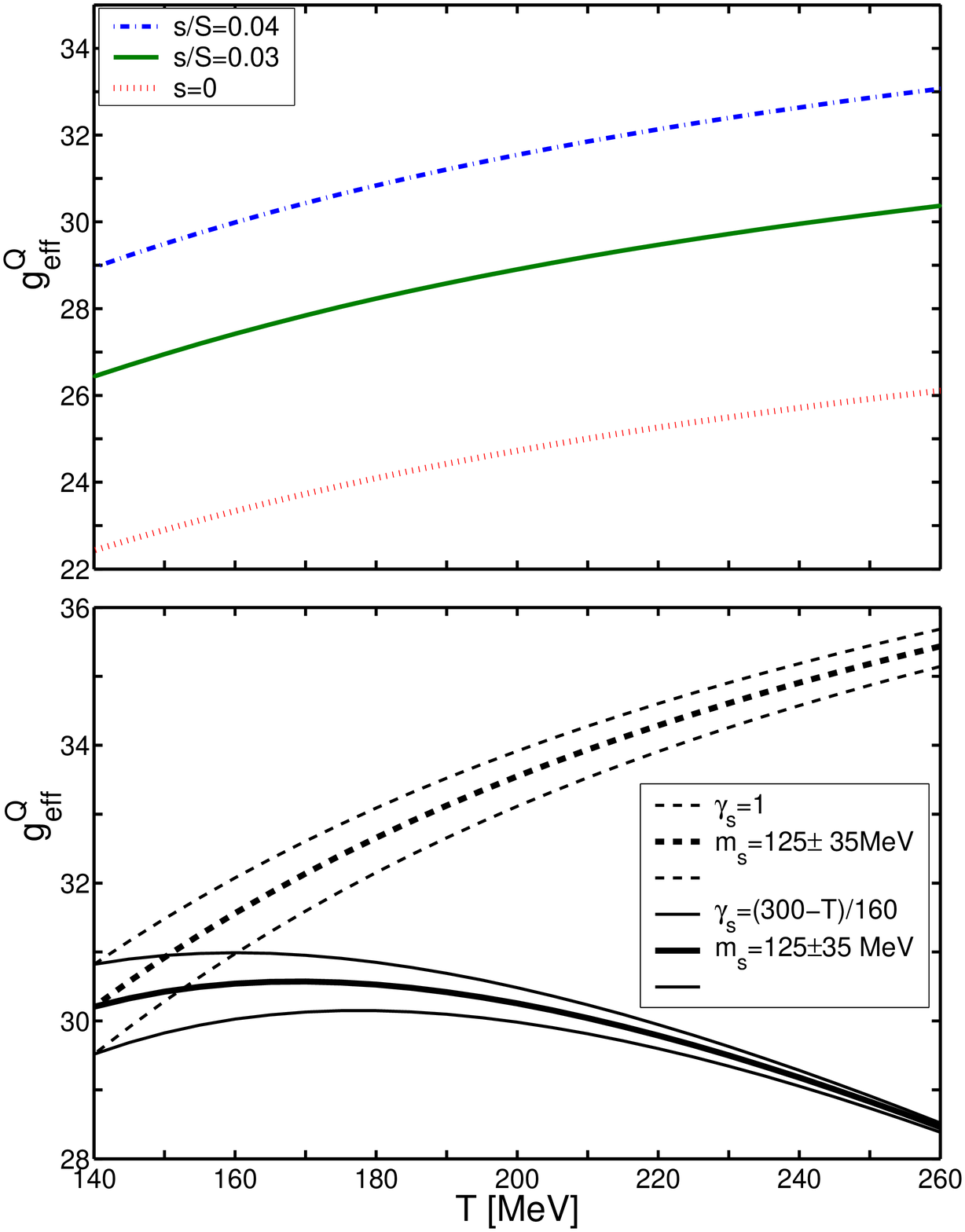}
\caption{(color on-line)
\small{
The Stefan-Boltzman degrees of freedom $g_{\rm eff}$   based on
entropy content of QGP, as function of temperature $T$.
Upper frame:   fixed $s/S$, the solid line with dots (green)
is for a system  with fixed strangeness per entropy $s/S=0.03$,
while dot-dashed (blue) line is for  $s/S=0.04$. The dotted (red)
line is for 2-flavor QCD $s/S=0$ ($u,d,G$ only);
The bottom frame shows dashed (black) line
2+1-flavor QCD with  $m_s=125\pm35$\,MeV (chemically equilibrated
$u,d,s,G$ system). The (thick, thin) solid lines are
for QGP in which strangeness contents is increasing as a function of
temperature, see text. }} \label{geff}
\end{figure}

We show the QGP degeneracy  in figure \ref{geff},  as a function
of $T\in [140,260]$ MeV,  top frame for
fixed $s/S=0, 0.03, 0.04$ (from bottom to top), and in the bottom
frame for the strangeness chemical  equilibrium, $\gamma_s=1$ (dashed)
and approach to chemical equilibrium cases (solid).
When we fix the specific strangeness content $s/S$  in the plasma
comparing different temperatures we find  that in all cases
$g_{\rm eff}^Q$  increases with $T$.
For $s/S=0$ we have a 2-flavor system (dotted line, red) and
the effective number of degrees of freedom $g_{\rm eff}^Q$
varies between 22 and 26.   The
solid line with dots (green) is for $s/S=0.03$, and the dot-dashed line (blue)
gives the result for  $s/S=0.04$.

%%%%%%%%%%%%%%%%%%%%%%%%%%%%%%%%%%%%%%%%%%%%%%%%%Fig 14/4
%%%%%%%%%%%%%%%%%%%%%%%%%%%%%%%%%%%%%%%%%%%%%%%%%%%%%%%%%%%%%%%

In the bottom panel of figure \ref{geff} we note that like for 2 flavors,
case (s=0), for the 2+1-flavor system ($\gamma_s=1$) $g_{\rm eff}^Q$ increases
with T (dashed line, black). $g_{\rm eff}^Q$ varies between 30 and 35.5.
The thin dashed lines indicate the range of uncertainty due to mass of the
strange quark, which in this calculation is fixed with upper curve
corresponding to $m_s=90$ MeV, and lower one $m_s=160$ MeV. The expected
decrease in value of $m_s$ with $T$ will thus have the effect to steepen
the rise in the degrees of freedom with $T$.

We now explore in a QGP phase  the effect of an increasing
strangeness fugacity with decreasing temperature. This study is
a bit different from the rest of this paper, where we consider
for comparison purposes hadronization for a range of temperatures
but at a {\em fixed} value of $s/S$. A variable $\gamma_s^{\rm Q}(T)$
implies a more sophisticated, and thus more model dependent
picture of plasma evolution. However, this offers us an important insight
about  $g_{\rm eff}^Q$.

We consider the function:
\begin{equation}\label{gamsmodel}
\gamma_s^{\rm Q}={300-T{\rm [MeV]} \over 160}.
\end{equation}
This is  consistent with the kinetic computation of
strange\-ness production~\cite{Letessier:2006wn}.
 At $T=140$ MeV  we have chemical equilibrium in the QGP phase,
while and at the  temperature $T=260$\,MeV  we  have
$\gamma_s=0.25$.
The result for   $g_{\rm eff}^Q$ is shown as a thick (black) line in figure \ref{geff},
with the range showing   strange quark mass range $m_s=125\pm35$\,MeV.
 We see that in a wide range of temperatures
we have $29.5<g_{\rm eff}^Q<30.5$.

The lesson is that  with
the growth of $\gamma_s^{\rm Q}$ with decreasing $T$
the entropy of the QGP is well described by
a constant value  $g_{\rm eff}^Q=30\pm0.5$. Since the entropy is (nearly) conserved
and $g_{\rm eff}^Q$ is (nearly) constant, Eq.\,(\ref{SdVdy}) implies that
 we can scale the system properties using the constraint $T^3dV/dy=$Const.
We stress again that these results  arises
in a realistic QGP with $2+\gamma_s^{\rm Q}$-flavors, but are model dependent
and of course rely on the lattice motivated description of
the  behavior of QGP properties.  On the other hand it is not surprising that
the rise of
strangeness chemical saturation with decreasing temperature compensates   the
`freezing'  of the  $q,G$-degrees of freedom
with decreasing temperature.

\subsection{Entropy content and chemical (non-)equilibrium}\label{ssentrocont}
%%%%%%%%%%%%%%%%%%%%%%%%%%%%%%%%%%%%%%%%%%%%%%%%%

We use as a reference a QGP state with
$dV/dy=800\,{\mathrm{fm}}^{3}$ at $T=200\,{\mathrm{MeV}}$,
see table \ref{VTN1}. We find  from Eq.\,(\ref{ent})
the Q and H phase particle multiplicity. The hadron multiplicity stated
is what results after secondary resonance decays. The  total hadron
multiplicity after hadronization and resonance decays
was calculated using on-line SHARE 2.1~\cite{Torrieri:2004zz}.
If a greater (smaller) yield of final state hadrons is observed at LHC,
the value of $dV/dy$ need to be revised up (down). In general
expansion before hadronization
will not alter $dS/dy$. We can  expect that as $T$ decreases,
$V^{1/3}$ increases. Stretching the validity of Eq.\,(\ref{SdVdy}) to
low temperature  $T=140$ MeV, we see the result in the second line of table \ref{VTN}.

%%%%%%%%%%%%%%%%%%%%%%%%%%%%%%%%%%%%%%%%%%%%%%%%%%%%%%TAble II
\begin{table}
\centering
\caption{Reference values of  volume, temperature, entropy, particle multiplicity}
 \label{VTN1}
\begin{tabular}{|c|c|c|c|c|}
  \hline
 $dV/dy$ $[\mathrm fm^{3}]$ & $T$[MeV]  & $dS^{\mathrm {Q}}/dy$ & $dN^{\mathrm {Q}}/dy $& $dN^{\mathrm {H}}/dy$\\
  \hline
  $800$& $200$ &10,970&2,700&5,000\\
  $2300$& $140$ &10,890&2,700&4,500\\
\hline
\end{tabular}
\end{table}
%%%%%%%%%%%%%%%%%%%%%%%%%%%%%%%%%%%%%%%%%%%%%%%%%%%%%%%%%%%%%%%%

For QGP, in general  the entropy content is higher than
in a comparable volume of chemically equilibrated
hadron matter,  because of the liberation of color degrees of freedom in the
color-deconfined phase. The total entropy has to be
conserved during transition between QGP and HG phases, and
thus after hadronization, the excess of entropy is
observed in excess particle multiplicity, which can be interpreted as a
signature of deconfinement~\cite{Letessier:1992xd,Letessier:1993hi}.
The dynamics of the transformation of QGP into HG determines how this
additional entropy manifests itself.

The comparison of entropy in both phases is temperature
dependent but in the domain of interest i.e.
$140<T<180$ MeV  the entropy density follows:
\begin{equation}
\sigma^{\mathrm{Q}} \gtrsim 3 \sigma^{\mathrm{H}}.
\end{equation}
Since the total entropy $S$ is conserved or slightly increases,
in the hadronization process some key parameter must grow in the
hadronization process. There are two options:\\
a) either the volume changes:
\begin{equation}
3 V^{\mathrm{H}} \gtrsim V^{\mathrm{Q}};
\label{volch}
\end{equation}
or \\
b) the phase occupancies change, and since $n_i\propto \gamma_i^{2,3}, i=q,s$ in
hadron phase
\begin{equation}
\gamma^{\mathrm{H}}_q\simeq \sqrt{3}, \ \ \ \gamma^{\mathrm{H}}_s/\gamma^{\mathrm{H}}_q\gtrsim 1.
\end{equation}
In a slow, on hadronic time scale, transition, such as is the case in the
early Universe, we can expect that case a) prevails. In
high energy  heavy ion collisions, there is
no evidence in the experimental results   for the long coexistence of hadron and
quark phases which is required for volume growth. Consequently, we have
$V^{\mathrm{H}}\sim V^{\mathrm{Q}}$ and
a large value of $\gamma^{\mathrm{H}}_q$ is required
to conserve entropy. The value of
$\gamma^\mathrm{H}_q$ is restricted by
\begin{equation}
\gamma^\mathrm{cr}_q\cong\exp(m_{\pi}^0/2T). \label{bcon}
\end{equation}
This value $\gamma^\mathrm{cr}_q$  is near to maximum
allowed value, which arises at condition of Bose-Einstein
condensation of pions. We will discuss quantitative results for
$\gamma^{\mathrm{H}}_q$ (and $\gamma^{\mathrm{H}}_s$) below
in subsection  \ref{noneqSec}.

%%%%%%%%%%%%%%%%%%%%%%%%%%%%
\section{Strangeness in Hadronization}\label{strSec}
%%%%%%%%%%%%%%%%%%%%%%%%%%%%
\subsection{Strangeness abundance in QGP and HG}\label{StrAbund}
%%%%%%%%%%%%%%%%%%%%%%%%%%%%%%%%%%%%%%%%%%%%%%%%%
The efficiency of strangeness production depends on energy and collision
centrality of heavy ions collisions. The increase, with value of
centrality (participant number), of
per-baryon specific strangeness yield indicates presents of
strangeness production mechanism acting beyond the first collision
dynamics. The thermal gluon fusion to strangeness can explain this
behavior~\cite{Koch:1986ud},
and a model of the flow dynamics at RHIC and LHC suggests
that the QGP approaches chemical equilibrium but also can
exceed it at time of hadronization~\cite{Letessier:2006wn}.

The strangeness yield in  chemically equilibrated  QGP is usually
described as an ideal Boltzman gas. However, a significant correction
is expected due to perturbative QCD effects. We implement this
correction based on   comments below Eq.\,(\ref{gsk}). We use here the expression:
\begin{equation}
\frac{dN_{s}^{\rm Q}}{dy}=\gamma_s^Q\left(1-\frac{\alpha_s}{\pi}\right)n^{\rm eq}_s\frac{dV}{dy}.\label{str}
\end{equation}
with the Boltzman limit density Eq.\,(\ref{BolzDis}), and mass $m_s=125$\,MeV,
$g_s=6, \lambda_s=1$. The QCD correction corresponds to discussion
of entropy in subsection \ref{sseentr}

We obtain strange quarks phase space occupancy
$\gamma^{\mathrm{H}}_s$ as a function of temperature from
condition of equality of the number of strange quark and antiquark
pairs in QGP and HG. Specifically,
in the sudden QGP hadronization,  quarks recombine and we expect that the
strangeness content does not significantly change.
For heavier flavors across the phase boundary
this condition  Eq.\,(\ref {flcons}) is very well satisfied,  for strangeness the
fragmentation effect adds somewhat to the yield,
\begin{equation}\label{sconshad}
 \frac{dN_{s}^{\rm H}}{dy} \gtrsim \frac{dN_{s}^{\rm Q}}{dy}.
\end{equation}
Using the equality of yields we underestimate slightly the value of strangeness
occupancy that results. We recall that we also conserver entropy  Eq.\,(\ref {Scons})
which like strangeness can in principle grow in hadronization,
 \begin{equation}  \label{encon}
 \left.{s\over S }\right|_{\mathrm H} \gtrsim
     \left.{s\over S }\right|_{\mathrm Q} .
\end{equation}
using   Eq.\,(\ref {Scons}) we underestimate the value of $ \gamma^{\mathrm{H}\,2}_q$.

Counting all
strange particles, the number of pairs is:
\begin{eqnarray}
\frac{dN_{s}^{\rm H}}{dy}=\frac{dV}{dy}\left[\right.
 &&\hspace*{-0.4cm}  {{\gamma^{\mathrm{H}}_{s}}}
  \left(\gamma^{\mathrm{H}}_qn^{\rm eq}_{K}
             +
      \gamma^{\mathrm{H}\,2}_qn^{\mathrm{eq}}_Y \right)\nonumber \\
+   && \hspace*{-0.4cm}
 \gamma^{\mathrm{H}\,2}_s(2\gamma^{\mathrm{H}}_qn^{\rm eq}_\Xi+n^{\rm eq}_\phi+P_sn^{\rm eq}_\eta) \nonumber \\
+  3 && \hspace*{-0.4cm}
        \gamma^{\mathrm{H}\,3}_sn^{\rm eq}_\Omega
\left. \right],
\label{gammas}
\end{eqnarray}
where $n^{\mathrm{eq}}_i$ are densities of strange hadrons
(mesons and baryons) calculated  using Eq.\,(\ref{dist}) in chemical
equilibrium. $P_s$ is the strangeness content of the $\eta$.
The way we count hadrons is to follow strangeness content, for  example
$n^{\rm eq}_K=n^{\rm eq}_{K^+}+n^{\rm eq}_{K^0}=n^{\rm eq}_{K^{-}}+n^{\rm eq}_{\bar{K}^0}$.
We impose  in our calculations $\bar{s}=s$.
The pattern of this  calculation follows an
established approach,  SHARE 2.1~\cite{Torrieri:2004zz} was  used in detailed evaluation.

\subsection{Strangeness per entropy $s/S$}
%%%%%%%%%%%%%%%%%%%%%%%%%%%%%%%%%%%%%%%%%%%%%%%%%
Considering that both strangeness, and entropy, are conserved in the
hadronization process, a convenient variable to consider as fixed
in the hadronization process, is the ratio
of these conserved quantities $s/S$. In chemical
equilibrium we expect that   in general such a ratio must be
different for different phases of matter from which particles
are produced~\cite{Kapusta:1986cb,Letessier:1993nz,Letessier:2006wn}.

\begin{figure}%[!b]
\centering
\includegraphics[width=8cm,height=8cm]{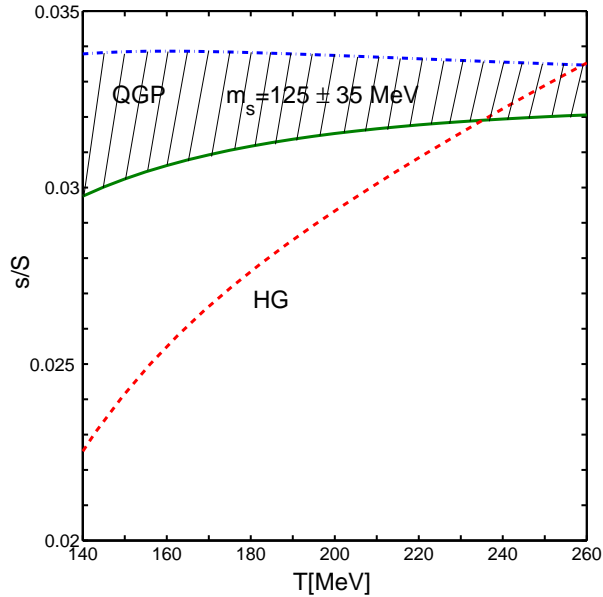}
\caption{(Color on line)
\small{Strangeness to entropy ratio $s/S$ as function of
temperature T, for the QGP (green, solid line for $m_s=160$ MeV,
blue dash-dot line for $m_s=90$ MeV) with $k=1$, see Eq.\,(\ref{str});
and for HG (light blue,dashed line) phases for
$\gamma_q=\gamma_s=\lambda_q=\lambda_s=1$
in both phases. }} \label{sS}
\end{figure}

We compare QGP and HG specific per entropy strange\-ness content
in figure {\ref{sS}}. We  show as function of temperature $T$
 the $s/S$ ratios for chemically equilibrated QGP and HG phase.
  For the QGP the entropy $S$ in QGP is calculated
as described in section \ref{entroSec}, and  we use $k=1$ in Eq.\,(\ref{str}).
 The shaded area
shows the range  of masses of strange
quarks, considered,  results for $m_s=90$ MeV (upper (blue) dash-dotted line)
and $m_s=160$ MeV ((green) solid line) form the boundaries. The central
QGP value is at about $s/S=0.032$.

%%%%%%%%%%%%%%%%%%%%%%%%%%%%%%%%%%%%%%%%%%%Fig 5
%%%%%%%%%%%%%%%%%%%%%%%%%%

The  short-dashed (light blue) line shows the hadron phase $s/S$ value found
using SHARE 2.1 . For HG  near to usual range of hadronization
temperature $T\simeq 160$\,MeV we find $s/S\simeq 0.025$. In general formation of
QGP implies and increase by 30\% in $s/S$. Both HG and QGP phases have a similar
specific strangeness content at $T=240$--260\,MeV, however it is
not believed that a HG at such high temperature would be a stable
form of matter.  This HG to QGP dissociation, or QGP hadronization depends on the
degree of strangeness equilibration in plasma~\cite{Rafelski:2005md},
and other dynamical factors.

In the QGP the value of $s/S$ for the range of realistic hadronization
temperature $140<T<180$ MeV is in general
 larger than in HG.  This implies that generally, the abundance of strange hadrons
produced in hadronization   over saturates   the strange
hadron  phase space, if QGP state had  reached (near) chemical  equilibrium. Moreover,
since we are considering the ratio $s/S$ and find in QGP a value greater than in
HG, for chemical equilibrium in QGP the hadronization process will lead to
$\gamma^{\rm H}_s/\gamma^{\rm H}_q>1$.

One can wonder if we have not overlooked some dynamical or microscopic
effect which could adjust the value of $s/S$ implied by QGP to the value
expected in HG. First we note
 that the fast growth of the volume $V$ cannot change $s/S$. Moreover, any
additional strangeness production in hadronization would enhance the
over-abundance  recorded in the resulting HG.  Only a highly significant entropy
production at fixed strangeness yield in the hadronization process could
bring the QGP $s/S$ ratio down, masking  strangeness over-saturation.
A  mechanism for such entropy production in hadronization
is unknown, and moreover, this  would further entail an
unexpected and high hadron multiplicity
excess.

One could of course argue that the perturbative QCD properties in the QGP are
meaningless, the entropy in QGP is much higher at given temperature.
However, the properties of QGP have been checked against the lattice
results, and the use of lowest order expressions is justified in
these terms~\cite{Letessier:2003uj}. Moreover, the value of $s/S$ is
established way before hadronization.

\subsection{Wr\'oblewski ratio $W_s$}
At this point it is appropriate to look  at another observable proposed
to study strangeness yield, the Wr\'oblewski ratio~\cite{Wroblewski:1985sz}:
\begin{equation}\label{WRs}
W_s\equiv \frac{2\langle \bar s s\rangle}{\langle \bar u u\rangle+\langle \bar d d\rangle}.
\end{equation}
$W_s$ compares the number of newly produced strange quarks to 
the produced number of light quarks. In an equilibrated   deconfined phase
$W_s$ compares the number of active strange quark 
degrees of freedom to the number of light quark degrees of freedom. 

The ratio $s/S$ compares the strange quark degrees of freedom to all degrees of 
freedom available in QGP. Therefore as function of $T$  the ratios  $s/S$ and $W_s$ can behave
differently: Considering the limit $T\to T_c$ a constant $s/S$ indicates 
that the reduction of $s$-degrees of freedom    goes hand in hand with  
the `freezing' of gluon degrees of freedom, which precedes the   `freezing' of light quarks.
This also  implies that for $T\to T_c$  in general $W_s$ diminishes.   
The magnitude of $m_s$, the strange quark mass decisively enters the  limit  $T\to T_c$.

For $T>> T_c$ the ratio $W_s$ 
can be evaluated comparing the rates of production of light and strange quarks,
using the fluctuation-dissipation theorem~\cite{Gavai:2002kq} , which allows 
to relate rate of quark production to 
 quark susceptibilities $\chi_i$  (see Eqs. (11) and (12) in~\cite{Gavai:2002kq}):
\begin{equation}\label{WRsc}
W_s\simeq R_\chi= \frac{2\chi_s}{\chi_u+\chi_d}.
\end{equation}
An evaluation of $R_\chi$ as function  of temperature  in lattice QCD  
has been  achieved~\cite{Gavai:2006ap}.  
For $T\simeq 2.5T_c$    the result obtained, 
 $  W_s\to R_\chi\simeq 0.8$,  is in agreement with the expectation for equilibrium QGP with 
nearly free quarks, with mass of strangeness having a small but noticeable  significance. 
With decreasing $T$, the ratio $R_\chi\to 0.3$ for  $  m_s=T_c$. However, this value
of $m_s$ is too large, the physical value should be nearly half as large, which  
would result in a greater $R_\chi$. Moreover, for $T\to T_c$ the 
relationship of $  W_s$ to $ R_\chi$, Eq. (\ref{WRsc})  is in question in that the greatly reduced 
rate of production of strangeness may not be
satisfying the conditions required in Ref.~\cite{Gavai:2002kq}.   

Comparing the observables $s/S$  and $W_s$ we note that the
experimental measurement requires in both cases a  detailed
analysis of   all particles produced. At lower reaction energies there
is additional complication in evaluation of $W_s$ due to the need to 
subtract the effect of quarks brought into the reaction region.  Turning
to the theoretical computation of   $s/S$  and $W_s$ we note 
that the thermal lattice QCD evaluation of $s/S$ is possible without any approximation, 
even if the actual computation of entropy near the phase boundary 
is a challenging task.  On the other hand, the lattice computation
of  $W_s$  relies on production rate of strangeness being sufficiently fast, 
which cannot be expected near to the phase boundary. Moreover, the 
variable    $s/S$ probes all QGP degrees of freedom, while  $W_s$ probes
only quark degrees of freedom. We thus conclude that  $s/S$ is both 
more accessible  theoretically and experimentally, 
and perhaps more QGP related  observable, as compared to $W_s$, 
since it comprises the gluon degrees of freedom.

%%%%%%%%%%%%%%%%%%%%%%%%%%%%%%%%%%%%%%%%%%%%%%%%%%%%%%%%%%%%%%%%%%%%%
\subsection{Strangeness chemical non-equilibrium}\label{noneqSec}
%%%%%%%%%%%%%%%%%%%%%%%%%%%%%%%%%%%%%%%%%%%%%%%%%
In order that in fast hadronization there is continuity of
strangeness   Eq.\,(\ref{sconshad}), and entropy, Eq.\,(\ref{encon})
the hadron phase  $\gamma^\mathrm{H}_s\ne 1$ and $\gamma^\mathrm{H}_q\ne 1$.
We have  to solve for  $\gamma^\mathrm{H}_s$ and  $\gamma^\mathrm{H}_q$
simultaneously  Eqs.\,(\ref{sconshad},\ref{encon}).

\begin{figure}  %[!b]
\centering
\includegraphics[width=7.3cm,height=8.1cm]{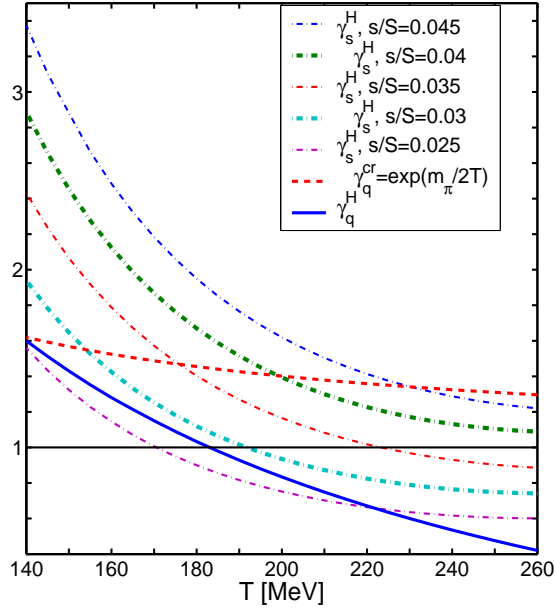}
\caption{(color on-line)
\small{Phase space occupancy as a function of $T$:
$\gamma^{\mathrm{H}}_q$ (blue, solid line),
$\gamma^\mathrm{H}_s$ (dash-dotted lines, from top to bottom)  for $s/S=0.045$,
for  $s/S=0.04$ (thick line), $s/S=0.035$, $s/S=0.03$ (thick line), $s/S=0.025$;
$\gamma^\mathrm{cr}_q$ (red, dashed line).}} \label{gq}
\end{figure}

In figure~{\ref{gq}} we show  as a function of $T$ the strange phase space occupancy
$\gamma^{\mathrm{H}}_s$,
obtained for several values of  $s/S$ ratio (from top to bottom 0.045, 0.04, 0.035, 0.03, 0.025)
evaluated for $S^{\rm Q}=S^{\rm H}$. The solid
line shows $\gamma^{\rm H}_q$  for $s^{\rm Q}=s^{\rm H}$ and $S^{\rm Q}=S^{\rm H}$.
The maximum allowed value Eq.\,(\ref{bcon}) is
shown dashed (red).

%%%%%%%%%%%%%%%%%%%%%%%%%%%%%%%%%%%%%%%%%%%%%%%%%%Fig 6
%%%%%%%%%%%%%%%%%%%%%%%%%%%%%%%%%%%%%%%%%%%%%%%%%%%%%%%%%%%%%%%

\begin{figure} %[!b]
\includegraphics[width=7.7 cm,height=14.2cm]{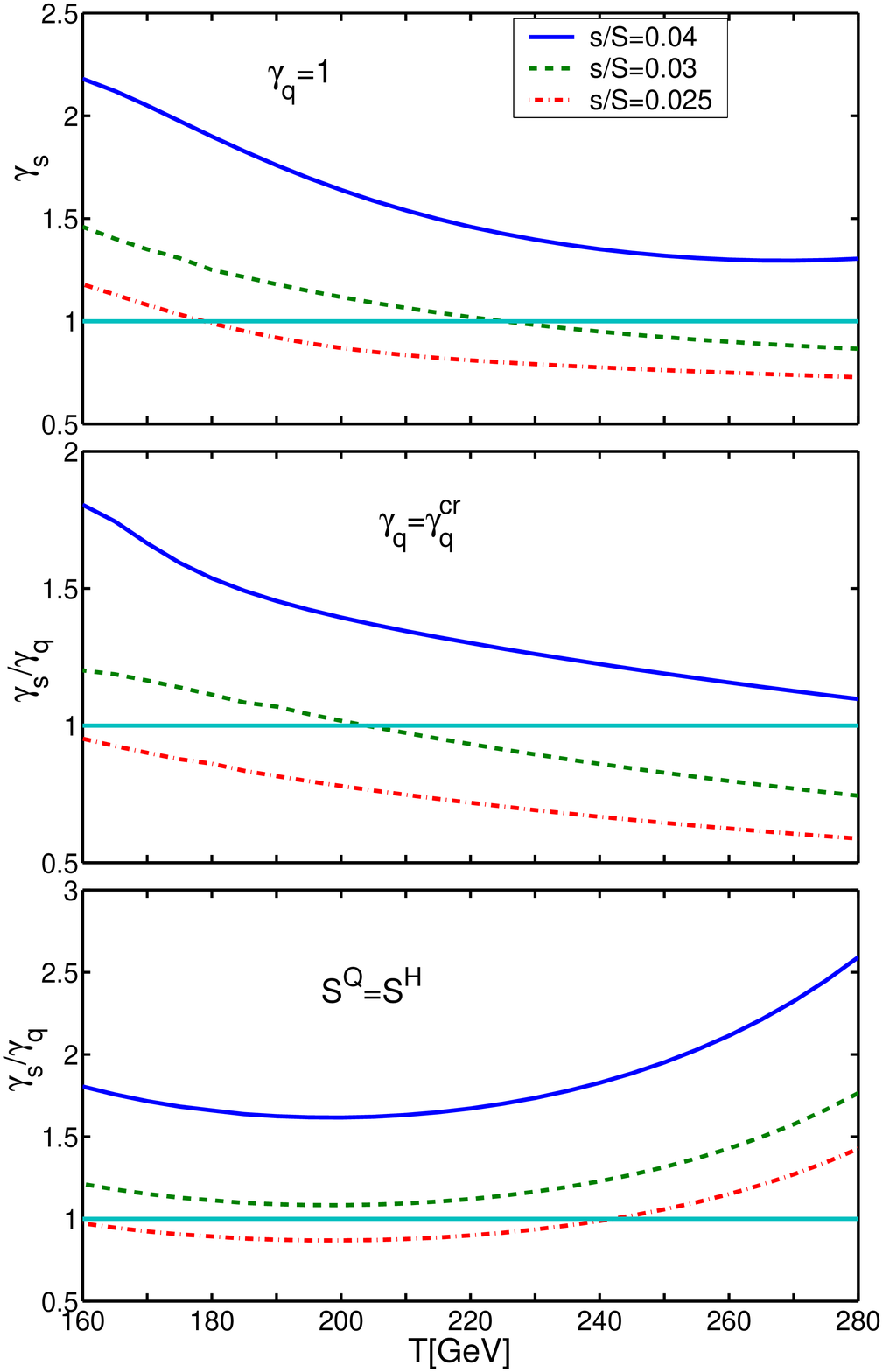}
\centering
\caption{(color on line)
\small{$\gamma^{\mathrm{H}}_s/\gamma^{\mathrm{H}}_q$ (=$\gamma^{\mathrm{H}}_s$  at
 $\gamma^{\mathrm{H}}_q=1$) as a function of
hadronization temperature $T$. Top frame:  $\gamma^{\mathrm{H}}_q=1$,
middle frame:   $\gamma^{\mathrm{H}}_q=\gamma^{cr}_q$, and bottom frame:
$S^H=S^Q$. Lines,  from top to bottom: $s/S=0.04$
(blue, solid line), $s/S=0.03$ (green, dashed line), $s/S=0.025$
(red, dash-dot line) }} \label{gseq} %\label{gsgq}
\end{figure}

In figure \ref{gseq} we show results for $\gamma^{\mathrm{H}}_s/\gamma^{\mathrm{H}}_q$
(where $\gamma^{\mathrm{H}}_q=1$ we show $\gamma^{\mathrm{H}}_s$ ) .
We consider the three cases: $\gamma_q=1$ , $\gamma^{\rm H}_q=\gamma^{cr}_q$,
and entropy conservation $S^H=S^Q$  for  $s/S=0.045$,
$s/S=0.04$, $s/S=0.035$, $s/S=0.03$, $s/S=0.025$ (dash-dot lines)
(lines from top to bottom).
We see that except  in case that strangeness were to remain well below
chemical equilibrium in QGP ($s/S\simeq 0.03$), the abundance of heavy
flavor hadrons we turn to momentarily will be marked by an
overabundance of strangeness, since practically
in all realistic conditions we find
$\gamma^{\rm H}_s>\gamma^{\rm H}_q$.

%%%%%%%%%%%%%%%%%%%%%%%%%%%%%%%%%%%%%%%%Fig 23/9+24/10/=7
%%%%%%%%%%%%%%%%%%%%%%%%%%

\begin{figure}% [!b]
\includegraphics[width=9cm,height=9cm]{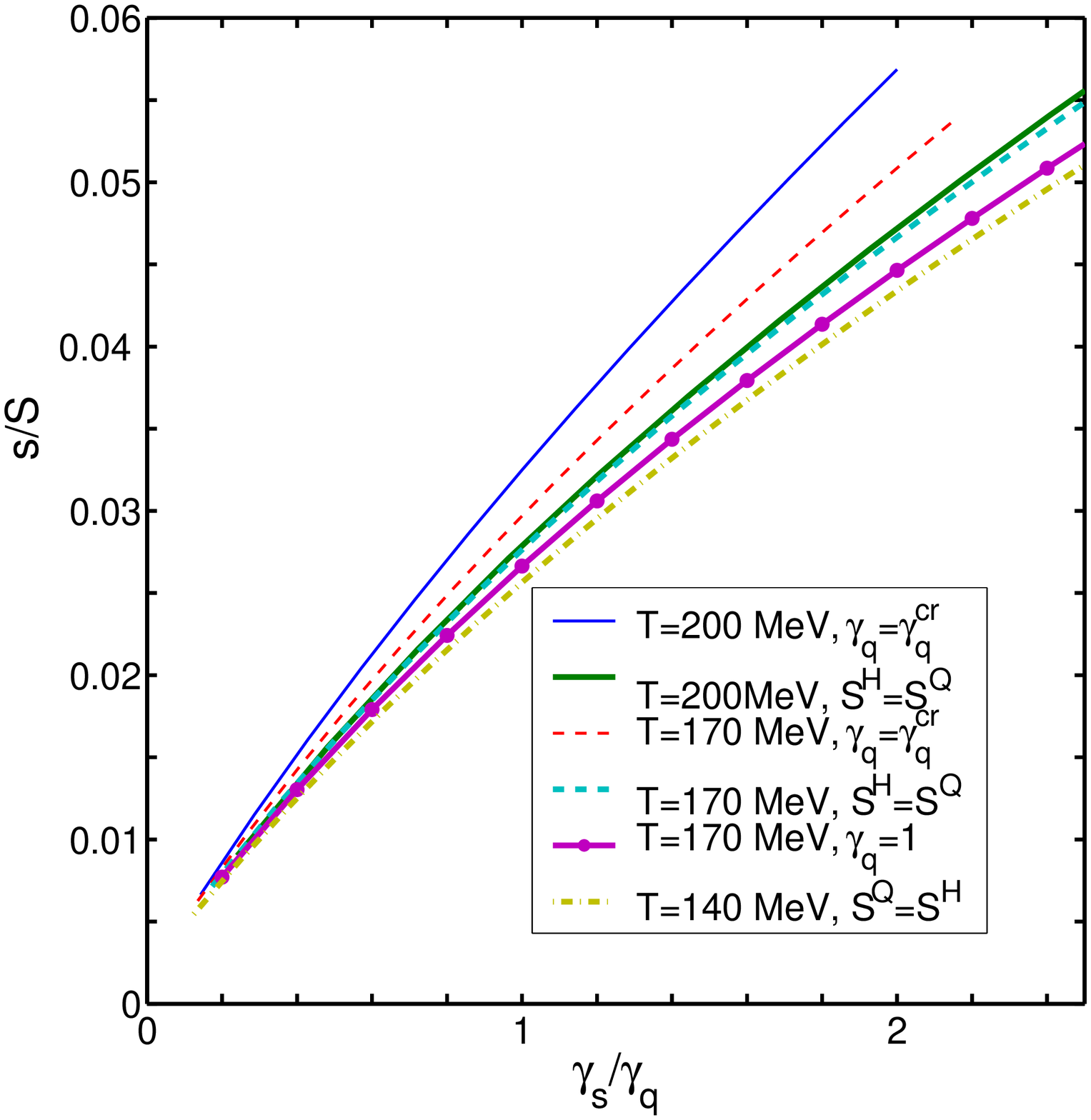}
\centering
\caption{(color on line) \small{Strangeness to entropy ratio, $s/S$,
as a function of $\gamma_s/\gamma_q$ for $T=200$ MeV,
$\gamma_q=0.083$, $S^H=S^Q$(solid line), $T=170$ MeV,
$\gamma_q=1.15$, $S^H=S^Q$ (dashed line), $T=140$ MeV,
$\gamma_q=1.6$, $S^H=S^Q$
(dash-dotted line); $\gamma_q = 1$ (dot
marked solid); $\gamma_q=\gamma_q^{cr}$: $T=200$ MeV (thin solid
line), $T=170$ MeV (thin dashed line).}}\label{sSrg}
\end{figure}

In figure~{\ref{sSrg}} we show  $s/S$ ratio as function of
$\gamma^{\rm H}_s/\gamma^{\rm H}_q$. The solid line is  for $T=200$
MeV, $\gamma^{\rm H}_q=0.83$, $S^Q=S^H$, dashed line for $T=170$
MeV, $\gamma^{\rm H}_q=1.15$  $S^Q=S^H$ and dash-dot line for
$T=140$ MeV, $\gamma^{\rm H}_q=1.6$ MeV, $S^Q=S^H$. We also consider
$\gamma_q=1$ case for $T=170$ MeV (dot marked (purple) solid line).
In this case strangeness content $\gamma_s/\gamma_q$ is higher than
for $S^H=S^Q$ with the same $T$ and $s/S$.  In the  limit
$\gamma^{\rm H}_q=\gamma_q^{cr}$, Eq.~\ref{bcon}, ($T=200$ MeV,
solid, thin line; $T=170$ MeV, dashed line) the strangeness content
$\gamma^{\rm H}_s/\gamma^{\rm H}_q$ is minimal for given $T$ and
$s/S$.

%%%%%%%%%%%%%%%%%%%%%%%%%%%%%%%%%%%%%%%%Fig 8  
%%%%%%%%%%%%%%%%%%%%%%%%%%

%%%%%%%%%%%%%%%%%%%%%%%%%%%%%%%%%%%%%%%%%%%%%%%%%%%%%%Fig 9
\begin{figure} %[!t]
\centering
\includegraphics[width=9cm,height=8cm]{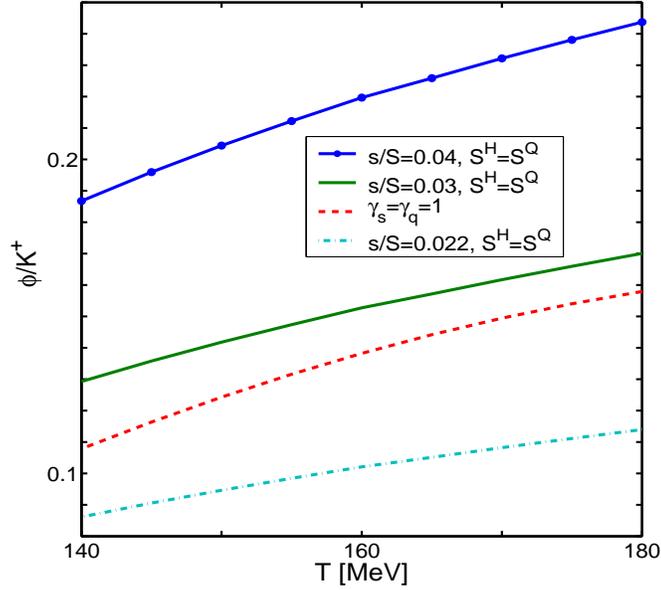}
\caption{(Color on line) The ratio  $\phi/{\rm K}^+$ as a function of $T$.
Dashed line (red) is for chemical equilibrium.
Solid line with dots (green)  $s/S=0.03$, solid line (blue)
$s/S=0.04$, dash-dot line (per)
is for $s/S=0.022$.} \label{phiK}
\end{figure}
%%%%%%%%%%%%%%%%%%%%%%%%

These results suggest that it is possible to measure the value of $s/S$ irrespective
of what the hadronization temperature may be, as long as the main yield dependence is
on the ratio ${\gamma^{\rm{H}}_s}/{\gamma^{\rm{H}}_q}$.
Indeed,  we find that the ratio $\phi/K^+$;
\begin{equation}
 { \phi\over K^+} =\frac{\gamma^{\rm{H}}_s}{\gamma^{\rm{H}}_q}
                   \frac{ n^{\rm eq}_{\phi}}{n^{\rm eq}_{K^+}},
\label{gr}
\end{equation}
is less sensitive to hadronization temperature compared to its
 strong dependence on the value of $s/S$.
In figure~\ref{phiK} we show the total hadron phase space
 ratio $\phi/K^+$ as function of $T$  for several
$s/S$ ratios, and for $\gamma^{\rm{H}}_{s,q}=1$ (chemical equilibrium, dashed (red) line).
The $K^+$ yield contains the contribution from
the decay of $\phi$ into kaons which is a noticeable correction.

We record in table \ref{s/S} for given $s/S$ and volume $dV/dy$
the corresponding total yields of strangeness,
which may be a useful guide in consideration of the consistency of experimental
results with what we find exploring heavy flavor hadron abundance.

%%%%%%%%%%%%%%%%%%%%%%%%%%%%%%%%%%%%%%%%%%%%%%%%%%%%%% Table III
\begin{table}
\centering
\caption{Specific and absolute strangeness yield for different reaction volumes at $T=200$ MeV.
 \label{s/S}}
\begin{tabular}{|c|c|c|c|}
  \hline
    $s/S$&$ds/dy$&$dV/dy$ $[\mathrm fm^{-3}]$&T [MeV]\\
  \hline
$0.045$ &$550$ &$1000$& $200$\\
$0.04$ & $360$ &$800$& $200$\\
$0.035$ &$250$ &$700$& $200$\\
$0.03$ & $165$ &$600$& $200$\\
$0.025$& $106$ &$500$& $200$\\
$0.022$& $83$ &$500$& $200$\\
\hline
\end{tabular}
\end{table}
%%%%%%%%%%%%%%%%%%%%%%%%%%%%%%%%%%%%%%%%%%%%%%%%%%%%%%%%%%%%%%%%

\subsection{Phase space occupancy $\gamma^{\mathrm{H}}_{c}$
and $\gamma^{\mathrm{H}}_{b}$}\label{gamcvalSec}
%%%%%%%%%%%%%%%%%%%%%%%%%%%%%%%%%%%%
 The first step in order to determine the  yields of heavy flavor
hadronic particles is the determination of the phase space occupancy
$\gamma^{\mathrm{H}}_c$ and $\gamma^{\mathrm{H}}_b$.
$\gamma^{\mathrm{H}}_c$ is obtained from equality of
number of these quarks (i.e. of  quark and anti quark  pairs) in QGP
and HG.  The yield constraint is:
\begin{equation}
\frac{dN_{c}}{dy}=\frac{dV}{dy}\left[{\gamma^{\mathrm{H}}_{c}}n^{c}_{\mathrm{op}}
+
\gamma^{\mathrm{H}\,2}_{c}(n^{c\,eq}_{\mathrm{hid}}
+
2\gamma^{\mathrm{H}}_qn^{\mathrm{eq}}_{ccq}
+
2\gamma^{\mathrm{H}}_sn^{\mathrm{eq}}_{ccs})\right]; \label{gammacb}
\end{equation}
where open `op'  charm yield is:
\begin{equation}
n^{c}_{\mathrm{op}}\!=\gamma^{\mathrm{H}}_qn^{\mathrm eq}_{D}+\gamma^{\mathrm H}_sn^{\mathrm eq}_{Ds}+
{\gamma^{\mathrm{H}\,2}_q}\,n^{\rm eq}_{qqc}+{\gamma^{\mathrm{H}}_s}{\gamma^{H}_q}n^{\mathrm eq}_{sqc}+
{\gamma^{\mathrm{H}\,2}_s}\,n^{\mathrm{eq}}_{ssc}.
\end{equation}
Here $n^{\mathrm{eq}}_{D}$ and $n^{\mathrm{eq}}_{Ds}$ are densities of
$D$ and $D_s$ mesons, respectively, in chemical equilibrium, $n^{\mathrm{eq}}_{qqc}$
is equilibrium density of baryons with one charm and two light quarks,
$n^{\mathrm{eq}}_{ssc}$ is density of baryons with one
charm (or later on one bottom quark) and two strange quarks ($\Omega^0_c$,
$\Omega^0_b$) in chemical equilibrium and $n^{\rm eq}_{\mathrm{hid}}$
 is equilibrium particle density with both, a charm (or bottom) and an anticharm
(or antibottom) quark (C=0, B=0, S=0).
The equilibrium densities can be calculated using Eq.(\ref{dist}).
$\gamma^{\mathrm{H}}_c$ can now  be obtained from Eq.(\ref{gammacb}).

Similar calculations can be done for $\gamma^{\mathrm{H}}_b$. The
only difference is that we need to add number of $B_c$ mesons to the
right hand side of Eq.(\ref{gammacb}),
\begin{equation}
\frac{dN_{Bc}}{dy}=\gamma^{\mathrm{H}}_b\gamma^{\mathrm{H}}_c n^{\mathrm{eq}}_{Bc}
\frac{dV}{dy}.
\end{equation}
$n^{\mathrm{eq}}_{Bc}$ is density in chemical equilibrium of $B_c$.  In the
calculation of $\gamma^{\mathrm{H}}_c$ the contribution of term with $n_{Bc}$
is very small and we did not consider it above.

The value of $\gamma_c$ is in essence controlled by the open single
charm mesons and baryons. For this reason we do not consider the
effect of exact charm conservation. The relatively small effects due
to canonical phase space of charm are leading to a slight
up-renormalization of the value of $\gamma_c$ so that the primary
$dN_c/dy$ yield is preserved. This effect   enters into the yields
of multi-charmed and hidden charm hadrons, where the compensation is
not exact and there remains slight change in these yields. However,
the error made considering the high yield of charm is not important.
On the other hand for multi-bottom and hidden bottom hadrons the
canonic effect can be large, depending on actual bottom yield, and
thus we will not discuss in this paper yields of these hadrons,
pending extension of the methods here developed to include canonical
phase space effect.  

%%%%%%%%%%%%%%%%%%%%%%%%%%%%%%%%%%%%%%%%%%%%%%%%%
%\begin{figure}[!hbt]
%\centering
%\includegraphics[width=8cm,height=14cm]{gcgq.eps}
%\caption{(Color on line)
%\small{$\gamma^\mathrm{H}_{c}\gamma^\mathrm{H}_q/N_c$
%ratio as a function of $\gamma^\mathrm{H}_s/\gamma^\mathrm{H}_q$ for
%T=0.14 GeV (blue, solid line), T=0.16 GeV (green, dashed line) and T=0.18 GeV
%(red,  dash-dot line). Results   for $\gamma^\mathrm{H}_q$
%are from Fig.{\ref{gs}}.}} \label{gcgq}
%\end{figure}
%%%%%%%%%%%%%%%%%%%%%%%%%%%%%%%%%%%%%%%%%

%In a good approximation, see Eq.\,(\ref{gammacb}),
%the value of $\gamma^{\mathrm{H}}_{c,b}$ scales with  the
%total yields $dN_{c,b}/dy$, except for immaterial corrections
%from the multi-heavy hadrons. Therefore,
%we show in figure~{\ref{gcgq}}  the relative dependence
%$\gamma^{\mathrm{H}}_b{\gamma^{\mathrm{H}}_q}/N_b$ (top) and
%$\gamma^{\mathrm{H}}_c{\gamma^{\mathrm{H}}_q}/N_c$ (bottom) on the chemical
%non-equilibrium ratio
%${\gamma^{\mathrm{H}}_s}/{\gamma^{\mathrm{H}}_q}$ for several temperatures
%(from top to bottom) $T=140,160,180$ MeV.  We note
%the monotonic decrease with
%${\gamma^{\mathrm{H}}_s}/{\gamma^{\mathrm{H}}_q}$.
%This is due to the fact that
%with an increase of ${\gamma^{\mathrm{H}}_s}/{\gamma^{\mathrm{H}}_q}$ there is more
%phase space for the available $c$ quarks to bind to. This decrease of the
%value of the heavy flavor fugacity with ${\gamma^{\mathrm{H}}_s}/{\gamma^{\mathrm{H}}_q}$
%influences significantly the heavy flavor hadron yields of interest to us.

\begin{figure}
\centering
\includegraphics[width=9cm,height=12cm]{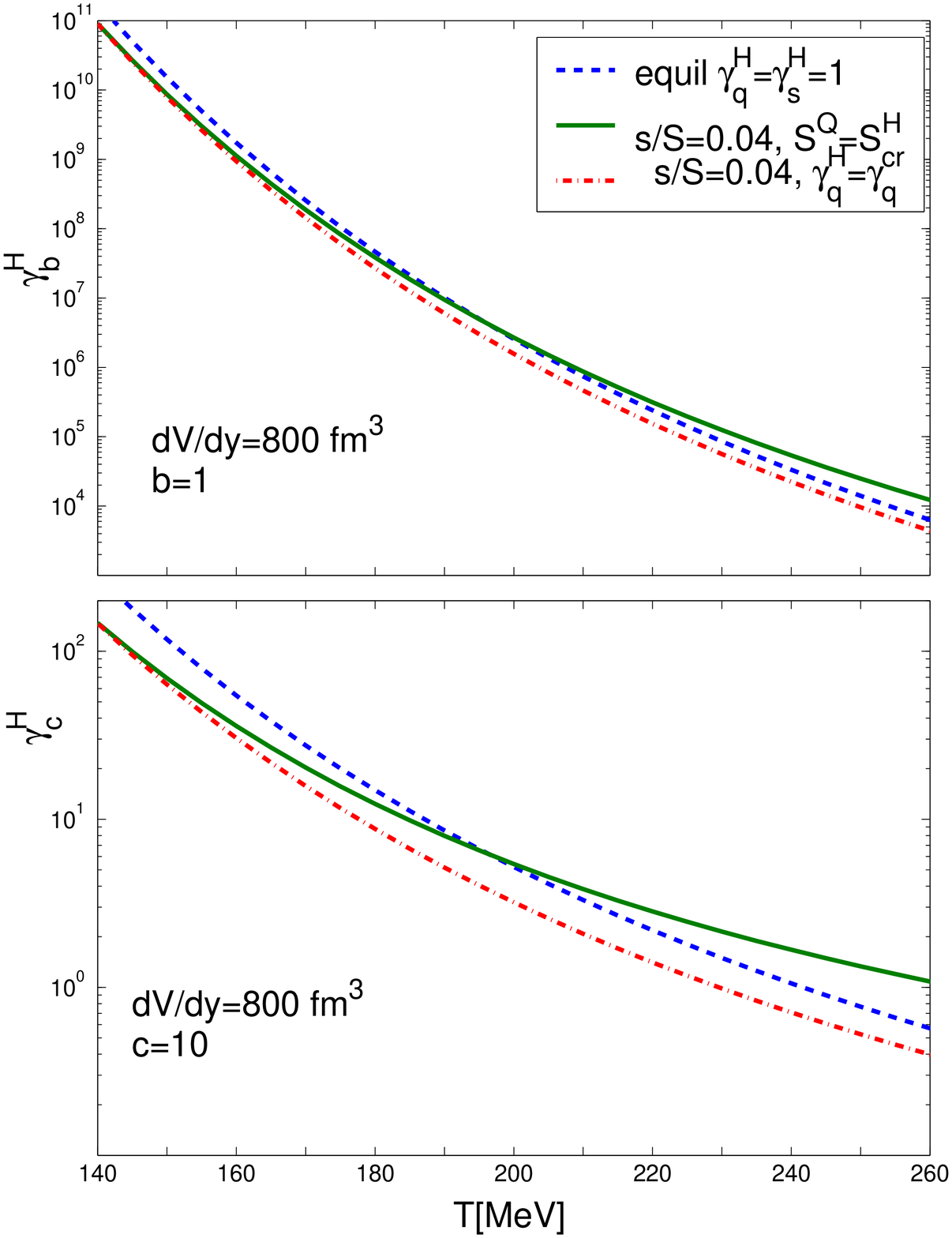}
\caption{(Color on line)
\small{$\gamma^{\mathrm{H}}_b (b=1)$ (upper panel),
and $\gamma^{\mathrm{H}}_c$ (c=10) (lower panel), as functions of
temperature of hadronization T. The solid lines are non-equilibrium for $s/S=0.04$ with $S^Q=S^H$,
dashed lines are
equilibrium case $\gamma_s$=$\gamma_q$=1 and dot-dash lines are for $s/S=0.04$ with maximal
value of $\gamma_q$($\gamma_q = \gamma^{cr}_q$) ($dV/dy=800$ $\mathrm{fm^3}$).}}
\label{gammaall}
\end{figure}

We consider in figure~\ref{gammaall} the temperature dependence of both
$\gamma^{\mathrm{H}}_b$ (top) and $\gamma^{\mathrm{H}}_c$ (bottom)
 for the heavy flavor yield given in Eqs.\, (\ref{nc},\ref{nb}).
In the non-equilibrium case (solid lines) the
space occupancy $\gamma^{\mathrm{H}}_s$ is obtained from
Eq.\,(\ref{gammas}) and $\gamma^{\mathrm{H}}_q$ is chosen to keep
$S^{\mathrm{H}}=S^{\mathrm{Q}}$.
$\gamma^{\mathrm{H}}_{c(b)}$
depend  on $\gamma_s$ and $\gamma_q$: the value of $N_s$ in Eq.(\ref{gammas})
is chosen to have $s/S=0.04$ after hadronization, the
corresponding $\gamma^{\rm H}_q$ and $\gamma^{\rm H}_s$ are shown
in figures~\ref{gq} and \ref{gseq} .
Since applicable  $\gamma^{\rm H}_q$ may depend on hadronization dynamics and/or details of  equation of state of
QGP, we show   charm quark phase space occupancies also for maximum possible
value of $\gamma^{\rm H}_q\to \gamma_q^{cr}$), also considered at $s/S=0.04$  for all hadronization temperatures.
We can compare our results with the chemical equilibrium (dashed lines)
setting $\gamma^{\mathrm{H}}_s = \gamma^{\mathrm{H}}_q=1$ in
Eq.\,(\ref{gammacb}). At hadronization condition $T=160\pm20$ MeV
temperatures we see in figure~\ref{gammaall} a significant (considering
the fast changing logarithmic scale) difference between the chemical equilibrium,
and non-equilibrium (s/S=0.04) results.

%%%%%%%%%%%%%%%%%%%%%%%%%%%%%%%%%%%%% Figure 10
%%%%%%%%%%%%%%%%%%%%%%%%%%

%%%%%%%%%%%%%%%%%%%%%%%%%%%%%%%%%%%%% Figure 11
\begin{figure}
\centering
\includegraphics[width=8cm,height=8 cm]{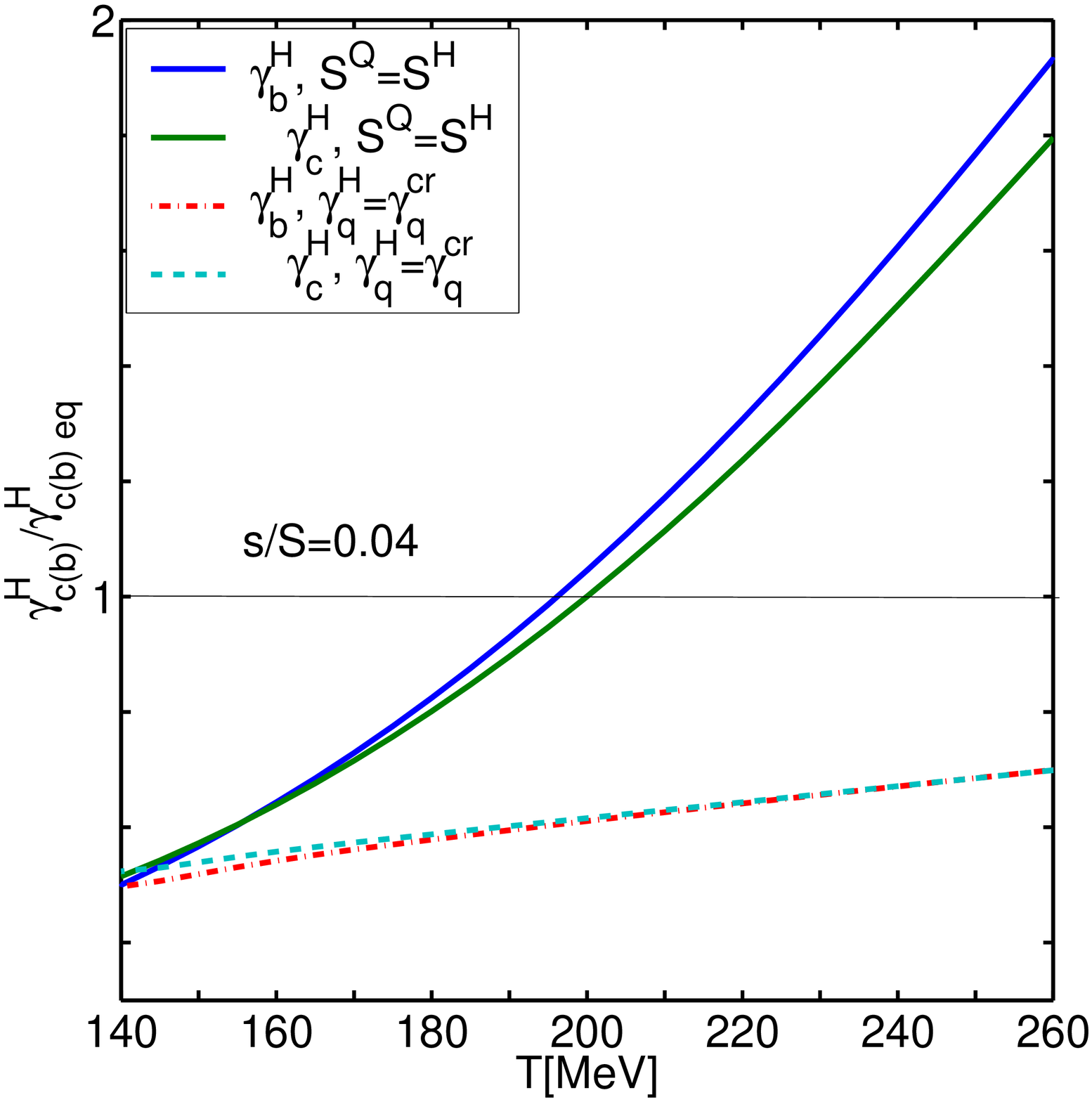}
\caption{(Color on line)
\small{$\gamma^{\mathrm{H}}_b/\gamma^{\mathrm{H}}_{b\,eq}$
and $\gamma^{\mathrm{H}}_c/\gamma^{\mathrm{H}}_{c\,eq}$, as functions of
temperature of heavy flavor hadronization T. The solid with dot marks line is
for $\gamma^{\mathrm{H}}_b/\gamma^{\mathrm{H}}_{b\,eq}$ with
$s/S=0.04$, solid line is for $\gamma^{\mathrm{H}}_c$ with $s/S=0.04$,
dot-dash and dashed lines are for $s/S=0.04$ with maximal value of
$\gamma_q\to \gamma^{cr}_q$  for $\gamma^{\mathrm{H}}_b/\gamma^{\mathrm{H}}_{b\,eq}$
and for $\gamma^{\mathrm{H}}_c/\gamma^{\mathrm{H}}_{c\,eq}$, respectively.}}
\label{rg}
\end{figure}
%%%%%%%%%%%%%%%%%%%%%%%%%%

In figure~\ref{rg} we show  the ratio $\gamma^{\rm H}_{c(b)}/{\gamma^{\rm H}_{c(b)\,eq}}$ as a
function of hadronization temperature $T$. This helps us understand when the
presence of chemical nonequilibrium is most noticeable. This is especially
the case should heavy flavor hadronization occur at the same  temperature $T=140$--170 MeV
as is  obtained  for non-heavy  hadrons, and/or when the entropy content of light hadrons
is maximized with $\gamma_q^{\rm H}\to \gamma_q^{\rm cr}$.  When no additional entropy
is formed in hadronization, that is   $S^H=S^Q$,  $\gamma^{\rm H}_{c(b)}/{\gamma^{\rm H}_{c(b)\,eq}}$
exceeds unity for  $T>200$ MeV, at which point   the heavy flavor hadron yields
exceed the chemical equilibrium expectations.  In general we find that heavy hadron yields
if produced at normal hadronization temperature would be effectively suppressed, compared to
statistical equilibrium results, by the high strangeness yield. This happens since the phase space is
bigger at $\gamma^{\rm H}_{s,q}>1$, and thus a smaller $\gamma^{\rm H}_{c,b}$ is required to reach a given heavy flavor yield.

$\gamma^{\mathrm{H}}_b$ and $\gamma^{\mathrm{H}}_c$ are
nearly proportional to  $dN_{b,c}/dy$, respectively. The deviation
from the proportionality is due to the abundance of multi-heavy hadrons
and it is small. To estimate this effect more quantitatively we first evaluate:
\begin{equation}
\gamma^{\mathrm{H}}_{c0} =
\frac{dN_{c}}{dy}/\left(\frac{dV}{dy}n^c_{\mathrm{open}}\right),
\label{linsol}
\end{equation}
{\it i.e.\/} the value expected in absence of multi heavy hadrons.
Next we compare with the result when  we take into account
the last three terms in Eq.\,(\ref{gammacb}).
The influence of these therms depend not only
on $dN_c/dy$ but also on $dN_c/dy/dV/dy$.
For fixed $dV/dy=800$ $\mathrm{fm^{-3}}$ in the range of
$dN_c/dy=(5,30)$, we find that $\gamma^{\mathrm{H}}_c/N_c$
(and therefore yields of open charm hadrons)
changes at temperature $T=140$ MeV  by $\sim 6 \%$ for the $s/S=0.04$. For the chemical
equilibrium case $\gamma_s=\gamma_q=1$, $\gamma^{\mathrm{H}}_c/N_c$
changes up to $15 \%$ at the same conditions.
For the particles with hidden charm or 2 charm quarks the yields
are proportional $\gamma_i^2$,  therefore  changes in their yields will be
about twice larger. For RHIC $N_c<3$ and $dV/dy=600$ $\mathrm{fm^{-3}}$
the dependence of yields on $N_c$ is much smaller.

The multiplicity $dN_c/dy$ can also influence   $\gamma^{\mathrm{H}}_b$, since as
we noted it also includes a term
proportional to $\gamma^{\mathrm{H}}_cn^{\mathrm{eq}}_{Bc}$. In the
range of $N_c=(5,30)$, $\gamma_b/N_b$ changes at temperature
$T=0.14$ MeV by $\sim 0.5 \%$ for $s/S=0.04$. Since the mass of $b$-quark is much
larger than that of $c$-quark, the effect due to multi-bottom
states is negligible.

%%%%%%%%%%%%%%%%%%%%%%%%%%%%%%%%%%%%%%%%%%%%%%%%%%%%%%%%%%%%%%%%%%%%%%%%%%%%%%%%%%%%%%%%%%%%%%%%%%%%%%%%%%%%%%%%%%%%%%%%%%%%%%
\chapter{HEAVY FLAVOR HADRONS IN STATISTICAL HADRONIZATION OF STRANGENESS AND ENTROPY RICH QGP}\label{hfhad}
%%%%%%%%%%%%%%%%%%%%%%%%%%%%%%%%%%%%%%%%%%%%%%%%%%%%%%%%%%%%%%%%%%%%%%%%%%%%%%%%%%%%%%%%%%%%%%%%%%%%%%%%%%%%%%%%%%%%%%%%%%%%
\section{Introduction}
A relatively large number of hadrons containing charmed  and bottom
quarks  are expected to be produced in heavy ion (AA) collisions at
the Large Hadrons Collider (LHC).
Because of their large mass $c,\bar c, b, \bar b$ quarks are produced
predominantly in primary
parton-parton collisions~\cite{Geiger:1993py},
at RHIC~\cite{Cacciari:2005rk}, and thus
even more so at LHC. These heavy flavor
quarks participate in the
evolution of the dense QCD matter from the beginning. In view of
the recent  RHIC results it can be
hoped that their momentum distribution could reach approximate
thermalization within the dense QGP phase~\cite{vanHees:2004gq}.

In the calculations in this chapter we assume the same  evolution stages as present in the beginning of dissertation 
introduction, except the kinetic. Here we assume that this phase does not have influence on charm and bottom hadrons.

It is important to observe that in the presence of deconfined
QGP phase heavy hadrons containing more than one heavy quark are made from
heavy quarks created in different initial NN collisions. Therefore yields of these
hadrons are expected to be  enhanced as compared  to yields seen in single
NN collisions~\cite{Schroedter:2000ek, Becattini:2005hb}.
We note that the Bc($b\bar c,\bar b c$) and $J/\Psi (c\bar c)$
and more generally all bound
$c\bar{c}$ states yields were calculated before in the kinetic
formation and dissociation models~\cite{Schroedter:2000ek, Thews:2005fs}.
Our present work suggests that it is important to account for the binding
of heavy flavor with strangeness, an effect which depletes the eligible
supply of heavy flavor quarks which could form
Bc($b\bar c,\bar b c$) and $J/\Psi (c\bar c)$~\cite{Kuznetsova:2006hx}.

Enhanced production yield of multi-heavy hadrons can be considered to be
an indicator of the presence of deconfined QGP phase for reasons
which are analogue to those of multi-strange (anti) baryons~\cite{Koch:1986ud}.
Considering that we have little doubt that QGP is the state of matter formed in
the very high energy AA interactions, the study of yields of multi-heavy
hadrons is primarily explored in this work in order to falsify, or  justify,
features of the statistical hadronization model (SHM) employed or the model
itself in the context of formation of the heavy flavor hadrons.

For example, differing from others recent studies which assume    that
the hadron yields after hadronization
are in chemical equilibrium~\cite{Becattini:2005hb,Andronic:2003zv},
we form the yields based on abundance of $u,d,s$ quark pairs
as these are  available
at the chemical freeze-out (particle formation) conditions in
the quark-gluon phase. This approach is justified by the
expectation that in a fast break-up
of the QGP formed at RHIC and LHC the
phase entropy and strangeness will be nearly   conserved
during the process of hadronization.
We will investigate in quantitative terms
how such  chemical non-equilibrium yields, in the conditions we explore  well above the
chemical equilibrium abundance,  influence
the expected yields of  single, and multi-heavy flavor hadrons.

In the order
to evaluate the yields of final state hadrons we enforce
conservation of entropy, and the flavor $s,c,b$ quark pair number
during phase transition or transformation. The faster the
transition, the less likely it is that there is significant
change in strange quark pair yield. Similarly, any entropy
production is minimized when the entropy rich QGP breakup
into the entropy poor HG occurs  rapidly. The entropy conservation
constraint fixes the final light quark yield. We assume a fast
transition between QGP and HG phases, such that all hadron yields
are at the same physical conditions as in QGP breakup.

In the evaluation of heavy particle yields we form ratios involving
 as normalizer the total heavy flavor yield, and for yields
of particles with two heavy quarks we use as normalizer the product
of total yields of corresponding heavy flavors  such that the
results we consider is as little  as possible dependent on the unknown total yield
of charm and bottom at RHIC and LHC.  The order of magnitude  of the remaining
 dependence on heavy flavor yield
is set by the  ratio of yield of all particles with two heavy quarks to
yield of particles with one heavy quark. This  ratio depends  on the density
of heavy flavor at hadronization,  $(dN_c/dy)/(dV/dy)$.
The results we present for LHC are obtained for an assumed
charm and bottom quark multiplicity:
\begin{eqnarray}
 {\frac{dN_c}{dy}}\equiv c&=&10  , \label{nc}\\
 {\frac{dN_b}{dy}}\equiv b&=&\ 1  .  \label{nb}
\end{eqnarray}
and $dV/dy=800$ $\mathrm{fm}^3$ at $T=200$ MeV. Theoretical
cross sections of $c$ and $b$ quarks production for RHIC and LHC can
be found in~\cite{Bedjidian, Anikeev:2001rk}. In certain situations
we will explore how variation of the baseline yields Eq.\,(\ref{nc})
and  Eq.\,(\ref{nb}) impact the results. In particular among the
yields of multi-heavy hadrons, this influence  can be noticeable,
see discussion in the end of section \ref{gamcvalSec}. We note that
the number of $b$ quarks can not change during expansion, because of
large mass $m_b>>T$. It is nearly certain that all charm in QGP at
RHIC is produced in the first parton collisions, for further
discussion of LHC see Ref.\cite{Letessier:2006wn} -- it appears that
for all practical purposes also in the more extreme thermal
conditions at LHC charm is produced in the initial parton
interactions.

In order to form
physical intuition about the prevailing conditions in the QGP phase at time of
hadronization, we also evaluate the  heavy quark  chemical reference density,
that is the magnitude of the chemical occupancy factor in QGP, considering the
pre-established initial  yields of $c$ and $b$ from parton collision.
For this purpose we use  in the deconfined QGP phase:
\begin{eqnarray}
m_c &=& 1.2\ \ \mathrm{GeV},  \nonumber\\
m_b &=& 4.2.\ \ \mathrm{GeV}  \nonumber
\end{eqnarray}
We also take  ${\lambda}_i=1, i=u,d,s$ for all light flavors,
since the deviation from particle-antiparticle
yield symmetry is rather small and immaterial in the present discussion.

When computing the yields of charmed (and bottom) mesons we will distinguish
only strange and non-strange abundances, but not charged with non-charged
(e.g. $\mathrm{D}^-(\bar c d)$ with $\mathrm{D}^0(\bar c u)$). We assume that
the experimental groups reporting results, depending on which types of D-meson
were observed, can infer the total yield (charged+non-charged) which we present.
We treat in similar way other heavy hadrons, always focusing on the heavy and
the strange  flavor content and not distinguishing the light flavor content.

This chapter is organized as follows: we  use the  elements of the SHM model introduced in section 
\ref{sthad} to evaluate heavy flavor hadron
yields.
This allows us to discuss the relative
yields of strange and non-strange heavy mesons in section \ref{RelCharSec},
and we show how this result relates the value of the strangeness chemical (non-)equilibrium
parameters. In this context, we also propose a multi-particle ratio as
a measure of the hadronization temperature, and explore how a multi-temperature,
staged, freeze-out would impact the relevant results.

We  turn to discuss the heavy flavor hadron yields
for given bulk QGP constraints in section \ref{heavyFlSec}, where we
also  compare when
appropriate to the strangeness and light quarks
chemical equilibrium results. We use the
charm and bottom quark phase space occupancy parameters
(subsection \ref{gamcvalSec}) and turn in subsection \ref{cbMesYielSec}
to discussion of the yields of single heavy mesons, which we follow
with discussion of yields of single heavy baryons in subsection \ref{BarYieSec}.
In last subsection \ref{MultiSec} we present the expected yields of the multi-heavy hadrons, in so
far these can be considered in the grand canonical  approach.
We conclude our work with a brief summary in section \ref{concSec}.

\section{Relative Charmed  Hadron Yeilds}\label{RelCharSec}
\subsection{Determination of  $\gamma_s/\gamma_q$}\label{RelGamSec}
%%%%%%%%%%%%%%%%%%%%%%%%%%%%%%%%%%%%%%%%%%%%%%%%%%%%%%%%%%%%%%%%%%%%
We have seen  considering $s/S$ and also $s$ and $S$ individually
across the phase limit that in general one would expect
chemical non-equilibrium in hadronization
of chemically equilibrated QGP.
We first show that this result
matters for the relative charm meson yield ratio $D/D_s$, where
$D_s(c\bar{s})$ comprises all mesons of type $(c\bar{s})$ which
 are   listed in the bottom section of table \ref{openbc}, and
 $D(c\bar{q})$ comprise yields of all $(c\bar{q})$states  listed in the top
section of table~\ref{openbc}. This ratio is formed
based on the assumption that on the time scale of strong interactions
the family of strange-charmed mesons can  be distinguished
from the family non-strange charmed mesons.

%%%%%%%%%%%%%%%%%%%%%%%%%%%%%%%%%%%%%%%%%%% Table I
\begin{table}
\centering
\caption{Open charm, and bottom, hadron states we considered. States
in parenthesis either need confirmation or have not been observed experimentally,
in which case  we follow the values of Refs.\,\cite{Cheu:2004zc,Matsuki:1997da}.
We have charm-bottom symmetry required for certain observables.
\label{openbc}}
\begin{tabular}{|c|c|c|c|c|c|c|c|}
  \hline
    &hadron&  &M[GeV]&hadron&  &M[GeV]&g\\
  \hline
   &$D^0(0^-)$&$c\bar{u}$ & 1.8646&$B^0(0^-)$&$b\bar{u}$&5.279&1\\
   &$D^+(0^-)$ &$c\bar{d}$ & 1.8694&$B^+(0^-)$&$b\bar{d}$&5.279&1\\
   &$D^{*0}(1^-)$&$c\bar{u}$&2.0067&$B^{*0}(1^-)$&$b\bar{u}$&5.325&3\\
   &$D^{*+}(1^-)$&$c\bar{d}$&2.0100&$B^{*+}(1^-)$&$b\bar{d}$&5.325&3\\
  $D$&$D^0(0^+)$ &$c\bar{u}$&2.352&$B^0(0^+)$&$b\bar{u}$&5.697&1\\
   &$D^+(0^+)$&$c\bar{d}$&2.403&$B^+(0^+)$&$b\bar{d}$&5.697&1\\
   &$D^{*0}_1(1^+)$&$c\bar{u}$&2.4222&$B_1^{*0}(1^+)$&$b\bar{u}$&5.720&3\\
   &$D^{*+}_1(1^+)$&$c\bar{d}$&2.4222&$B_1^{*+}(1^+)$&$b\bar{d}$&5.720&3\\
   &$D^{*0}_2(2^+)$&$c\bar{u}$&2.4589&$B^{*0}_2(2^-)$&$b\bar{u}$&(5.730)&5\\
   &$D^{*+}_2(2^+)$&$c\bar{d}$&2.4590&$B^{*+}_2(2^+)$&$b\bar{d}$&(5.730)&5\\
  \hline
   &$D_s^+(0^-)$&$c\bar{s}$&1.9868&${B_s^0}(0^-)$&$s\bar{b}$&5.3696&1\\
   &$D^{*+}_s(1^-)$&$c\bar{s}$&2.112&${B^{*0}_s}(1^-)$&$s\bar{b}$&5.416&3\\
  $D_s$&$D^{*+}_{sJ}(0^+)$&$c\bar{s}$&2.317&${B^{*0}_{sJ}}(0^+)$&$s\bar{b}$&(5.716)&1\\
   &$D^{*+}_{sJ}(1^+)$&$c\bar{s}$&2.4593&${B^{*0}_{sJ}}(1^+)$&$s\bar{b}$&(5.760)&3\\
   &$D^{*+}_{sJ}(2^+)$&$c\bar{s}$&2.573&${B^{*0}_{sJ}}(2^+)$&$s\bar{b}$&(5.850)&5\\
 \hline
\end{tabular}
\end{table}
%%%%%%%%%%%%%%%%%%%%%%%%%%%%%%%%%%%%%%%%%%%%%%%%%%%%%%%%%%

%%%%%%%%%%%%%%%%%%%%%%%%%%%%%%%%%%%%%%%%%%%%%%%Fig 1
\begin{figure}
\centering
\includegraphics[width=8cm,height=8cm]{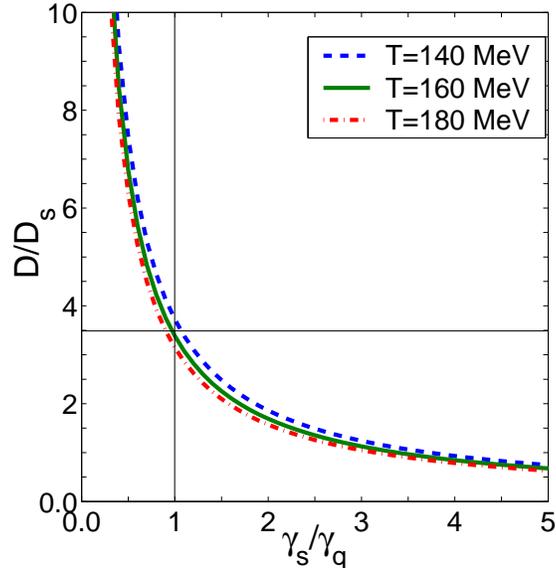}
\caption{(Color on line) \small{$D/D_s$ ratio as a function of
$\gamma^\mathrm{H}_s/\gamma^\mathrm{H}_q$ for $T = 140$ MeV (blue,
dashed line), $T = 160$ MeV (green, solid line) and $T= 180$ MeV
(red, dash-dot line)}. } \label{rDsDg}
\end{figure}
%%%%%%%%%%%%%%%%%%%%%%%%

%%%%%%%%%%%%%%%%%%%%%%%%%%%%%%%%%%%%%%%%%%%%%%%%%%%%%%Fig 2
\begin{figure}[!hbt]
\centering
\includegraphics[width=8cm,height=8cm]{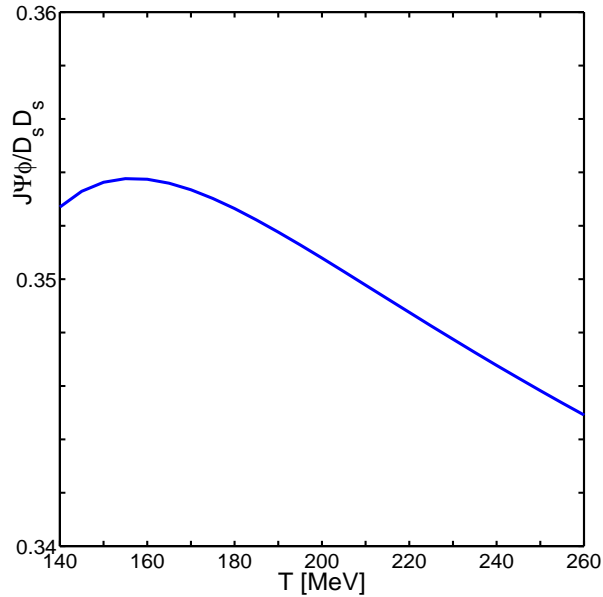}
\caption{(Color on line)
\small{$J\!/\!\Psi\,\phi/D_s{\overline{D_s}}$ ratio  as a function of hadronization temperature
T.}} \label{JpsiD}
\end{figure}
%%%%%%%%%%%%%%%%%%%%%%%%

The yield ratio $D/D_s$ calculated using Eq.\,(\ref{dist}) and
Eq.\,(\ref{BolzDis}) is shown in figure~\ref{rDsDg}. Using
Eq.\,(\ref{dist1})  we see that this ratio is inverse proportional to
${\gamma^{\rm H}_s}/{\gamma^{\rm H}_q}$  and weakly dependent on
$T$:
\begin{equation}
\frac{D}{D_s}\approx\frac{\gamma^{\rm H}_q}{\gamma^{\rm{H}}_s}\frac{\Sigma_i
g_{Dsi}m_{Dsi}^{3/2}\exp(-m_{Dsi}/T)}{\Sigma_i
g_{Di}m_{Di}^{3/2}\exp(-m_{Di}/T)}=f(T)\frac{\gamma^{\rm H}_q}{\gamma^{\rm H}_s}.
\end{equation}
A deviation of
${\gamma_s}/{\gamma_q}$
from unity in the range we will see in section \ref{noneqSec}
leads to a noticeable difference in the ratio $D/D_s$.
We show in figure~\ref{rDsDg} results for $T=140, 160, 180$ MeV.
In this temperature range the effect due to
${\gamma^{\rm H}_s}/{\gamma^{rm H}_q}\ne 1$
is the dominant contribution to the variation of this relative yield.

\subsection{Check of statistical hadronization model}\label{singHadSec}
%%%%%%%%%%%%%%%%%%%%%%%%%%%%%%%%%%%%%%%%%%%%%%%%%%%%%%%%%%%%%%%%%%%%
We next   construct a heavy flavor particle ratio that depends on
hadronization temperature  only. To cancel the fugacities and  the
volume we consider the ratio $J\!/\!{\Psi}\phi/D_s\overline{D_s}$ in
figure~{\ref{JpsiD}}. Here $J\!/\!\Psi$ yield  includes the  yield
of $(c\bar c)$ mesons decaying into the $J\!/\!\Psi$.  All phase
space occupancies cancel since $J/\Psi \propto
\gamma^{\mathrm{H}\,2}_c$, $\phi \propto \gamma^{\rm H\,2}_s$, $D_s
\propto \gamma^{\mathrm{H}}_c\gamma^{\rm H}_s$ and similarly
$\overline{Ds}\propto \gamma^{\mathrm{H}}_c\gamma^{\rm H}_s$. When
using here the particle $D_s(\bar c s)$ and antiparticle
$\overline{D_s}(c \bar s)$    any chemical potentials present are
canceled as well. However, for the LHC and even RHIC environments
this refinement is immaterial.

\begin{figure}
\centering
\includegraphics[width=9cm,height=9cm]{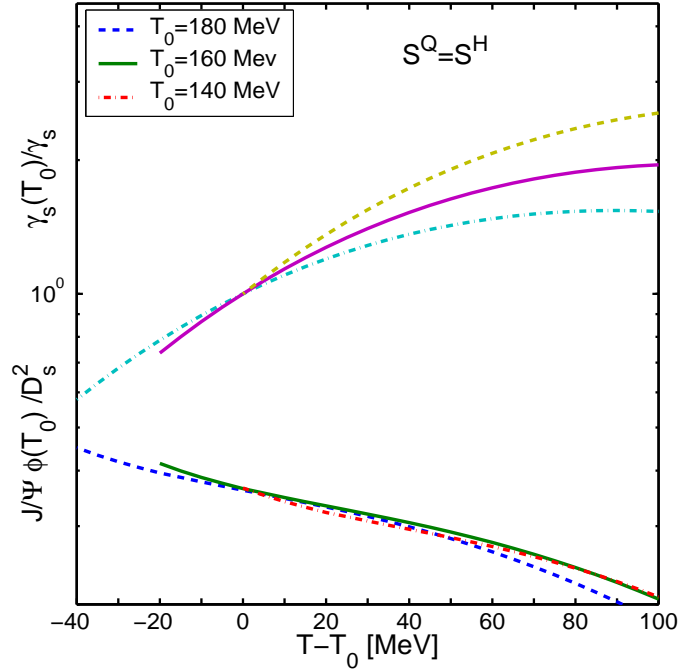}
\caption{(Color on line) \small{$J\!/\!\Psi\,\phi(T_0)/D_sD_s$ ratio
is evaluated at two temperatures, $T $ for  heavy flavor hadrons,
and $T_0$ for  $\phi$ as a function of $T-T_0$, with three values of
$T_0=140,160,180$ MeV is considered  with $S^H=S^Q$. }}
\label{JpsiD2T}
\end{figure}

This ratio  $J\!/\!{\Psi}\phi/D_s\overline{D_s}$, turns out to be
practically constant, within a rather wide  range of  hadronization
temperature $T$, see figure~{\ref{JpsiD}}.
The  temperature range we study $140<T<280$ MeV allows us to consider
an early freeze-out of different hadrons. To be sure
of the temperature independence  of $J\!/\!{\Psi}\phi/D_s\overline{D_s}$
we next consider the possibility that
hadronization temperature $T$ of charmed hadrons is higher than
hadronization temperature $T_0$ of $\phi$. We study this
question by exploring the sensitivity of the ratio
 $J\!/\!{\Psi}\phi/D_s\overline{D_s}$ to the two temperature
freeze-out in figure \ref{JpsiD2T}, see bottom three
lines for $T_0=180,160,140$ MeV  with $\gamma_q$ from condition $S^Q=S^H$, see figure~\ref{gq}.  If charmed hadrons hadronize later, $T>T_0$, and $T-T_0<60$
the change in $J\!/\!{\Psi}\phi/D_s\overline{D_s}$ ratio is small (about 20\%).
If this were to be
measured as experimental result,
\begin{equation}
{\frac{J\!/\!{\Psi}\phi} {D_s\overline{D_s}}}\simeq 0.35,
\end{equation}
one could not but conclude that
all particles involved are formed by mechanism of statistical hadronization.

%%%%%%%%%%%%%%%%%%%%%%%%%%%%%%%%%%%%%%%%%%%%%%%%%%%%%%Fig 3
%%%%%%%%%%%%%%%%%%%%%%%%

This interesting result can be understood, considering the behavior
of the $\gamma^{\rm H}_s(T_0)/\gamma^{\rm H}_s(T)$ ratio, which
increases rapidly  with increasing $T-T_0$ (see the top three lines
in figure \ref{JpsiD2T}). This ratio almost compensates the change
in $\phi$-yield, an effect we already encountered in the context of
the results we show below in figure \ref{phiK}. For large
$T-T_0$ the ratio ${J\!/\!{\Psi}\phi/ D_s\overline{D_s}}$ begins to
decrease more rapidly because $\gamma_s$ increases for $S^H=S^Q$,
see figure~(\ref{gq}).

%. and $J\!/\!{\Psi}\phi(T_0)/D_s\overline{D_s} \propto
%(n^{\rm eq}_{\phi}(T_0)\gamma_s(T_0)/(n^{\rm eq}_{\phi}(T)\gamma_s(T)))^2$
%if consider $J\!/\!{\Psi}\phi/D_s\overline{D_s}(T)=$const. From this
%figure we can conclude that $\phi(T)$, especially for $T < 200$, is
%almost constant. That is because $\gamma_s$ is approximately
%proportional to $1/n^{\rm eq}_K$ and $m_{\phi} \approx 2m_K$

%%%%%%%%%%%%%%%%%%%%%%%%%%%%%%%%%%%%%%%%%%%%%%%%%%%%%%%%%%%%%%%
\section{Yields of heavy flavored hadrons}\label{heavyFlSec}
%%%%%%%%%%%%%%%%%%%%%%%%%%%%%%%%%%%%%%%%%%%%%%%%%%%%%%%%%%%%%%%

%%%%%%%%%%%%%%%%%%%%%%%%%%%%%%%%%%%%%%%%%%%%%%%%%%%%%%%%%%%%%%%
\subsection{D, Ds, B, Bs meson yields}\label{cbMesYielSec}
\begin{figure}
\centering
\includegraphics[width=9cm,height=12cm]{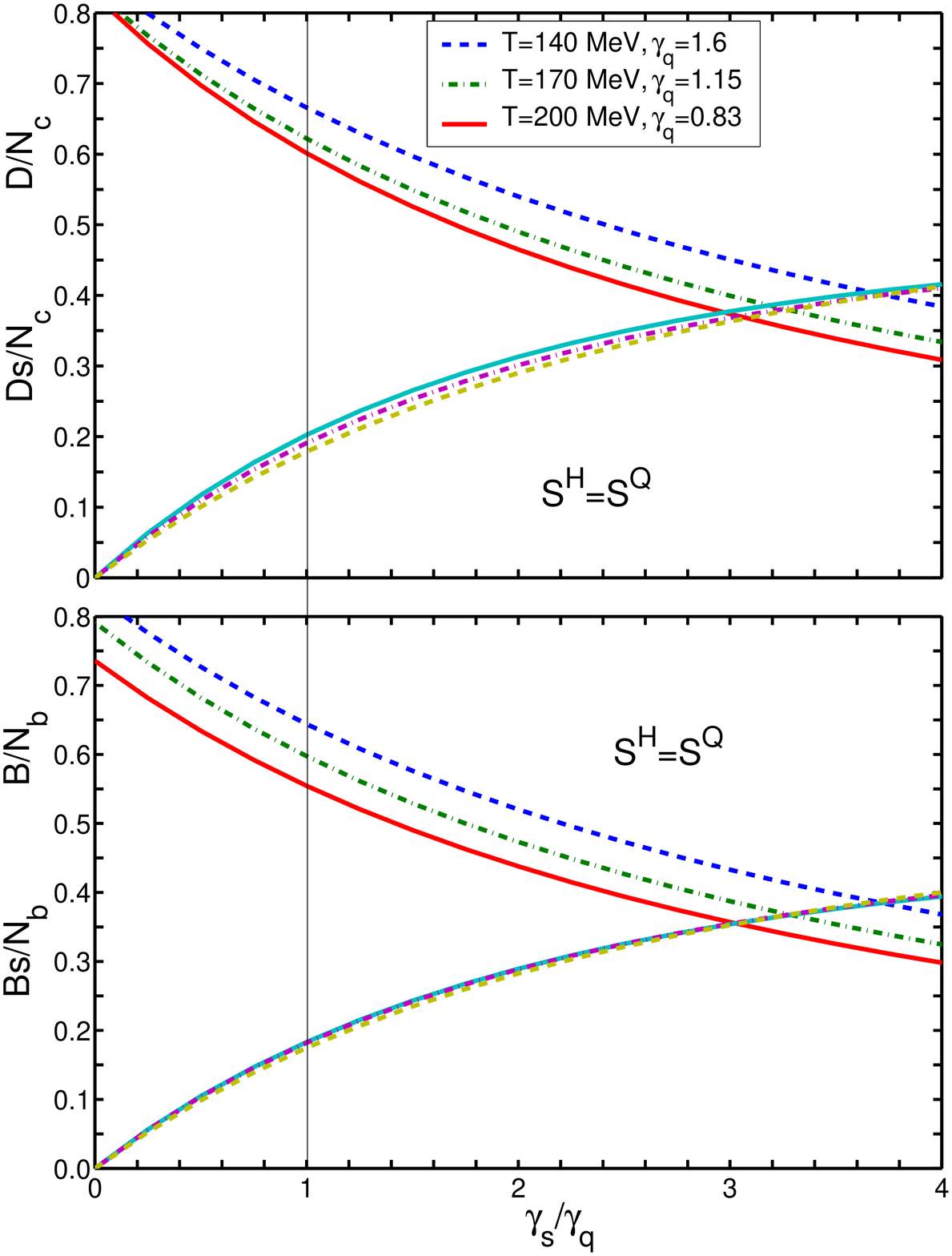}
\caption{(Color on line) \small{Upper panel, fractional charm meson
yield, and lower panel, fractional bottom meson yields as a function
of $\gamma_s/\gamma_q$ ratio for fixed hadronization temperature
$T$. Upper lines in each panel are for D(B) mesons, solid line is
for $T=200\,MeV$, $\gamma_q=1.1$, dashed line is for $T=170$ MeV,
$\gamma_q=1.15$ and dash-dot line is for $T=140$ MeV, $\gamma_q=0.83$ $(S^H=S^Q)$.
}} \label{mesrg}
\end{figure}

%%%%%%%%%%%%%%%%%%%%%%%%%%%%%%%%%%%%%%%%%%%%%%%%%%%%%%%%%%%%%%%
In next sections we will mostly consider particles yields after
hadronization and we will omit superscript H in $\gamma$s.
Considering Eq.\,(\ref{dist}),
 we   first obtain $\gamma_c$
as a function of $\gamma_s/\gamma_q$ ratio and T. Substituting this
$\gamma_c$ and appropriate  equilibrium hadron densities into
Eq.\,(\ref{dist}) we  obtain yields of $D(B)$ and $D_s(B_s)$, as
functions of the $\gamma_s/\gamma_q$ ratio, at fixed temperature,
which are shown on figure~\ref{mesrg}. In   the upper panel we show
the fractional yields of charmed $D/N_c$ and $D_s/N_c$ mesons, and
in the lower panel  $B/N_b$ and $B_s/N_b$ for $T=200$ MeV
(solid line), $T=170$ MeV (dashed line), $T=140$ MeV (dash-dot
line). Fractional yield means that these yields are normalized by
the total number of charm quarks $N_c$ and, respectively bottom
quarks $N_b$, and thus tell us how big a fraction of available heavy
flavor quarks binds to non-strange and strange heavy mesons,
respectively. Using figure~\ref{sSrg}  the  ratio
$\gamma_s/\gamma_q$ can be  related to the $s/S$ ratio. $\gamma_q$
was chosen to conserve entropy during hadronization process, see
figure~\ref{gq}. In general the heavy non-strange mesons yield
decreases and strange heavy meson yield increases with
$\gamma_s/\gamma_q$. The yields $D,B$ and $D_s,B_s$ are sum over
exited states of $D,B$ and $D_s,B_s$ respectively, see table
\ref{openbc} for the `vertical tower' of resonances we have
included.

%%%%%%%%%%%%%%%%%%%%%%%%%%%%%%%%%%%%%%%%%%%%%%%%%%%%%%Figure 12
%%%%%%%%%%%%%%%%%%%%%%%%%%%%%%%%%%%%%%%%%%%%%%%%%%%%%%%%%%%%%%%

Using $\gamma_c$, $\gamma_s$, $\gamma_q$ at a given $T$
(see figures~\ref{gq}, \ref{gseq}, \ref{gammaall}) we have now all the inputs
required to compute absolute and relative particle yields of all
heavy hadrons which we can consider  within the grand canonical
phase space. When we consider chemical equilibrium case, we  use
naturally $\gamma_s=\gamma_q=1$.

\begin{figure}
\centering
\includegraphics[width=9cm,height=12cm]{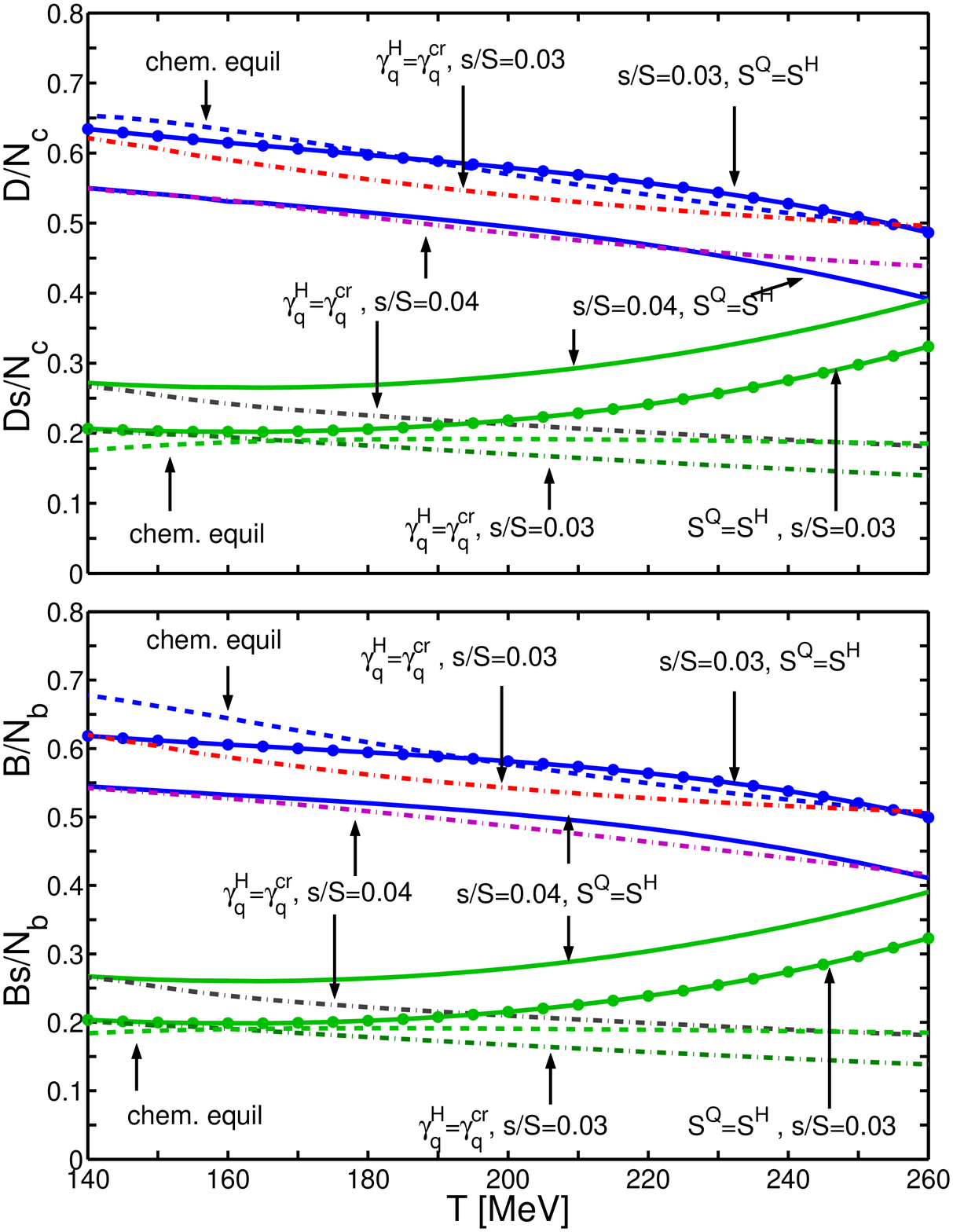}
\caption{(Color on line) \small{Upper panel, fractional charm meson
yield, and lower panel, fractional bottom meson yields. Equilibrium
(dashed lines) and non-equilibrium for $s/S=0.03$ (point marked
solid line) and $s/S=0.04$ (solid line) for $D/N_c$ (blue lines,
upper panel); $D_s/N_c$ (green lines, upper panel); for $D/N_c$ and
$D_s/N_c$ with $s/S=0.03$ and $s/S=0.04$ for $\gamma_q =
\gamma^{cr}_q$ (dash-dotted lines); $B/N_b$ (solid line, lower
panel); and $B_s/N_b$ (point marked solid line, lower panel), for
$B/N_b$ and $B_s/N_b$ with $s/S=0.03$ and $s/S=0.04$ for $\gamma_q =
\gamma^{cr}_q$ (dash-dot lines); as a function of $T$.}}
\label{Dmes}
\end{figure}

In figure~\ref{Dmes} we consider  the yields shown in
figure~\ref{mesrg} as a functions of hadronization temperature. The
dashed blue and green lines were obtained for chemical equilibrium
yields of $D$ and $D_s$ respectively. The extreme upper and lower
lines are for fractional $D$ and $D_s$ yields with $s/S=0.03$ (dot
marked, blue and green lines, respectively), while the central lines
are for $s/S=0.04$ (solid, blue and green lines). Also we show
fractional yields for maximal possible value $\gamma_q\to
\gamma^{cr}_q$, see figure~\ref{gq} for $\gamma^{cr}_q(T)$)
(dash-dot lines) and Eq.\,(\ref{bcon}).

%%%%%%%%%%%%%%%%%%%%%%%%%%%%%%  figure 13
%%%%%%%%%%%%%%%%%%%%%%%%%%%

%%%%%%%%%%%%%%%%%%%%%%%%  figure 14
\begin{figure}
\centering
\includegraphics[width=9cm,height=12cm]{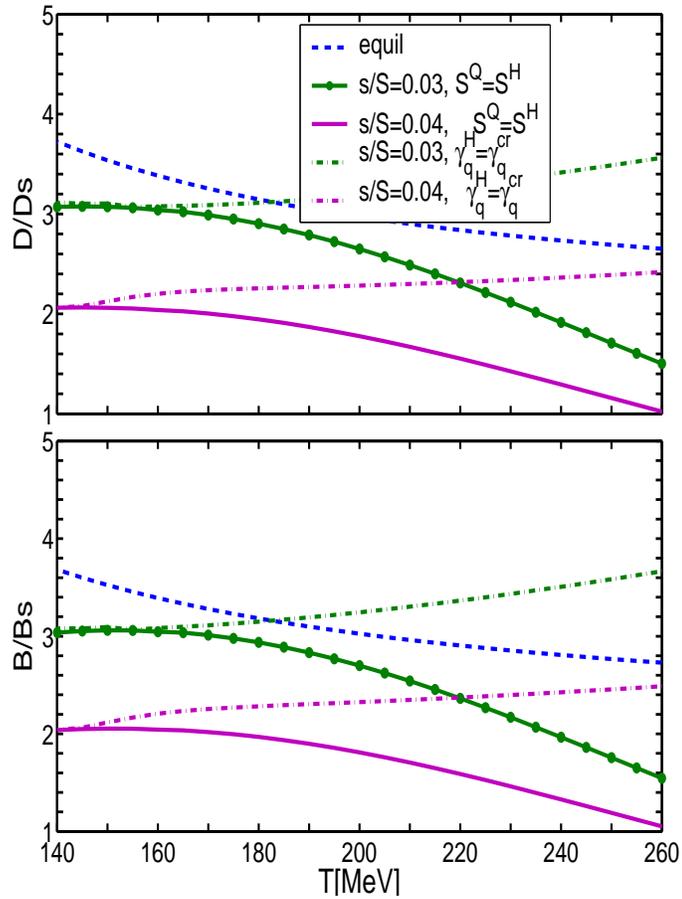}
\caption{(Color on line) \small{ Ratios $D/Ds$ (upper panel) and
$B/Bs$ (lower panel) are shown as a function of $T$ for different
$s/S$ ratios and in chemical equilibrium. Solid line is for
$s/S=0.04$, dash-dot line is for $s/S=0.03$, dashed line is for
$\gamma^{\mathrm{H}}_s=\gamma_s^{\mathrm{H}}=1$.}} \label{Dratio}
\end{figure}
%%%%%%%%%%%%%%%%%%%%%%%%%%%

We note that there is considerable symmetry at fixed $T$ between the
fractional yields of charmed, and bottom mesons,  for the same
condition of $s/S$. The chemical equilibrium results show
significant difference between strange and non-strange heavy mesons.
In the case of chemical equilibrium, for the considered very wide
range of hadronization temperatures  ${D_s}/N_c\simeq
{B_s}/N_b\simeq 0.2$ are nearly constant.
 A significant deviation from
this result would suggest the presence of chemical non-equilibrium
mechanisms of heavy flavor meson production.

The yields of $D_s/N_c$ and ${B_s}/N_b$ are very similar,
and similarly so for $D/N_c$ and $B/N_b$. Thus the relative
yield of either of these mesons measures the relative yield of charm to
bottom participating in the statistical hadronization process:
\begin{equation}
{D_s\over B_s}\simeq {D\over B}={N_c\over N_b}
\end{equation}
This is a very precise result, which somewhat depends on the
tower of resonances included, and thus in particular on the
symmetry in the heavy quark spectra between charmed and bottom states
which we imposed.

It is useful to reconsider here the ratio $D/D_s$ ($B/B_s$) which is
 proportional to $\gamma_q/\gamma_s$, see figure~\ref{rDsDg} for
$D_s/D$ presented as a function of $\gamma_s/\gamma_q$.
We consider this ratio now as a function of $T$, the upper panel
in figure \ref{Dratio} is for charm, the lower for bottom.
We see that there is considerable symmetry in the relative yields
between charmed and bottom mesons with upper and lower panels
looking quasi-identical. Except for accidental values of $T$ where
the equilibrium results (blue, dashed lines) cross the fixed $s/S$
results, there is considerable deviation in these ratios expected
from chemical equilibrium.
 For LHC with
$s/S=0.04$ this ratio is always noticeably smaller than in chemical equilibrium
(solid purple line is for $S^Q=S^H$ and purple, dash-dot line is for
$\gamma_q=\gamma_q^{cr}$). Even for
RHIC-like conditions with $s/S=0.03$   % (with $dN_c/dy=2$)
this ratio is smaller than in chemical equilibrium for all temperatures when
entropy conservation in hadronization is assumed, $S^Q=S^H$ (dot marked
solid, green line).

%%%%%%%%%%%%%%%%%%%%%%%%%%%%%%%%%%%%%%%%%%%%%
\subsection{Heavy baryon yields}\label{BarYieSec}
\begin{figure}
\centering
\includegraphics[width=9cm,height=14cm]{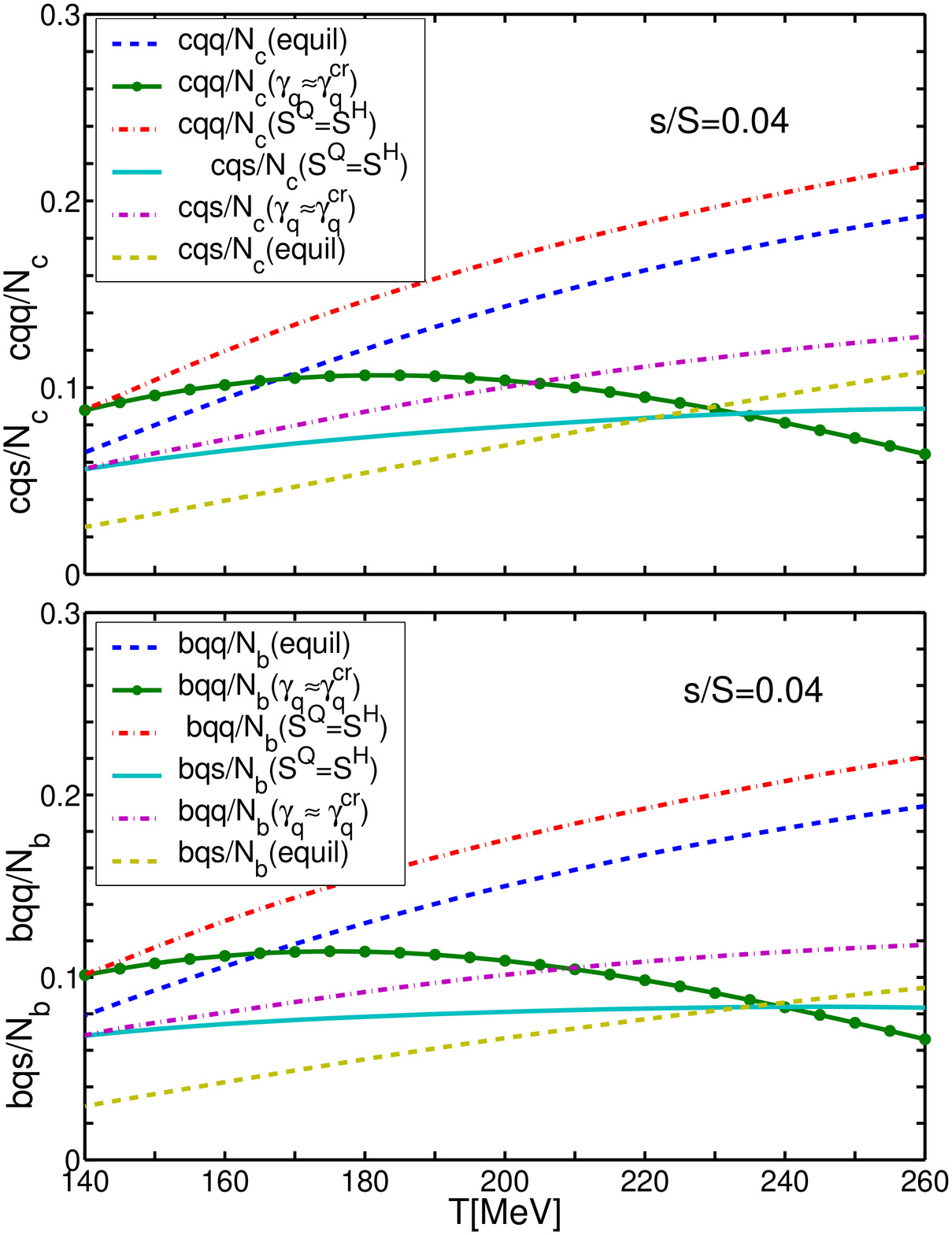}
\caption{(Color on line) \small{Equilibrium (dashed lines),
$s/S=0.04$, $S^Q=S^H$ (solid lines), $s/S=0.04$, $\gamma_q=\gamma^{cr}_q$, (the upper
panel) upper lines for each type are for ratio $(\Lambda_c+\Sigma_c)/N_c$ and lower lines are for $\Xi_c/N_c$ (upper panel) and
(lower panel) upper lines of each type are for $(\Lambda_b+\Sigma_b)/N_c$ and lower lines are for $\Xi_b/N_b$ as functions of T.}}
 \label{bar}
\end{figure}

%%%%%%%%%%%%%%%%%%%%%%%%%%%%%%%%%%%%%%%%%%%%%
As was the case comparing charm to bottom mesons we also establish a
symmetric set of charmed and bottom baryons, shown in the
table~\ref{baryonbc}. Many of the bottom baryons are result of theoretical
studies and we include that many states to be sure that both charm
and bottom are consider in perfect symmetry to each other. In
figure~\ref{bar} (upper panel) we show hadronization temperature
dependencies of yields of baryons with one charm quark normalized to
charm multiplicity $N_c$. We show separately yields of baryons
without strange quark $(\Lambda_c+\Sigma_c)/N_c$, and with one
strange quark S=1 ($\Xi_c/N_c$). We show two cases for
$s/S=0.04$ with conserved entropy at hadronization
$S^{\mathrm{Q}}=S^{\mathrm{H}}$ (solid lines) and with maximum
possible entropy value $\gamma_q=\gamma^{cr}_q$ (dash-dot lines).
The chemical equilibrium case $\gamma_q=\gamma_s=1$ is also shown
(dashed lines). The upper lines of each type are for
$(\Lambda_c+\Sigma_c)/N_c$, the lower lines are for $\Xi_c/N_c$.
A similar result is presented for bottom baryons in the lower panel of
figure~\ref{bar}. We note that the result for bottom baryons is more
uncertain since most baryon masses entering are not experimentally
verified.

%%%%%%%%%%%%%%%%%%%%%%%%%%%%%%%%%%%%%%%%%%% table IV
\begin{table}
\centering
\caption{Charm and bottom baryon states  considered. States
in parenthesis are not known experimentally and have been
adopted from references~\cite{Albertus:2003sx}.\label{baryonbc}}
\begin{tabular}{|c|c|c|c|c|c|c|}
  \hline
  hadron&  &M[GeV]&hadron&  &M[GeV]&g\\
\hline
  $\Lambda_c^+(1/2^+)$&udc&2.285&$\Lambda_b0(1/2^+)$&udb&5.624&2\\
  $\Lambda_c^+(1/2^-)$&udc&2.593&$\Lambda_b0(1/2^-)$&udb&(6.000)&2\\
  $\Lambda_c^+(3/2^-)$&udc&2.6266&$\Lambda_b0(1/2^-)$&udb&(6.000)&2\\
  $\Sigma_c^+(1/2^+)$&qqc&2.452&$\Sigma^0_b(1/2^+)$&qqb&(5.770)&6\\
  $\Sigma_c^{*}(3/2^+)$&qqc&2.519&$\Sigma^{0*}_b(3/2^+)$&qqb&(5.780)&12\\
  $\Xi_c(1/2^+)$&qsc&2.470&$\Xi_b(1/2^+)$&qsb&(5.760)&4\\
  $\Xi_c^{'}(1/2^+)$&qsc&2.5741&$\Xi^{'}_b(1/2^+)$&qsb&(5.900)&4\\
  $\Xi_c(3/2^+)$&qsc&2.645&$\Xi^{'}_b(3/2^+)$&qsb&(5.900)&8\\
  $\Omega_c(1/2^+)$&ssc&2.700&$\Omega_b(1/2^+)$&ssb&(6.000)&2\\
  $\Omega_c(3/2^+)$&ssc&(2.700)&$\Omega_b(3/2^+)$&ssb&(6.000)&4\\
  \hline
\end{tabular}
\end{table}
%%%%%%%%%%%%%%%%%%%%%%%%%%%%%%%%%%%%%%%%%%%%%%%%%%%%%%%%%%

We note that the results shown  figure~\ref{bar} imply
that under LHC conditions at least  15\% of heavy flavor can
be bound in heavy baryons, but possibly 30\%. For large
$\gamma_q=\gamma^{cr}_q > 1$ we see increase in
$(\Lambda_c+\Sigma_c)/N_c$ yields compared to chemical equilibrium
and especially compared to  entropy conserved hadronization $S^Q=S^H$.
This is so since    yields are
proportional to $\gamma_q^2$, $\gamma_s \gamma_q $. This results to
relative suppression the $D_s/N_c$ (see figure~\ref{Dmes}).

%%%%%%%%%%%%%%%%%%%%%%%%%%%%%%%%%%%   figure 15
%%%%%%%%%%%%%%%%%%%%%%%%%%%%%%%%%%%

%%%%%%%%%%%%%%%%%%%%%%%%%%%%%%%%%%% figure 16
\begin{figure}
\centering
\includegraphics[width=8cm,height=8cm]{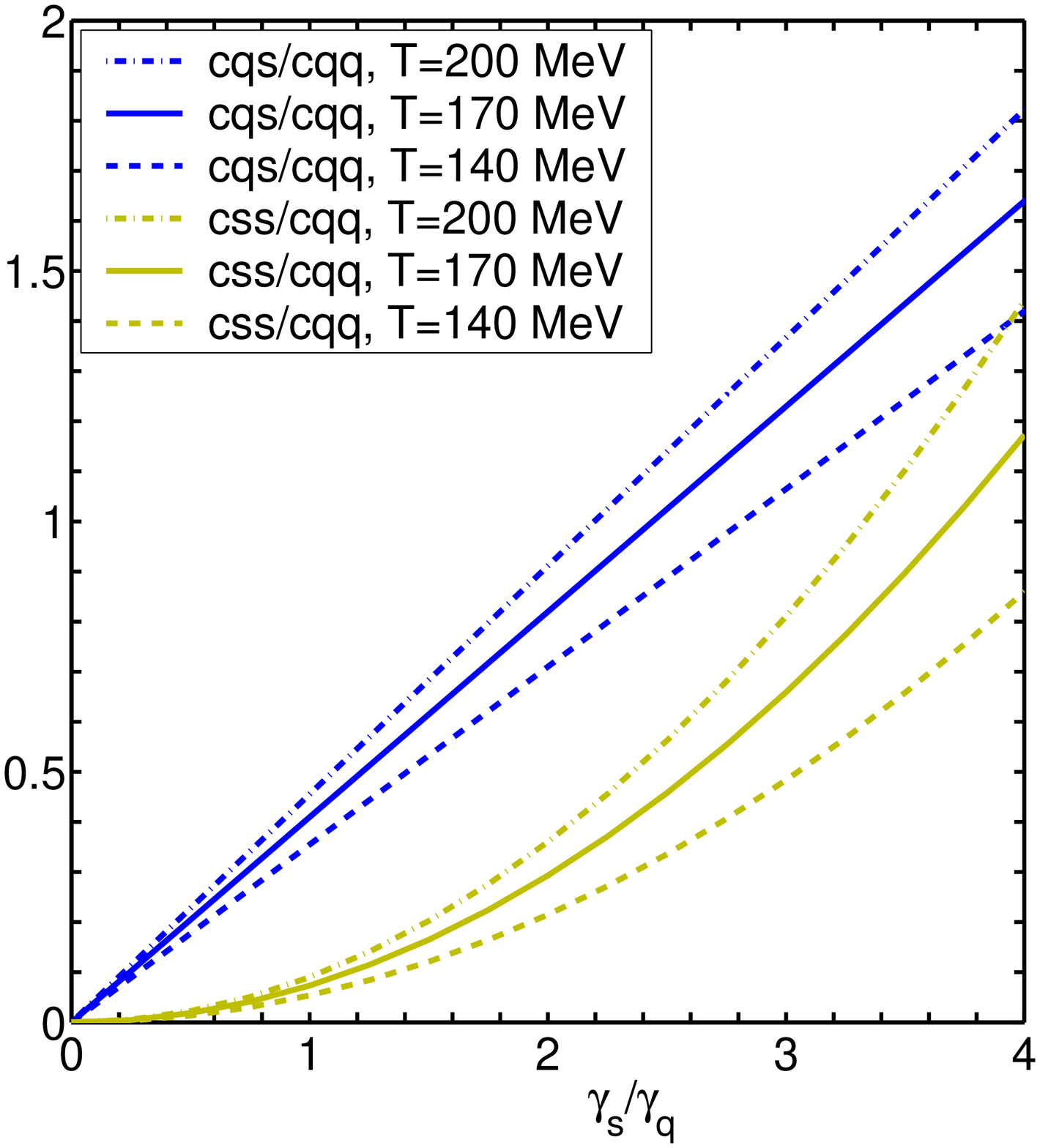}
\caption{(Color on line)
\small{The ratios $cqs/cqq=\Xi_c/(\Lambda_c+\Sigma_c)$ (upper lines)
and $css/cqq=\Omega_c/(\Lambda_c+\Sigma_c)$ (lower lines)
for $T=200$ MeV (dash-dot line), $T=170$ MeV (solid line) and
$T=140$ MeV (dashed line) as functions of $\gamma_s/\gamma_q$} .}
\label{barratio}
\end{figure}
%%%%%%%%%%%%%%%%%%%%%%%%%%%%%%%%%%%

In figure~\ref{barratio} we show ratio
$cqq/cqs=(\Lambda_c+\Sigma_c)/\Xi_c$ as a function of
$\gamma_s/\gamma_q$ for $T=200$ MeV (dash-dot line),
$T=170$ MeV (solid line) and $T=140$ MeV (dashed line).
This dependence is linear,  the slope   depends only on
hadronization temperature $T$.  The $\gamma_s/\gamma_q$ ratio can be converted to $s/S$ ratio using figure~\ref{sSrg}.

\begin{figure}
\centering
\includegraphics[width=8cm,height=8cm]{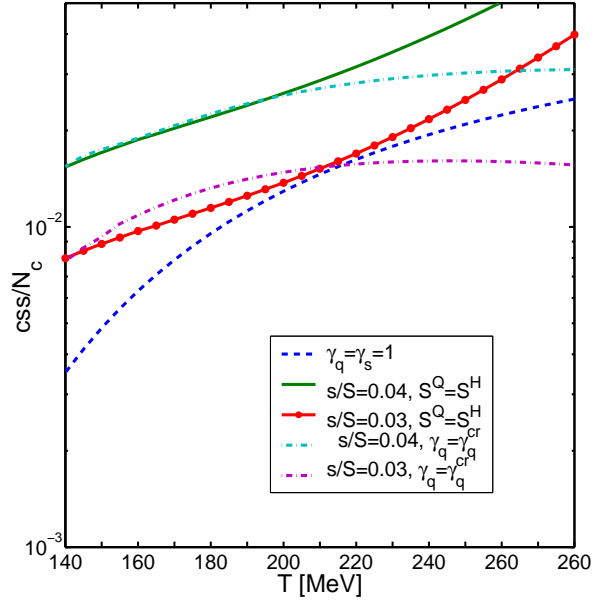}
\caption{(Color on line)
\small{$\Omega_c(css)/N_c$ as function of $T$: dashed line for chemical equilibrium;
solid lines are
 for $S^{\mathrm{Q}}=S^{\mathrm{H}}$, dashed dotted lines are for $\gamma_q=\gamma^{cr}_q$: both for
$s/S=0.03$  and $s/S=0.04$ (upper lines).}}
\label{barss}
\end{figure}

The yield  of multi-strange charmed baryon, $\Omega_c(css)$
is, similar to the light multi-strange hadrons, much more sensitive
to chemical non-equilibrium. In figure~\ref{barss} we see a
large increase in fractional yield of $\Omega_c(css)/N_c$
for $s/S=0.04$ and $S^{\mathrm{Q}}=S^{\mathrm{H}}$ (solid line) compared
to the chemical equilibrium (dashed line) expectation
for the entire considered range of hadronization  temperature. As expected,
this yields increase  with $T$. This also means that higher formation
temperature can be invoked to explain an unusually high yield.
We expect that at LHC more than one percent of total
charm yield will  be found in the $\Omega_c(css)$  state.

%%%%%%%%%%%%%%%%%%%%%%%%%%%%%% 17
%%%%%%%%%%%%%%%%%%%%%%%%%%%

%%%%%%%%%%%%%%%%%%%%%%%%%%%%%%%%%%%%%%%%%%%
\subsection{Yields of hadrons with two heavy quarks}\label{MultiSec}
%%%%%%%%%%%%%%%%%%%%%%%%%%%%%%%%%%%%%%%%%%%
We consider multi-heavy hadrons listed in the table~\ref{multihiddennbc}.
The yields we will compute are now more model dependent since
we cannot completely reduce the result, it either remains dependent on
the reaction volume $dV/dy$, or on the total charm(bottom) yields $dN/dy$.
For example the yields of hadrons  with two heavy quarks are
approximately proportional to $1/(dV/dy)$ because $\gamma^H_{b,c}$ for heavy
quarks is proportional to $1/dV/dy$, see Eq.\,(\ref{gammacb}):
\begin{eqnarray}
\frac{dN_{hid}}{dy} &\propto& \gamma^{H\,2}_c \frac{dV}{dy} \propto
\frac{1}{dV/dy},\\
\frac{dN_{Bc}}{dy} &\propto& \gamma^H_c \gamma^H_b \frac{dV}{dy}
\propto \frac{1}{dV/dy}.
\end{eqnarray}
Moreover, unlike it is the case for single heavy hadrons,  the
canonical correction to grand-canonical phase space does not cancel out
in these states, adding to the uncertainty.

%%%%%%%%%%%%%%%%%%%%%%%%%%%%%%%%%%%%%%%%%%% table V
\begin{table}
\centering
\caption{Hidden charm and multi heavy hadron states
considered. States in parenthesis are not known experimentally
\label{multihiddennbc}}
\begin{tabular}{|c|c|c|c|}
  \hline
  hadron&  &mass(GeV)&g\\
  \hline
   $\eta_c (1S)$&$c\bar{c}$&2.9779&1\\
   $J\!/\!\Psi (1S)$&$c\bar{c}$&3.0970&3\\
   $\chi_{c0}(1P)$&$c\bar{c}$ &3.4152&1\\
   $\chi_{c1}(1P)$&$c\bar{c}$ &3.5106&3\\
   $h_c(1P)$&$c\bar{c}$ &3.526&3\\
   $\chi_{c2}(1P)$&$c\bar{c}$ &3.5563&5\\
   $\eta_c(2S)$&$c\bar{c}$&3.638&1\\
   $\psi(2S)$&$c\bar{c}$&3.686&3\\
   $\psi$&$c\bar{c}$&3.770&3\\
   $\chi_{c2}(2P)$&$c\bar{c}$&3.929&5\\
   $\psi$&$c\bar{c}$&4.040&3\\
   $\psi$&$c\bar{c}$&4.159&3\\
   $\psi$&$c\bar{c}$&4.415&3\\
   $B_c$&$b\bar{c}$&6.27&1\\
   $\Xi_{cc}$&ccq&3.527&4\\
   $\Omega_{cc}$&ccs&(3.660)&2\\
\hline
\end{tabular}
\end{table}
%%%%%%%%%%%%%%%%%%%%%%%%%%%%%%%%%%%%%%%%%%%%%%%%%%%%%%%%%%

\begin{figure}%[!t]
\centering
\includegraphics[width=9cm,height=12cm]{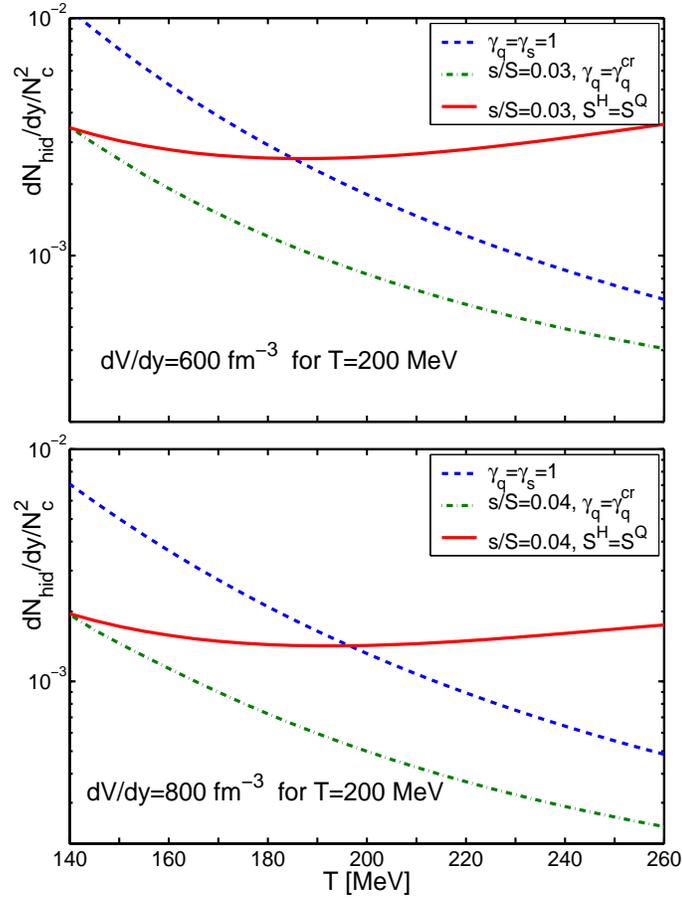}
\caption{(Color on line) \small{$c\bar{c}/N^2_c$  yields as a
function of hadronization temperature T, at
  $dV/dy=600$ $\mathrm{fm^{-3}}$ for $T=200$ MeV, $s/S=0.03$  (upper panel),
$dV/dy=800$ $\mathrm{fm^{-3}}$ for $T=200$ MeV,  $s/S=0.04$ (lower panel).
Results shown are for  $S^Q=S^H$  (solid lines),   for $\gamma_q=\gamma_q^{cr}$  (dash-dot lines),
and    for chemical equilibrium case (dashed lines, $s/S$ is not fixed).}} \label{cc}
\end{figure}

Thus the result we present must seen as a guiding the eye and
demonstrating a principle.
In figure~\ref{cc} we show the yield of hidden charm $c\bar{c}$ mesons
(see table~\ref{multihiddennbc}) normalized by the  square
of charm multiplicity $N_c^2$ as a
function of hadronization temperature $T$. We consider again cases with
$s/S=0.03$ (upper panel) and $s/S=0.04$ (lower panel),
 solid line is for $S^H=S^Q$, dot-dash line is for $\gamma_q=\gamma^{cr}_q$,
and dot-dash line is for $\gamma_q=\gamma_q^{cr}$.
The chemical equilibrium $c\bar{c}$ mesons yields are
shown (dashed lines on both panels) for two different values of
$dV/dy=600$\,fm$^3$ for $T = 200$ MeV (upper panel) and
$dV/dy=800$\,fm$^3$ for $T = 200$ MeV (lower panel).

%%%%%%%%%%%%%%%%%%%%%%%%% figure 18
%%%%%%%%%%%%%%%%%%%%%%%%%%%%%%

The yield of $c\bar{c}$ mesons is much smaller for $s/S=0.04$ than
in equilibrium for the same $dV/dy$ for large range of hadronization
temperatures. For $s/S=0.03$ the effect is similar, but suppression
is not as pronounced. For $\gamma_q=\gamma_q^{cr}$ suppression
the yield of hidden charm particles is always
smaller than equilibrium value. This suppression occurs due to
competition with the yield of strange-heavy mesons,
and also, when $\gamma_q>1$, with
heavy baryons with two light quarks.
The enhanced yield of $D, D_s$ and heavy baryons in effect depletes
the pool of available charmed quark pairs, and fewer hidden charm
$c\bar{c}$ mesons are formed. For particles with two heavy quarks
the effect is larger than for hadrons with one heavy quark and light
quark(s).

\begin{figure}%[!t]
\centering
\includegraphics[width=8cm,height=8cm]{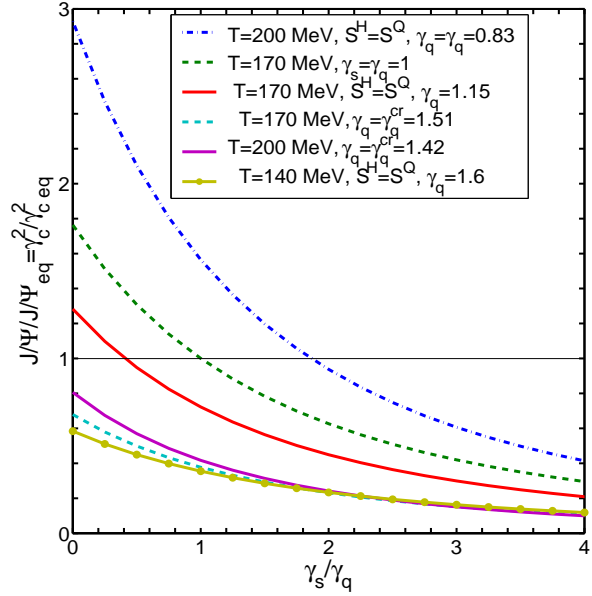}
\caption{(Color on line) \small{Ratio $J\!/\!{\Psi}/J\!/\!\Psi_{eq}=\gamma^2_c/\gamma^2_{c\,eq}$ as a function
of $\gamma^{\rm H}_s/\gamma^{\rm H}_q$ at fixed value of $\gamma^{\rm H}_q$ and if required, entropy
conservation.
Shown are:   $T=200$ MeV at $\gamma_q=0.83$ (dot-dash line) and at
$\gamma_q=\gamma^{cr}_q=1.42$ (lower solid line (purple) );
$T=170$ MeV at $\gamma_q=1$(upper dashed line) , at  $\gamma_q=1.15$, (upper solid line (red)), and at
$\gamma_q=\gamma^{cr}_q=1.51$, (lower dashed line); and
 $T=140$ MeV, $\gamma_q=1.6$}} \label{jpsrg}
\end{figure}

In  figure~\ref{jpsrg} we compare the $J\!/\!\Psi$ yield to the chemical equilibrium
yield $\Psi/J\!/\!\Psi_{eq}$, as a function of $\gamma^{\rm H}_s/\gamma^{\rm H}_q$,
each line is at a fixed value $\gamma^{\rm H}_q$.
This ratio is:
\begin{equation}
\frac{J\!/\!{\Psi}}{J\!/\!{\Psi}_{\rm eq}}=\frac{N_{hid}}{N_{hid\,eq}}=\frac{\gamma_c^2}{\gamma_{c\,{\rm eq}}^2}.
\end{equation}
$J\!/\!{\Psi}/J\!/\!{\Psi}_{\rm eq}$  always decreases when $\gamma_s/\gamma_q$ increases.
For $\gamma_q = \gamma_{cr}$ $J\!/\!\Psi/J\!/\!\Psi_{\rm eq}$
is smaller than unity  even when $\gamma_s\to 0$, because of large phase space occupancy of light quarks.
$J\!/\!{\Psi}/J\!/\!{\Psi}_{\rm eq}>1 $ for small $\gamma_q$ and small $\gamma_s/\gamma_q$ .
This ratio decreases with $\gamma_s/\gamma_q$ grow. This $J\!/\!{\Psi}/J\!/\!{\Psi}_{\rm eq}$ ratio behaviour is similar to 
experimental results for SPS enerdies shown in figure~\ref{stjpsi}~\cite{Becattini:2005yj}. At SPS energies the chemica potential the chemical
potential has influence on result. However it have not to change the effect qualitively. The reason of $J\!/\!\Psi$ suppression in this case can be the same as in our model.  

%%%%%%%%%%%%%%%%%%%%%%%%% figure 19
%%%%%%%%%%%%%%%%%%%%%%%%%%%%%%

\begin{figure}%[!t]
\centering
\includegraphics[width=9cm,height=9cm]{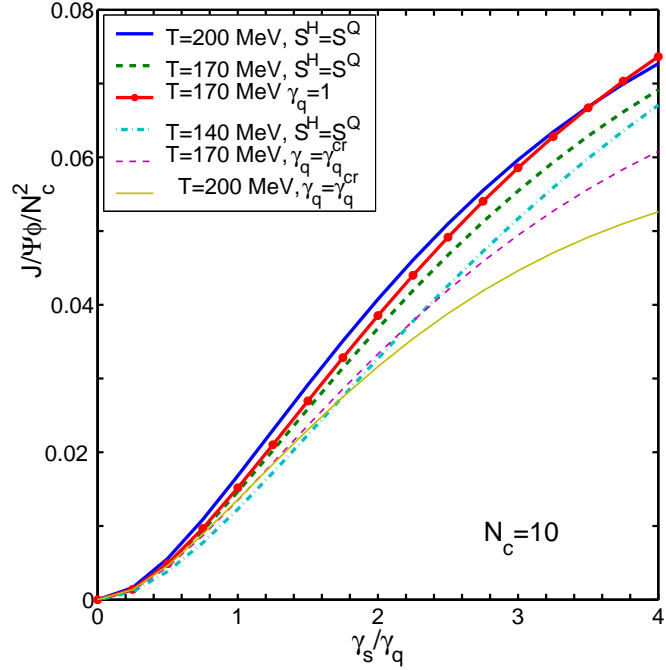}
\caption{(Color on line) \small{$J/\Psi\phi/N^2_c$ states yields as a
function of $\gamma_s/\gamma_q$ ratio for $T=200$ MeV, $S^Q=S^H$ (solid line)
and $\gamma_q=\gamma_q^{cr}$ (solid thin line),
$T=170$ MeV: $S^Q=S^H$ (dashed line), $\gamma_q=\gamma_q^{cr}$ (thin dashed line)
 and $\gamma_q=1$ (solid line with dot marker); for
$T=140$ MeV, $S^Q=S^H$ (dash-dot line)}} \label{jpsphrg}
\end{figure}

 Considering  the product of $J\!/\!\Psi$ and $\phi$
yields normalized by $N^2_c$ we eliminate nearly all the uncertainty
about the yield of charm and/or hadronization volume. However,
we tacitly assume that both $J\!/\!\Psi$ and $\phi$ hadronize at the
same temperature.  In figure~\ref{jpsphrg} we show $J/\Psi\phi/N^2_c$
as function of $\gamma_s/\gamma_q$. There is considerable  difference
to the ratio considered in figure~\ref{JpsiD}.  We see mainly dependence on
 $\gamma_s/\gamma_q$. As before, see
 section \ref{singHadSec} $J\!/\!\Psi$ is the sum of all states
$c\bar{c}$ from table~\ref{multihiddennbc} that can decay to
$J\!/\!\Psi$. We show results for $T=200$
MeV (solid lines), $T=170$ MeV (dashed line) and $T=140$ MeV
(dash-dot line). The $\gamma_q$, for each $T$,  is fixed by   entropy
conservation condition during hadronization (figure~\ref{gq}) (thick lines) or by
$\gamma_q=\gamma_q^{cr}$ (thin lines). For $T=140$ MeV these lines
coincide.   $T=170$ MeV, $\gamma_q=1$ case is also shown (solid
line with dot markers). The  $s/S$ values, which correspond  to given
$\gamma_s/\gamma_q$ ratio can be found in figure~\ref{sSrg}.
Figure~\ref{jpsphrg} shows  that despite the yield $\phi/(dV/dy)$ increasing as $(\gamma_s/\gamma_q)^2$,
$J\!/\!{\Psi}\phi/N^2_c$ is increasing as $\gamma_s/\gamma_q$ considering
compensation effects.

%%%%%%%%%%%%%%%%%%%%%%%%% figure 20
%%%%%%%%%%%%%%%%%%%%%%%%%%%%%%

\begin{figure}%[!t]
\centering
\includegraphics[width=9cm,height=12cm]{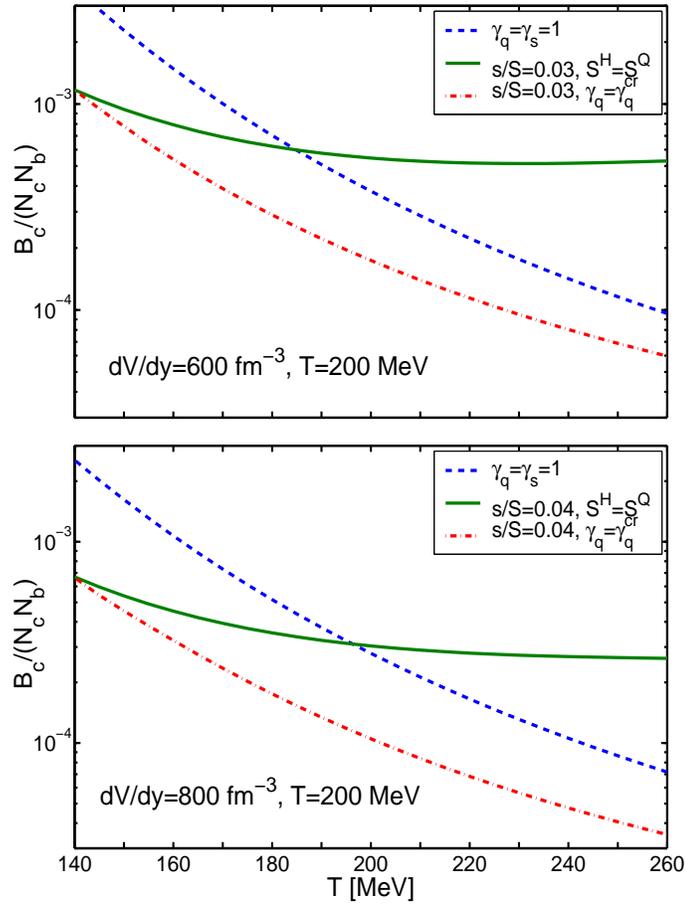}
\caption{(Color on line) \small{Bc mesons yields as function of T for
chemical equilibrium case with $dV/dy=600$ $fm^{-3}$ for $T=200$ MeV
(the upper panel, dashed line), for $s/S=0.03$ with $dV/dy=600$
$fm^{-3}$ for $T=200$ MeV (the upper panel, solid line), for
chemical equilibrium case with $dV/dy=800$ $fm^{-3}$ for $T=200$
MeV (the lower panel, dashed line) for $s/S=0.04$ $dV/dy=800$
$fm^{-3}$ for $T=200$ MeV (lower panel, solid line)}} \label{bc}
\end{figure}

A similar situation,
as in figure~\ref{jpsrg} for hidden charm,
arises for the $B_c$ meson yield, see
figure~\ref{bc}, where $B_c/N_cN_b$ ratio is shown as a function of
hadronization temperature $T$, for the same strangeness yield cases
as discussed for the hidden charm meson yield.  Despite
suppression in strangeness rich environment, the  $B_c$ meson yield
continues to be larger than the yield of $B_c$ produced in single NN
collisions, where the scale yield is at the level of
$\sim{10^{-5}}$, see cross sections for $b\bar{b}$ and $B_c$
production in~\cite{Anikeev:2001rk} and in \cite{Chang:2003cr}, respectively.

%%%%%%%%%%%%%%%%%%%%%%%%%%%%%%%figure 21
%%%%%%%%%%%%%%%%%%%%%%%%%%%%%%%%%

\begin{figure}
\centering
\includegraphics[width=9cm,height=12cm]{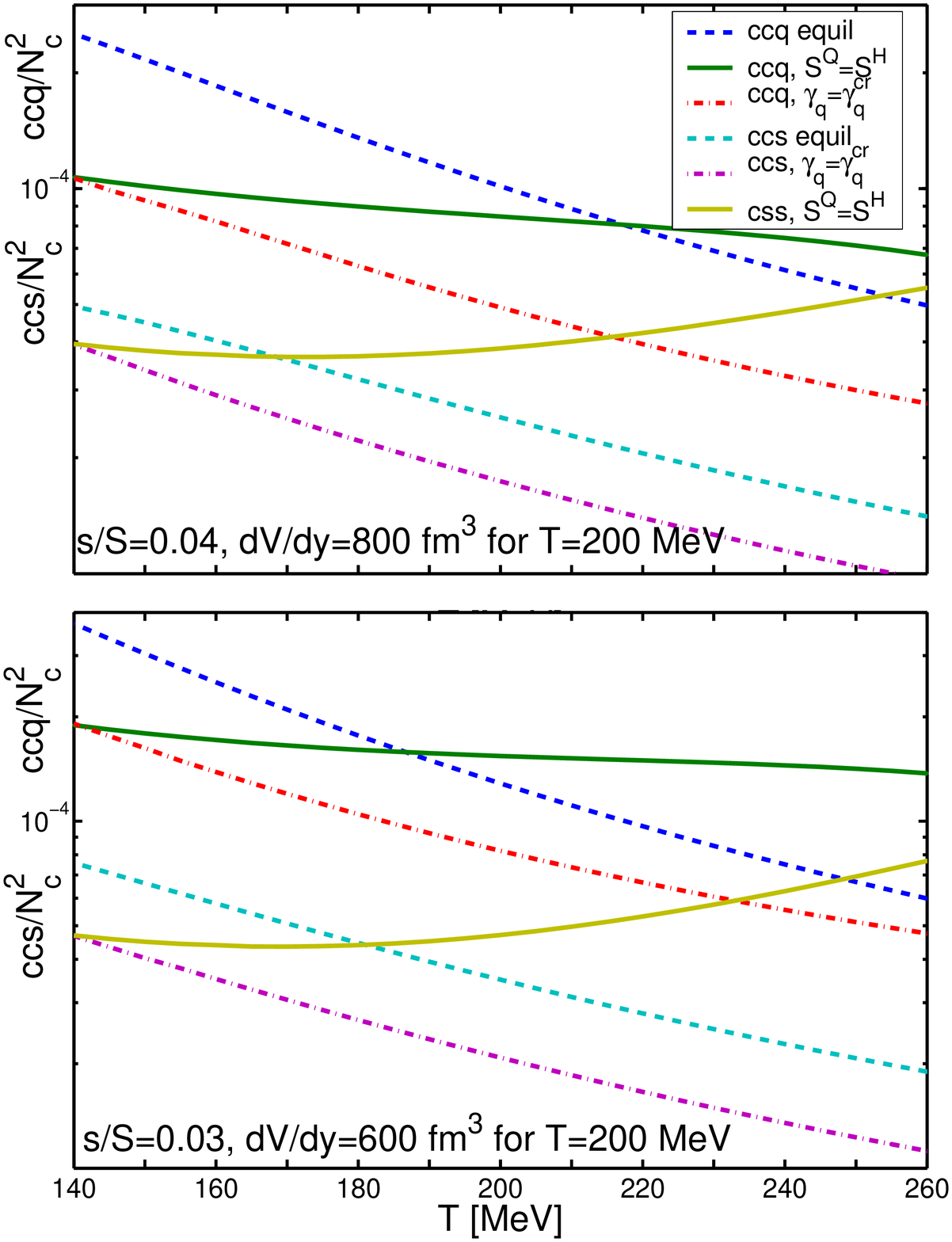}
\caption{(Color on line) \small{$ccq/N_c^2$ (upper lines in each panel) and $ccs/N_c^2$ (lower lines
in each panel) baryon  yields as a function of $T$. Upper panel:
chemical equilibrium case with $dV/dy=800$ $\mathrm{fm^{-3}}$ for $T=200$
MeV(dashed line), $s/S=0.04$ with
$dV/dy=800$ $fm^{-3}$ for $T=200$ MeV: $S^H=S^Q$ (solid
line) and $\gamma_q=\gamma_q^{cr}$ (dash-dot line); and lower panel: chemical equilibrium case with
$dV/dy=600$ $\mathrm{fm^{-3}}$ for $T=200$ MeV (dashed line), and $s/S=0.03$
$S^Q=S^H$ (solid line) and $\gamma^q=\gamma_q^{cr}$ (dash-dot line).}}
\label{qcc}
\end{figure}

In figure~\ref{qcc} we show $N_c^2$ scaled yields of $ccq$ and $ccs$ baryons
as a function of temperature. Upper panel shows aside of the equilibrium case
(dashed lines) the yields for  $s/S=0.04$  with $S^H=S^Q$ (solid lines) and with
 $\gamma_q=\gamma_q^{cr}$ (dash-dot line) for
$dV/dy = 800\,\mathrm{fm^{-3}}$ for $T=200$ MeV.
Lower panel is for $dV/dy = 600\,\mathrm{fm^{-3}}$ and  $s/S=0.03$.
For the $ccq$ baryons the chemical nonequilibrium suppression effect is
similar to what we saw for $c\bar c$ and Bc mesons. Equilibrium yield is much
larger than non-equilibrium for $T<230$ MeV when $s/S=0.04$ and $S^H=S^Q$,
and for $T<190$ MeV when $s/S=0.03$ and $S^H=S^Q$.
In case $\gamma_q=\gamma_q^{cr}$, the yield of $ccq$ is
always smaller than equilibrium. The yield of $ccs$ baryons
has similar suppression, but it becomes larger than equilibrium
for smaller temperatures and yield enhancement for higher T is
 larger for $S^H=S^Q$ then in case of $ccq$ because of large number of strange quarks.

%%%%%%%%%%%%%%%%%%%%%%%%%%%%%%% figure 22
%%%%%%%%%%%%%%%%%%%%%%%%%%%%%%%

\begin{figure}
\centering
\includegraphics[width=9cm,height=12cm]{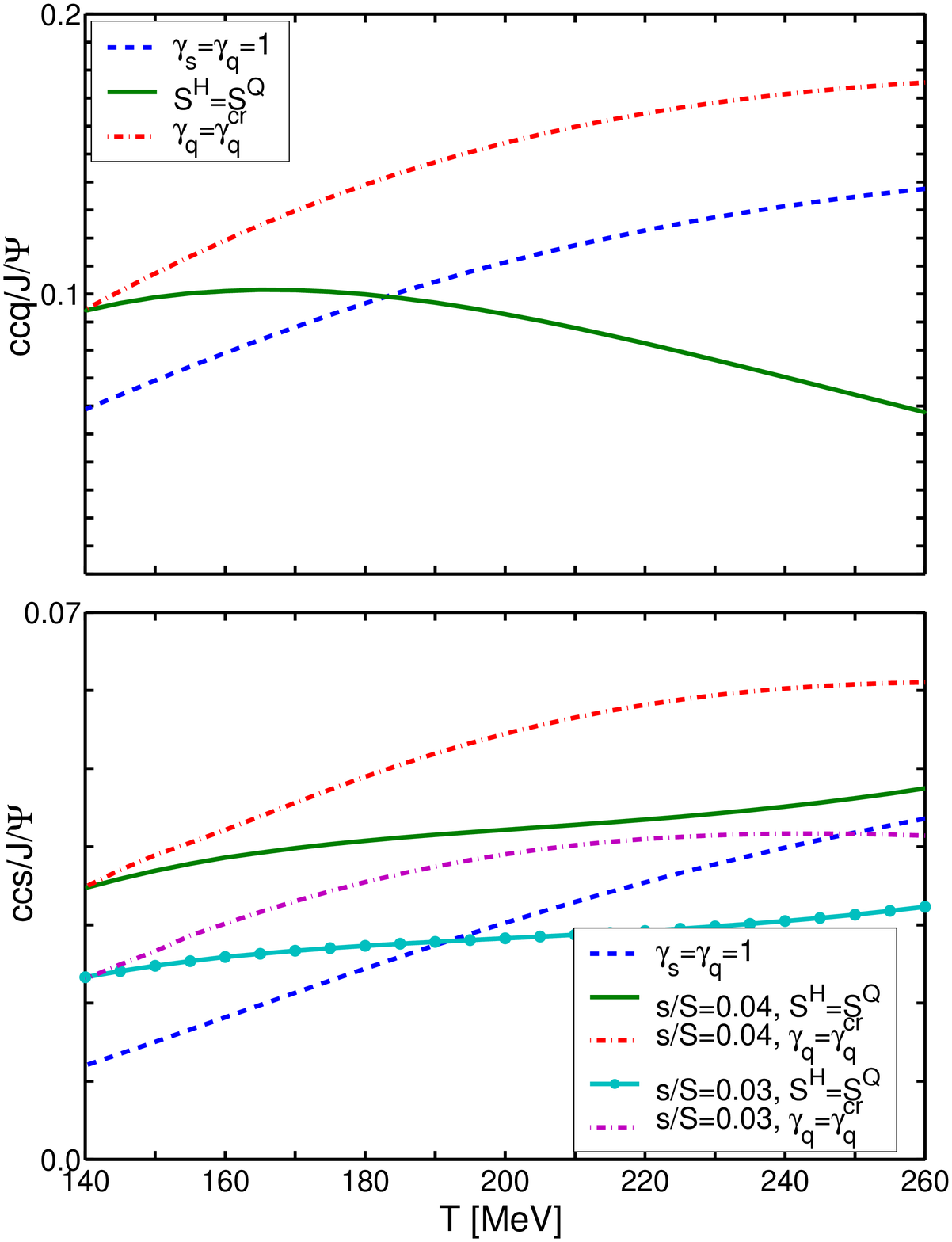}
\caption{(Color on line) \small{$ccq/J\!/\!\Psi$ (upper panel) and $ccs/J\!/\!\Psi$ (lower  panel) ratios as a function of $T$.
Upper panel:
chemical equilibrium case  (dashed line), $S^H=S^Q$ (solid
line) and $\gamma_q=\gamma_q^{cr}$ (dashed-dot line); and lower panel: chemical equilibrium case with
(dashed line), $s/S=0.04$: $S^Q=S^H$ (solid line with dot marker) and $\gamma_q=\gamma_q^{cr}$ (thin dash-dot line); $s/S=0.03$
(solid line) and $\gamma_q=\gamma_q^{cr}$ (thin dash-dot line).}}
\label{qccjpsi}
\end{figure}

In the figure~\ref{qccjpsi} we show ratios $ccq/J\!/\!\Psi$ (upper panel) and $ccs/J\!/\!\Psi$ (lower panel)
as a function of hadronization temperature. These ratios do not
depend on $dV/dy$.  $ccq/J\!/\!\Psi\propto \gamma_q$ does not depend
on $s/S$. For $ccq/J\!/\!\Psi$ ratio we show three cases: chemical
equilibrium $\gamma_s=\gamma_q=1$ (dashed line), $S^H=S^Q$ (solid
line) and $\gamma_q=\gamma_q^{cr}$ (dash-dot line). For
$ccs/J\!/\!\Psi$ ($ccq/J\!/\!\Psi \propto \gamma_s$) we show
chemical equilibrium case (dashed line), $s/S=0.04$: $S^H=S^Q$
(solid line with point marker) and $\gamma_q=\gamma_q^{cr}$ (thin
dash-dot line); $s/S=0.03$: $S^H=S^Q$ (solid line) and
$\gamma_q=\gamma_q^{cr}$ (thin dash-dot line). The overall all
yields of double charmed (strange and non-strange) baryons and
anti-baryons is clearly larger than the yield of $J\!/\!\Psi$.

%%%%%%%%%%%%%%%%%%%%%%%%%%%%%%% figure 23
%%%%%%%%%%%%%%%%%%%%%%%%%%%%%%%

%%%%%%%%%%%%%%%%%%%%%%%%%%%%%%%%%
\section{Conclusions}\label{concSec}
%%%%%%%%%%%%%%%%%%%%%%%%%%%%%%%%%

We have considered here in some detail the abundances
of heavy flavor hadrons within the statistical hadronization model.
While we compare the yields to the expectations based on
chemical equilibrium yields of light and strange quark pairs,
we present results based on the hypothesis that
the QGP entropy and QGP flavor yields determine
the values of phase space occupancy $\gamma^\mathrm{H}_i$ $i=q,s,c,b$,
which are of direct interest in study of the heavy hadron yields.

For highest energy heavy ion
collisions the range of  values discussed  in literature is
$1\le \gamma^{\mathrm{H}}_q\le 1.65$ and
$0.7\le \gamma^{\mathrm{H}}_s/\gamma^{\mathrm{H}}_q\le 1.5$. However
$\gamma^{\mathrm{H}}_c$ and $\gamma^{\mathrm{H}}_b$
values which are much larger than unity arise. This is
due to the need to describe the large primary parton based
production, and considering that the   chemical
equilibrium yields   are suppressed by the factor $\exp(-m/T)$.

Our work is based on the grand canonical treatment of phase space.
 This approach is valid for charm hadron production at LHC,  since
the  canonical corrections, as we have discussed, are
 not material.  On the other hand, even at LHC 
the much smaller  yields of  bottom heavy hadrons are
subject to canonical suppression. The value of the parameter $\gamma_b^{H}$
obtained at a   fixed bottom yield  $N_b$, using either  the canonical, or the grand canonical methods,
are different, see e.g.  Eq.\,(15) in \cite{Rafelski:2001bu}.  Namely, to obtain  a given yield $N_b$ 
in canonical approach, a greater value of $\gamma_b^{H}$ is needed 
in order to compensate the canonical suppression effect. 
However, for any individual single-$b$ hadron, 
the  relative yields, i.g. $B/B_s$ do  not depend on $\gamma_b^{H}$
and thus such ratios are not influenced by  canonical suppression.  Moreover, as long as 
the yield of single-$b$ hadrons dominates the total  bottom yield: 
$N_b\simeq B+B_s+\Lambda_b+\ldots$, also the $N_b$ scaled yields of
hadrons comprising one $b$-quark i.e. ratios such as $B/N_b$, $B_s/N_b$, $B_c/N_b$, etc,  
are not sensitive to the value of   $\gamma_b^{H}$ and  can be obtained
within either the canonical, or grand canonical method. 
 On the other hand for $b\bar{b}$ mesons and multi-$b$ baryons the
canonical effects should be considered. Study of the
yields of these particles is thus postponed.

We address here  in particular how the yields of heavy hadrons are influenced by
$\gamma^{\mathrm{H}}_s/\gamma^{\mathrm{H}}_q\ne 1$ and $\gamma_q \ne 1$. The actual values
of $\gamma^{\mathrm{H}}_s/\gamma^{\mathrm{H}}_q$ we use are related to
the strangeness per entropy yield $s/S$ established in the QGP phase.
Because the final value $s/S$ is established well before hadronization,
and the properties of the hadron phase space are well understood,
the resulting $\gamma^{\mathrm{H}}_s/\gamma^{\mathrm{H}}_q$ are well
defined and turn out to be quite different from unity in the range of
temperatures in which we expect particle freeze-out to occur.
We consider in some detail the effect of QGP hadronization on
the values of $\gamma^{\mathrm{H}}_s$ and $\gamma^{\mathrm{H}}_q$.

One of first results we present (figure~\ref{JpsiD}) allows a test of the
statistical hadronization model for heavy flavor:
 we show that the yield ratio
$c\bar c$ $s\bar s$/($c\bar s$ $\bar c s$) is nearly independent
of temperature and it is also nearly constant when the $\phi$ is
allowed to freeze-out later (figure~\ref{JpsiD2T}), provided that the condition of
production is at the same value of strangeness per entropy $s/S$.

We studied in depth how the (relative) yields of strange and non-strange
charmed mesons vary with strangeness content. For a chemically
equilibrated QGP source, there is considerable shift of the yield
from non-strange $D$ to the strange $D_s$
 for $s/S=0.04$ expected at LHC.
The expected fractional yield $D_s/N_c \simeq {B_s}/N_b\simeq 0.2$
when one assumes $\gamma^{\mathrm{H}}_s=\gamma^{\mathrm{H}}_q=1$,
 the expected
enhancement of the strange heavy mesons is at the level of 30\%
when $s/S=0.04$, and greater when greater strangeness yield is
available.

As the result we find a relative suppression of the
multi-heavy hadrons, except when they contain strangeness. This suppression depends on both factors $\gamma_s$ and $gamma_q$. When phase space occupancy of
light and strange quark is relatively high the probability for charm quarks to make hadrons with strange quarks increases and probability to find the second charm quark
among light and strange quarks decreases. Therefor the $c\bar c$ yield suppression increases when $\gamma_s/\gamma_q$ ratio increases for constant $\gamma_q$. This result
is qualitatively in agreement with experimental results obtained for SPS energies~~\cite{Becattini:2005yj}.

On the other hand, the yield of $c\bar{c}/N_c^2 \simeq 2 10^{-3}$ is found to be
almost independent on  hadronization
temperature when  entropy at hadronization is conserved. That is because for larger $T$ $\gamma_q$ decreases. The suppression effect decreasees, compared to SHM and
become even negative for $T>200$ MeV, resulting to the $c \bar c$ yield almost independent on temperature. 
We don't know exactly equation of state in QGP and so the value of
 $\gamma_q$ which is needed to conserve the entropy may be different.
 If $\gamma_q$ is larger for higher temperatures, suppression of
$c{\bar{c}}$ is larger for a fixed $s/S$.
The same result is found for $B_c \approx 5-6\,10^{-4} N_cN_b $,
that  yield remains considerably larger (by a factor 10 --- 100) compared to the scaled
yield in single nucleon nucleon collisions.

We have shown that the study of heavy flavor hadrons will provide
important information about the nature and properties of the QGP
hadronization. The yield of Bc($b\bar c$) mesons remains enhanced
while the hidden charm $c\bar c$ states encounter another suppression
mechanism, compensating for the greatly enhanced production due to
large charm yield at LHC.

%%%%%%%%%%%%%%%%%%%%%%%%%%%%%%%%%%%%%%%%%%%%%%%%%%%%%%%%%%%%%%%%%%%%%%%%%%%%%%%%%%%%%%%
\chapter{CHEMICAL EQUILIBRATION INVOLVING DECAYING PARTICLES AT FINITE TEMPERATURES}\label{dpeq}
%%%%%%%%%%%%%%%%%%%%%%%%%%%%%%%%%%%%%%%%%%%%%%%%%%%%%%%%%%%%%%%%%%%%%%%%%%%%%%%%%%%%%%%%%%%%%%%
%%%%%%%%%%%%%%%%%%%%%%%%%%%%%%%%%%%%%%%%%%%
\section{Introduction, motivation, overview}
%%%%%%%%%%%%%%%%%%%%%%%%%
In this chapter we consider relativistic master population equation and equations for reaction rates similar to considered
in section~\ref{partpr}, but extended to the case of ${\rm two-body}  \leftrightarrow {\rm one-body}$ reaction~\ref{123}.
To our surprise, we realized that such ${\rm two-body}  \leftrightarrow {\rm one-body}$ reaction  has so far NOT  been
addressed   in the relativistic context in literature.
This study begun with the question at which temperature in the expanding early Universe the reaction
$\gamma +\gamma \leftrightarrow  \pi ^{0}$
`freezes' out, that is the $\pi^0$ decay overwhelms the production rate and the yield
falls out from chemical equilibrium yield. This reaction has  the lowest threshold,
one pion mass,  with two thermal particles available to reach it. Thus
this reaction  should be still operational at a relatively  low temperature when all other
hadron production reactions cease to be effective. 

In chapter~\ref{eeg} we will also consider this reaction in $e^+e^-\gamma$ plasma, created by high intensity laser pulse. 
Aside of cosmological implications,
insights from this study are clearly of
relevance to the general understanding of quark gluon plasma and
hadron gas evolution in relativistic heavy ion collision. For
example this study allows us to consider the chemical yields arising in
reactions such as $\rho \leftrightarrow \pi \pi $, $\pi
^{0}\leftrightarrow \gamma \gamma $, $\Delta \leftrightarrow N\pi $,
and so on~\cite{Kuznetsova:2008zr}., 
chapter~\ref{respr}.

We recall here that the reaction~\ref{123} % \qquad m_{1}+m_{2}\le m_{3}
considered   in the rest frame of the decaying particle $m_{3}$ implies the constraint $m_1 +m_2\le m_{3}$ since
\begin{equation}
m_3^2=(p_1+p_2)^2=(m_1 +m_2)^2 +2(E_1E_2-m_1m_2-\vec p_1\cdot \vec p_2)\ge (m_1 +m_2)^2 ,\label{123c}
\end{equation}%
considering that the condition $E_1^2E_2^2\ge (m_1m_2+\vec p_1\cdot \vec p_2)^2 $ implies
$(m_1|\vec p_2|-m_2|\vec p_1)^2\ge 2m_1m_2\vec p_1\cdot \vec p_2-2m_1m_2| \vec p_1|\,| \vec p_2|$ which is always
 true. The equality  sign corresponds to the case that
$m_1+m_2=m_3$ for which the reaction rate vanishes by virtue of vanishing phase space.
This text book exercise shows that the raction Eq.\,(\ref {pigg}) is possible, has a `good' phase space size,
and it invites to evaluate the rates of the processes of interest
 in the rest frame of particle `3',  boosting, as appropriate,
from/to laboratory frame. To do this effectively  we will need to formulate the master population
equations in explicitely covariant fashion.

This constraint  Eq. (\ref{123c})  forbids many reactions. For example, the hydrogen formation
 $p+e\to $H is forbidden since for a bound state $m_H<m_p+m_e$. Thus
there must be  a  second particle in the final state,  the  electron capture involves  either a radiative emission,
$p+e\to $H$+\gamma$ or a surface/third atom, which picks  the   recoil momentum.
The situation would be different if there were `resonant' intermediate state of relative long lifespan
with energy above  ionization threshold.  Such `doorway' resonances are available in many important
physical processes, including e.g. the $d+t\to \alpha +n $ fusion.

The general kinetic master equation approach to reactions of type~\ref{1234}
for the yield (chemical) equilibration in nuclear
and particle physics  has been studied  frequently  in the context
of heavy ion reactions\cite{Koch:1986ud}.  However, the simpler situation
was not considered in this framework, and the adaptation is not trivial given
novel quantum and relativistic effects involving
particle decay.

At temperatures $T\simeq m$ ($\hbar=c=k=1$)
the  particle number present   follows rapidly  the relativistic statistical phase
space. Due to the conservation of
energy and momentum, reaction Eq. (\ref{123}) is subject to a particular kinematical
constraint.

In the present work we present for the first time
the dynamics of reaction Eq. (\ref{123}) in a microscopic description
of particle production and the
associated decay reactions  within the
frame work of kinematical master equation obtained from the
Boltzmann-Uehling-Uhlenbeck (BUU) equation under thermal bath\cite{BUU}.

We will ask questions such as \textquotedblleft Will decaying particles reach
chemical equilibrium, and if so, how fast? Does the presence of background
thermal particles stimulate or slow down reaction rate?\textquotedblright .
In the foillowing section \ref{2} we write down the kinetic equations for time evolution of number
density $n$ of decaying particle and equations for invariant rates. We show
that the time variation of density of particle $3$ is
\begin{equation}
\frac{dn_{3}}{dt}=\left( \frac{\Upsilon _{1}\Upsilon _{2}}{\Upsilon_{3}} -1\right)
               \frac{dW_{3\rightarrow 12}}{dVdt},  \label{fe}
\end{equation}%
where ${dW_{3\rightarrow 12}}/{dVdt}$ is the decay rate of particle $3$ and
$\Upsilon _{i}$ is fugacity for the particle $i$. Here the number density
$n_{i}$ of particle $i$ in thermal (kinetic) but not necessarily in
the chemical equilibrium is given by:
\begin{equation}
n_{i} =\frac{1}{(2\pi )^{3}}\int d^{3}p_{i}f_{b/f}(p_{i}),  \label{n} \\
\end{equation}%
$f_{b/f}$, defind by Eq.~\ref{bf},
 is the covariant form of the usual Bose or Fermi distribution
function defined in the rest frame of the thermal bath, and describes the
corrresponding quantity in a general reference frame where the thermal bath
has the relative velocity defined by $u^{\mu }$, see Eq.\ref{4v}. Note that the
distribution function $f$ is a Lorentz scalar but spatial density $n_{i}$ is
not.

The particle $3$ attains the chemical equilibrium when the following
condition among fugacities is satisfied:
\begin{equation}
\Upsilon _{1}\Upsilon _{2}=\Upsilon _{3}.  \label{equilcon1}
\end{equation}%
This, as expected, is equivalent to the Gibbs condition for the chemical
equilibrium. In section \ref{3} we evaluate invariant rate using decay time in vacuum
in rest frame of decaying particle, and discuss the behavior of the average
decay rate of an unstable particle in the presence of thermal bath. In
section \ref{4}, we apply our formalism to three examples: \\
a) relaxation time of formation $\rho $ meson through $\pi +\pi \leftrightarrow \rho $ in
a baryon-free hot hadronic gas, where mesons are considered in thermal and
chemical equilibrium;\\
b) the decay and production relaxation time of resonance $\Sigma(1385)$ in reaction $\Sigma(1385) \leftrightarrow \Lambda \pi$ in hot hadronic gas;\\
c) $\pi^0$ equilibration  the  reaction $\gamma +\gamma \leftrightarrow \pi
^{0}$ in thermal $e^+e^-\gamma$ plasma or early universe\\
However, one should note that at sufficiently low temperatures the
local density of  $\pi ^{0}$ is too low to apply the methods of
statistical physics.

%%%%%%%%%%%%%%%%%%%%%%%%%%%%%%%%%%%%%%%%%%%

\section{Kinetic equations for decaying particles}\label{2}

\subsection{Decaying particle density evolution equation}
%%%%%%%%%%%%%%%%%%%%%%%%%%%%%%%%%%%%%%%%
Consider an unstable particle, say $3$, which decays uniquely into other two
particles,
\begin{equation}
3\rightarrow 1+2  \label{2eq}
\end{equation}
in the vacuum. In a dense and high temperature thermal ambient  phase   particles
$2 $ and $3$ are present, and the inverse reaction:
\begin{equation}
1+2\rightarrow 3  \label{1eq}
\end{equation}
can occur to produce the particle $3$. If we assume that the abundance of
particle $3$ changes solely by  thermal production of particle $3$ by
particles $1$ and $23$ and its decay via Eqs.(\ref{1eq}, \ref{2eq}), then we write the
time variation of the number density as:
\begin{equation}
\frac{dn_{3}}{dt}=\frac{dW_{12\rightarrow 3}}{dVdt}-\frac{dW_{3\rightarrow
12}}{dVdt},  \label{popeq}
\end{equation}
where $dW_{12\rightarrow 3}/dVdt$ is the production rate per unit volume of
particle $ 3$ via Eq.(\ref{1eq}) and $dW_{3\rightarrow 12}/dVdt$ is the decay rate of
particle $3$ per unit volume.

In a normal situation, the abundance of particles $1$ and $2$ are determined
by the other processes which produces these particles. For example, consider
the reaction $\rho \leftrightarrow \pi \pi $ in dense hot matter formed the heavy ion collisions.
Then pions can be easily created by inelastic collisions of other hadrons
and thus in principle we have to deal with multi-component systems. However, there exists
special situation where the total abundances of particles $1$ and $2$ are
initially determined,  and in the following the time variation of number densities of particles
$1$ or $1$ is established by the above reactions In such cases we
have:
\begin{equation}
\frac{dn_{1,2}}{dt}=\frac{dW_{3\rightarrow 12}}{dVdt}-\frac{dW_{12\rightarrow 3}}{dVdt}.  \label{popeq23}
\end{equation}

In the following, we assume that the system is spatially homogeneous and all
of the particles are in thermal equilibrium. Furthermore, we consider that
the interaction time among particles is short enough so that all the
dynamical information can be obtained from the single particle distribution
function $f\left( p\right)$ for each particle. In a thermal equilibrium,
this function is specified completely by 2 parameters, $T$ the temperature
and $\Upsilon$ the fugacity. In this paper, we assume that the thermal
back-ground is inert, so that we keep $T$ constant, but the fugacity
$\Upsilon $ changes in time through the chemical reactions so does the
density of each component of the gas.

The thermal production rate ${dW_{12\rightarrow 3}}/{dVdt}$ and the decay
rate of the particle $3$ under the thermal background
${dW_{3\rightarrow 12}}/{dVdt}$ can then be expressed using these distribution functions for each
of particles involved in the reaction.

\subsection{Decay and production rates}

%%%%%%%%%%%%%%%%%%%%%%%%%
According to the boson or fermion nature of the particle $1,$ we have to
consider different cases. If the particle $1$ is boson, then there are two
different cases of the decay and production mode:
\begin{eqnarray}
&&\mathrm{boson_{3}\longleftrightarrow boson_{1}+boson_{2},} \\
&&\mathrm{boson_{3}\longleftrightarrow fermion_{1}+\overline{fermion_{2}}.}
\end{eqnarray}
On the other hand, if the particle $3$ is fermion it should decay into a
boson and a fermion:
\begin{equation}
\mathrm{fermion_{3}\longleftrightarrow boson_{1}+fermion_{2}.}
\end{equation}
The Lorentz invariant transition probability per unit time and unit volume
corresponding to the process (\ref{1eq}) can be expressed as
\begin{eqnarray}
\frac{dW_{12\rightarrow 3}}{dVdt} &=& \frac{1}{1+I}
  \frac{g_{1}}{(2\pi )^{3}}  \int \frac{d^{3}p_{1}}{2E_{1}}f_{b,f}(\Upsilon _{1},p_{1})
 \frac{g_{2}}{(2\pi )^{3}}\int \frac{d^{3}p_{2}}{2E_{2}}f_{b,f}(\Upsilon _{2},p_{2})\int \frac{
d^{3}p_{3}}{2E_{3}\left( 2\pi \right) ^{3}}\nonumber\\
       && \times  
\left( 2\pi \right) ^{4}\delta ^{4}\left( p_{1}+p_{2}-p_{3}\right)
\frac{1}{g_{1}g_{2}}\sum_{\mathrm{spin}}\left\vert \langle p_{1}p_{2}
 \left\vert M\right\vert p_{3}\rangle \right\vert ^{2}\left(1\pm f_{b,f}(\Upsilon_{3},p_{3})\right) ,  \label{pp}
\end{eqnarray}
where $I=1$ for the reaction of two indistinguishable particles $1$ and $2$, and $I=0$ if
 $1$ and $2$ are distinguishable. The factor $1/(g_{1}g_{2})$ and the summation are
due to averaging over all initial spin states. The last factor accounts for
the enhancement or hindrance of the final state phase due to the quantum
statistical effect, as is introduced first by Uehling and Uhlenbeck~\cite{BUU}. The
sign $^{\prime }+^{\prime }$ is for the case when the particle $3$ is boson
and $^{\prime }-^{\prime }$ when it isa fermion. It is clear that Eq. (\ref{pp})
is manifestly Lorentz invariant and it can be used in any frame of reference.

Now we write in the  same way the decay rate of the process (\ref{2}), per unit volume
we have:
\begin{eqnarray}
\frac{dW_{3\rightarrow 12}}{dVdt} &=&\frac{g_{3}}{(2\pi)^{3}}
\int \frac{d^{3}p_{3}}{2E_{3}}f_{b,f}(\Upsilon _{3},p_{3})
\int \frac{d^{3}p_{1}}{2E_{1}(2\pi)^3}
\int \frac{d^{3}p_{2}}{2E_{2}(2\pi) ^{3}}
(2\pi)^{4}\delta ^{4}\left( p_{1}+p_{2}-p_{3}\right) \times  \notag \\
&&\frac{1}{g_{3}}\frac{1}{1+I}
\sum_{spin}\left\vert \langle p_{3}\left\vert M\right\vert p_{1}p_{2}\rangle \right\vert ^{2}
\left( 1\pm f_{b,f}(\Upsilon_{1},p_{1})\right) \left( 1\pm f_{b,f}(\Upsilon _{2},p_{2})\right)
\label{pd}
\end{eqnarray}%
Here, the one particle state is normalized as
\begin{equation}
\langle p^{\prime }|p\rangle =2p^{0}\left( 2\pi \right) ^{3}\delta
^{3}\left( \vec{p}^{\prime }-\vec{p}\right) .
\end{equation}%
The quantum statistical effects on the final state affects the decay rate,
compared to the free-space case. The decay process is simulated by the
presence of thermal background particles $1$  and $2$ if they
are both mesons, and hindered if they are both fermions.

Note that the pure thermal production rate ${dW_{12\rightarrow 3}}/{dVdt}$
of the particle $3$ is related to its decay rate ${dW_{3\rightarrow 12}}/{dVdt}$
through the time-reversal relation of the transition matrix element
which can be shown in the following. Using Eq.({\ref{FBrel}}),
we can rewrite the Eq.(\ref{pp}) as
\begin{eqnarray}
&&\frac{dW_{12\rightarrow 3}}{dVdt}=\frac{1}{1+I}\Upsilon _{3}^{-1}
\int\frac{d^{3}p_{1}}{2E_{1}(2\pi)^3}
\int \frac{d^{3}p_{2}}{2E_{2}(2\pi)^3}
\int \frac{d^{3}p_{3}}{2E_{3}(2\pi)^3}
\left( 2\pi \right)^{4}\delta ^{4}\left( p_{1}+p_{2}-p_{3}\right) \times   \notag \\
&&\sum_{spin}\left\vert \langle p_{3}\left\vert M\right\vert
p_{1}p_{2}\rangle \right\vert ^{2}f_{b}(\Upsilon _{1},p_{1})f_{b,f}(\Upsilon
_{2},p_{2})f_{b,f}(\Upsilon _{3},p_{3})\exp (u\cdot p_{3}/T);  \label{drg23}
\end{eqnarray}
and for the Eq.(\ref{pd}), using energy-momentum conservation $ p_{1}+p_{2}=p_{3}$ we obtain
\begin{eqnarray}
&&\frac{dW_{3\rightarrow 12}}{dVdt}=
\frac{1}{1+I}\Upsilon_{1}^{-1}\Upsilon_{2}^{-1}
\int \frac{d^{3}p_{1}}{2E_{1}\left( 2\pi \right) ^{3}}
\int \frac{d^{3}p_{2}}{2E_{2}\left( 2\pi \right) ^{3}}
\int \frac{d^{3}p_{3}}{2E_{3}\left( 2\pi \right) ^{3}}
\left( 2\pi \right) ^{4}\delta ^{4}\left( p_{1}+p_{2}-p_{3}\right)
\times
\notag \\
&&\sum_{spin}\left\vert \langle p_{1}p_{2}\left\vert M\right\vert p_{3}\rangle \right\vert ^{2}
f_{b}(\Upsilon _{1},p_{1})f_{b,f}(\Upsilon_{2},p_{2})f_{b,f}(\Upsilon _{3},p_{3}) \exp (u\cdot p_{3}/T);
 \label{drg1}
\end{eqnarray}
Using the time reversal symmetry of the transition matrix element,
\begin{equation}
\left\vert \langle p_{3}\left\vert M\right\vert p_{1}p_{2}\rangle\right\vert ^{2}
 =\left\vert \langle p_{1}p_{2}\left\vert M\right\vert p_{3}\rangle \right\vert ^{2},
\end{equation}
we find:
\begin{equation}
\frac{dW_{12\rightarrow 3}}{dVdt}\frac{1}{\Upsilon_{1}\Upsilon_{2}}=
\frac{dW_{3\rightarrow 12}}{dVdt} \frac{1}{\Upsilon_{3}} \label{pdr} = R_{12 \leftrightarrow 3}
\end{equation}
which is the detailed balance relation for the process of formation and
decay of unstable particle. Therefore chemical equilibrium
${\Upsilon_{1}\Upsilon _{2}}={\Upsilon _{3}}$ corresponds to equal decay and
production rates as we expected. Using this relation, Eq.(\ref{popeq}) can
be written in the form of Eq.(\ref{fe}).

Given a thermal bath with a fixed temperature $T $, we wish that the change of number density
is related directly to the change of fugacity. This is achieved by defining the decay time  by
\begin{equation}
\tau _{3}\equiv \frac{dn_{3}/d\Upsilon _{3}}{A},  \label{tau}
\end{equation}
where:
\begin{equation}
A=\frac{1}{\Upsilon _{3}}\frac{dW_{3\rightarrow 12}}{dVdt} \label{A}
\end{equation}
Therefore, from Eq.(\ref{fe})
the time derivative of the fugacity of the particle $3$ is:
\begin{equation}
\dot{\Upsilon}_{3}=({\Upsilon _{1}\Upsilon _{2}}-{\Upsilon _{3}})\frac{1}{\tau _{1}}.  \label{U1_dot}
\end{equation}

For the case where the abundances of $1$ and $2$ are determined only from
the reactions $\left( 3\leftrightarrow 1+2\right) $, then analogous
expressions for particles $1$, $2$ are obtained by introducing $\tau _{i}$
for each particles as
\begin{equation}
\tau _{i}=\frac{dn_{i}/d\Upsilon _{i}}{A}.  \label{taui}
\end{equation}

\section{Calculations of invariant decay (production) rate} \label{3}

\subsection{General case}
%%%%%%%%%%%%%%%%%%%%%%%%%%%%%%%%%%%%%%%%%%%%%%%%%%%%%%%%%%

The vacuum decay width of particle  $3$ in its own rest frame is found in   textbooks. In our notation:
\begin{align}
\frac{1}{\tau _{0}}& =\frac{1}{2m_{3}}\frac{1}{1+I}
\int \frac{d^{3}p_{1}}{2E_{1}\left( 2\pi \right) ^{3}}
\int \frac{d^{3}p_{2}}{2E_{2}\left( 2\pi \right)^{3}}
\left( 2\pi \right) ^{4}\delta ^{4}\left( p_{1}+p_{2}-p_{3}\right)\frac{1}{g_3}
\sum_{spin}\left\vert \langle p_{1}p_{2}\left\vert M\right\vert p_{3}\rangle \right\vert ^{2}  \notag \\
& =\frac{1}{2m_{3}g_3}\frac{1}{4\left( I+1\right) \left( 2\pi \right) ^{2}}
\int\frac{d^{3}p}{E_{1}E_{2}}\delta (E_{1}+E_{2}-m_{3})\sum_{spin}\left\vert
\langle \vec{p},-\vec{p}\left\vert M\right\vert m_{3}\rangle \right\vert ^{2}
\notag \\
& =\frac{1}{2m_{3}g_3}\frac{1}{4(I+1)}\frac{p}{\pi m_3}\sum_{spin}\left\vert
\langle \vec{p},-\vec{p}\left\vert M\right\vert m_{3}\rangle \right\vert ^{2} \label{VacTau}
\end{align}
Here $p=p_1=p_2$ and $E_{1,2} =\sqrt{p^2+m^2_{1,2}}$
are the magnitude of the momentum and, respectively, the energy, of particles $1$ and $2$ in the
rest frame of the particle $3$. From energy conservation:
\begin{equation}
E_{1,2}=\frac{m_{3}^{2}\pm (m_{1}^{2}-m_{2}^{2})}{2m_{3}}, \quad
p^2= E_{1,2}^{2}-m_{1,2}^{2} =\frac{m_3^2}{  4} -\frac{m_i^2+m_2^2}{2}+\frac{(m_1^2-m_2^2)^2}{4m_3^2}. \label{encon}
\end{equation}

We denote   by $\tau_3^{\prime }$    the decay rate of the particle $3$ in the  rest frame of   the
thermal bath,  $E_{3}$ and $p_{3}$ are  the corresponding energy and the
momentum.  The thermal decay rate per unit volume $dW_{3\rightarrow 1+2}/dVdt$
should then be the average (over the inverse of this life time) in
the thermal bath frame:
\begin{equation}
\frac{dW_{3\rightarrow 1+2}}{dVdt}=\frac{g_3}{\left( 2\pi \right) ^{3}}\int
d^{3}p_{3}f_{b,f}(\Upsilon _{3},p_{3})\frac{m_{3}}{E_{3}}\frac{1}{\tau^{\prime }_3},  \label{Decay1}
\end{equation}%
where $E_{3}\tau ^{\prime }_3/m_3$ is the decay time of the particle $3$ with
moment $p_{3}$.

Comparing this expression Eq.(\ref{Decay1}) with the complete Eq.(\ref{drg1}), we conclude
that the in medium, at finite temperature $T$,  decay rate $\tau_3^{\prime }$  is given by:
\begin{align}
\frac{1}{\tau ^{\prime }_3}& =\frac{1}{2m_{3}}\frac{1}{1+I} \int \frac{
d^{3}p_{1}}{2E_{1}\left( 2\pi \right) ^{3}}\int \frac{d^{2}p_{2}}{2E_{2}\left(2\pi \right) ^{3}}
 \left( 2\pi \right) ^{4}\delta ^{4}\left(p_{1}+p_{2}-p_{3}\right)\times  \notag \\
&
\frac{1}{g_3}\sum_{spin}\left\vert \langle p_{1}p_{2}\left\vert
M\right\vert p_{3}\rangle \right\vert ^{2} f_{b,f}(\Upsilon _{1},p_{1})f_{b,f}(\Upsilon _{2},p_{2})
\Upsilon_{1}^{-1}\Upsilon _{2}^{-1}\exp (u\cdot p_{3}/T),  \label{tau23}
\end{align}%
which is a Lorentz invariant form. We note that  $ u\cdot p_{3}=E_3$ denotes the energy of the
particle $3$ in the rest frame of the bath.

Using the vacuum rest-frame decay time, Eq.(\ref {VacTau}), we find that  Eq.(\ref{tau23}) takes the form:
\begin{equation}
\frac{1}{\tau ^{\prime }_3}=\frac{1}{\tau _{0}}\frac{e^{E_{3}/T}}{2}\Phi (p_{3}).
\label{tau-tau0}
\end{equation}%
 The function $\Phi (p_{3})$ is:
\begin{equation}
\Phi \left( p_{3}\right) =\int_{-1}^{1}d\zeta
\frac{\Upsilon_{1}^{-1}}{\Upsilon _{1}^{-1}e^{\left( a_{1}-b\zeta \right) }\pm 1}
\frac{\Upsilon_{2}^{-1}}{\Upsilon _{2}^{-1}e^{\left( a_{2}+b\zeta \right) }\pm 1}.
\label{phif}
\end{equation}
with
\begin{equation}
a_{1}=\frac{E_{1}E_{3}}{m_{3}T}, \quad
a_{2} =\frac{E_{2}E_{3}}{m_{3}T}, \quad
b =\frac{pp_{3}}{m_{3}T}\quad {\rm and}\quad
\zeta =\cos \theta =\cos (\vec{p}_{2}\wedge \vec{p}_{1}).
\end{equation}
%and $\zeta =\cos \theta =\cos (\vec{p}_{2}\wedge \vec{p}_{1})$.

The integral $\Phi(p_3)$ can be evaluated analitically.
The integrant in this equation ($\zeta=x$) is even, therefore
\begin{equation}
\Phi(p_{\pi})=2\int_0^{1}dx
\frac{\Upsilon_{1}^{-1}}{\Upsilon _{1}^{-1}e^{\left( a_{1}-bx \right) }\pm 1}
\frac{\Upsilon_{2}^{-1}}{\Upsilon _{2}^{-1}e^{\left( a_{2}+bx\right) }\pm 1}%\\ \nonumber
=\int_0^{1}dx\frac{e^{bx}}{\Upsilon _{1}^{-1}e^{a_1}\pm e^{bx}}\frac{2e^{-a_2}\Upsilon _{1}^{-1}}{e^{bx}\pm \Upsilon _{2} e^{-a_2}}
\end{equation}
Introducing $y=e^{bx}$ this integral can be written as

\begin{eqnarray}
\Phi(p_{\pi})&=&\frac{2e^{-a_2}}{b}\int_1^{e^b}dy\frac{\Upsilon_{1}^{-1}}{(\Upsilon_{1}^{-1}e^{a_1}\pm y)(y\pm \Upsilon_{2}e^{-a_2})}
\nonumber\\
\nonumber\\
&=&\frac{2}{b(e^{a_1+a_2}-\Upsilon_{1}\Upsilon_{2})}\int_1^{e^b}dy\left(\frac{1}{(\Upsilon_{1}^{-1}e^{a_1}\pm y)}+\frac{1}{(y\pm \Upsilon_{2}e^{-a_2})}\right)
\nonumber\\
\nonumber\\
\nonumber\\
&=&\frac{2}{b(e^{a_1+a_2}-\Upsilon_{1}\Upsilon_{2})}\ln\left(\frac{(\Upsilon_{2}^{-1}e^b \pm e^{-a_2})(\Upsilon_{1}\pm e^{a_1})}{(\Upsilon_{1}e^b\pm e^{a_1})(\Upsilon^{-1}_{2} \pm  e^{-a_2})}
\right).\label{phiab}
\end{eqnarray}
The result is
\begin{equation}
 \Phi(p_3)=\frac{1}{b(e^{a_1+a_2} \pm \Upsilon_1\Upsilon_2)}\ln\frac{\left(e^{-a_2} \pm \Upsilon_2^{-1}e^b\right)
\left(e^{a_1} \pm \Upsilon_1\right)}{\left(e^{-a_2}\pm \Upsilon_2^{-1}\right)\left(e^{a_1} \pm \Upsilon_1e^b\right)}
. \label{phia1a2b}
\end{equation}
We note that in the non-relativistic limit ($m_{3}\gg T, p_3$), this quantity
tends to
\begin{equation}
\Phi (0) = 2\frac{\Upsilon _{1}^{-1}\Upsilon_{2}^{-1}}
 {(\Upsilon_{1}^{-1}e^{E_{1}/T}\pm 1)(\Upsilon _{2}^{-1}e^{E_{2}/T}\pm 1)}.
\label{philim}
\end{equation}
Finally, the average particle $3$ decay rate per unit volume in a thermally
equilibrated system is given by
\begin{equation}
\frac{dW_{3\rightarrow 1+2}}{dVdt}=\frac{g_3}{\left(2\pi^{2}\right) }
\frac{m_{3}}{\tau _{0}}\int_{0}^{\infty }\frac{p_{3}^{2}dp_{3}}{E_{3}}
\frac{e^{E_{3}/T}}{\Upsilon_{3}^{-1}e^{E_{3}/T}\pm 1}\Phi (p_{3}),  \label{Decay1-final}
\end{equation}

%%%%%%%%%%%%%%%%%%%%%%%%%%%%%%%%%%%%%%%%%%%%%%%%%%%%%%%%%%%%%%%%%%%%%%%%%%%%%%
\subsection{Decay and production rates in Boltzmann limit}
%%%%%%%%%%%%%%%%%%%%%%%%%%%%%%%%%%%%%%%%%%%%%%%%%%%%%%%%%%%%%%%%%%%%%%%%%%%%%%

The equations become much simpler in case of Boltzmann limit when we can
omit unity in distributions Eq.(\ref{bf}). This is possible when
\begin{equation}
\Upsilon _{i}^{-1}e^{u\cdot p_{i}/T}\gg 1,  \label{Boltzmann}
\end{equation}%
that is, when $\Upsilon _{i}\ll 1$ or $T\ll m_{1}/2$. The condition $T\ll m_{1}/2$
comes from fact that the minimal energy of lighter particles is
$m_{1}/2$ in the particle $1$ rest frame. In this limit the decay time in the
particle $1$ rest frame from Eq.(\ref{tau23}) $\tau ^{\prime }\rightarrow\tau _{0}$
 so that from Eq.(\ref{tau}) we have for the average decay rate
$\tau $ in the reference frame (the rest frame of the bath) as
\begin{eqnarray}
\tau_3^\prime  &\approx &\tau _{0}
\frac{\int_{0}^{\infty }p^{2}dp \,e^{{E_{3}}/{T}}}
        {\int_{0}^{\infty }p^{2}dp \, e^{{E_{3}}/{T}}{m_{3}}/E_{3}} \\[0.3cm]
&=&\frac{\tau _{0}}{\left\langle 1/\gamma \right\rangle} = \tau _{0}\frac{K_{2}(m_{1}/T)}{K_{1}(m_{1}/T)}.  \label{taubl}
\end{eqnarray}
As we see in Eq.(\ref{Decay1}), the average decay time $\tau_3^\prime $ in lab frame
is proportional to the  (inverse) average of Lorentz factor of particle
$3$. We will discuss this effect next    in quantitative manner, and
note that the ratio of $\tau_3^\prime  $ to $\tau _{0}$ is shown
in figure {\ref{taurho}} as dash-dot line. For $T\ll m_{3}$ this ratio goes to
unity because the Lorentz factor becomes $1$. For large $T $, the
rate increases because of the larger average energy of particle $3$
or equivalently the larger average Lorentz factor. Therefore, if we
have small number of all particles ($\Upsilon _{i}\ll 1$) so that
Eq.(\ref{Boltzmann}) is yet valid for $T>m_{3},$ the average
particle life time increases with $T$ due to relativistic effects.
%%%%%%%%%%%%%%%%%%%%%%%%%%%%%%%%%%%%%%%%%%%%%%%%%%%%%%%%%%%%%%%%%%%%%%%%%%%%%%
\section{Examples} \label{4}

\subsection{Production of $\protect\rho $ mesons via $\protect\rho %
\leftrightarrow \protect\pi \protect\pi $ process}

Here we consider example of $\rho $ meson thermal decay and production:
\begin{eqnarray}
&&\rho ^{0}\leftrightarrow \pi ^{+}+\pi ^{-},  \label{ropi++} \\
&&\rho ^{\pm }\leftrightarrow \pi ^{\pm }+\pi ^{0}.  \label{ropi+0}
\end{eqnarray}%
We consider the pions to be in chemical equilibrium with chemical potential
$\mu _{\pi}=0$. In this case all particles are bosons and $m_{1}=m_{2}$ and
in integral (\ref{phif}) we have $E_{1}=E_{2}=m_{\rho }/2$ in $\rho$ rest
frame. Integrant in $\Phi(p)$ function is symmetric function. Then we can write
\begin{equation}
\Phi \left( p_{\rho }\right) =2\int_{0}^{1}d\zeta
\frac{\Upsilon _{\pi }^{-2}}{\Upsilon _{\pi }^{-1}e^{\left( a-b\zeta \right) }-1}
\frac{1}{\Upsilon_{\pi }^{-1}e^{\left( a+b\zeta \right) }-1}.  \label{phif2}
\end{equation}
where
\begin{eqnarray}
a &=&\frac{\sqrt{m_{\rho }^{2}+p_{\rho }^{2}}}{2T};\\
b &=&\frac{\sqrt{1-4m_{\pi }^{2}/m_{\rho }^{2}}p_{\rho }}{2T}.
\end{eqnarray}
The integral (\ref{phif2}) can be evaluated in this case as
\begin{equation}
\Phi (p_{\rho })=\frac{2\Upsilon _{\pi }^{-2}}{b(\Upsilon _{\pi
}^{-2}e^{2a}-1)}\left( b+\ln \left( 1+\frac{ \Upsilon _{\pi
}\left(e^{(b-a)}-e^{-(a+b)}\right) }{\left(1-\Upsilon_{\pi }e^{b-a}\right) }\right) \right).  \label{phiab}
\end{equation}%
Then we substitute $\Phi$ into Eq.(\ref{Decay1}) and using Eq.(\ref{pdr})
we can calculate $\rho $ decay and production rates. To calculate $\tau$
we use definition (\ref{tau}).

%%%%%%%%%%%%%%%%%%%%%%%%%%%%%%
\begin{figure}[tbp]
\centering \includegraphics[width=8.6cm,height=8.5cm]{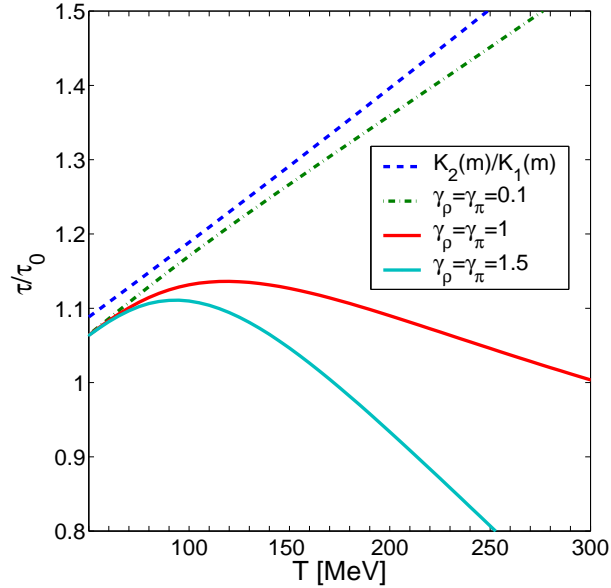}
\caption{\small The ratio $\protect\tau_3 /\protect\tau _{0}$
as a function of temperature $T$   in the
reaction $\rho\leftrightarrow \pi\pi$. The  dashed line is
for Boltzmann limit showing just time dilation. Nearly this limit arises (dot-dashed line)
for   $\Upsilon_{\rho}=\Upsilon_{\pi}=0.1$. Solid lines are   for $\protect\Upsilon_{\rho} = \Upsilon_{\pi}=1$ (top, red)
and   $\Upsilon_{\rho} =\Upsilon_{\pi}= 1.5 $  (bottom). }\label{taurho}
\end{figure}
%%%%%%%%%%%%%%%%%%%%%%%%%%%

In figure {\ref{taurho}} we presents $\rho $ decay time in lab frame as a
function of temperature $T$ for $\Upsilon _{\rho }=\Upsilon _{\pi }=1$,
solid line, for $\Upsilon _{\rho }=\Upsilon _{\pi }=1.5$, solid with dot
marker, $\Upsilon _{\rho }=\Upsilon _{\pi }=0.1$, dashed line and dash-dot
line is for Boltzmann limit Eq.(\ref{taubl}). We consider range of
temperatures between $50$ and $300$ MeV which includes quark gluon plasma
hadronization temperature ($\approx$ 140 -180 MeV). We show case $\Upsilon _{\rho }=\Upsilon _{\pi
}=0.1$ to check transition to Boltzmann limit. We can see that for this case
result is close to Boltzmann approximation for our range of $T$ as it is
expected. In case $\Upsilon _{\rho }=\Upsilon _{\pi }=1$ we have chemical
equilibrium. In this case and for $\Upsilon _{\rho }=\Upsilon _{\pi }=1.5$
for small $T\ll m_{\rho }/2$ we have ratio $\tau /\tau _{0}$ near Boltzmann
limit, near unity. For such small $T$, when Boltzmann limit is applied,
decay time $\tau $ doesn't depend on $\Upsilon $. When $T$ increases quantum
effects take place then $\tau $ begin to decrease with increase T. The larger $\Upsilon$ the
faster $\tau$ decreases with temperature.

The case with $\Upsilon_{\pi}=\Upsilon_{\rho} \approx 1.5$ can take place after quark qluon plasma
hadronization. Light hadrons multiplicities has to be above chemical equilibrium for hadronization
temperature smaller than $180$ MeV to conserve entropy during hadronization.

%%%%%%%%%%%%%%%%%%%%%%%%%%%%%%%%%%%%%%%%%%%%%%%%%%%%%%%%%%%%%%%%%%%%%%%%%%%%%%%%%%%%%%%%%%%%%%%%%%%%%%%%%%%%
\subsection{Baryon resonance ($\Sigma(1385)$) lifespan calculations in dense hadronic gas}\label{chapter 2.3}
%%%%%%%%%%%%%%%%%%%%%%%%%%%%%%%%%%%%%%%%%%%%%%%%%%%%%%%%%%%%%%%%%%%%%%%%%%%%%%%%%%%%%%%%%%%%%%%%%%%%%%%%%%%%%%%%

In this subsection we consider the effect of oversuturated pion component in hadronic gas and the effect of the motion of the decaying
resonance with respect to the thermal rest frame on its lifespan and 
then also on resonance production relaxation time, considering example $\Sigma(1385) \leftrightarrow \Lambda\pi$.

%%%%%%%%%%%%%%%%%%%%%%%%%%%%%%  Figure 3
\begin{figure}[tbp]
\centering \includegraphics[width=8.5cm,height=8.5cm]{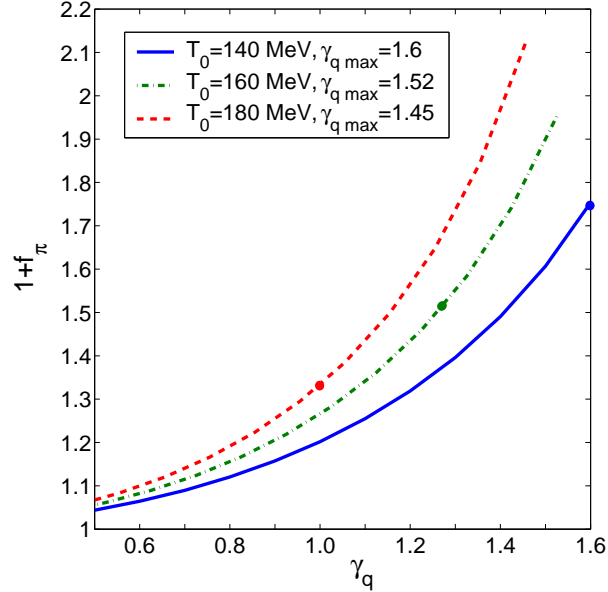}
\vskip -0.31cm
\caption{\small (color on line) The Bose enhancement factor $1+f_{\pi}(E^{*}_1)$
in $\Sigma(1385)$ rest frame
as a function of light quark fugacity $\gamma_q$ for the
reaction $\Sigma(1385) \leftrightarrow \Lambda \pi$ at $T=140$ MeV (blue, solid line),
at $160$ MeV (green, dash-dot line) and $180$ MeV (red, dashed line). The dots
show  the initial value of  fugacities  for the three possible
hadronization cases.} \label{bose140}
\end{figure}
%%%%%%%%%%%%%%%%%%%%%%%%%%%

For the temperatures of interest
(hadronization of QGP and below) $m_{\Lambda}$ and $m_{\Sigma} >>
T$. In this case with sufficient accuracy we can  rewrite function $\Phi(p_3)$ as
\begin{equation}
\Phi(p_3)\simeq \frac{1}{be^{E_{3}/T}}\ln \frac{\left(e^{a_1+b} - \Upsilon_{\pi}\right)}{\left(e^{a_1-b} -\Upsilon_{\pi}\right)}.
\label{phia1a2bfnr}
\end{equation}
Here fugacities for $\Lambda$ and $\pi$  correspond to those for
particles 1 and 2, respectively. 
There are no significant medium  effects upon decay rate of
$\Sigma(1385)$ and $\Lambda$  resonances. However the pions have
energy $E^{*}_2 = 250$ MeV (Eq.(\ref{encon})) in the $\Sigma$ rest
frame and the Bose enhancement effect is possible in the
oversaturated   hadronic gas after QGP hadronization.

For the  low temperatures considered here we can assume that
$\Sigma$ resonances almost do not move.  Thus the enhancement effect
in the thermal bath  frame is close to the enhancement in the
$\Sigma(1385)$ rest frame. The decay  rate increases by Bose
enhancement factor $1+f_{\pi}$ (here $f_{\pi}=f_{\pi}(E^{*}_2, T)$).
In figure \ref{bose140} we show Bose enhancement factor as a
function of light quark fugacity $\gamma_q$ for temperature
$T_0=140$ MeV (blue, solid line), $T_0=160$ MeV (green, dash-dot
line), $T_0=180$ MeV (red, dashed line). The large dots show Bose
enhancement factor for our initial $\gamma_q$ determined from
entropy conservation in  fast hadronization. The fugacity
$\gamma_q=1.6$ is close to maximum expected value at $T_0=140$ MeV.
The maximum fugacities for each temperature correspond to Bose -
Einstein singularity. The Bose enhancement effect is largest for
maximum $\gamma_q$ and it diminishes  for small $\gamma_q$. At fixed
entropy the greatest enhancement is for smallest ambient
temperature, see the dot on solid line in figure \ref{bose140}.

%%%%%%%%%%%%%%%%%%%%%%%%%%%%%% Figure 4
\begin{figure}
\centering \includegraphics[width=9.cm,height=9.cm]{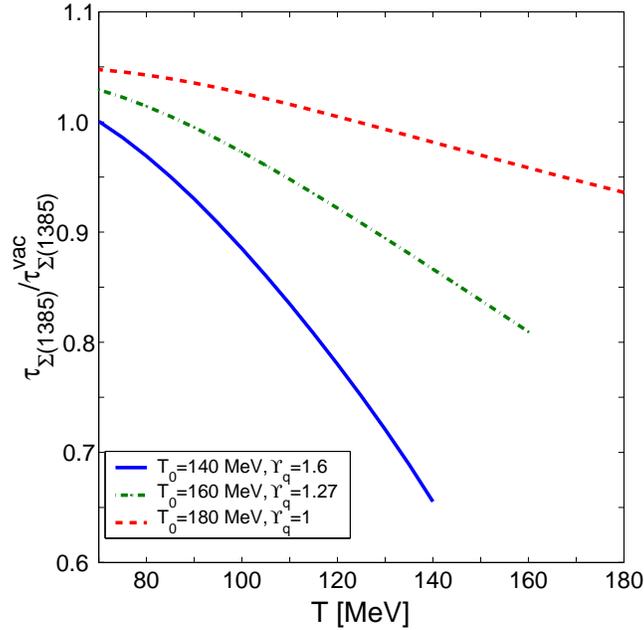}
\vskip -0.31cm
\caption{\small    (color on line)
The ratio  of the in medium lifespan  $\protect\tau_{3}$  with the vacuum
lifespan $\tau _{0}$ as a function of temperature $T$ for the reaction
$\Sigma(1385)\leftrightarrow \Lambda\pi$. The  dashed (red) line is for
hadronization at $T_0=180$ MeV, $\gamma_q=1.0$; the dot-dashed line (green)
for hadronization at $160$ MeV, $\gamma_q=1.27$; solid line (blue) is for
hadronization at $140$ MeV and   $\gamma_{q}= 1.6$.}\label{tausig}
\end{figure}
%%%%%%%%%%%%%%%%%%%%%%%%%%%

In figure~\ref{tausig} we show the corresponding decrease in the
lifespan, the ratio $\protect\tau_{3}/\protect\tau _{0}$ as a
function of temperature $T$ in the reaction
$\Sigma(1385)\leftrightarrow \Lambda\pi$. We consider temperature
range from corresponding hadronization temperature until $T=70$ MeV.
We assumed, that $\Upsilon_{\pi}$ is a constant. Fugacities of heavy
resonances do not influence the result. The lowest $\tau_3/\tau_0$
ratio is for $\gamma_q=1.6$ at $T_0=140$ MeV when we have maximum
value of $\gamma_q$ for given temperature. If we compare this value
of $\tau_3/\tau_0 = 0.65$ with inverse Bose enhancement factor
$1/(1+f_{\pi}(E_2^*, T)) = 0.54$ for this $T$ and $\gamma_q$ (see
figure \ref{bose140}) we see that these values are near to each
other (difference is about 20\% ) as expected for $m_{\Sigma}>>T$.
For smaller $T$, $\gamma_q$ decay time goes to its vacuum value.

The same calculations are applicable for heavier $\Sigma^*$.  When
the difference of mass of the initial and final state resonance
decreases, the Bose enhancement effect increases, since it involves
small momenta. The largest effect is for reaction $\Sigma(1670)
\leftrightarrow \Lambda(1520) + \pi$. On the other hand, for the
reactions which satisfy condition $m_3-(m_1+m_2)>300$ MeV the enhancement
effect becomes very small. 

%%%%%%%%%%%%%%%%%%%%%%%%%%%%%%%%%%%%%%%%%%%%%%%
\subsection{Thermal Production of $\protect\pi ^{0}$}\label{pi0pr}

As mentioned in the Introduction, it is interesting to examine the mean life
time of $\pi^{0}$ in the end of hadronic gas stage of the universe where
the temperature becomes several tens of MeV. Then the $\pi^0$ production in two photons fusion,
Eq.(\ref{ggpi0}) determines the abundance of $\pi ^{0}$.

The difference with previous example is that the photons are
massless and they are in chemical equilibrium
($\Upsilon_{1}=\Upsilon_2=1$). Then we can rewrite function
(\ref{phia1a2b}) as
\begin{equation}
\Phi (p_{\rho })=\frac{2}{b(e^{2a}-1)}\left( b+\ln \left(1+\frac{\left(
e^{(b-a)}-e^{-(a+b)}\right)}{\left(1-e^{b-a}\right) }%
\right) \right).  \label{phiab1}
\end{equation}%
with
\begin{eqnarray}
a &=&\frac{\sqrt{m_{\pi^0 }^{2}+p_{\pi^0 }^{2}}}{2T}; \\
b &=&\frac{p_{\pi^0}}{2T}.
\end{eqnarray}%
Again we use Eq.(\ref{Decay1}) and (\ref{pdr})
we can calculate $\pi_0$ decay and production rates. To calculate $\tau$
we use definition (\ref{tau}).

%%%%%%%%%%%%%%%%%%%%%%%%%%%%%%
\begin{figure}[tbp]
\centering \includegraphics[width=8.6cm,height = 8.5cm]{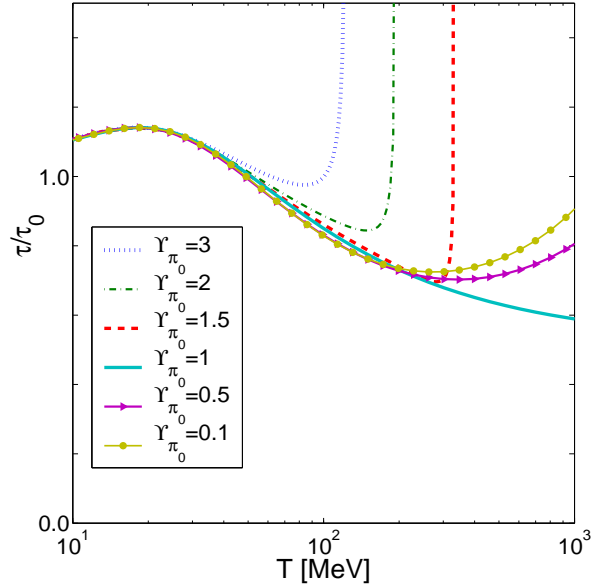}
\caption{{\protect\small {The ratio $\protect\tau /\protect\tau
_{0}$ for $\pi^0$ decay/production as a function of temperature $T$.
Dotted, blue line is for $\protect\Upsilon_{\pi^0}=3$, dash-dot,
green line is for $\protect\Upsilon_{\pi^0}=2$, dashed, red line is
for $\protect\Upsilon_{\pi^0}=1.5$, solid, turquoise line is for
$\Upsilon_{\pi^0}=1$, purple solid line with triangle markers is for
$\Upsilon_{\pi^0}=0.5$, brown solid line with dot markers is for
$\Upsilon_{\pi^0}=0.1$.}}} \label{taupi0}
\end{figure}
%%%%%%%%%%%%%%%%%%%%%%%%%%%

In figure \ref{taupi0} we show ratio of $\pi^0$ decay time in the
presence of thermal particles to the decay time in vacuum in $\pi_0$
rest frame: $\tau_{\pi^0}/\tau^0_{\pi^0}$. In this figure the large
range of temperature is shown $10-10^3$ MeV. For
$\Upsilon_{\pi^0}=1$ the ratio $\tau_{\pi^0}/\tau^0_{\pi^0}$ the
temperature dependence is similar to that for $\rho$ decay,
considered in previous chapter. It increases at first until
relativistic effects become noticeable. Then, after $T\approx
$,$\tau$ goes slowly down with temperature, when quantum in-medium
effect becomes important. Range of change of $\tau$ is not large for
this large range of temperature. The smallest
$\tau_{\pi^0}/\tau^0_{\pi_0}$ is about 0.6 at $T=10^3$ and the
maximal value of this ratio is about 1.2.

The cases when $\Upsilon_{\pi} \ne 1$ are different from those considered for $\rho$ decay because photons
are stay in chemical equilibrium. When $\Upsilon_{\pi_0}<1$, $\Upsilon_{\pi_0}=0.5$ and  $\Upsilon_{\pi_0}=0.1$ (purple line with dots and brown line with tringles, respectively), the $\tau$ also begins to decrease slowly after $T \approx 20$ MeV because of quantum effect from photons distribution. Then for $T>300$ it is slowly increases because relativistic effects becomes slightly dominant.

When $\Upsilon_{\pi_0}>1$, $\Upsilon_{\pi_0}=1.5$ (red, dashed line), $\Upsilon_{\pi_0}=2.0$ (green, dash-dot line) and $\Upsilon_{\pi_0}=3.0$ (blue, dotted line), there is Bose - Eistein critical point when
\begin{equation}
T=m_{\pi^0}/\log(\lambda).
\end{equation}
$dn_{\pi^0}/d{\lambda}$ is increasing faster near this critical point than $\pi^0$ production rate. Decay relaxation time $\tau_{\pi^0}$ goes sharply up, diverges, near critical point.

\begin{figure}
\centering
\includegraphics[width=8.6cm,height=8.5cm]{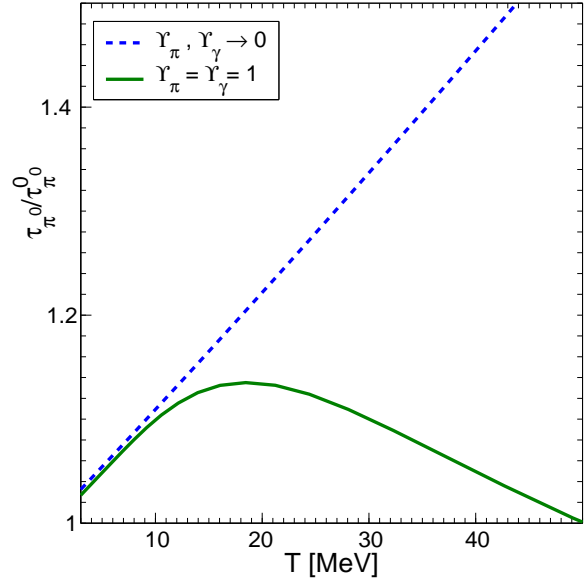}
\caption{\small{The ratios $\tau_{\pi^0}/\tau^0_{\pi_0}$ as
functions of temperature $T$ for relativistic Boltzmann limit  (blue, dashed
line) and for quantum distribution in chemical equilibrium, $\Upsilon_{\pi}=\Upsilon_{\gamma}=1$
(green, solid line).}} \label{taupi0app}
\end{figure}

In figure \ref{taupi0app} we show $\tau_{\pi^0}$ for more realistic temperatures. This temperature range can be interesting for early universe 
evolution and for $e^+e^-\gamma$ plasma created by laser pulse. It turns out that there are
both relativistic and quantum effects which contribute and they (nearly) cancel at this range of temperature.
The relativistic effect arises because $\tau_{\pi^0}$ in Eq.(\ref{taupi0}) is in lab frame while the
known $\tau^0_{\pi^0}$ is in the pion rest frame. In the relativistic
Boltzmann limit the correction is obtained considering the related time dilation effect Eq.~(\ref{taubl}).
We find that this effect implies that
$\tau_{\pi^0}$ in the lab frame increases with temperature. This effect is shown by dashed
 (blue) line   in  figure \ref{taupi0app}. Furthermore,
with increasing  temperature quantum distribution functions for photons and   for
the  produced particle need to be considered. This leads to the result shown as solid line
(green)  in  figure \ref{taupi0app}. Thus in general $ \tau_{\pi^0} >  \tau_{\pi^0}^0$, by up to 14\%.

%%%%%%%%%%%%%%%%%%%%%%%%%%%%%%
%%%%%%%%%%%%%%%%%%%%%%%%%%%

\section{Conclusions and Discussion}

In this chapter, we examined in detail the kinetic master equation for the
process involving formation of an unstable particle through the reaction Eq.(%
\ref{123}) in a relativistically covariant fashion. Assuming that
all the particles in the process are in thermal equilibrium, we
calculate the thermal averaged decay and formation rate of the
unstable particle based on the BUU equation. Using the time reversal
symmetry, we show that the time evolution of the density of the
unstable particle as Eq.(\ref{fe}).
Therefore in chemical equilibrium particles fugacities are connected by Eq.(%
\ref{equilcon1}) as expected. We have explicit the thermal decay rate
of unstable particle, obtaining Eq.(\ref{Decay1-final}), which is
our principal result.

Using the formalism developed above, we examined the general properties of
the thermal particle decay/production rate. We see that for $T\ll m_{i}$
where the Boltzmann limit can be applied, the decay width is reduced to $%
\Upsilon _{1}/\tau _{0}$ and production width is $\Upsilon _{2}\Upsilon
_{3}/\tau _{0}$. For larger values of $T$ but $\Upsilon _{i}\ll 1$ so that
the Boltzmann approximation is valid, then decay width and production width
tend simply to $\Upsilon _{1}/\tau $ and $\Upsilon _{2}\Upsilon _{3}/\tau $,
respectively, where $\tau $ is essentially proportional to average Lorentz
factor and doesn't depend on $\Upsilon _{i}$. When some of $m_{i}/T$ and $%
\Upsilon _{i}$ are about unity or larger we see dependence of $\tau $ on $%
\Upsilon _{i}$.

We applied our formalism to $3$ examples, $\rho \leftrightarrow \pi +\pi $,
$\Sigma(1385) \leftrightarrow$ and $\pi ^{0}\leftrightarrow \gamma +\gamma .$ 
The first and second processes can take
place both in a hot hadronic gas created by the heavy ion collisions and in
the expanding early Universe. In particular for the heavy ion reaction case,
our analysis, coupled to the hydrodynamical expansion of the system will
furnish additional information of the dynamics of the system. We will study baryon resonances
evolution in heavy ions collisions in next chapter.
The relaxation time for $\pi^0$ decay remains close (within 50\%)  to relaxation
time in vacuum for large temperature range. In chapter~\ref{eeg} we will apply this for $\pi^0$
evolution in $e^+e^-\gamma$ plasma, created by the intensive laser pulse and in early universe. 
%%%%%%%%%%%%%%%%%%%%%%%%%%%%%%%%%%%%%%%%%%%%%%%%%%%%%%%%%%%%%%%%%%%%%%%%%%

\chapter{RESONANCE PRODUCTION IN HEAVY IONS COLLISIONS} \label{respr}
%%%%%%%%%%%%%%%%%%%%%%%%%%%%%%%%%%%%%%%%%%%%%%%%%%%%%%%%%%%%%%%%%%%%%%%%%%%%

\section{Introduction}
Hadron resonances are produced copiously in the quark-gluon plasma  (QGP)
fireball break up into hadrons (hadronization, chemical freeze-out)
e.g. at RHIC~\cite{Adams:2006yu,Salur:2006jq, Markert:2007qg,Witt:2007xa}.
Within the  statistical hadronization
model (SHM) approach~\cite{Torrieri:2004zz, Torrieri:2006xi},
the initial  yields are described by  chemical fugacities $\Upsilon$, and
hadronization temperature$\,T$. The production of heavy resonances
is suppressed exponentially in $m/T$.  Once  formed, resonances
decay. If this occurs inside matter, detailed balance requires also production
of resonances, called `regeneration' and/or `back-reaction'.

If the chemical freeze-out  occurs much earlier than thermal, the initially produced
resonances are practically invisible due to rescattering
of decay products~\cite{Rafelski:2001hp}. The observed yield of resonances
is fixed by the physical conditions prevailing at the  final  breakup of the fireball,
 at which time last scattering occurs, this is the  `kinetic freeze-out'.
The present work addresses two  questions:\\
a)  how observable  resonance
yield depends on the difference between chemical freeze-out temperature (e.g. point of hadronization of QGP)
 and  the kinetic freeze-out temperature;\\
b) how this yield depends on the degree of initial chemical non-equilibrium at hadronization. \\
One can see this work as an effort to improve on the concept of chemical freeze-out for
the case of resonances: given the relatively fast  reactions   their yield remains sensitive to  the
conditions prevailing  between chemical and thermal freeze-out, even if this time is just 1\,fm/c.

Hadron resonances are observed in a surprisingly large yield when a
quark-gluon plasma  (QGP) fireball breaks up into
hadrons~\cite{Markert:2002xi,Adams:2006yu,Salur:2006jq,Markert:2007qg,Witt:2007xa,Abelev:2008yz}.
This is unexpected, since  the invariant mass signature formed from
decay products could be erased by  rescattering of the strongly
interacting decay products~\cite{Rafelski:2001hp}.
In order to describe evolution of the resonance abundance
one can  perform a microscopic transport simulation of the expanding system.
In this approach the  regeneration of resonances
was previously studied by Bleicher and collaborators~\cite{Bleicher:2002dm, Bleicher:2003ij, Vogel:2006rm}.
There are many detailed features of particle interactions to resolve in a microscopic model description
and thus it seems appropriate to   simplify the situation. We study  resonance decay and regeneration  
using the momentum integrated  population master equations, and assuming  hydrodynamic expansion
inspired model of fireball dynamics with conserved  entropy content.
In all our considerations we presume that
the yield of pions $\pi$  is so large
that we can assume it to be unaffected by any of the reactions we
consider, thus we fix pion yield in terms of fugacity and temperature values.
As a result, the final short lifespan resonance yield can be considerably different from statistical
hadron gas (SHG) benchmark expectation. 

The other result, we obtain here, is that the long lived  resonances, such as
$\Lambda(1520)$, can be considerably suppressed in their yield. This effect is
amplified for the case when the initial  hadron fugacities, and thus
particle yields,  are  above chemical equilibrium. This situation is
expected for a hadronizing QGP phase. The low
$\Lambda(1520)$ yield  has been  reported
both in RHIC and SPS experiments~\cite{Markert:2002xi,Adams:2006yu}.

Here we consider two models. The first simplified model we apply for $\Delta(1232)/N_{tot}$ and $\Sigma(1385)/\Lambda_{tot}$ ratios calculation.
In this model we consider only one dominant (fastest) reaction for each resonance:
\begin{eqnarray}
\Delta(1232) \leftrightarrow N+ \pi; \label{dnp}\\
\Sigma(1385) \leftrightarrow \Lambda + \pi.  \label{L1115L}
\end{eqnarray}
Then, for the `fast' baryon resonances considered here  we keep the sum of  yields  constant:
\begin{eqnarray}\label{partcon}
\Delta+ N&=&\Delta_0+N_0\equiv N^{\rm tot}_0 ={\rm Const.}, \\
\Sigma (1385)+  \Lambda & = &\Sigma_0(1385) + \Lambda_0\equiv \Lambda^{\rm tot}_0  ={\rm Const.} . \nonumber
\end{eqnarray}
The  baryon  annihilation,  strangeness exchange such as $N+K\leftrightarrow \Lambda+\pi$  reactions,
and population exchanges with   higher resonances
are assumed not to have a material impact within
the time scale during which the temperature drops from chemical to kinetic freeze-out condition.

The experimentally observable hyperon yield  appearing in our final result  is
\begin{equation} \label{Ltot}
\Lambda_{\rm tot}=\Sigma (1385)+ \Lambda+\Sigma^0(1193)  + Y^*
\end{equation}
due to experimentally inseparable  $\Sigma^0(1193) \to \gamma+\Lambda$ decay and
the decay of further hyperon resonances $Y^*$. Similarly, when
we refer to  $N_{\rm tot}$ we include   baryon resonances in the count.

In this model we also do not take into account medium effect on the reaction rates. It will be added in the second model.
These effect  is described in section \ref{chapter 2.3}.

The second model we apply for $\Lambda(1520)/\Lambda_{tot}$ and $\Sigma(1385)/\Lambda_{tot}$ ratios.
In this model we include many reactions, which we will describe in section~\ref{chapter 2.1}. In case of $\Lambda(1520)$ (which is suppressed) it is necessary to consider a few reactions when
for $\Sigma(1385)$ the resut does not change much compared to first model.
The resonance suppression, or enhancement, mechanism works as follows.
In thermal hadronic gas the reaction (\ref{123}),
can occur  in both directions: the resonance decay $3\to 1+2$, and
the back-reaction (regeneration) resonance formation $1+2\to 3$.
When the reaction  goes with the same rate in both directions,  we
have chemical  detailed balance, e.g. particles yields do not change
in this period of temporal evolution of the system. This does not
necessarily mean that we have a chemical equilibrium. Instead it may
be a transient condition for which none of the three particles is
equilibrated chemically - we will show when this can happen.

In the study of resonance decay and regeneration we are using the
momentum integrated  population master equations. We assume  a
fireball  expansion model governed by hydrodynamic inspired flow
with conserved  entropy content. In our considerations we presume
that the yield of pions $\pi$ is so large that we can assume it not
to be materially affected by any of the reactions we consider. Thus
we fix pion yield in terms of an ambient fugacity and temperature
value, and in essence the total (per unit rapidity at RHIC) yield is
fixed since we conserve entropy.

An important assumption implied  below is that the rapidly expanding
hadron system maintains for the relevant particles a fully thermal
(Boltzmann) momentum distribution.
To describe the evolution of hadron abundances in the kinetic phase
we  track in time the yields of single strange hadrons after their initial formation.
This is implemented in terms of time dependence of
the chemical fugacities $\Upsilon(t)$, and  the time dependence of the
hadronization temperature $\,T(t)$.

We  look in detail at three potential evolution scenarios:\\
a) a high temperature  breakup
at $T_0\simeq 180$ MeV where the entropy content of the equilibrated QGP and HG-phase are similar;\\
b) the   $T_0\simeq 160$ MeV case where  chemical non-equilibrium among produced hadrons is already
required; and \\
c) at $T_0\simeq 140$ MeV  which is favored by descriptions of stable hadron production, and  in which case a
strong chemical non-equilibrium situation arises.

For the late stage of the expansion, at relatively low density
 the assumption of thermal momentum distribution
may not be anymore  fully satisfied. In particular pions of high
momentum could  be escaping from the fireball.  For this reason we
will consider here a second scenario, which we call ``dead
channel''.  In this scenario we assume that the reaction (\ref{pdr})
goes mainly in the direction of resonance 3 decay and the resonance
formation  is switched off for
\begin{equation}
m_3-(m_1+m_2)>300\,{\rm MeV}. \label{dchcon}
\end{equation}

Without a complete kinetic model including equilibration and
particle emission we do not know the exact energy in condition
(\ref{dchcon}) and timescale (during expansion) for which Boltzmann
distribution is violated and dead channels appear.  It is possible
that reality lies between the two cases (kinetic Boltzmann
distribution and  dead-channels) considered here which, in our
opinion, are the two most extreme limits.

In section~\ref{simplmod} we calculate $\Delta(1232)/N_{tot}$ and $\Sigma(1385)/\Lambda_{tot}$ 
considering reactions~(\ref{dnp}) and (\ref{L1115L}). 
However, in section ~\ref{chapter 2.1} we investigate many further reactions in which
resonances $\Lambda(1520)$ an $\Sigma(1385)$ participate. Thus we are obliged to develop a completely
numerical evolution, for which the  analytical study of
$\Sigma(1385)/\Lambda^0$ provides a benchmark check of our approach.
 We discuss the temporal evolution of HG
particle fugacities $\Upsilon(t)$ in section~\ref{chapter 3.1}. In
section~\ref{chapter 3.2} we present results for the evolution of
particle $\Sigma(1385)$, $\Lambda(1520)$ multiplicities during
kinetic phase. In section~\ref{chapter 3.3} we obtain the observable
`ob' ratios $\Lambda(1520)_{\rm ob}/\Lambda_{\rm tot}$ and
$\Sigma(1385)_{\rm ob}/\Lambda_{\rm tot}$.
We discuss our results in section~\ref{chapter 4}
\section{Short lived resonances $\Delta(1232)$ and $\Sigma(1385)$ (simplified model)}~\label{simplmod}
\subsection{$\Delta$ multiplicity evolution equation}

%\subsection{Equilibrium conditions for $\Delta$ density}

In the following we will be referring explicitly to  the
$\Delta$  yield   governed by $ c\tau_\Delta\equiv 1/\Gamma_\Delta=1.67$ fm.
All equations apply equally to   $\Sigma(1385)  $ yield
(partial decay width $\Gamma_{\Sigma\to \Lambda}\simeq 35$ MeV) and we  will
compare our results   with experiment for this case.
We note that even though the $\Sigma(1385)  $ decay width is much smaller than $\Gamma_\Delta$, the
number of reaction channels and particle densities available lead to a significant
effect for $\Sigma(1385)$, comparable to our finding for $\Delta$.

The  evolution in time  of the $\Delta$ (or $\Sigma (1385)$) resonance yield is described
 by the process of resonance formation in scattering and decay, population equation (\ref{popeq23}),
where particle 3 is ${\Delta(1323)}$ or $\Sigma(1385)$, particle 1 is the ground 
state $N$ or $\Lambda$, and particle 2 is a pion.  
Allowing for Fermi-blocking and Bose enhancement in the final state, the two in-matter rates ${dW_{N\pi \rightarrow \Delta}}/{dVdt}$ and ${dW_{\Delta
\rightarrow N \pi}}/{dVdt}$ are described by
Eq.~(\ref{pp}) and (\ref{pd}). 
 The distribution
functions for $\Sigma$, $\Delta$, $N$, $\Lambda$ are Fermi and Bose for pions, Eq. (\ref{bf}).

Using detailed balance equation (\ref{pdr}) the master equation,  Eq.(\ref{popeq23}), can now be cast into the form:
\begin{equation}
\frac{1}{V}\frac{dN_{\Delta}}{dt}=
\left(\frac{\Upsilon_{\pi}\Upsilon_{N}}{\Upsilon_{\Delta}}-1\right)
    \frac{dW_{\Delta\rightarrow N \pi}}{dVdt}.\label{ddup}
\end{equation}
This is a rather intuitive and simple result, yet only recently the $1\leftrightarrow 2$ population master equations
have been considered~\cite{KuznKodRafl:2008}.
Equation (\ref{ddup}) implies for $dN_\Delta/dt=0$ the chemical equilibrium condition:
\begin{equation}
\Upsilon^{\rm eq}_{\pi} \Upsilon^{\rm eq}_{N}=\Upsilon^{\rm eq}_{\Delta}. \label{equilcon}
\end{equation}
This equation is solved by the global chemical equilibrium
$\Upsilon^{\rm eq}_{\pi} =\Upsilon^{\rm eq}_{N}=\Upsilon^{\rm eq}_{\Delta}=1$.
However, there are also other, transient, equilibrium states possible,
given a  prescribed value of e.g. the background
pion abundance, $\Upsilon^{\rm eq}_{\pi} \ne 1$.  When the initial state is formed
away from transient equilibrium condition, we recognize that for
$\Upsilon_{\Delta}<\Upsilon_{\pi} \Upsilon_{N}$  the
$\Delta$ production is dominant, and conversely,
for $\Upsilon_{\Delta}>\Upsilon_{\pi} \Upsilon_{N}$  the $\Delta$ decay
dominates.

We now introduce into the population master equation (\ref{ddup})
the effective lifespan,  $\tau_\Delta$ aiming to find an equation
similar to classic radioactive decay population equation. We define
the in medium $\Delta$-lifespan to be:
\begin{equation}\label{Delt}
\tau_\Delta\equiv \frac{\Upsilon_\Delta}{V} \frac{dN_\Delta/d\Upsilon_\Delta}{dW_{\Delta\to N\pi}/dVdt}.
\end{equation}
We recognize that in the Boltzmann limit this corresponds to the ratio of equilibrium yield to the rate per unit time
at which  the equilibrium is approached.
We obtain  for  Eq.(\ref{ddup}):
\begin{equation}\label{Ndelta}
\frac{dN_{\Delta}}{dt}=\left (\Upsilon_{\pi}\Upsilon_{N}- \Upsilon_{\Delta} \right)
              \frac{dN_{\Delta}}{d\Upsilon}\frac 1 {\tau_\Delta}.
\end{equation}

In case that
the ambient temperature does not vary with time, and thus only populations evolve due to change in
fugacities, we have  $dN/dt=dN/d\Upsilon\, d\Upsilon/dt$ and
 the following dynamical equation for the fugacity arises:
\begin{equation}\label{dUpsdt}
\tau_\Delta\frac{d\Upsilon_\Delta}{dt}=\left (\Upsilon_{\pi}\Upsilon_{N}- \Upsilon_{\Delta} \right).
\end{equation}
This is `classical' population equation form where the fugacity plays the role of the classical densities.
When the dynamical values of  $\Upsilon_i(t)$ are used in the quantum Bose/Fermi distributions, the
effects of blocking, and stimulated emission are explicit.

If we instead  were to introduce the lifespan by  $\tilde
\tau_\Delta\equiv (N_\Delta/V)/(dW_{\Delta\to N\pi}/dVdt)$,  this
implies for all particles (Bose, Fermi, Boltzmann)  the classical
population equation, e.g. $dN_\Delta/dt
=(\Upsilon_{\pi}\Upsilon_{N}/\Upsilon_{\Delta}
-1)N_\Delta/\tilde\tau_\Delta$, and the quantum effects are now
hidden in the  definition of $\tilde \tau$. Both definitions
coincide for the case of a dilute system, and differ most for dense
systems. In the limit of very dilute, vacuum system, the relaxation
time is the same as the lifespan of the particles. The computed
yields of particles as function of time are  not dependent on the
finesse of the relaxation time definition.

We now set up for semi-analytical solution of master equation (\ref{Ndelta}).
For multiplicities ${\Delta}$ and ${N}$ considering the small yield and $m\gg T$
we will use the  Boltzmann distribution:
\begin{eqnarray}
\frac {N_{\Delta} }V&=&\Upsilon_{\Delta}\frac{T^3}{2\pi^2}g_{\Delta}x_{\Delta}^2K_2(x_{\Delta}), \\
\frac {N_{N} }V&=&\Upsilon_{N}\frac{T^3}{2\pi^2}g_{N}x_{N}^2K_2(x_{N}),
\end{eqnarray}
where $x_{\Delta,N }=m_{\Delta,N }/T$, $K_2(x)$ is Bessel function.
Considering that  fugacities, temperature  and volume vary in
time, we   rewrite the left hand side of Eq.(\ref{Ndelta}):
\begin{equation}
\frac{\ dN_{\Delta}}{N_\Delta d\tau} = \frac{\ d\Upsilon_{\Delta}}{\Upsilon_{\Delta}d\tau } +
 \frac{d\ln(x_{\Delta}^2K_2(x_{\Delta}))}{dT}\dot{T} + \frac{d(VT^3)}{VT^3\ \ d\tau}.
\label{Ups}
\end{equation}
We changed from $t$ to $\tau$ to make explicit the fact that we work in fluid-element co-moving frame
and thus do not consider the effect of flow on the volume time dependence.

Combining  Eq.(\ref{Ndelta}) with Eq.(\ref{Ups}) we obtain
\begin{equation}
\frac{d\Upsilon_{\Delta}}{d{\tau}} =
\left({\Upsilon_{\pi}\Upsilon_{N}}-\Upsilon_{\Delta}\right)\frac{1}{\tau_{\Delta}}
+\Upsilon_{\Delta}\frac{1}{\tau_T}+\Upsilon_{\Delta}\frac{1}{\tau_S} ,
\label{Ups2}
\end{equation}
%where
\begin{eqnarray}
\frac{1}{\tau_{T} } &=&  -\frac{d\ln({x_{\Delta}}^2K_2(x_{\Delta}))}{dT} \dot{T}. \label{Teq}\\
\frac{1}{\tau_{S} } &=&  -\frac{d\ln( VT^3  )}{dT} \dot{T}. \label{Seq}
\end{eqnarray}
The last term is negligible, $\tau_S\gg \tau_\Delta, \tau_T$ since pions dominate and we have near conservation of
entropy which for massless particles would in fact imply $VT^3=$Const.

Since entropy must be (slightly) increasing, while $T$ is decreasing with time, $\tau_S>0$. Similarly,
$\tau_T>0$, since the temperature decreases with time, and $x^2K_2(x), x=m/T$ increases with $T$ :
\begin{equation}
x^2K_2(x) \approx \sqrt{0.5 \pi}x^{3/2}\exp(-x);
\end{equation}
Therefore:
\begin{equation}\label{1tT}
\frac{1}{\tau_T} \approx -\frac{m_{\Delta}}{T}\left( 1-\frac 3 2 \frac T m_{\Delta}\ldots \right)\frac{\dot{T}}{T}.
\end{equation}

We now evaluate the magnitude of $\tau_T$   invoking a model of matter expansion
of the type used e.g. in~\cite{Letessier:2006wn}, where the longitudinal
and transverse expansion is considered to be independent.
In this model
\begin{equation}
\frac{dV}{dy}=\pi R_{\perp}^2(\tau)\tau,
\end{equation}
where $\tau$ is the proper time in the
local volume element,  this  is exact for a 1-d ideal hydro flow.
The growth of the transverse dimension
can be generically described by
\begin{equation}
R_{\perp}(\tau) = R_0 + \int^\tau_{\tau_0} v(\tau')d{\tau'}, \label{rperp}
\end{equation}
where we take velocity
\begin{equation}
v(\tau) = v_{max}\frac{2}{\pi}\arctan(4(\tau-\tau_0)/\tau_c), 
\end{equation}
where $v_{max} \approx 0.5-0.8c$ ($c$ is speed of light and we take $c=1$), relaxation time $\tau_c \approx 0.5$ fm, $\tau_0 \approx 0.1-1$ fm.

In the proper rest frame of the outflowing matter,
\begin{equation}
\frac{dS}{dy}\propto T^3\frac{dV}{dy}= \pi R^2_{\perp}(\tau)T^3\frac {dz}{ dy}\simeq {\rm Const.}. \label{volt1}
\end{equation}
We will use ${dz}/{dy}\simeq \tau$. 

The growth of the transverse dimension
can be generically described by Eq. (\ref{rperp}).

From Eq. (\ref{rperp}) and Eq.(\ref{volt1}) by elementary evaluation we obtain:
\begin{equation}\label{DTT}
\frac{\dot {T}}{T} = -\frac{1}{3}\left( \frac{2\,(v\tau/R_{\perp  })  + 1}{  \tau}\right).
\end{equation}

\begin{figure}
\centering
\includegraphics[width=8.3cm]{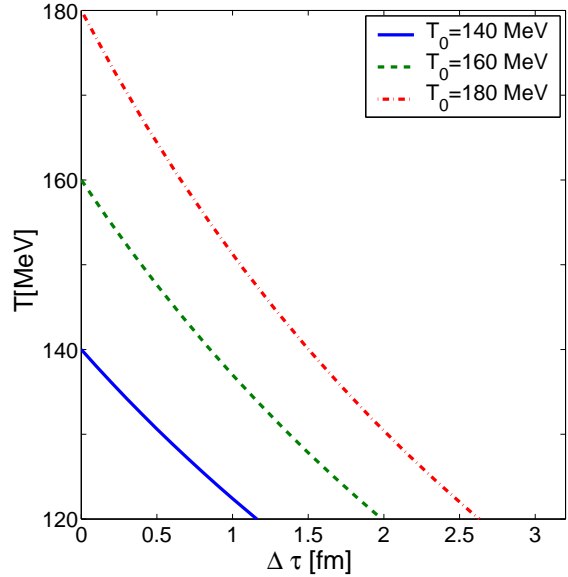}
\caption{\small{Temperature $T$ as function of $\delta \tau$, the proper time interval between
chemical and thermal freeze-out or  chemical
freeze-out temperature (from top to bottom)
 $T=180,\, 160,\, 140$ MeV and thermal  freeze-out $T\ge 120$ MeV.} }
\label{Ttaut}
\end{figure}

Equation (\ref{DTT}) evaluated near hadronization condition is yielding the magnitude of $\tau_T$, see Eq.(\ref{1tT}).
If the maximum expansion velocity is practically instantly achieved, $v\tau/R_\perp\simeq 1$. This leads to maximum
value of $\dot T /T\simeq -1/\tau$. However if a more realistic profiles are assumed, $\dot T /T$ is diminished
in magnitude as much as 30\%. We thus conclude that
$$\frac {0.5}{\tau_h}\, \frac{m_{\Delta}}{T}< \frac 1 {\tau_T} < \frac {0.7}{\tau_h}  \frac{m_{\Delta}}{T}$$
which for hadronization time $\tau_h<10$ fm can compete with the width of the $\Delta$-resonance, $1/t_\Delta\simeq 120$ MeV.
As this shows, the details of the expansion model are not critical for the results we obtain. In actual calculations we employ
$v(\tau)$ described in~\cite{Letessier:2006wn}, where we assume that the expansion is already at maximum velocity
at the time of chemical freeze-out. The resulting dependence $T(\tau)$ after chemical freeze-out is shown in figure \ref{Ttaut}.
We note that the time between chemical and thermal freeze-out $\Delta \tau$ is not longer than about 2.5fm/c, and can be as short
as  1fm/c. However, even such a short scattering period is enough to alter the visible yields of strong resonances, in fact most 
pronounced effect we find in the latter case, since the longer time allows a greater degree of chemical equilibration.

%%%%%%%%%%%%%%%%%%%%%%%%%%%%%%
%%%%%%%%%%%%%%%%%%%%%%%%%%%

We now can solve  Eq.(\ref{Ups2}). Employing Eq.(\ref{partcon}) we have:
\begin{equation}
\frac{d\Upsilon_{\Delta}}{d\tau} + \tilde\Gamma (\tau)\Upsilon_{\Delta} = q(\tau), \label{Ups3}
\end{equation}
%where
\begin{eqnarray}
\tilde\Gamma(\tau) &=& \left[1+\Upsilon_{\pi}\frac{N^{\infty}_{\Delta}}{N^{\infty}_{N}}\right]\frac{1}{\tau_{\Delta}} - \frac{1}{\tau_T}, \\[0.3cm]
q(\tau) &=& \Upsilon_{\pi}\frac{N_0^{\rm tot}}{N^{\infty}_{N}}
\frac{1}{\tau_{\Delta}},
\end{eqnarray}
where $N^{\infty}_{\Delta}$ and $N^{\infty}_{N}$ are densities of
$\Delta$ and $N$ resonances with $\Upsilon_{\Delta}=\Upsilon_N=1$.
The solution of Eq.(\ref{Ups3}) is elementary:
\begin{eqnarray}
\Upsilon_{\Delta} (\tau)= \left(\Upsilon^0_{\Delta} +
     \int_{\tau_{h}}^{\tau}q\,e^{ \int_{\tau_{h}}^{\tau'}\tilde\Gamma d\tau''}d\tau' \right )e^{ -\int_{\tau_h}^{\tau }\tilde\Gamma  d\tau' }
\end{eqnarray}
where $\tau_h$ is initial expansion time at hadronization, and $\tau_h<\tau<\tau_{\rm max}$, upper time limit chosen
to yield  $T_ {\rm max}= 120$ MeV, i.e. $\tau_{\rm max}\simeq 8$ fm.

%%%%%%%%%%%%%%%%%%%%%%%%%%%%%%%%%%%%%%%%%%%%%%%%%%%%%%%%%%%%%%%%%%%%%%%%%%%%%%%%%%%%
\subsection{Results for $\Delta(1232)$ and $\Sigma(1386)$ resonance multiplicities}
%%%%%%%%%%%%%%%%%%%%%%%%%%%%%%%%%%%%%%%%%%%%%%%%%%%%%%%%%%%%%%%%%%%%%%%%%%%%%%%%%%%%

In order to evaluate the  final $\Delta$ multiplicity we need also to know initial
particles densities  right after hadronization which we consider for RHIC
head-on Au--Au collisions at $\sqrt{s_{\rm NN}}=200$ GeV.
We introduce  the  initial hadron yields  inspired by a picture of a rapid
hadronization of QGP  with all hadrons produced with
yields governed by entropy and strangeness
content of QGP by quark recombination. In this model
the yields  of mesons and baryons are controlled by
the  constituent  quark fugacity $\gamma_q$:
\begin{equation}
\Upsilon^0_{\pi}=\gamma_q^{2}; \qquad \label{upinpi}
\Upsilon^0_{\Delta, N}=\gamma_q^{3}. %\label{upindN}
\end{equation}
Thus for $\gamma_q>1$ we always have the initial condition 
\begin{equation}
\left.\frac{\Upsilon_1
\Upsilon_2}{\Upsilon_3}\right\vert_{t=0}=\gamma_q^2 \ge 1\,\label{incon}
\end{equation}
and the yield of $\Delta$ will increase  in the time evolution.

For each
entropy content of the QGP fireball, the corresponding fixed background value of $\gamma_q$ can
be found once hadronization temperature is known, see section \ref{noneqSec}.  For $T=140$ MeV pions
form a  nearly fully  degenerate Bose gas with $\gamma_q\simeq 1.6$.
In the following discussion, aside of this  initial condition,  we also consider the value pairs
$T=150{\rm\, MeV},\,\gamma_q=1.42$,
 $T=160{\rm\, MeV},\,\gamma_q=1.27$,
  $T=170{\rm\, MeV},\,\gamma_q=1.12$   and
$T=180$ MeV with $\gamma_q=1$.

We assume in this section that  $m\gg T$ the density  $\Delta$ is relatively low,  thus there is no significant dependence of $1/{\tau_{\Delta}}$ (the same for $\Sigma(1385)$) on
$T$   and $\Upsilon_{\Delta}$; in essence $\tau_\Delta=\hbar /\Gamma_\Delta$
takes the free space value $\tau_\Delta\simeq \hbar/120\rm{MeV}$. Although from section~\ref{chapter 2.3} we know that noticeabale but not very large large effect on $\tau_{\Sigma(1385)}$ from dense pion gas exits.  We will take this effect into account 
in section , where more detailed model is presented.
 
 As already noted, we do not need to follow the evolution in time for
the pion yield, which is fixed by conservation of entropy per unit
rapidity, as incorporated in Eq.\,(\ref{DTT}). Thus it is
(approximately) a constant of motion. This can be seen recalling
that the  entropy per pion is nearly 4 within the domain of
temperatures considered. Thus the conservation of entropy implies
that pion number is conserved. With $VT^3\simeq \mathrm{Const.}$,
this further implies that during the expansion
$$\Upsilon_\pi=\gamma_q^2=\mathrm{Const.},$$
which we keep at the initial value.

%%%%%%%%%%%%%%%%%%%%%%%%%%%%%%
\begin{figure}
\centering
\includegraphics[width=8.3 cm,height=8.3 cm]{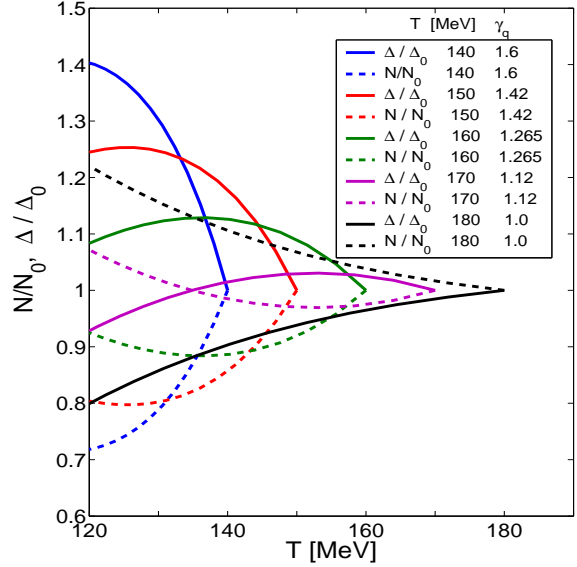}
\caption{\small{The ratio ${\Delta}/{\Delta}^0$ (solid lines) and $N/N^0$ (dashed lines)
 as functions of temperature $T$ for select given pairs of values $T,\gamma_q$,
see text and figure box for details.}}
\label{nd}
\end{figure}
%%%%%%%%%%%%%%%%%%%%%%%%%%%

In figure \ref{nd} we present results for ratios
${\Delta}/{\Delta}_0$ (solid lines) and $N/N_0$ (dashed lines) as
functions of temperature $T$, beginning from the presumed initial hadronization
temperature $T$ through  $T_ {\rm max}= 120$ MeV.
${\Delta}_0$ and $N_0$ are the initial yields obtained at each hadronization
temperature.  For $T < 180$ MeV, initially $\Upsilon_\Delta < \Upsilon_N\Upsilon_\pi$,
thus   based on our prior discussion, we expect that
the master equation leads to an initial increase in the yield of resonances. However,
as temperature drops,  due to the dynamics of the expansion   the increasing yield
of $\Delta$ turns over, and a final net
increase of resonance yield is observed  for $T\le 160$ MeV. We note that for $T \ge 180$ MeV there is 
a  continuous depletion of resonance yield. The nucleon yields
move in opposite direction to the $\Delta$-resonance.

This behavior can be understood in qualitative manner as follows:
  The total number $\Delta + N$ is conserved therefore $\Delta$ multiplicity increases and $N$
multiplicity decreases until they reach transient chemical equilibrium
($dN_{\Delta}/d\tau=0$),  corresponding to the maximum point seen for  $\Delta$ in figure \ref{nd}.
There is also influence of expansion:  even if for some temperature
the transient equilibrium condition (\ref{equilcon}) is reached,  the system cannot
stay in this equilibrium, $\Upsilon_{\Delta}$ and $\Upsilon_{N}$ are increasing to conserve
total number of particles. $\Upsilon_{\Delta}$ increases faster
because of larger $\Delta$'s mass.  After
$\Upsilon_{\Delta}$ becomes larger than $\Upsilon_{\pi}\Upsilon_{N}$
$\Delta$ decay begins to dominate and  their multiplicity is decreasing. The special case at
hadronization temperature $T=180$ MeV where, $\Upsilon_i=1$ and equilibrium
condition is satisfied initially. As expansion sets in,   ${\Delta}$ is decreasing
because $\Upsilon_{\Delta} >  \Upsilon_{N}$ (recall that here $\Upsilon_{\pi}=1$).  
In the SHM evaluation of yields one assumes that all ratios seen in figure \ref{nd}  are unity.

\begin{figure}
\centering
\includegraphics[width=8.3 cm,height=8.3 cm]{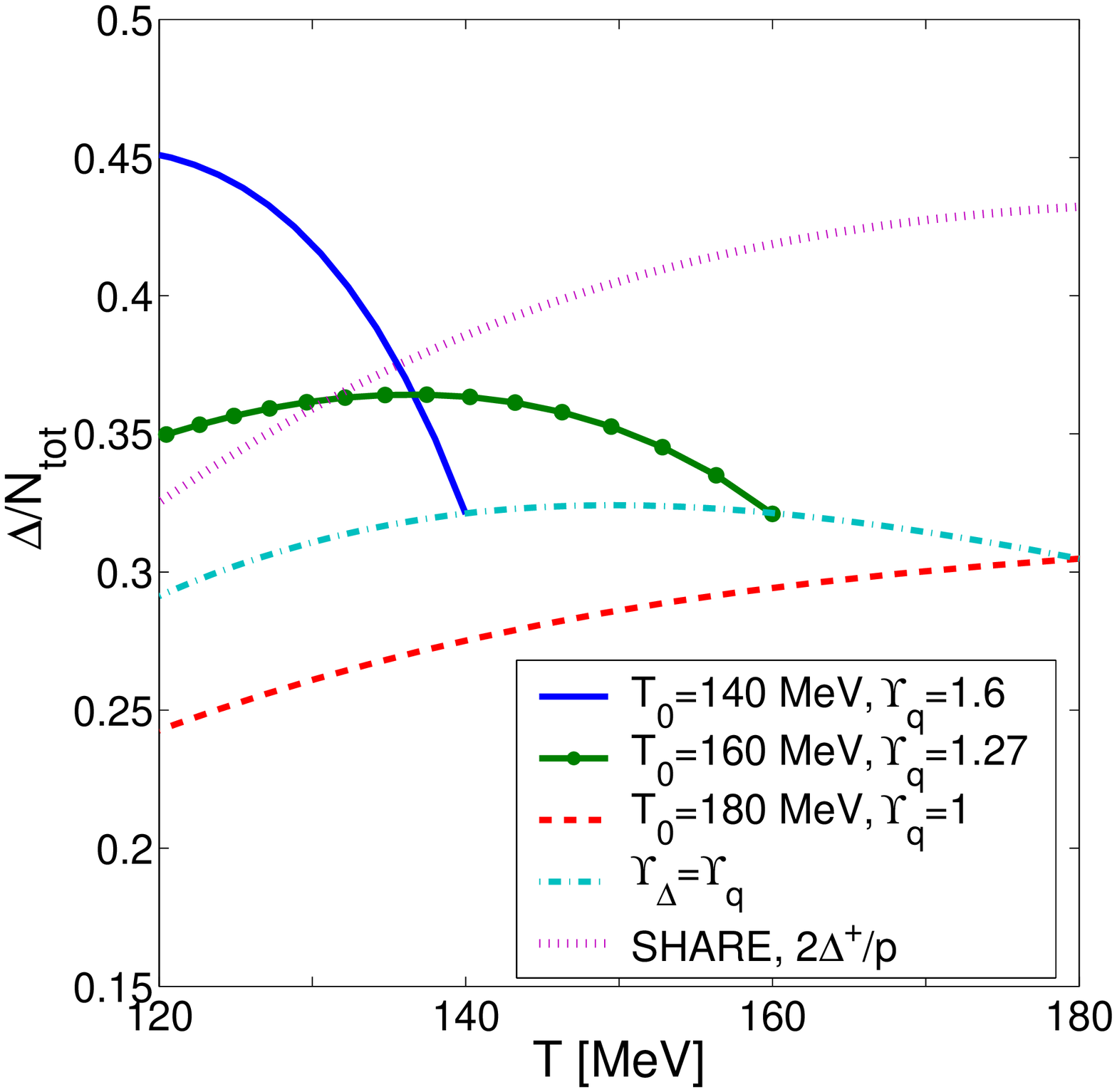}\\[-0.6cm]
\includegraphics[width=8.3 cm,height=8.3 cm]{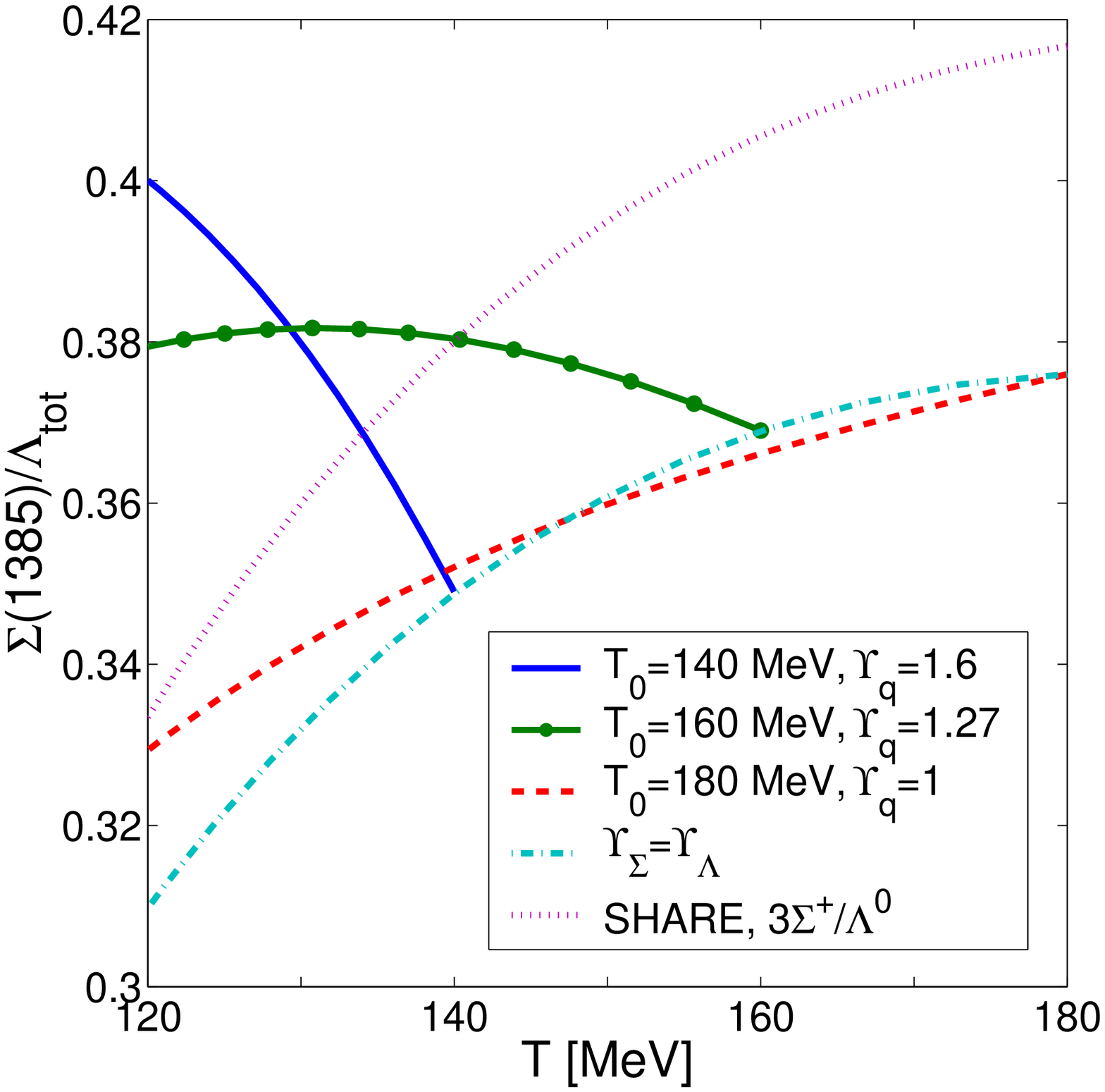}
\caption{\small{Relative resonance yield, for (top)
${\Delta}/N_{\rm tot}$ and (bottom)
${\Sigma(1385)/\Lambda_{\rm tot}}$ as a functions of freeze-out
temperature, for hadronization temperatures
$T_0=140,160,180$ MeV, see box and text for details.
The dotted brown line gives the expected SHM chemical equilibrium  result.}}
\label{ndNtot}
\end{figure}

The initial hadronization  yields which we used as
reference in   figure \ref{nd} are not accessible to measurement. Therefore,
we  consider in  figure \ref{ndNtot} the fractional yield   ${\Delta}/N_{\rm tot}$ (top frame), again as 
a function of temperature $T$. 
The results  for hadronization temperatures $T_0=140$ (solid
blue line), $T_0=160$ (dash-dot green line) and $T_0=180$ MeV (dashed brown
line) are shown. $N_{\rm tot}$ is fixed by hadronization
condition and is not a function of time, as discussed.  Thus the observable final rapidity nucleon yield
corresponds to the initial  value at hadronization. Note that 
up to strange and multi  strange baryon contribution,  $N_{\rm tot}$
is the  total baryon (rapidity) yield. 

%%%%%%%%%%%%%%%%%%%%%%%%%%%%%%
%%%%%%%%%%%%%%%%%%%%%%%%%%%

Since in this study  we have considered a  subset of all relevant baryon resonances
our chemical equilibrium reference yield (line for $\Upsilon_{\Delta}=\Upsilon_{N}$) is not the same as
the corresponding reference line for the full statistical hadronization model (SHM)
evaluation,  obtained using SHARE2, and presented as    $2\Delta^+/p$ (upper frame) and
 $3\Sigma^+/\Lambda^0$ (lower frame).
 The SHARE2-SHM value  $\Delta^{++}/p\simeq 0.2$ at $T\simeq 160$ MeV
is consistent with the STAR d--Au results~\cite{Abelev:2008yz}. Also,  comparing
our with the SHARE2  result we note that SHARE2  yield
 is larger at chemical freeze-out. The magnitude of
the difference in the yields at time of chemical freeze-out provides a
measure of the magnitude of the corrections we can expect to arise in
the full treatment at thermal freeze-out and/or systematic error for these
yields.

The nature of these effects is  different for the two yield cases considered: the
presence of heavier  resonances which cascade by way of $\Delta$ leads to an increase
of the thermal freeze-out yield. The correction is thus nearly as much as we see the SHARE2 yield higher at
chemical freeze-out.
For $\Sigma(1385) $  the difference  with SHARE2 arises from a difference  of contributions
of partial decays producing $\Lambda_{\rm tot}$, thus the correction is multiplicative
factor which does not change, but is uncertain in magnitude due to lack of
knowledge about the branching ratios.

We believe that  the  $\Delta$ and  $\Sigma(1385) $ yields are  underestimated  by about 15\% -- 35\%.
(bigger effect for hadronization at higher $T$).
This implies that depending on hadronization temperature a relative yield range
$0.16<\Delta^{++}/p=0.5 {\Delta}/N_{\rm tot}<0.26$ arises,
and similarly (see lower frame in figure \ref{ndNtot})   $0.35<\Sigma(1385)/\Lambda_{\rm tot}<0.43$
with the {\it higher} relative yield
corresponding to the {\it lower} hadronization temperature. One of the key results of this work
is the narrow range for $\Sigma(1385)/\Lambda_{\rm tot}$, and the
fact that the initial chemical non-equilibrium  effect leads to a
reversal of  the SHM model situation:   the relative yields of massive resonances decreases
with decreasing hadronization temperature.

In order to compare with the experimental results we
note that the data presented~\cite{Adams:2006yu,Salur:2006jq}
are for  charged $\Sigma(1385)$,
particle and antiparticle channels,
$(\Sigma^\pm(1385)+\overline{\Sigma^\pm(1385)}/(\Lambda_{\rm tot}+\overline{\Lambda_{\rm tot}})\simeq 0.29$.
This result needs to be multiplied with 3/2 to be comparable to results presented here which include $\Sigma^0(1385)$.
Multiplying the  value   for hadronization at $T=140$ MeV with thermal
freeze-out at $T=120$ MeV, and  allowing for contribution by heavier resonances as indicated by SHARE2
our result is in perfect agreement with~\cite{Adams:2006yu,Salur:2006jq}
However, given the narrow range of results we find, it seems that the high yield of $\Sigma(1385)$, seen
the error ${\cal O}(20\%)$ is nearly compatible with the entire range of chemical freeze-out
temperatures here considered -- the low $T$ chemical freeze-out is favored by 1.5  s.d. over high $T$.

The reader should take note that the `thermal' model result presented in Ref.~\cite{Adams:2006yu}
corresponds to initial high temperature freeze-out in chemical equilibrium   which is
 unobservable, since the high $T$ hadronization resonance decay products have no chance
to escape into free space. Thus this comparison of this model with experiment is flawed.
The evolved yield is shown as (red) dashed line in figure \ref{ndNtot}, and is found 25\%
below the value measured. The reason this happens is that the high $T$ chemical freeze-out
happens near chemical equilibrium and the yields follow closely the chemical equilibrium 
yield described by temperature, thus it is the
{\em thermal freeze out temperature which in this case controls the final observable
resonance yield}.

%%%%%%%%%%%%%%%%%%%%%%%%%%%%%%
\begin{figure}
\centering
\includegraphics[width=17 cm, height=12 cm]{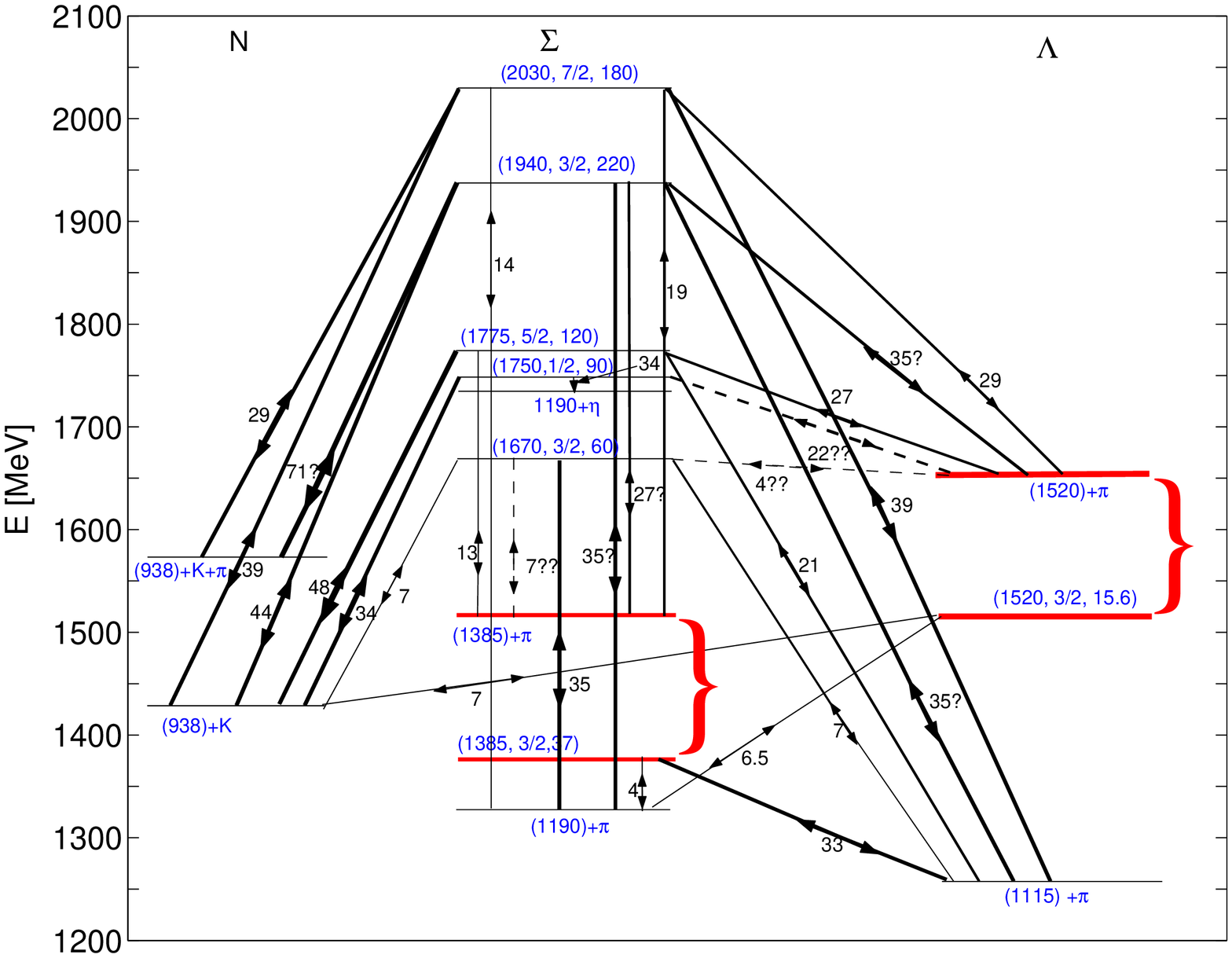}
\vskip -0.51cm
\caption{\small{ (color on line)  Reactions scheme for $\Lambda(1520)$ and
$\Sigma(1385)$  population evolutions.} } \label{Lam1520}
\end{figure}
%%%%%%%%%%%%%%%%%%%%%%%%%%%

\section{Suppression of $\Lambda(1520)$ and Enhancement of $\Sigma(1385)$}
\subsection{Reactions scheme for $\Lambda(1520)$ and $\Sigma(1385)$}\label{chapter 2.1}

In figure~{\ref{Lam1520}} we show the scheme of reactions  which all
have a noticeable effect on $\Lambda(1520)$ yield after the chemical
freeze-out kinetic phase. The format of this presentation is
inspired by nuclear reactions schemes. On the vertical axis the
energy scale is shown in MeV. There are three classes of particle
states, which we denote from left to right as "$N$" (S=0 baryon),
"$\Sigma$" ($S=-1, I=1$ hyperon) and "$\Lambda$" ($S=-1, I=0$
hyperon).
 Near each particle  bar   we state (on-line in blue) its mass,
and/or angular momentum and/or total width in MeV. The states
$\Lambda(1520)$ and $\Sigma(1385)$ are shown along with the location
in energy of $\Lambda(1520)+\pi$ and $\Sigma(1385)+\pi$
respectively, both entries are connected by the curly bracket, and
are highlighted (on-line in red). The  inclusion of the $\pi$-mass
is helping to see the kinetic threshold energy of a reaction. The
lines connecting the $N,\Sigma,\Lambda$ columns are indicating the
reactions we consider in the  numerical computations. All reactions
shown in figure~{\ref{Lam1520}} can go in both directions, as shown
by the double arrows placed next to the numerical value of the
partial decay width $\Gamma_i$ in MeV.

$\Lambda(1520)$ decays with a total decay width of about 15.6 MeV,
with two main channels:
\begin{eqnarray}
 \Sigma +\, \pi  \leftrightarrow \Lambda(1520) , \quad\Gamma \approx 6.5\, {\rm MeV}; \\\notag
    N + K  \leftrightarrow \Lambda(1520) ,\quad\Gamma \approx 7\phantom{.5}\, {\rm MeV}.
\end{eqnarray}
However, $\Lambda(1520)$ reacts with several
heavier $\Sigma^*$-resonances,
($\Sigma^*\equiv \Sigma(1670)$, $ \Sigma(1750)$, $ \Sigma(1775)$, $\Sigma(1940)$, $\Sigma(2030)$):
\begin{equation}
   \Lambda(1520)+\pi \leftrightarrow \Sigma^*, \label{LS*}
\end{equation}
and these reactions have a larger reaction strength
 shown  in  figure~{\ref{Lam1520}}.   $\Lambda(1520)$ nearly behaves
like a `stable' hadronic particle since:\\
a) it is dominantly coupled to heavier resonances; \\
b) its natural lifespan is  larger than the hadronic reaction rate.\\

 Hereto we note that (several) $\Sigma^*$ involved in Eq.\, (\ref{LS*}) participate   in further reactions:
\begin{eqnarray}
&&   \Lambda(1115) +\pi  \leftrightarrow \Sigma^*; \label{L1115S*}\\   %\notag
&&   \Sigma(1190)+\pi \leftrightarrow  \Sigma^*;  \\  %\notag
&&    N + K  \leftrightarrow \Sigma^*;  \\
&&    \Sigma(1385) + \pi  \leftrightarrow \Sigma^*; \label{S1385S*}\\ %\notag
&&   \Delta + K  \leftrightarrow \Sigma(1940, 2030) ;    \\ %\notag
&&   N + K(892)  \leftrightarrow \Sigma(1940);   \\ %\notag
&&   \Sigma   + \eta\leftrightarrow  \Sigma(1750).
\end{eqnarray}
All  reactions shown above can excite $\Sigma^*$ resonances. Since
the mass of $\Lambda(1520)$ is near to the
$\Sigma^*$ mass, the yield of  $\Lambda(1520)$ is effectively depleted
by the   reaction chain
\begin{equation}
\Lambda(1520)+\pi  \rightarrow  \Sigma^* \rightarrow N+{\rm K},\  {\rm etc} .\label{LS*N}
\end{equation}
The balancing
two step back-reaction can also occur, especially once $\Lambda(1520)$ has been  depopulated.
Thus   a dynamical reduced   detailed balance yield of  $\Lambda(1520)$ would result
if the system were at fixed volume rather than  expanding.

The multiplicity of $\Sigma(1385)$ is mostly determined by its dominant
decay and production in the reaction~(\ref{L1115L})
and to a lesser extent by  the reaction
\begin{equation}
  \Sigma(1190) + \pi \leftrightarrow \Sigma(1385)  . \label{S1385S}
\end{equation}
The resonance $\Sigma(1385)$ participates further in reactions with
heavier $\Sigma^*$; see reaction (\ref{S1385S*}), but strength of
these interactions is smaller than for similar reactions with
$\Lambda(1520)$ and smaller than the decay width of $\Sigma(1385)$.
Thus we find that the  influence of these reactions on
$\Sigma(1385)$ yield is small. Another reason for a reduced
effective depletion rate of  $\Sigma(1385)$ is that a lesser
fraction of this resonance is needed to excite $\Sigma^*$. Thus in
such a reaction the depopulation effect decreases  because of a
larger mass difference between $\Sigma(1385)$ and $\Sigma^*$ in
comparison with  $\Lambda(1520)$ and $\Sigma^*$.

\begin{figure}
\centering
\includegraphics[width=17 cm, height=12 cm]{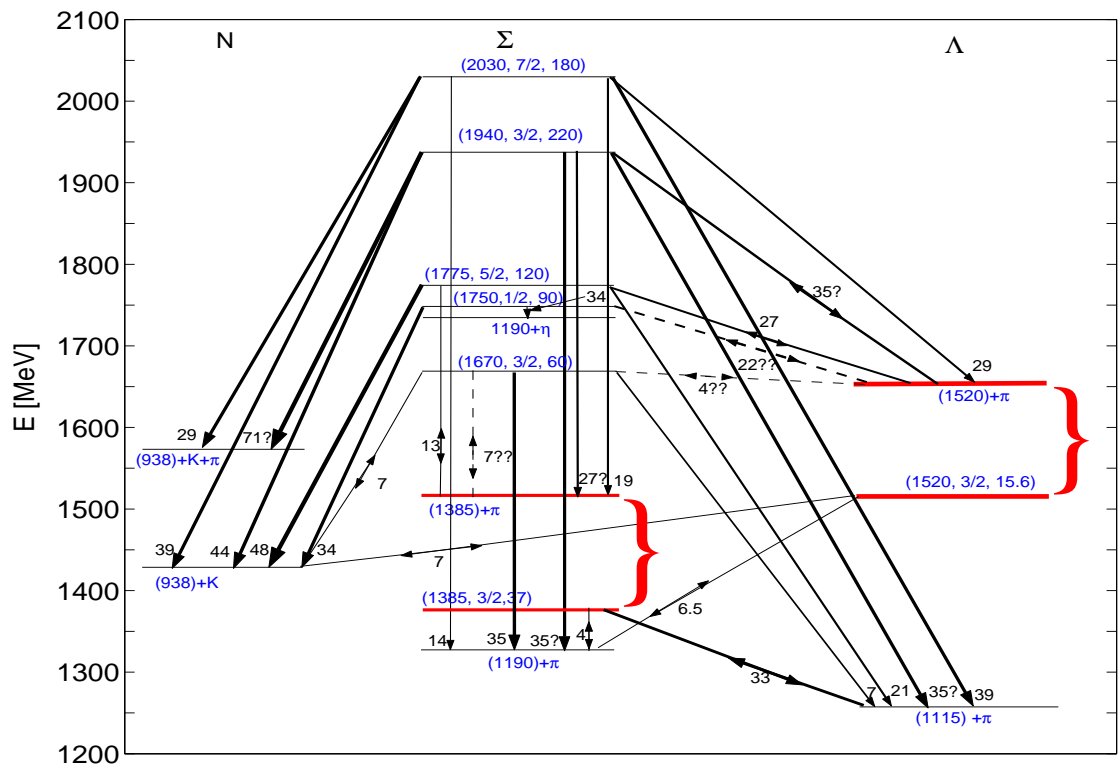}
\vskip -0.51cm \caption{\small{ (color on line)  Reactions scheme
for $\Lambda(1520)$ and $\Sigma(1385)$ interactions in the ``dead
channel'' model.} } \label{Lam1520dc}
\end{figure}

The reactions scheme for $\Lambda(1520)$ reactions with dead
channels is shown in figure~\ref{Lam1520dc}. The difference between
figure~\ref{Lam1520}  and figure~\ref{Lam1520dc} is that some of the
reaction lines have single-directional arrows, as is stipulated by
the condition Eq.\,(\ref{dchcon}).

%%%%%%%%%%%%%%%%%%%%%%%%%%%%%%
%%%%%%%%%%%%%%%%%%%%%%%%%%%\ni

%%%%%%%%%%%%%%%%%%%%%%%%%%%%%%%%%%%%%%%%%%%%%%%%%%%%%%%%%%%%%%%%%%%%%%%%%%%%%
\subsection{Resonances densities, time evolution equations}\label{chapter 2.2}
%%%%%%%%%%%%%%%%%%%%%%%%%%%%%%%%%%%%%%%%%%%%%%%%%%%%%%%%%%%%%%%%%%%%%%%%%%%%%%

The  evolution in time  of the resonance yield is described by a
master equation, similar to (\ref{popeq23}), where in general multichannel case the all processes of resonance formation in
scattering is balanced by all natural resonance decay channels:
\begin{equation} \label{delev}
\frac{1}{V}\frac{dN_{3}}{dt}=\sum_i\frac{dW^i_{{1+2 \rightarrow
3}}}{dVdt}-\sum_j\frac{dW^j_{{3 \rightarrow 1 +2}}}{dVdt},
\end{equation}
where subscripts $i$, $j$ denote different reactions channels when
available. We further allow different subscripts $i$, $j$ for the
case where there are dead channels. Thus ${dW^i_{1+2 \rightarrow
3}}/{dVdt}$ and ${dW^j_{{3 \rightarrow 1 + 2}}}/{dVdt}$ are
invariant rates (per unit volume and time) for particle $3$
production and decay respectively.   In case all reactions occur in
both directions the total number of fusion channels is the same as
the total number of decay channels.

Allowing for Fermi-blocking and Bose enhancement in the final state,
where  by designation   particles $1$  and $3$ are fermions (heavy baryons) and
particle $2$ is a boson (often light pion) we have Eq.(~\ref{pp}) for resonance production
and Eq.~(\ref{pd}) for resonanse decay rate.

Using detailed balance Eq.\,(\ref{pdr}) we obtain  for fugacity
$\Upsilon_3$ the  evolution
equation:
\begin{equation}
\frac{d\Upsilon_{3}}{d{\tau}} =
\sum_i{\Upsilon^i_{1}\Upsilon^i_{2}}\frac{1}{\tau^i_{3}}
+\Upsilon_{3}\left(\frac{1}{\tau_T}+\frac{1}{\tau_S}-\sum_j\frac{1}{\tau^j_3}\right),
\label{Ups2}
\end{equation}

Using detailed balance Eq.\,(\ref{pdr}) we obtain  for fugacity
$\Upsilon_3$ the  evolution
equation~\cite{KuznKodRafl:2008,Kuznetsova:2008jt}:
\begin{equation}
\frac{d\Upsilon_{3}}{d{\tau}} =
\sum_i{\Upsilon^i_{1}\Upsilon^i_{2}}\frac{1}{\tau^i_{3}}
+\Upsilon_{3}\left(\frac{1}{\tau_T}+\frac{1}{\tau_S}-\sum_j\frac{1}{\tau^j_3}\right),
\label{Ups2}
\end{equation}
where  characteristic time constants of temperature $T$ and entropy $S$ evolution, $\tau_{T}$ 
and $\tau_{S}$, are from Eq. (\ref{Teq}) and (\ref{Seq}).

The entropy term is negligible, $\tau_S\gg \tau_3, \tau_T$ since we
implement near conservation of entropy. We implement this in the way
which would be exact for massless particles taking  $VT^3=$Const..
Thus there is some entropy growth in HG evolution to consider, but
it is not significant. In order to evaluate  the magnitude of
$\tau_T$ we use the relation between Bessel functions of order 1 and
2 (not to be mixed up with particles 1,2)
${d}\left(z^2K_2(z)\right)/{dz}=-z^2K_1(z)$. We obtain
\begin{equation}
\frac{1}{\tau_{T}} = - \frac{K_1(x_{3})}{K_2(x_3)}x_3 \frac{\dot{T}}{T}, \label{Teq2}\\
\end{equation}
$\tau_{T}>0$. We  invoke a model of matter expansion. For a static system with $\tau_T \to
0$ we see that Eq.\,(\ref{Ups2}) has  transient stable population
points whenever
 \begin{equation}
\sum_i\Upsilon_1^i\Upsilon_2^i\frac{1}{\tau_3^i}-\Upsilon_3\sum_j\frac{1}{\tau_3^j}=0.
\label{stable}
 \end{equation}

Next we address the functional dependence on time of
$\Upsilon_{1},\Upsilon_{2}$.   In the equation for $\Upsilon_{1}$ we
have terms which compensate what is lost/gained in  $\Upsilon_{3}$
see Eq.\,(\ref{Ups2}). Further we have to allow that particle `1'
itself plays the role of particle 3 (for example this is clearly the
case for $\Lambda(1520)$). That allows a chain of populations
relations as follows:
 \begin{equation}
(1'+2' \leftrightarrow 1) + 2 \leftrightarrow  3, \label{dpr2}
 \end{equation}
Then we  obtain:
\begin{eqnarray}
\frac{d\Upsilon_{1}}{d{\tau}}\!\! &=& \!\!
\Upsilon_{3}\sum_k\frac{1}{\tau^k_3}\frac{dN^k_3/d{\Upsilon^k_3}}{dN_1/d{\Upsilon_1}}
-\sum_n{\Upsilon_{1}\Upsilon^n_{2}}\frac{1}{\tau^n_{3}}\frac{dN^n_3/d{\Upsilon^n_3}}{dN_1/d{\Upsilon_1}}\nonumber\\
&+& \Upsilon_{1}\left(\frac{1}{\tau_T}+\frac{1}{\tau_S}-\sum_j\frac{1}{\tau^j_1}\right)
+\sum_i{\Upsilon^i_{1'}\Upsilon^i_{2'}}\frac{1}{\tau^i_{1}}
\label{Ups3}
\end{eqnarray}
The ratios of derivative of $N_i$ seen in the first line are due to
the definition of relaxation time Eq.\,(\ref{Dect}). The system  of
equations for baryons closes with the equation for $\Upsilon_{1'}$
 \begin{eqnarray}
 \frac{d\Upsilon_{1'}}{d{\tau}} \!\! &=  &\!\!
\Upsilon_{1}\sum_k\frac{1}{\tau^k_1}\frac{dN^k_1/d{\Upsilon^k_1}}{dN_{1'}/d{\Upsilon_{1'}}}
-\sum_n{\Upsilon_{1'}\Upsilon^n_{2'}}\frac{1}{\tau^n_{1}}\frac{dN^n_1/d{\Upsilon^n_1}}{dN_{1'}/d{\Upsilon_{1'}}}\nonumber\\
&+&  \Upsilon_{1'}\left(\frac{1}{\tau_T}+\frac{1}{\tau_S}\right).
\label{ups4}
 \end{eqnarray}
In the present setting  $\Upsilon_{2=\pi}=$Const. by virtue of
entropy conservation (see discussion below) and the same applies to
the case $2'=\pi$. However, if either particle $2$ or  $2'$ is a
kaon, we need to follow the equation for $\Upsilon_{2,2'=K}$ which
is analogous  to equation for particle $1$ or $1'$.\

The evolution equations can be integrated once we determine the {\em
initial} values of particle densities (fugacities) established at
hadronization/chemical freeze-out. We determine these for RHIC
head-on Au--Au collisions at $\sqrt{s_{\rm NN}}=200$ GeV. We
introduce  the  initial hadron yields  inspired by a picture of a
rapid hadronization of QGP in which quarks combine into final state
hadrons. For simplicity we assume here that  the net baryon yield at
central rapidity  is negligible. Thus the baryon-chemical and
strangeness potentials vanish. The initial yields  of mesons ($q\bar
q, s\bar q$) and baryons ($qqq, qqs$) are controlled aside of the
ambient temperature $T$ by the  constituent light quark fugacity
$\gamma_q$ and the strange quark fugacity~$\gamma_s$.

The strangeness pair-yield in QGP is maintained in transition to HG.
This fixes the initial value of $\gamma_s$. In fact, since we
investigate here relative chemical equilibrium reactions our results
do not depend significantly  on the exact initial value $\gamma_s$
 and/or strangeness content. The entropy
conservation at hadronization fixes $\gamma_q$. For hadronization
temperature $T(t=0)\equiv T_0=180$ MeV,  $\gamma_q=1$. However, when
$T_0<180$ MeV, $\gamma_q>1$ in order to have entropy conserved at
chemical freeze-out. At $T_0=140$ MeV $\gamma_q=1.6$ that is close
to maximum possible value of $\gamma_q$, defined by Bose-Einstein
condensation condition \cite{Kuznetsova:2006bh}.

For reactions, such as  shown in Eq.\, (\ref{123}), we have (lower
index defines particle considered, where $Y\equiv \Sigma, \Lambda$
is a hyperon)
\begin{equation}
\Upsilon^0_{(1=Y)}=\gamma_q^{2}\gamma_s,\qquad %\label{upindN}
\Upsilon^0_{(2={\rm K)}}=\gamma_q\gamma_s; \qquad \label{upinpi1}
\end{equation}
$\Upsilon^0_{(1=N)}$ and $\Upsilon^0_{(2=\pi0)}$ are defined by Eq. $\ref{upinpi}$,
where the particle 1 in reaction (\ref{123}) is a baryon and particle 2
is a meson. The particle 3 is always a strange baryon:
\begin{equation}
\Upsilon^0_{(3=Y)}=\gamma_q^{2}\gamma_s, %\label{upindN}
\end{equation}
As a consequence initially the pair of particles 1,2 reacts into 3.
We have again condition~(\ref{incon}) satisfied.

%%%%%%%%%%%%%%%%%%%%%%%%%%%%%%%

%%%%%%%%%%%%%%%%%%%%%%%%%%%%%
\section{Numerical results}\label{chapter 3}
%%%%%%%%%%%%%%%%%%%%%%%%%%%%%%
\subsection{Evolution of fugacities}\label{chapter 3.1}
%%%%%%%%%%%%%%%%%%%%%%%%%%%%%%

%%%%%%%%%%%%%%%%%%%%%%%%%%%%%%%%%%%%%%%%%%%%%% Figure 5
\begin{figure*}
\centering
%\hspace*{-.3cm}\includegraphics[width=7.1 cm]{Upsil140.eps}\hspace*{-1.5cm}
%\includegraphics[width=7.1 cm]{Upsil160.eps}\hspace*{-1.5cm}
%\includegraphics[width=7.1 cm]{Upsil180.eps}
\hspace*{-.3cm}\includegraphics[width=16.5 cm]{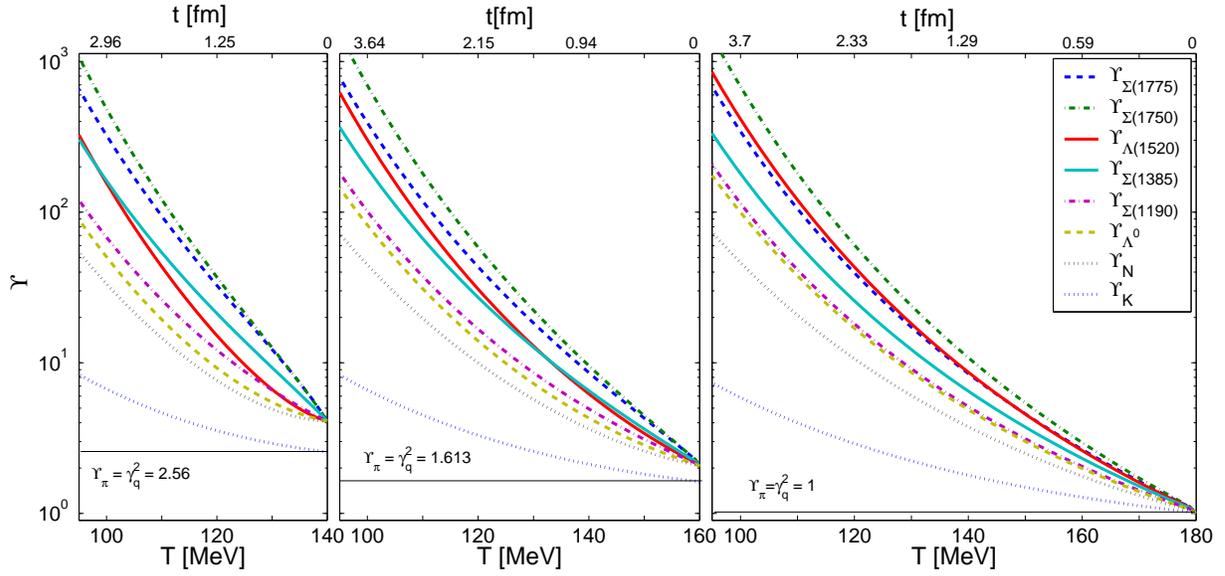}
\caption{\small{ (color on line) The fugacities $\Upsilon$  for selected particles are shown as a function of
temperature $T(t)$, for $T_0=140$ MeV on the left, for $T_0=160$ MeV in the middle and
for $T_0=180$ MeV, on the right. See text for further details.}} \label{Upsil}
\end{figure*}
%%%%%%%%%%%%%%%%%%%%%%%%%%%%%%%%%%%%%%%%%%%%%%%%%%%%%%
In order to evaluate the $\Lambda(1520)$ and $\Sigma(1385)$
multiplicities we must  integrate  Eq.\,(\ref{Ups2}), or
Eq.\,(\ref{Ups3}), or Eq.\,(\ref{ups4}) for each particle involved
in figure~\ref{Lam1520}, and perform similar operations for
reactions with dead channels in figure~\ref{Lam1520dc}. This system
of equations includes equations for $\Lambda(1520)$, $\Sigma(1385)$,
five equations for $\Sigma^*$s, equations for K(892) and $\Delta$
and equations for ground states $\Lambda(1115)$, $\Sigma(1190)$, N,
K. All reactions in figures~{\ref{Lam1520}} are included. We solve
this system of equations numerically, using classical fourth order
Runge-Kutta method.

%%%%%%%%%%%%%%%%%%%%%%%%%%%%%%%%%%%%%%%%%%%%%%Fig 6
\begin{figure}
\centering
\includegraphics[width=8 cm, height=10 cm]{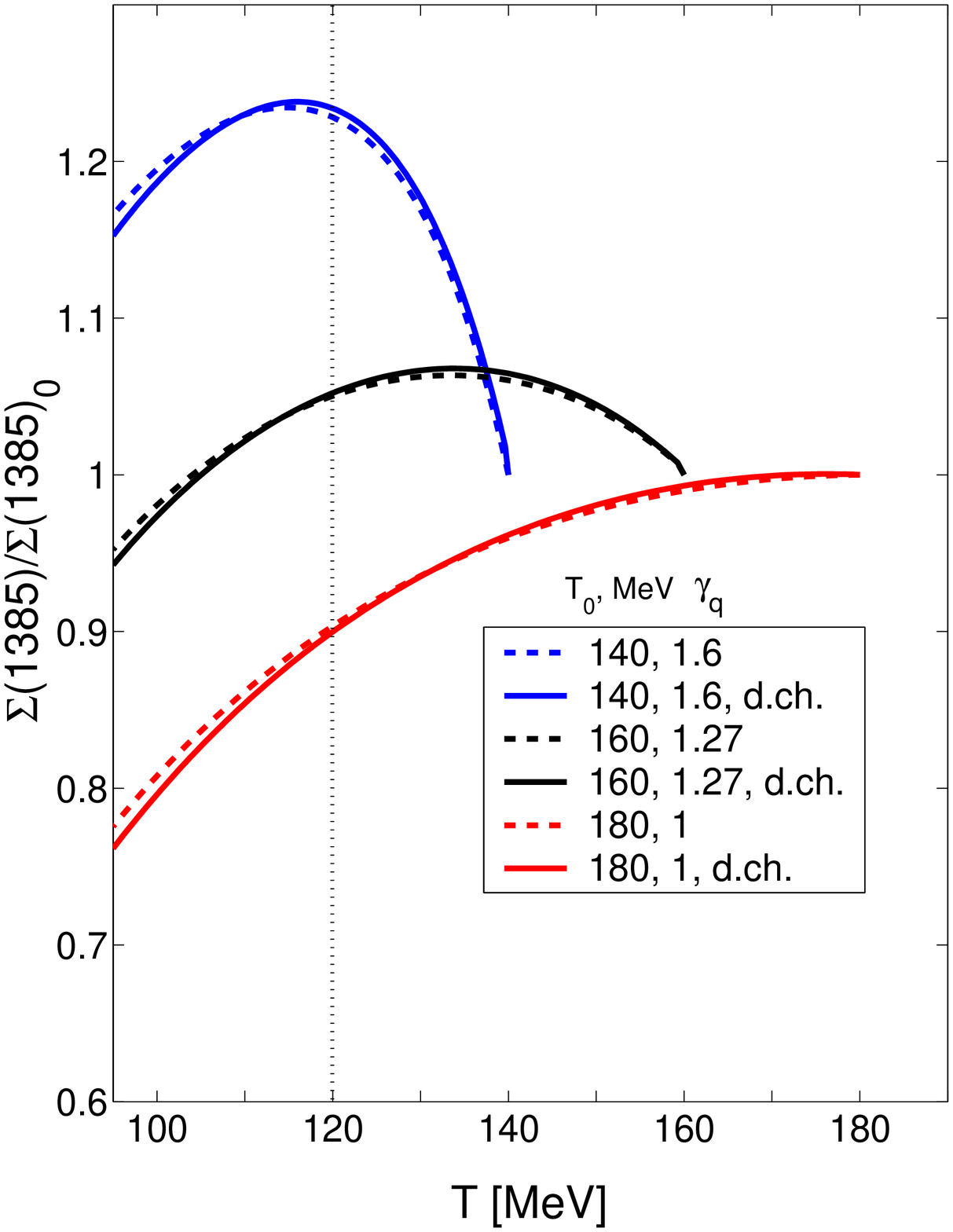}
\includegraphics[width=8 cm, height=10 cm]{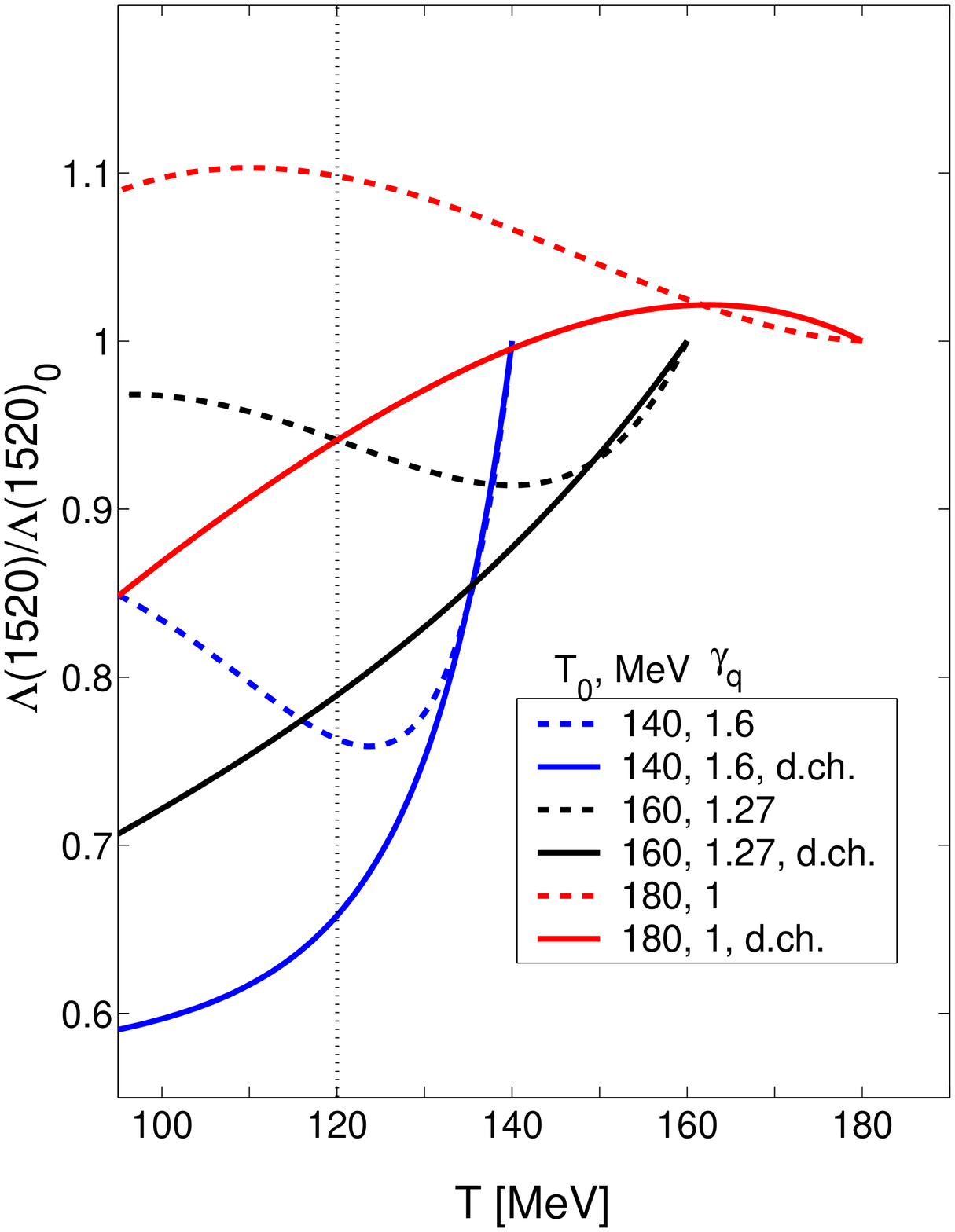}
\caption{\small{The ratio
$\Sigma(1385)/\Sigma(1385)_0$ on left
and $\Lambda(1520)/\Lambda(1520)_0$ on right
as a functions of temperature $T(t)$ for
different initial hadronization temperatures $T_0=140$, $160$
and $180$ MeV (blue/bottom, black/middle and red/top lines, respectively). Solid lines
are for calculations with dead channels, dashed lines are for
calculations without dead channels.}} \label{Lam1520r}
\end{figure}
%%%%%%%%%%%%%%%%%%%%%%%%%%%%%%%%%%%%%%%%%%%%%%%%%%%%%%

%%%%%%%%%%%%%%%%%%%%%%%%%%%%%%%%%%%%%%%%%%%%%%Fig 7
\begin{figure}
\centering
\includegraphics[width=8 cm, height=10 cm]{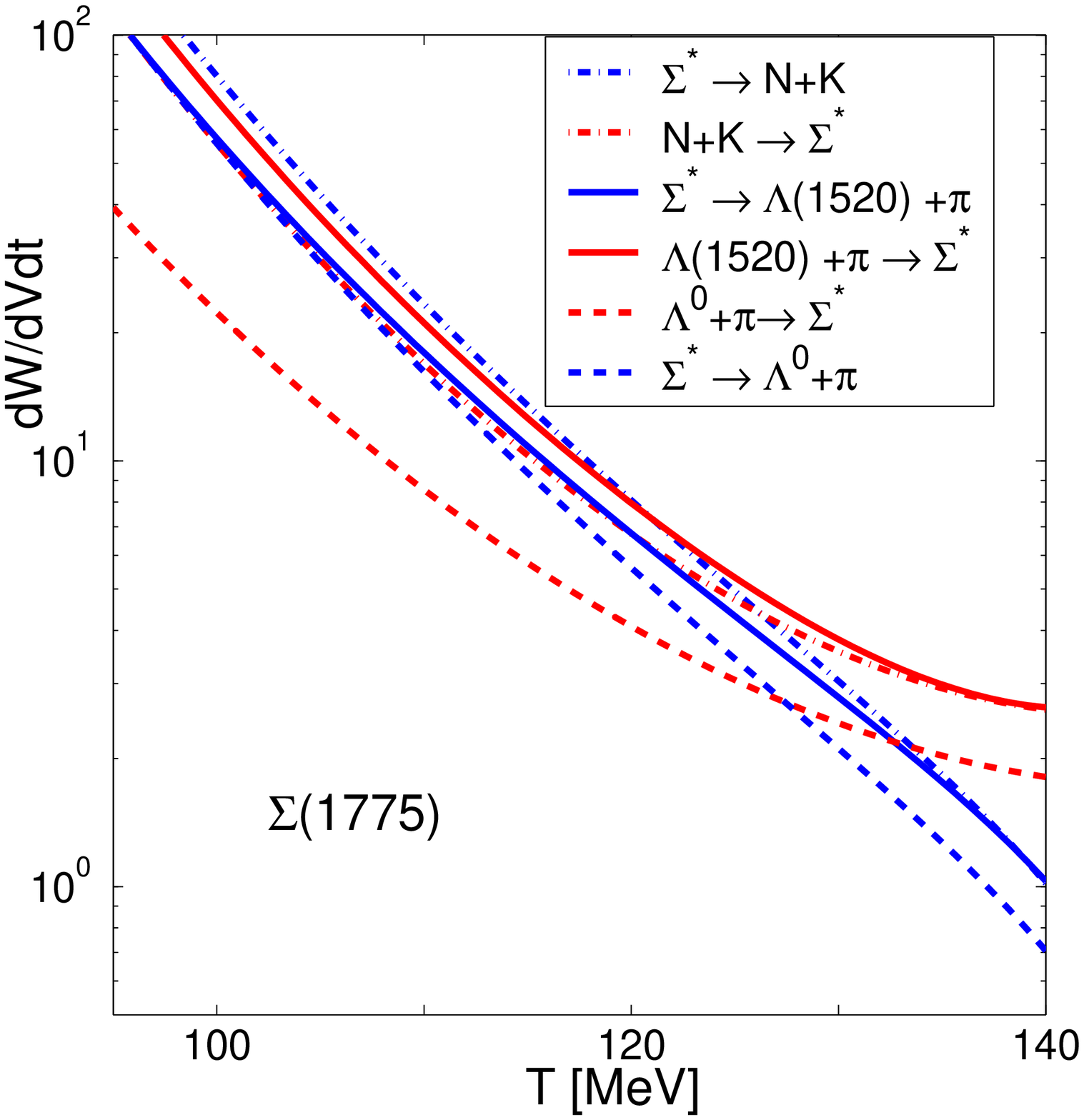}
\includegraphics[width=8 cm, height=10 cm]{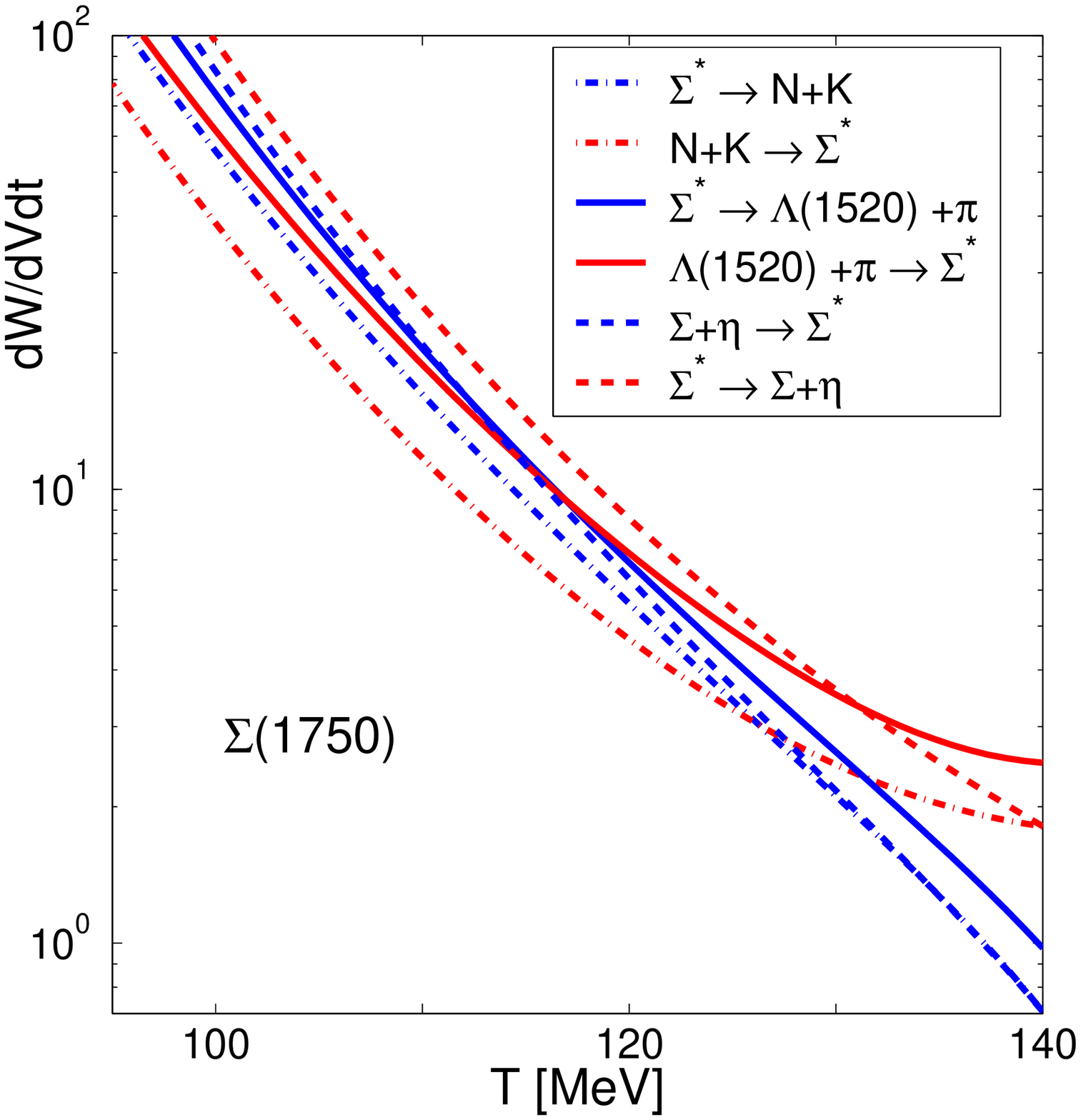}
\caption{\small{The rates for main channels of $\Sigma(1775)$ (on
the left) and  $\Sigma(1750)$ (on the right) decay and production as
a functions of temperature $T$  in the case when all reactions go in
both directions and $T_0=140$ MeV. Solid lines  are for reaction
$\Sigma^* \leftrightarrow \Lambda(1520) + \pi$; dash-dot lines are
for reaction $\Sigma^* \leftrightarrow N+K$; dashed lines are for
reaction $\Sigma(1775) \leftrightarrow \Lambda^0+\pi$ on the left
and $\Sigma(1750) \leftrightarrow \Sigma+\eta$ on the right; blue
and  red lines are for decay and backward fusion reaction,
respectively.}} \label{rates1775}
\end{figure}
%%%%%%%%%%%%%%%%%%%%%%%%%%%%%%%%%%%%%%%%%%%%%%%%%%%%%%

Particle fugacities, except $\Upsilon_{\pi}$, change rather rapidly.
Figure~\ref{Upsil} shows the computed $\Upsilon(t)$ as a function of
temperature $T(t)$. We present here the  scenario in which  all
reactions evolve in both directions, for the initial condition
$\gamma_s=\gamma_q$. The time, corresponding to the temperature
shown at the bottom, is shown at the top of figure~\ref{Upsil}, in
each frame. On the left we have hadronization at 140 MeV, in the
middle at 160 and to the right at 180 MeV. Each frame has the same
scale size for temperature unit, not time. For $\Upsilon_{\Sigma^*}$
we show two possible evolution examples, for $\Sigma_{1750}$
(dash-dot dark line) and $\Sigma(1775)$ (dashed line). These
resonances have significant influence on the $\Lambda(1520)$ yield.
The solid lines are for $\Upsilon_{\Lambda(1520)}$ (upper, red line)
and $\Upsilon_{\Sigma(1385)}$ (lower, light blue line). The dash-dot
and dashed light lines are for $\Upsilon_{\Sigma(1190)}$ and
$\Upsilon_{\Lambda^0}$, respectively. The upper dotted line is for
$\Upsilon_{N}$ and lower dotted line is for $\Upsilon_K$.

An important feature is that the $\Upsilon$s of massive hadron
(resonances) increase very fast when $T$ decreases.  This is so
since in absence of a rapid re-equilibration  reactions,
multiplicity of given resonance must  be conserved. Then, according
to Eq.~(\ref{BolzDis}) $\Upsilon_i \propto 1/K_2(m_i/T)$, and thus
for large $m_i$ $\Upsilon_i \propto \exp(m_i/T)$. We would expect
$\Upsilon_i>\Upsilon_j$, when $m_i > m_j$, and $T$ decreases. This
behavior is just like we found for the case of large charm fugacity
~\cite{Kuznetsova:2006bh}. However, because of the decay and
regeneration reactions there are some deviations from this
expectation in figure~\ref{Upsil}.

For $T_0=180$ MeV in most cases $\Upsilon_3>\Upsilon_1\Upsilon_2$
$(t>0)$. Massive resonances decay to lower mass particles. The
result is defined by resonance mass, its decay width and decay
products. For example $\Upsilon_{\Sigma(1775)}$ is smaller  than
$\Upsilon_{\Sigma(1750)}$ and $\Upsilon_{\Lambda(1520)}$, because of
its large decay width. Therefore excitation of $\Sigma(1775)$ by
$\Lambda$ slightly dominates over $\Sigma(1775)$ decay to
$\Lambda(1520)$ even in this case, when for most resonances the
decay is dominant. For smaller initial hadronization temperatures
$\Upsilon_{\Lambda(1520)}$ becomes smaller than
$\Upsilon_{\Sigma(1775)}$, and  even smaller than
$\Upsilon_{\Sigma(1385)}$ in some range of temperatures. This
suppression occurs because of $\Sigma(1775)$, and others $\Sigma^*$
regeneration. Because of large $\Upsilon_{\pi}$,
$\Upsilon_{\Sigma(1775)} < \Upsilon_{\Lambda(1520)}\Upsilon_{\pi}$,
the $\Sigma(1775)$ production by $\Lambda(1520)$ is dominant in the
full range of $T$  considered here.

%%%%%%%%%%%%%%%%%%%%%%%%%%%%%%%%%%%%%%%%%%%%%%%%%%%%%%%%%%%%%%%%%%%%%%%%%%%%%%%%%%%%%%%
\subsection{Final $\Lambda(1520)$ and $\Sigma(1385)$ multiplicities}\label{chapter 3.2}
%%%%%%%%%%%%%%%%%%%%%%%%%%%%%%%%%%%%%%%%%%%%%%%%%%%%%%%%%%%%%%%%%%%%%%%%%%%%%%%%%%%%%%

In this section  we consider the evolution of the multiplicity of \
resonances $\Lambda(1520)$, $\Sigma(1385)$, $\Sigma(1775)$ during
the kinetic phase. We use the Boltzmann yield limit. By the symbol $X(T)$  we refer to a
particular resonance, and $X_0$ is the initial multiplicity for that
resonance. The dynamic yield of this resonance may be expressed as
\begin{equation}
\frac{X(T)}{X_0} = \frac{\Upsilon_X(t)T(t)^3K_2(m_X/T(t))}
            {\Upsilon_{X\,0}T_0^3K_2(m_X/T_0)}
\end{equation}

Figure~\ref{Lam1520r} shows this yield as a function of T(t) for
$X=\Sigma(1385)$ (left) and $X=\Lambda(1520)$ (right). We consider
three   initial conditions, temperature $T_0 = 140, 160, 180$ MeV,
with corresponding $\gamma_q =1.6, 1.27, 1.0$, respectively. The
solid lines correspond for the model with dead channels and dashed
one are for case when all reactions are symmetric  in both
directions. The thin dotted vertical line at $T=120$ MeV marks the
kinetic freeze-out temperature, assumed before
in~\cite{Kuznetsova:2008zr}. The main result is that the resulting
relative yields for $\Lambda(1520)$ and $\Sigma(1385)$ behave
qualitatively different from  each other. In particular, as the
temperature decreases, for the case $T_0 =140$ MeV we observe a
strong yield suppression for $\Lambda(1520)$, and a strong
enhancement for $\Sigma(1385)$ (as compared to initial SHM yields).

To  better understand the mechanism of $\Lambda(1520)$ suppression,
we analyze in some detail the  case of $\Sigma(1775)$ and
$\Sigma(1750)$ decay and production rates $dW/dVdt$. We assume here
that these reactions can go in both directions. In figure
\ref{rates1775} we show the reactions rates for the principal
channels of decay and production as a functions of temperature T for
$\Sigma(1775)$ (left) and $\Sigma(1750)$ (right), for the case of
initial temperature  $T_0=140$ MeV which provides the largest
$\Lambda(1520)$ suppression. Solid lines are for the reaction
$\Sigma \leftrightarrow \Lambda(1520) + \pi$, dash-dot lines are for
reaction $\Sigma \leftrightarrow N+K$, dashed lines are for reaction
$\Sigma \leftrightarrow \Lambda^0+\pi$. Two set of lines are
presented for the  decay (on-line blue) and backward fusion reaction
(on-line red), respectively.

As temperature decreases, all rates $dW/dtdV$ are increasing
rapidly. This is mainly because fugacities $\Upsilon$  increase
nearly exponentially when number of particles is conserved, see
figure \ref{Upsil}. We see that at the beginning of the kinetic
phase all reactions go in the direction of $\Sigma(1775)$
production, since $\Sigma(1775)$ production rate is larger than its
decay rate for all channels. Then at  first $\Sigma(1775)
\leftrightarrow \Lambda^0+\pi$   decay rate    becomes dominant over
$\Sigma(1775)$ production rate in this channel, followed by the same
for $\Sigma(1775)\leftrightarrow N+K$ channel.

For the reaction  $\Sigma(1775) \leftrightarrow \Lambda(1520) + \pi$
backward reaction is always dominant. As  result, during the
kinetic phase always more $\Lambda(1520)$ resonances are excited
into $\Sigma(1775)$ than they are produced by $\Sigma(1775)$ decay.
The reason for this is the decay of $\Sigma(1775)$ to the other channels,
as long as   $\Upsilon_{\Sigma(1775)}<\Upsilon_{\Lambda(1520)}\Upsilon_{\pi}$.
The lighter is the total mass of decay products,  the earlier the
decay reaction becomes dominant. This is due to the fact that
the fugacity of $\Upsilon$ for heavier particles increases faster
with expansion. Therefore, the decay rate becomes dominant earlier,
when the difference between initial and final mass is larger.
The  net result is $\Lambda(1520)$ suppression by $\Sigma(1775)$ excitation.

\begin{figure}
\centering
\includegraphics[width=8 cm, height=10 cm]{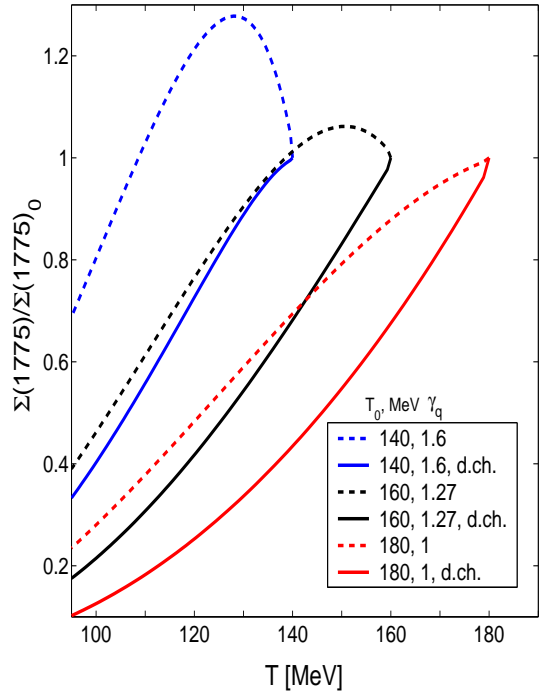}
\caption{\small{The ratio
$\Sigma(1775)/\Sigma(1775)_0$
as a functions of temperature $T(t)$ for
different initial hadronization temperatures $T_0=140$, $160$
and $180$ MeV (blue/bottom, black/middle and red/top lines), respectively. Solid lines
are for calculations with dead channels, dashed lines are for
calculations without dead channels.}} \label{Sigm1775r}
\end{figure}

In figure~\ref{Sigm1775r} we show the yield of $\Sigma(1775)$
normalized by its initial yield at hadronization:
$\Sigma(1775)/\Sigma(1775)_0$ as a function of $T(t)$. Like in the
other figures above, solid lines are for the dead channels and
dashed lines are for case when reactions go in both directions,
solid (blue) lines are for $T_0=140$ MeV, solid (black) lines for
$T_0=160$ MeV, and solid (red) lines are for $T_0=180$ MeV. Each of
the lines can be identified by their initial $T$-value. We see that
when all reactions go in both direction the ratio
$\Sigma(1775)/\Sigma(1775)_0$ increases at first similar to
$\Sigma(1385)/\Sigma(1385)_0$ and $\Delta(1230)/\Delta(1230)_0$
ratios~\cite{Kuznetsova:2008zr}.

Compared to these ratios, $\Sigma(1775)/\Sigma(1775)_0$ ratio
reaches its maximum value earlier, and after the maximum, the yield
of $\Sigma(1775)$ decreases faster. The reason for this behavior is
that the mass of $\Sigma(1775)$ is larger. The phase space occupancy
$\Upsilon_{\Sigma(1775)}$, and therefore its decay rates,  increase
faster than the fugacity and decay rates for $\Sigma(1385)$ and
$\Delta(1230)$. Therefore decays $\Sigma(1775)$ to some channels and
its total decay rate become dominant earlier (see figure
\ref{rates1775}). Although the total decay width of $\Sigma(1775)$
is approximately the same as for $\Delta(1230)$, the maximum value
of this ratio is smaller.

Said differently, the maximum yield of $\Sigma(1775)$ does not have
time to reach the value as high as that for $\Delta(1230)$. We thus
learn that the time evolution of the yield of resonances with large
decay width depends not only on their decay width, but also on mass
difference between initial and final states. Similar time evolution
occurs for the other $\Sigma^*$, which quantitatively depends on
their mass, decay products masses and decay width.

For most $\Sigma^*$s, the decay products in the channel
$\Lambda(1520)+\pi$ are heavier than the decay products in others
channels, which are thus favored by phase space.  For most
resonances in our range of temperature, the decay into
$\Lambda(1520)+\pi$ remains weak. The exception is $\Sigma(1750)$
which decays also to $\Sigma + \eta$, see figure~\ref{rates1775}.
($m_{\Sigma}+m_{\eta} > m_{\Lambda(1520)}+m_{\pi}$). $\Sigma(1750)$
begins to decay dominantly to $\Lambda^0(1520)$ at relatively low
temperature $T=116$ MeV, and continues to be produced by
$\Sigma+\eta$ fusion.

As a result,  allowing  all reactions to go in both directions, the
ratio $\Lambda(1520)/\Lambda(1520)_0$ has a minimum. This is
specifically due to  $\Sigma(1750)$) decay back to $\Lambda(1520)$
at small temperatures as described above. However,  when we  satisfy
Eq.(\ref{dchcon})  for dead channels the only decay occurs in the
beginning of kinetic the dead-channel model  phase.
 In that case the $\Upsilon_{\Sigma^*}$s are smaller,
and the rate of reaction $\Lambda(1520)+\pi\rightarrow \Sigma^*$
exceeds the rate for backward reaction by  larger amount, compared
to the scenario without dead channels. This amplifies the effect of
$\Lambda(1520)$ suppression. In this case, $\Sigma^*$  decay to
lighter hadrons right after they are produced by $\Lambda(1520)$. We
can see that for $T_0 = 140$ MeV and $T_0 = 160$ MeV $\Lambda(1520)$
yield is always decreasing in the here considered temperature range.

For $\Sigma(1385)$ multiplicity we find a result quite different from
$\Lambda(1520)$ behavior discussed here, but  similar to what we obtained in
\cite{Kuznetsova:2008zr} by a very different method in a smaller basis set of states.
In particular, the $\Sigma(1385)$
yield is enhanced, but the maximum value of
$\Sigma(1385)/\Sigma(1385)_0$ we find is  a few percent  higher, since
 we took into account the Bose enhancement of interaction rates,
reaction (\ref{S1385S}), and $\Sigma^*$ production. $\Sigma(1385)$ contribution to
$\Sigma^*$ production is small, compared to the influence of the first two effects.
The time (i.e. temperature) evolution of $\Sigma(1385)$ practically does not depend on
the presence of dead channels, and the   maximum enhancement of $\Sigma(1385)$
is even less sensitive. This in fact
indirectly confirms that  $\Sigma^*$ has a small influence on $\Sigma(1385)$ multiplicity.
Thus we confirm that: \\
a)   for $T_0=180$ MeV $\Sigma(1385)$   evolves with the system following
the ambient temperature;\\
b)    for $T_0=160$ MeV $\Sigma(1385)$  shows some increase in yield;\\
c)   for $T_0=140$ MeV there is a strong yield increase of
$\Sigma(1385)$.

While there is little sensitivity in the  yield of $\Sigma(1385)$ to issue of particle momentum distribution
(little difference between the two models considered, dashed and solid lines), the $\Sigma(1385)$ yield is highly
sensitive to initial  hadronization condition.  While for
$\Sigma(1385)$ the yield increases with decreased hadronization
temperature, for $\Lambda(1520)$ the opposite is true,  and in
particular the smallest final $\Lambda(1520)$ yield  corresponds to
the smallest hadronization temperature  for both models.

%%%%%%%%%%%%%%%%%%%%%%%%%%%%%%%%%%%%%%%%%%%%%%Fig 8
%%%%%%%%%%%%%%%%%%%%%%%%%%%%%%%%%%%%%%%%%%%%%%%%%%%%%%

%%%%%%%%%%%%%%%%%%%%%%%%%%%%%%%%%%%%%%%%%%%%%%Fig 9
\begin{figure*}
\centering
\includegraphics[width=8 cm, height=10 cm]{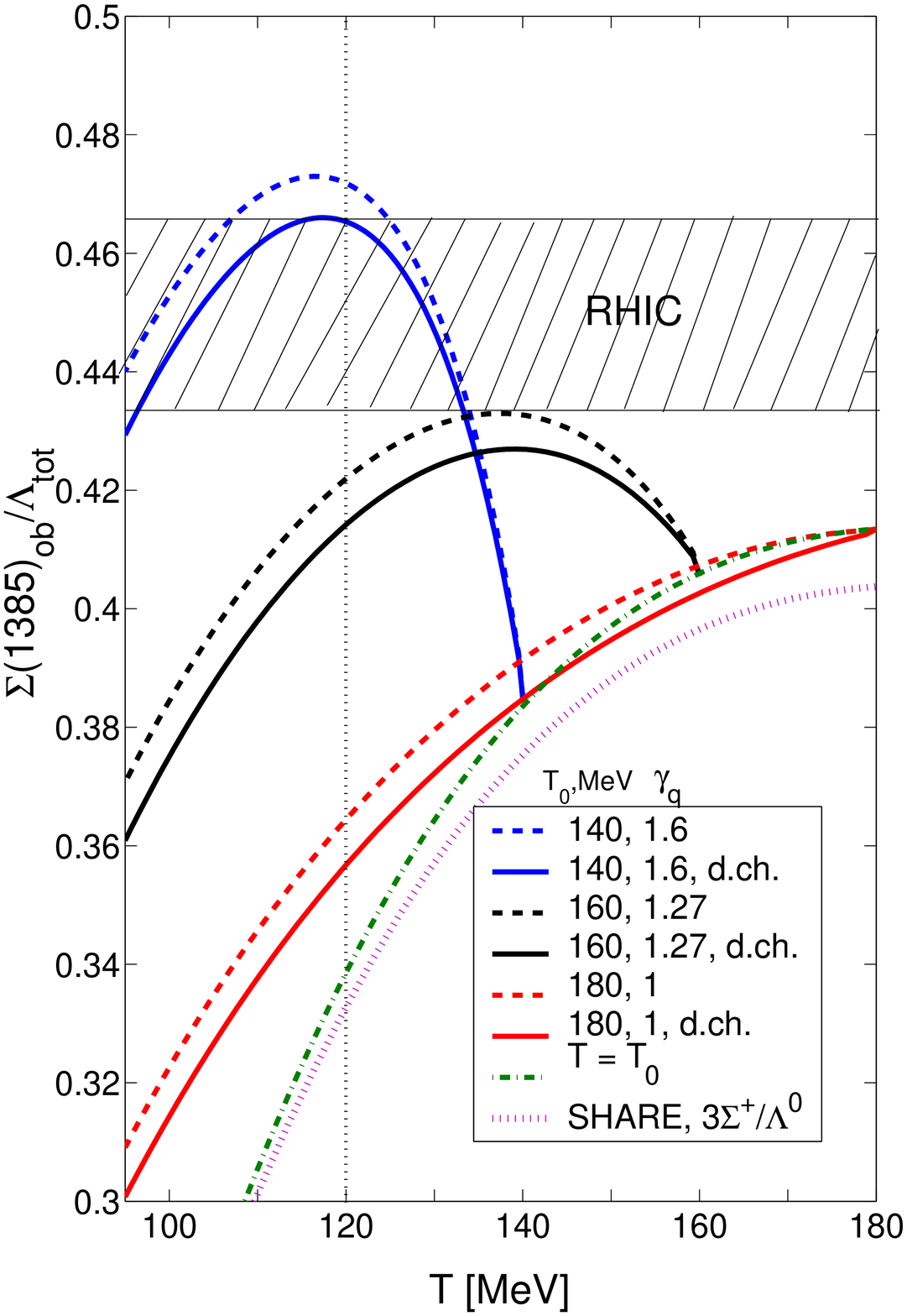}
\includegraphics[width=8 cm, height=10 cm]{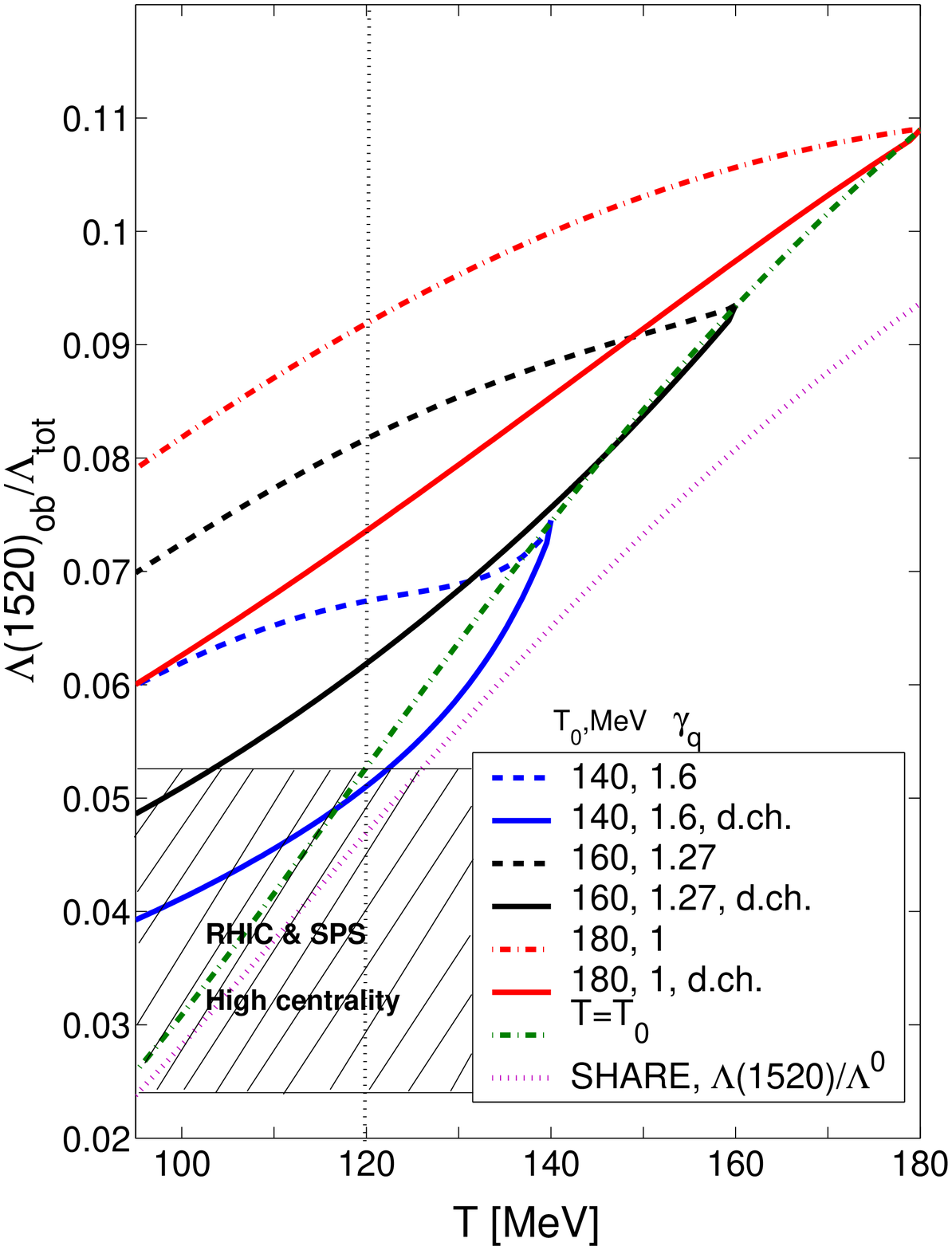}
\caption{\small{The ratios $\Sigma(1385)/\Lambda_{\rm tot}$ (on
left) and $\Lambda(1520)/\Lambda_{\rm tot}$ (on right) as a function
of temperature $T$ of final kinetic freeze-out, for different
initial hadronization temperatures $T_0=140$, $160$ and $180$ MeV
(blue, black and red lines, respectively). Dashed lines are for
calculations without dead channels, solid lines are for calculations
with dead channels.  The dotted purple line gives the expected SHM
chemical equilibrium  result. The dash-dot line is relative yield
result result for SHM with $T_0=T$.}} \label{lamltot}
\end{figure*}
%%%%%%%%%%%%%%%%%%%%%%%%%%%%%%%%%%%%%%%%%%%%%%%%%%%%%%

%%%%%%%%%%%%%%%%%%%%%%%%%%%%%
\subsection{Experimentally measurable resonance  ratios}\label{chapter 3.3}

The initial hadronization yields, which we used as a reference in
figure~\ref{Lam1520r}  in order to understand the physical behavior,
are not measurable. What is commonly used as a reference for the
yields of single strange hyperon resonances is the overall yield of
the stable $\Lambda^0(1115)$, without the weak decay feed from
$\Xi$. Aside of the initially produced particles, the experimental
yield of $\Lambda^0(1115)$  also includes resonances decaying during
the free expansion after kinetic freeze-out, in particular (nearly)
all decays of   $\Sigma(1385)$, and the experimentally inseparable
yield of $\Sigma^0(1193) \to \gamma+\Lambda^0$ decay and the decay
of any further hyperon resonances $Y^*$.

Thus we normalize our final result with the experimentally
observable final $\Lambda_{\rm tot}^0$ hyperon yield:
\begin{equation} \label{Ltot}
\Lambda_{\rm tot}= \Sigma^0(1193)  + 0.91 \Sigma(1385)+ \Lambda + Y^*.
\end{equation}
The factor $0.91$ shows that $91\%$ of end-state $\Sigma (1385)$
decays to $\Lambda$. We also included in  $\Lambda_{\rm tot}$
calculations decays of $\Xi^* \rightarrow \Lambda+K$, which makes
the result slightly dependent on $\gamma_s/\gamma_q$ ratio. We use
$\gamma_s/\gamma_q =1$, since this ratio value is expected at top
RHIC energy~\cite{Kuznetsova:2006bh}.

As noted,   $\Lambda(1520)$ and $\Sigma(1385)$
experimentally observable yields also include any decays which occur
in the free-streaming post-kinetic period. Thus we have:
\begin{eqnarray}
\Sigma(1385)_{\rm ob}=\Sigma(1385)+Y^*_{\Sigma(1385)},\\
\Lambda(1520)_{\rm ob}=\Lambda(1520)+Y^*_{\Lambda(1520)},
\end{eqnarray}
where $Y^*_{\Sigma(1385)}$ and $Y^*_{\Lambda(1520)}$ are hyperon
multiplicities at kinetic freeze-out temperature, and which decay to
$\Sigma(1385)$ and $\Lambda(1520)$, respectively. The multiplicities
$\Sigma(1385)$ and $\Lambda(1520)$ are taken at the moment of
kinetic freeze - out.

In  figure~\ref{lamltot} we present the fractional yields
${\Sigma(1385)}/\Lambda_{\rm tot}$ (left), and
${\Lambda(1520)}/\Lambda_{\rm tot}$ (right) as a function
of temperature of final kinetic freeze-out $T$.
The results  for the hadronization
temperatures $T_0=140$ (blue lines), $T_0=160$ (black
lines) and $T_0=180$ MeV (red lines) are shown. Solid lines
are for the case with dead channels and dashed lines are for the
case when all reactions are going in both directions.

In  figure~\ref{lamltot}  the green dash-dotted line is the result
when the kinetic freeze-out temperature $T$ coincides with the
hadronization temperature $T_0$. There is no kinetic phase in this
case, only   resonances decay after hadronization. This result is
similar to SHARE result (purple, dotted line). The  small difference
is mainly due to us taking into account the decays
\begin{equation}
\Sigma(1670, 1750) \rightarrow \Lambda(1520) + \pi,
\end{equation}
which are expected/predicted in~\cite{Cameron:1977jr}.
Similarly, for $\Sigma(1385)$ our results for $T_0=T$
are different from SHARE results because we include the decay:
\begin{equation}
\Sigma(1670) \rightarrow \Sigma(1385) + \pi,
\end{equation}
expected/predicted in~\cite{Prevost:1974hf}. These additional resonances
are part of current particle data set~\cite{Amsler:2008zz}.

For all initial
hadronization temperatures, as the freeze-out temperature decreases, the suppression for
$\Lambda(1520)_{\rm ob}/\Lambda_{\rm tot}$ ratio is larger than for
$\Lambda(1520)/\Lambda(1520)_0$ (at the same temperature $T$ of final kinetic freeze-out).
 This is particularly evident for dead channels and hadronization temperatures $T_0=160,\, 180$ MeV
(see figure~{\ref{Lam1520r}}).
The effect is due to $\Sigma(1775)$ suppression, as shown in
figure~\ref{Sigm1775r} (and similar for other $\Sigma^*$).
For $T_0=140$ MeV the additional suppression of $\Lambda(1520)$,
described above, is relatively small.

For $T_0=140$ MeV in the case without dead channels at final kinetic
freeze-out $T>120\,{\rm MeV}$, the final observed $\Lambda(1520)$
suppression is even smaller, compared to its suppression in the
kinetic phase at the same temperature  (see
figure~{\ref{Lam1520r}}). The reason is that yield of $\Sigma(1775)$
(and of the other $\Sigma^*$s) is much enhanced for this range of
temperatures see figure~\ref{Sigm1775r}. This additional
$\Sigma(1775)$ decays back to $\Lambda(1520)$. That results in a
smaller suppression at these temperatures.

The  above suppression effect   increases in magnitude  for higher
hadronization temperatures, since the suppression of $\Sigma(1775)$
and the sensitivity of $\Lambda(1520)_{\rm ob}$ multiplicity to
$\Sigma^*$ decays increase with temperature. However, when we
consider dead channels (see figure~\ref{lamltot}), the former effect
of $\Lambda(1520)$ suppression during evolution of  kinetic phase
increases for decreasing hadronization temperatures. Thus in the
combined effect, the  observable  relative suppression of
$\Lambda(1520)_{\rm ob}/\Lambda_{\rm tot}$, is approximately of the
same magnitude  for all hadronization temperatures $T_0$. However,
the initial hadronization yield of  $\Lambda(1520)$ is sensitive to
temperature, and decreases rapidly with $T$. Therefore only for $T_0
= 140 $ MeV, a kinetic freeze-out temperatures $\approx 95 -105$
MeV, and allowing for dead channels, the  ratio $\Lambda_{\rm
ob}(1520)/\Lambda_{\rm tot}$ reaches the experimental domain
$\Lambda_{\rm ob}(1520)/\Lambda_{\rm tot}<0.042\pm
0.01$~\cite{Adams:2006yu,Markert:2002xi} shown in
figure~\ref{lamltot} by dashed lines.

For the same initial conditions, that is
for $T_0=140$ MeV, we find the ratio $\Sigma(1385)/\Lambda_{\rm tot} \approx$
 0.45 at $T \approx 100$ MeV  (and for the entire range 95 -- 135 MeV,
in good agreement with experimental
data~\cite{Adams:2006yu,Salur:2006jq}). In~\cite{Kuznetsova:2008zr}
this value of $\Sigma(1385)/\Lambda_{\rm tot}$  is found at $T=120$ MeV,
which was in the reference the presumed lowest possible temperature of the final kinetic freeze-out.
Here we find that at $T=120$ MeV the ratio $\Sigma(1385)/\Lambda_{\rm tot}$ can be even higher
(about 0.47), which is due to the Bose enhancement of
in-medium $\Sigma(1385)$ production rate (see discussion following
figure~\ref{Lam1520r}).

%%%%%%%%%%%%%%%%%%%%%%%%%%%%%%%
%%%%%%%%%%%%%%%%%%%%%%%%%%%%%%
%%%%%%%%%%%%%%%%%%%%%%%%%%%%%
\section{Summary and conclusions}\label{chapter 4}

The resonant hadron states, considering their very  large decay and
reaction  rates, can interact beyond the chemical and thermal
freeze-out of stable particles. Thus the observed yield of
resonances is fixed by the physical conditions prevailing at a later
breakup of the fireball matter rather than the production of
non-resonantly interacting hadrons. Moreover,  resonances, observed
in  terms of the invariant mass signature,  are only visible when
emerging from a more dilute hadron system  given the ample potential
for rescattering of decay products. The combination of experimental
invariant mass method with a large resonant scattering makes the
here presented  population study of resonance kinetic  freeze-out
necessary. The evolution effects we find are greatly amplified at
low hadronization temperatures where greatest degree of initial
chemical equilibrium is present.

Our study quantifies the expectation that in a dense hadron  medium
narrow resonances are ``quenched''\cite{Rafelski:2001hp} that is,
effectively mixed with other states, and thus their observed
population is reduced. Since we follow here the particle density,
the effect we study is due to incoherent population  mixing of
$\Lambda(1520)$, in particular with $\Sigma^*$. This effect is
possible for particle densities out of chemical non-equilibrium.
However, this mixing can occur also at the amplitude (quantum
coherent) level. As the result the yield suppression effect  could
further increase, in some situations further improving the agreement
with experiment.

In first part of this chapter, 
we have presented master equation governing the evolution in time
of the $\Delta, \Sigma(1385)$ baryon resonance yield after QGP hadronization, allowing
for  resonance decay and production process.
We have shown  considering the properties of the master equation that if the
yield of hadrons is initially above chemical equilibrium,
the resonance population increases beyond the initial   yield.
Conversely, we find that in a physical system  in which the particle
multiplicities  of hadrons arise   below chemical equilibrium yields, a circumstance
expected below threshold to QGP formation, the final  yield of resonances is
suppressed by  the  dominance of the resonance decay process over back reaction  resonance production.

In a quantitative model we evolved the yields
after QGP hadronization allowing for initial chemical
non-equilibrium particle abundances,  and volume expansion assuring  entropy
conservation. We found, see   figure  \ref{ndNtot}, that the thermal freeze-out fractional resonance yield
differs significantly from the chemical-freeze out SHM expectation, with the scenario involving
high-$T$ hadronization resonance yield being depleted,
and low-$T$ hadronization yield  scenario further enhanced in relation to the total  yield.

The resonance enhancement effect  we presented
can only occur when the initial state is out of chemical equilibrium, and the decay/formation processes
are fast enough to compete
with  the hadron volume evolution. One would thus think that  `narrow', i.e. quasi-stable resonances
are not subject to the effects considered here. However, a special consideration
must be given to   narrow resonance which are  strongly coupled to more massive resonances
which can decay fast into other channels. An example is  $\Lambda(1520)$, which is considered in next section.
Aside of several specific predictions we made here,  there are three important  general consequences of our study: a)
the fractional yield of resonances $A^*/A$ can be considerably higher than expected naively
in SHM model of QGP hadronization,   b) since there is nearly
a factor of two difference in the final thermal freeze-out ratio in $\Delta/N_{\rm tot}$, while the SHM yields
  a more $T$ independent result, one can imagine the use of $\Delta/N_{\rm tot}$ as
a tool to distinguish the different hadronization conditions e.g. chemical non-equilibrium vs chemical equilibrium
a point noted in similar context before~\cite{Torrieri:2006yb}; and c) we have
shown that the relatively high yield of charged  $\Sigma^\pm(1385)$
reported by STAR is  well explained  by our considerations
with  hadronization at $T=140$ MeV being favored.

In second model in this chapter our results show that the observable ratio $\Lambda(1520)_{\rm
ob}/\Lambda_{\rm tot}$ can be suppressed by two effects. First
$\Lambda(1520)$ yield is suppressed   due to excitation of heavy
$\Sigma^*$s in the resonance scattering process. Moreover, the final
$\Lambda(1520)_{\rm ob}$ yield is suppressed, because $\Sigma^*$s,
which decay to $\Lambda(1520)$, are suppressed at the end of the
kinetic phase evolution by their (asymmetric) decays to lower mass
hadrons, especially when dead channels are present (see
figure~\ref{Sigm1775r}). As a result, fewer  of these hadrons can
decay to $\Lambda(1520)_{\rm ob}$ during the following free
expansion. A contrary mechanism operates for the resonances such as
$\Sigma(1385), \Delta(1230)$. These resonances can be so strongly
enhanced, that in essence most final states strange and non-strange
baryons come from a resonance decay.

We note that despite a   scenario dependent resonance formation or suppression,
the stable particle yields used in study of chemical freeze-out remain unchanged, since
all resonances ultimately decay into the lowest ``stable'' hadron.   Therefore after a
description e.g. within a statistical hadronization model  of the yields
of stable hadrons, the understanding of resonance yields is a second, and  separate task
which helps to establish the consistency of our physical understanding of the hadron
production process.

We conclude noting the key result of this study, that we can now understand the opposite behavior
of $\Lambda(1520)$ (suppression in high centrality reactions) and $\Sigma^(1385) $ (enhancement,
 and similarly $\Delta(1230)$) by considering their rescattering in matter. In order to explain both,
the behavior of the $\Lambda(1520)_{\rm ob}/\Lambda_{\rm tot}$
and $\Sigma(1385)/\Lambda_{\rm tot}$ ratios, one has to consider
$T=95-100$ MeV as the  favorite temperature of final kinetic
freeze-out  of hadron resonances, with   $T_0=140$ MeV being the
favored chemical freeze-out (hadronization, QGP break-up) temperature.
When there is little  matter  available to scatter, e.g. in peripheral  collisions,
  the average value of
$\Lambda(1520)_{\rm ob}/\Lambda_{\rm tot}$ ratio is higher,
approaching  the   expected chemical freeze-out  hadronization yield
for $T_0=140$ MeV. All these findings are in good agreement with
available experimental data.

\chapter{RELATIVISTIC $e^+e^-\gamma$ PLASMA CREATED BY LASER PULSE}  \label{eeg}
\section{Freeze-out condition of relativistic $e^-,e^+,\gamma$-plasma} \label{frcon}
\subsection{Introduction}
The formation of the relativistic ,
electron-positron-photon $ e^-, e^+, \gamma $  plasma  (EP$^3$, temperature $T$ in MeV range) in
the laboratory using ultra-short pulse lasers is one of the current topics
of interest and forthcoming experimental effort
~\cite{TajMou,TajMouBoul}.

For an  expanding  drop of plasma there is the freeze-out size $R$ where the particle 
density $\rho\propto 1/R^3$ decrease  allows the free-out-streaming of all particles,
since the scattering length $l\propto 1/(\sigma \rho)$ 
grows with $R^{3}$.
Here we consider this freeze-out condition  for  a relativistic $e^-,e^+,\gamma$ plasma. 
The conventional wisdom from keV temperature `fusion'
domain implies that an opaque plasma drop is not possible without 100's of MJ of energy. 

Here  we demonstrate a new temperature domain in which 
opaque plasma drops are possible for the energy content  
of $\cong 0.5$\,kJ  with a radius   in a range of 
$R=2  \div  10$~nm, at a temperature at the scale of MeV. 
This new and interesting plasma domain arises since for $T>m_e$ 
the density of electron-positron pairs  grows rapidly and the scattering 
length $l$ accordingly decreases rapidly. These physical conditions should 
become accessible in the foreseeable future upon the development 
of wavelength compression technology employing an  optical wavelength laser beam
reflected from a relativistic mirror, generated by a pulsed high intensity 
laser~\cite{Bulanov:2003zz}.

We evaluate mean free path length $l$ of photon in EP$^3$ plasma for
Compton scattering and pair production assuming thermal
equilibrium. By comparing this length with plasma size at constant energy content we determine at
what conditions plasma can be opaque and therefore in thermal
equilibrium. For energy 0.5 kJ we study the limits for plasma size and temperatures. Similarly, we can also find chemical equilibration conditions, considering reactions of pair production 
and annihilation. We evaluate photon free path  using the method of thermal Lorentz invariant reaction rate, which was used before
for strangeness production~\cite{Koch:1986ud},~\cite{Matsui:1985eu}.
This method allows us to take into account quantum effects in dense medium and
easy to use in the observer rest frame. Note that corresponding equations can also be used in the astrophysical plasma environment.

To create plasma drop with lager radius for
given energy, the plasma temperature must be decreased. However we have a limit on the lowest temperature and therefore an upper limit on plasma size, defined by opaqeness condition at fixed plasma energy. For example, we will show at section~\ref{plpr2} that fully 
plasma has maximum radius of ~7 nm at temperature ~2 MeV [and energy E=0.5kJ]. 
We found that to create low temperature ($T < 0.5$ MeV) opaque $EP^3$ we need to deposit much more energy to a larger volume.
This is due to the fact that the photon free path and, therefore plasma size, grow exponentially for this low temperature, when densities of photons with $E>m$ and electron-positron pair are small.

In susection~\ref{plpr1} of this section we discuss statistical properties of EP$^3$, including
master equation for electron - positron photon densities chemical equilibration
under assumption that the particles are in thermal equilibrium.
In subsection~\ref{plpr2} we calculate the photon mean free path in
plasma for Compton scattering and pair production and disscuss possible plasma size at given energy.
In subsection~\ref{conc1} there are summary and conclusions 

%%%%%%%%%%%%%%%%%%%%%%%%%%%%%%%%%%%%%%%%%%%%%%%%%%%%%%%%%%%
\subsection{Statistical properties of EP$^3$ plasma}\label{plpr1}
%%%%%%%%%%%%%%%%%%%%%%%%%%%%%%%%%%%%%%%%%%%%%%%%%%%%%%%%%%
Up to small QED interaction effects we can
use Fermi and Bose momentum distribution ~(\ref{bf}), respectively to describe
the particle content in the plasma
\begin{equation} \label{fstat}
f_{e^\pm}  = \frac{1}{\Upsilon_ee^{(u\cdot p_{e}\pm\nu_e)/T} +1},\quad
f_{\gamma}  =  \frac{1}{\Upsilon_{\gamma}e^{u\cdot p_{\gamma}/T} -1},
\end{equation}

When the electron chemical potential $\nu_e$ is small, $\nu_e\ll T$,
the number of particles and antiparticles is the same,
$n_{e^-}=n_{e^+}$. Physically, it means that the number of $e^+e^-$
pairs produced is dominating residual matter electron yield. Here we
will set $\nu_e=0$ , and will consider elsewhere the case for very
low density plasma where chemical potential may become important.
$\Upsilon_{e(\gamma)}$ is the fugacity of a given particle.

If plasma size is large enough, then plasma is opaque for photon electron
scattering or pair production:
\begin{eqnarray}
&&\gamma+\gamma \leftrightarrow e^+ + e^-; \label{ggee}\\
&&\gamma+e^{\pm}\leftrightarrow \gamma + e^{\pm}. \label{ge}
\end{eqnarray}

This plasma lives long enough and
electrons and positrons are in thermal and relative chemical
equilibrium with photons. The maximum photon density (black body
radiation) is reached when photon fugacity $\Upsilon_{\gamma}=1$.
Plasma is in chemical equilibrium, when $\Upsilon_{e}=1$, and all others
particles fugacities are $\Upsilon_{i}=1$. 
Under this condition the plasma density has maximum value at given $T$.

In our model we assume that the relative $e^+e^-$ pair and photon yields are
equilibrated by pair production and annihilation reactions. If we assume that
thermal equilibrium establishes faster than chemical then
the photon density evolution equations and chemical equilibrium conditions are
similar to those for muon production considered
in~\cite{Kuznetsova:2008jt}:
\begin{equation}
\frac{1}{V}\frac{dN_{\gamma}}{dt} = (\Upsilon_{e}^2
-\Upsilon_{\gamma}^2){R_{\gamma\gamma \leftrightarrow e^+e^-}} ,
\label{gammapr}
\end{equation}
where 
$$R_{\gamma\gamma\leftrightarrow e^+e^-} = \frac{1}{\Upsilon^2_{\gamma}} \frac{dW_{\gamma\gamma \rightarrow e^+e^-}}{dVdt}= \frac{1}{\Upsilon^2_{e}} \frac{dW_{e^+e^- \rightarrow \gamma\gamma}}{dVdt},$$ 
${dW_{\gamma\gamma \rightarrow e^+e^-}}/{dVdt}$ and ${dW_{e^+e^- \rightarrow \gamma\gamma}}/{dVdt}$ are Lorentz invariant
rates for pair production and annihilation reactions, respectively. 
Then the EP$^3$ plasma is in {\em relative chemical equilibrium} for
$\Upsilon_{\gamma}<1$ when
\begin{equation}
\Upsilon_{e} = \Upsilon_{\gamma}=\Upsilon. \label{eppeq}
\end{equation}
At $\Upsilon \to 1$ we achieve full chemical equilibrium.

We introduce pair production relaxation time defined by:
\begin{equation}
\label{taugg} \tau^{ch}_{\gamma\gamma} =
\frac{1}{2\Upsilon}\frac{dn_{\gamma}/d\Upsilon_{\gamma}}{R_{\gamma\gamma
\leftrightarrow e^+e^-}},
\end{equation}
then for the simplest case $\Upsilon(t)=$ const, $T(t)=$ const and
$R(t)=$ const the equation for $\Upsilon_{\gamma}$ is
\begin{equation}
\frac{d\Upsilon_{\gamma}}{dt} = (\Upsilon^2
-\Upsilon_{\gamma}^2)\frac{1}{2\Upsilon\tau^{ch}_{\gamma\gamma}}. \label{upgeq}
\end{equation}
The rates and relaxation times are discussed in depth
in~\cite{Kuznetsova:2008jt}.

If we introduce variable $\gamma = \Upsilon_{\gamma}/\Upsilon$, which
shows deviation from chemical equilibrium, the equation for $\gamma$
is
\begin{equation}
\frac{d\gamma}{dt} = (1 - \gamma^2) \frac{1}{2\tau^{ch}_{\gamma\gamma}}.
\label{muev2}
\end{equation}
The choice for the definition of relaxation time is made such that
particle multiplicity reaches magnitude of equilibrium value during the time interval on the order of relaxation time.
Note that the relaxation time for particle ($\gamma$ in this example)
production in two to two particles reaction increases by factor
$\Upsilon_i^{-1}$, where $i$ is the initial particle in reaction
(here $e^{\pm}$). The physical reason why one introduces
$\Upsilon_i^{-1}$ into the relaxation time is that the collision
rate drops by that factor due to reduced density in plasma.

In simple case, considered here, $e^+e^-$ pair production and annihilation are chemically 
equilibrated when
\begin{equation}
\tau^{ch}_{\gamma\gamma}<<\tau_{pl}, \label{cheqtau}
\end{equation}
where $\tau_{pl} \approx 0.1 \div 10$ fs
is lifespan of plasma. 

The plasma properties, such as particle density and energy density can be
evaluated using relativistic expressions:
\begin{equation}%{eqnarray}
n_i =\int g_i f_i(p)d^3p,\qquad E = \int \sum_i g_iE_if_i(p)d^3pV,
\label{Etherm}
\end{equation}%{eqnarray}
where $E_i=\sqrt{m_i^2+\vec p^{\,2}}$, $f_i(p)$ is the  momentum
distribution of the particle $i\in \gamma$, $e^{\pm},\mu^{\pm}$,
$\pi^0, \pi^{\pm}$ and $g_i$ its degeneracy: $g_i=1$ for $\pi^0$ and
$g_i=2$ for the other particles, which can contribute.

It is convenient to parameterize the equilibrium electron, positron
and photon $ e^-, e^+, \gamma$ plasma properties in terms of
the properties of the Stephan-Boltzmann law for  massless particles
(photons). We present energy of plasma in terms of the effective
degeneracy $g(T)$ comprising the count of all particles present at a
given temperature $T$. Energy at $\Upsilon =1$ is
\begin{equation}\label{SB}
E = g(T)  \sigma T^4,\qquad \sigma=\frac{\pi^2}{30}.
\end{equation}
At temperatures $T\ll m_e$ we only have truly massless photons and
$g(T) \simeq 2_\gamma$. Once temperature increases beyond $m_e$ we
find  $g \simeq 2_\gamma+(7/8)(2_{e^-}+2_{e^+})=5.5$ degrees of
freedom when $\Upsilon=1$. The factor 7/8 expresses the difference
in the evaluation of  Eq.\,(\ref{Etherm}) for the momentum
distribution of Fermion and Boson Eq.\,(\ref{fstat}), with Bosons
providing the reference  point at low $T$, where only massless
photons are present. In principle these particles acquire additional
in medium mass which reduces the degree of freedom count. However
this effect is compensated by collective `plasmon' modes. Thus we
proceed with naive counting of nearly free EP$^3$ components.

In classical case, $\Upsilon << 1$, we have for massless particles
($m/T \rightarrow 0$)
\begin{equation}
E = 3NT, \qquad N=\Upsilon\frac{g}{\pi^2}T^3V, \label{energ}
\end{equation}
where $g = 6$. There is no difference between Bose and Fermi
particles.

The densities and multiplicities of heavy particles ($m_i >> T$) can
be calculated using relativistic Boltzmann distribution:
\begin{equation}
\label{bolpi0}
\frac{N_{i}}{V} \equiv n_{i} = \Upsilon_{i}\frac{g_i}{2\pi^2}Tm_i^2
K_2(m_{i}/T),
\end{equation}
where subscript $i\in \pi,\mu$, $g_i$ is the degeneracy, $V$ is the
volume $K_2$ is the modified Bessel functions of integer order `2'.

\section{Mean free path of photon in $e^+e-\gamma$ plasma}\label{plpr2}

In order to be in thermal and chemical equilibrium plasma must be opaque for
the reactions, which establish this equilibrium. The major reactions which
may establish thermal and/or chemical equilibrium between photons and $e^+e^-$ pairs
are Compton scattering and pair production and annihilation.

The mean free path of photon to produce $e^+e^-$ pair is
\begin{equation}
l_{\gamma\gamma} = \frac{1}{n_{\gamma} \left\langle v\sigma_{\gamma\gamma\rightarrow e^+e^-}\right\rangle}
=\frac{n_{\gamma}}{\Upsilon_{\gamma}^2R_{\gamma\gamma}}, \label{lgg}
\end{equation}
where $v$ is relative velocity of interacting particles $\sigma_{\gamma\gamma\rightarrow e^+e^-}$ is
cross section.  For $1+2\rightarrow 3+4$ reactions thermally averaged
\begin{equation}
\left\langle v\sigma_{12\rightarrow 34}\right\rangle  = \frac{\Upsilon_1\Upsilon_2R_{12\leftrightarrow 34}}{n_1n_2}, \label{thcr}
\end{equation}
velocity $v=c$ for photons scattering (we take $c=1$).
Mean free path length $l_{\gamma\gamma}$ is in the order of magnitude of $\tau^{ch}_{\gamma\gamma}$. 
If size of plasma is on the order of magnitude of $c\tau_{pl}$. The condition of opaqueness for pair production 
is approximately the same as condition of chemical equilibration, Eq.(\ref{cheqtau}).

For Compton scattering mean free path is
\begin{equation}
l_{e\gamma} = \frac{1}{n_e\left\langle v\sigma_{e\gamma}\right\rangle}= \frac{n_{\gamma}}{\Upsilon_e\Upsilon_{\gamma}R_{e\gamma}}, \label{leg}
\end{equation}
where $R_{e\gamma}$ is Lorentz invariant Compton scattering rate.
The plasma drop is opaque when
\begin{equation}
l_{e\gamma}(l_{\gamma\gamma})<<R_{pl}. \label{opcon}
\end{equation}
The equations for cross sections for pairs production and annihilation in center of mass frame are
\cite{Jauch:1976}
\begin{eqnarray}
&&\sigma_{\gamma\gamma\rightarrow e^+e^-} = \frac{4\pi \alpha^2}{
m^2 x}{(-4/x-1)}{\sqrt{1-{4}/{x}}}+ \nonumber\\
&&\frac{4 \pi \alpha^2}{ m^2 x}\left(-\frac{8}{x^2}+4/x+1\right)\ln{\frac{1+\sqrt{1-4/x}}{1-\sqrt{1-4/x}}};
\label{crggee}\\
&&\sigma_{e^+e^-\rightarrow \gamma\gamma} =\frac{2\pi \alpha^2}{m^2 x}
\frac{(-4/x-1)}{\sqrt{1-4/x}}+ \nonumber\\
&&\frac{2\pi \alpha^2}{m^2
x{{(1-4/x)}}}\left(-\frac{8}{x^2}+\frac{4}{x}+1\right)
\ln{\frac{1+\sqrt{1-4/x}}{1-\sqrt{1-4/x}}};\label{creegg}
\end{eqnarray}
where $x=s/m^2$, $s=(p_1+p_2)^2$. Note that there is extra 1/2
factor in equation for pairs annihilation because we have two
identical particles or symmetrical wave function in final state. In
rate $R$ we add additional factor 1/2 when there are initial
identical particles. In backward reaction there is also factor 1/2
from cross section. Therefore rate is symmetrical in both reaction
directions.

The cross section for Compton scattering in electron rest frame is
\begin{eqnarray}
\sigma_{e^{\pm}\gamma} =
\frac{2\pi\alpha^2}{m^2}\left(\frac{1+\omega}{\omega^3}\right[\frac{2\omega(1+\omega)}{1+2\omega}-\ln{(1+2\omega)}\left]\right)\nonumber\\
+\frac{2\pi\alpha^2}{m^2}\left(\frac{\ln{1+2\omega}}{2\omega}-\frac{1+3\omega}{(1+2\omega)^2}\right);
\label{crcom}
\end{eqnarray}
where $\omega=E_{ph}/{m}$, $E_{ph}$ is photon energy.

%%%%%%%%%%%%%%%%%%%%%%%%%%%%%%Fig 1
\begin{figure}
\centering
\includegraphics[width=8.6cm,height=8.5cm]{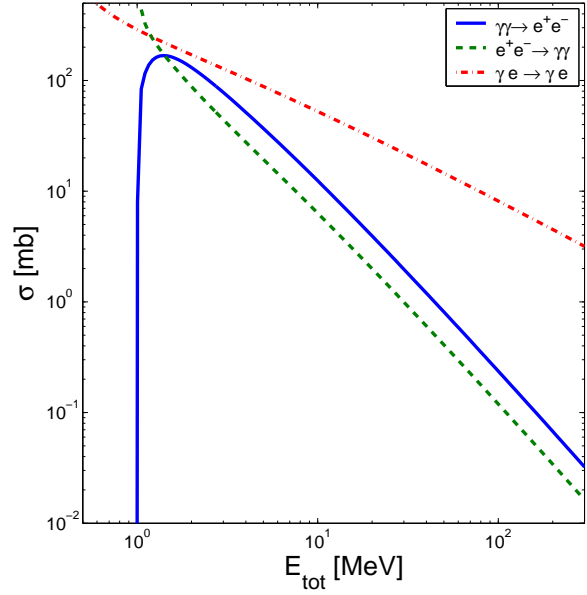}
\caption{\small{The cross sections for pairs production and annihilation in center of mass frame
and for Compton scattering in electron rest frame
shown as functions of total energy of interacting particles $E_{tot}$.}} \label{sigmaeg}
\end{figure}
%%%%%%%%%%%%%%%%%%%%%%%%%%%

In figure~\ref{sigmaeg} we show cross sections
(\ref{crggee})-(\ref{crcom}) as functions of total energy of
particles in reaction $E_{tot}$, $E_{tot}=s^{1/2}$ for pair
production and annihilation and $E_{tot}= E_{ph}+m_e$ for Compton
scattering. The decrease of cross sections with particles energy
increase can result that very high energy particle escape from
plasma and do not participate in heavier particle production. We do
not study this question in details here and we assume that this
effect is small, leaving it for future research.

The equation for rate $R_{\gamma\gamma}$ in Eq.(\ref{lgg}) is defined by Eq. (\ref{db}) and (\ref{pp})  in Introduction.
Similar invariant rate calculations were done before for strangeness
production in~\cite{Koch:1986ud} and \cite{Matsui:1985eu}. We perform
integration similar to in~\cite{Matsui:1985eu}. Here we
extend similar method to Compton scattering invariant rate
$R_{e\gamma}$ calculation.

The equation for Compton scattering rate is
\begin{eqnarray}
&&R_{\gamma e}=
\frac{g_eg_{\gamma}}{(2\pi)^8}\int\frac{d^{3}{p_{3}^{e}}}{2E_{3}^{e}} \int\frac{d^{3}{p_{1}^
{\gamma}}}{2E_{1}^{\gamma}}
\int\frac{d^{3}{p_{3}^{e}}}{2E_{4}^{e}}
\int\frac{d^{3}{p_{2}^{\gamma}}}{2E_{2}^{\gamma}}
\delta^{4}\left(p_{1}^{\gamma} + p_{3}^{e} -
p_{4}^{e} - p_{2}^{\gamma} \right) \nonumber\\
&&\times \sum_{\rm spin}\left|\langle
p_{1}^{\gamma}p_{3}^{e}\left| M_{\gamma e \rightarrow \gamma
e}\right|p_{2}^{\gamma}p_{4}^{\,e}\rangle\right|^{2}f_{\gamma}(p_{1\gamma})f_{\gamma}(p_{2\gamma})f_{e}(p_{3}^{e})f_{4}(p_{4}^{e})
\Upsilon_{\gamma}^{-2}\Upsilon_{e}^{-2}e^{u \cdot_e (p_{3}^{e} +
p_{1}^{\gamma} )/T},\label{comscat}
\end{eqnarray}
where $s = ({\bf p_3}+{\bf p_1})^2$, $t=({\bf p_1}-{\bf p_2})^2$, $u=({\bf p_3}-{\bf p_2})^2$, ($t+u+s=2m^2$)
(compared to pairs production $s=({\bf p_1}+{\bf p_2})^2$, $t=({\bf p_3}-{\bf p_1})^2$, $u=({\bf p_3}-{\bf p_2})^2$,
we cross $s$ and $t$). $g_e=4$ is $e^{\pm}$ degeneracy.
The matrix element for Compton scattering is
\cite{Aksenov:2009dy}
\begin{eqnarray}
|M_{\gamma e \rightarrow \gamma e}|^2
 =  64 \pi^2 \alpha^2 \left(\frac{m^2}{m^2-s}
 +\frac{m^2}{m^2-u}\right)^2- \nonumber\\
 16\pi^2 \alpha^2\left(\frac{4m^2}{m^2-s}+\frac{4m^2}{m^2-u} -
 \frac{m^2-u}{m^2-s} +\frac{m^2-s}{m^2-u}\right), \label{m2comp}
 \end{eqnarray}
where $g_{\gamma}$ and $g_{e}$ are photon and electron (positron)
degeneracies respectively. We define:
\begin{eqnarray}
{\bf q} = p_1+p_3;\,\,\,\,{\bf p}=\frac{1}{2}(p_1-p_3); \nonumber\\
{\bf q}^{\prime}=p_4 + p_2;\,\,\,\,{\bf p^{\prime}}=\frac{1}{2}(p_4-p_2);
\end{eqnarray}
then we have ${\bf q}^2=q_0^2-q^2 = q_0^{\prime\,2}-q^{\prime\,2}=s
\geq m^2$ and
\begin{eqnarray}
p_1 = \frac{{\bf q}}{2}+{\bf p};\qquad p_2 = -{\bf p}^{\prime}+\frac{1}{2}{\bf q}; \nonumber\\
p_3 = \frac{{\bf q}}{2}-{\bf p};\qquad p_4 = \frac{{\bf q}}{2}+{\bf p}^{\prime}. \label{p1234}
\end{eqnarray}

Using $$\int\frac{d^3p}{2E}=\int d^4p\delta^4(p^2-m^2)\theta(p_0)$$
and $p^2_{3,4}-m^2=0$ and $p_{1, 2}^2=0$, we obtain:
\begin{eqnarray}
&& {R_{e \gamma \rightarrow e\gamma}}= \frac{g_eg_{\gamma}}{(2\pi)^8} \int
d^4q\int d^4p\int d^4p^{\prime}
\delta\left({p}_1^2\right)\delta\left({ p}_3^2-m^2\right) \nonumber\\
&&\times\delta\left({p}^2_4-m^2\right)\delta\left({
p}_2^2\right)\theta\left({p}^0_1\right) \theta\left(
p_2^0\right)\theta\left({p}_3^0\right)
\theta\left({ p}_4^0\right)\nonumber\\
&&\times \sum|M_{\gamma e \rightarrow \gamma
e}|^2\Upsilon_{e}^{-2}f_{e}\left(p_3^0\right)f_{\gamma}\left(p_1^0\right)
\Upsilon_{\gamma}^{-2}f_{\gamma}\left(p_2^0\right)\nonumber\\
&& \times f_{e} \left(p_4^0\right)\exp{(q_0/T)}. \label{ratege}
\end{eqnarray}

The  integrals from Eq.(\ref{comscat}) can be evaluated in spherical
coordinates. The angle coordinates are chosen with respect to the
direction of 
$\overrightarrow{q}=\overrightarrow{p_3}+\overrightarrow {p_1}$:
$$q_{\mu}=(q_0,0,0,q),\,\,\,\,p_{\mu}=(p_0, p\sin\theta,0, p\cos\theta),$$
$$p_{\mu}^{\prime}=(p^{\prime}_0, p^{\prime}\sin\phi\sin\chi, p^{\prime}\sin\phi\cos\chi, p^{\prime}\cos\phi).$$

Using equations (\ref{p1234}) from delta functions we obtain equations:
\begin{eqnarray}
&&p_0^2-p^2 +\frac{s}{4} + p_0q_0 - pq\cos{(\theta)}= 0; \label{p1}\\
&&p_0^2-p^2+\frac{s}{4} - p_0q_0 + pq\cos{(\theta)}-m^2= 0; \label{p2}\\
&&p^{\prime\,2}_0-p^{\prime\,2} + \frac{s}{4} + p^{\prime}_0q_0 - p^{\prime}q\cos{(\phi)}-m^2 = 0; \label{ppr1}\\
&&p^{\prime\,2}_0-p^{\prime\,2}+\frac{s}{4}-p^{\prime}_0q_0 + p^{\prime}q\cos{(\phi)} = 0; \label{ppr2}
\end{eqnarray}
This system of equations is equivalent to (add and subtract pairs of
equations (\ref{p1}) and (\ref{p2}), (\ref{ppr1}) and (\ref{ppr2})):
\begin{eqnarray}
&&p_0^2-p^2 + \frac{s}{4} - \frac{m^2}{2} = 0; \label{p}\\
&&p_0q_0 - pq\cos{(\theta)} + \frac{m^2}{2} =0; \label{costh}\\
&&p^{\prime\,2}_0 - p^{\prime\,2}+\frac{s}{4} - \frac{m^2}{2} = 0; \label{ppr}\\
&&p^{\prime}_0q_0 - p^{\prime}q\cos{(\phi)} - \frac{m^2}{2} =0;
\label{cosphi}
\end{eqnarray}
then using  $\delta(f(x))=\sum_i 1/|f^{\prime}(x_i)|\delta(x-x_i)$, we can rewrite integral (similar to~\cite{Matsui:1985eu}) as
\begin{eqnarray}
&& {R_{e\gamma \rightarrow e\gamma}}=
\frac{2g_eg_{\gamma}}{(2\pi)^6 16} \int_{m_{e}}^{\infty}dq_0
\int_0^{\sqrt{q_0^2-s}}dq\int_{q_1}^{q_2}dp_0\int_{q^*_1}^{q^*_2}dp^{\prime}_0
\nonumber\\[0.4cm]
&&\times\int_0^{\infty}dp\int_0^{\infty}dp\prime\int^{1}_{-1}d(\cos{\theta})\int^{1}_{-1}d(\cos{\phi})\
\int_0^{2\pi}d{\chi} \nonumber\\
&&\times\sum|M_{e
\gamma \rightarrow e\gamma}|^2\delta\left(p-\left(p_0^2+\frac{s}{4}-\frac{m^2}{2}\right)^{1/2} \right)
 \nonumber\\
&&\times\delta
\left(p^{\prime}-\left(p^{\prime\,2}_0-\frac{m^2}{2}+\frac{s}{4}\right)^{1/2}\right)\delta\left(\cos{\phi}-\frac{q_0p^{\prime}_0}{qp^{\prime}}+\frac{m^2}{2qp^{\prime}}\right) \nonumber\\[0.4cm]
&&\times \delta\left(\cos{\theta}-\frac{q_0p_0}{pq}-\frac{m^2}{2qp}\right)f_{e}\left(\frac{q_0}{2}+p_0\right)f_{\gamma}\left(\frac{q_0}{2}-p_0\right)
\Upsilon_{e}^{-2}\nonumber\\
&&\times \Upsilon_{\gamma}^{-2}f_{\gamma}\left(\frac{q_0}{2}+p^{\prime}_0\right)f_{e}
\left(\frac{q_0}{2}-p^{\prime}_0\right)\exp{(q_0/T)}, \label{ratemu}
\end{eqnarray}
where
\begin{eqnarray}
&&q_{1,2}=-\frac{m^2q_0}{2s}\pm \frac{q}{2}{\left(1-\frac{m^2}{s}\right)};\\
&&q^*_{1,2}=\frac{m^2q_0}{2s}\pm \frac{q}{2}{\left(1-\frac{m^2}{s}\right)}.
\end{eqnarray}
$q_{1,2}$ and $q^*_{1,2}$ come from constrains $\cos{\theta}$, $\cos{\phi}<1$ and Eq.(\ref{p})-(\ref{cosphi}).

The integration over $p$, $p^{\prime}$, $\cos{\theta}$, $\cos{\phi}$
can be done analytically considering the delta-functions. The other
integrals can be evaluated numerically. In the order to simplify numerical integration we introduce dimensionless
variables:
\begin{eqnarray}
q = (q_0^2-m^2)^{1/2}z; \label{z}\\
p_0= -\frac{m^2q_0}{2s} + \frac{q}{2}\left(1-\frac{m^2}{s}\right)x, \label{x}\\
p_0^{\prime}=\frac{m^2q_0}{2s} + \frac{q}{2}\left(1-\frac{m^2}{s}\right)y \label{y},
\end{eqnarray}
$0<z<1$; $-1<x(y)<1$.

In these variables, using Eq.(\ref{p})-(\ref{cosphi}), we obtain for $u$ and $t$:
\begin{eqnarray}\label{ut2}
&&u = (p-p^{\prime})^2={m^2}-\frac{s}{2}+\frac{m^4}{2s}+\frac{s}{2}\left(1-\frac{m^2}{s}\right)^2(xy + \sqrt{(1-x^2)(1-y^2)}\sin{\chi});\nonumber\\
\nonumber\\
&&t = (p+p^{\prime})^2={m^2}-\frac{s}{2}-\frac{m^4}{2s}-\frac{s}{2}\left(1-\frac{m^2}{s}\right)^2(xy-\sqrt{(1-x^2)(1-y^2)}\sin{\chi}).
\end{eqnarray}
Then limits for $t$ are $0 > t >-s+2m^2-m^2/s$.

In new variables (\ref{z})-(\ref{y}) the equation (\ref{ratemu}) is
\begin{eqnarray}
&&R_{e\gamma \rightarrow e\gamma} = \frac{\alpha_s^2}{2^{8}\pi^4}\int_{m^2}^{\infty}dq_0
\exp(q_0)\int_0^1dzz^2(q_0^2-m^2)^{3/2}\int_0^{2\pi} d\chi
\int_{-1}^1 dx\int_{-1}^1dy\left(1-\frac{m}{s}\right)^2
 \nonumber\\
&& \times \sum|M_{\gamma e^{\pm}\rightarrow \gamma e^{\pm}}|^2 \Upsilon_{e}^{-2}f_{e^{\pm}}
\left(\frac{q_0}{2}+p_0\right)f_{\gamma}\left(\frac{q_0}{2}-p_0\right)\Upsilon_{\gamma}^{-2}f_{\gamma}\left(\frac{q_0}{2}+p^{\prime}_0\right)f_{e}
\left(\frac{q_0}{2}-p^{\prime}_0\right).
\end{eqnarray}

%%%%%%%%%%%%%%%%%%%%%%%%%%%%%%Fig 2
\begin{figure}
\centering
\includegraphics[width=8.6cm,height=8.5cm]{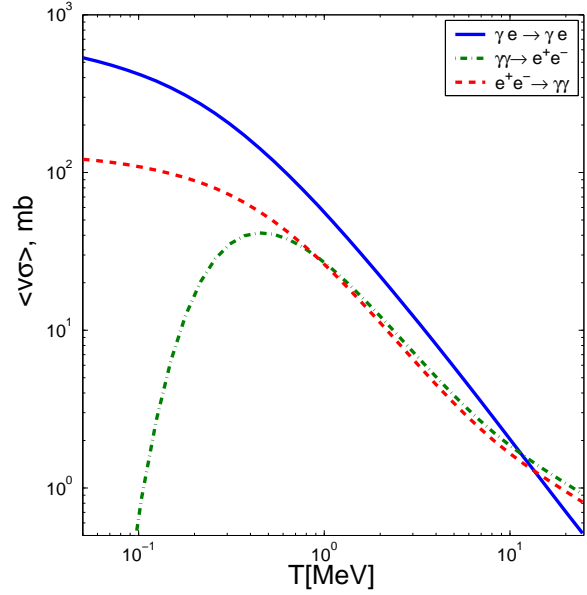}
\caption{\small{The thermal products $\left\langle v\sigma\right\rangle$ for Compton scattering (solid line), pair production (dash-dot line) and annihilation (dashed line) in observer rest frame shown as
a function of temperature $T$ at $\Upsilon=1$}} \label{sigcom}
\end{figure}
%%%%%%%%%%%%%%%%%%%%%%%%%%%

In figure~\ref{sigcom} the thermally averaged products $\left\langle v\sigma\right\rangle$ in observer frame, calculated using Lorentz invariant rates (Eq.(\ref{thcr})),
are shown for Compton scattering (solid line), pair production (dot-dash line) and pair annihilation (dashed line). For pair production 
and Compton scattering $v = c =1$. We see that at $T << m$ for Compton scattering $\left\langle v\sigma\right\rangle$ goes to Thompson limit:
$$\sigma = \frac{8\pi \alpha^2}{3m^2}=6.7\,10^2\,\rm mb.$$ For electrons production $\left\langle v\sigma\right\rangle$ starts to decrease at $T < m$ and goes to 0 with $T \rightarrow 0$. The density 
of photons with energy larger than threshold for pairs production drops in the tail of Boltzmann distribution with temperature decrease. For electrons
annihilation $\left\langle v\sigma\right\rangle$ stays finite at small temperature, $v \rightarrow 0$, because $\sigma$ diverges as $1/v$.  
%%%%%%%%%%%%%%%%%%%%%%%%%%%%%%Fig 1
\begin{figure}
\centering
\includegraphics[width=8.6cm,height=8.5cm]{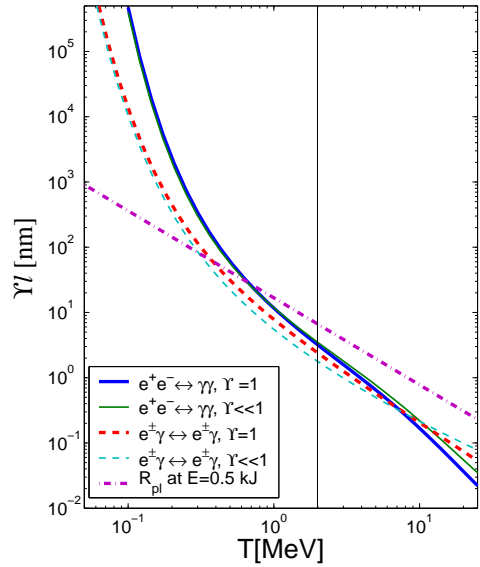}
\caption{\small{$l\Upsilon$ for Compton scattering and
pairs production at $\Upsilon=1$ (thick dashed and solid lines) and
$\Upsilon=0.1$ (thin dashed and solid lines) as functions of
temperature $T$; radius of equilibrium
($\Upsilon=1$) plasma at energy 0.5 kJ (dot-dash).}} \label{tauee1}
\end{figure}
%%%%%%%%%%%%%%%%%%%%%%%%%%%
In figure~\ref{tauee1} the mean free photon paths multiplied by $\Upsilon$ are shown for $e^+e^-$
pair production reaction $\Upsilon l_{\gamma\gamma}$ (Eq.(\ref{lgg})) and Compton scattering
$\Upsilon l_{e\gamma}$ (Eq.(\ref{leg})) are shown
as functions of temperature $T$ for $\Upsilon=1$ (thick dashed and solid lines, respectively) and 
$\Upsilon = 0.1$ (thin dashed and solid lines, respectively). Cases at $\Upsilon =1$ and $\Upsilon=0.1$ are shown to demonstrate the magnitude of the difference between these two cases due
to quantum effects.
For electron-positron pair production $l_{\gamma\gamma}$ is suppressed slightly at $\Upsilon=1$
due to Bose enhancement of reaction rate. Both effects Bose enhancement and Fermi blocking contribute to the Compton mean free path . At $\Upsilon<<1$, when there is no quantum effects from plasma, the free mean path of photon $\propto \Upsilon^{-1}$. 
We see that $l_{\gamma\gamma} < l_{e\gamma}$ at $T>8$ MeV for $\Upsilon = 1$ and at $T>10$ MeV for $\Upsilon << 1$  and 
therefore the thermal equilibrium in EP$^3$ is
established by reaction (\ref{ggee}) at $T > 8$ MeV, approximately at
the same time with chemical equilibrium of pairs and photons.
$l_{\gamma\gamma}$ drops fast when temperature increases:
\begin{equation}
l_{\gamma\gamma} \propto \frac{1}{\Upsilon T^2}.
\label{lee}
\end{equation}
At temperature range 1 MeV$<T<10$ MeV the temperature dependence of $l_{e\gamma}$ is a little slower than $1/T^2$.
In figure~\ref{tauee1} we also show the radius of EP$^3$ plasma drop with energy 0.5 kJ and $\Upsilon=1$ 
as a function of temperature. We see that at these conditions plasma loose
opaqueness at temperatures smaller than $2$ MeV (thin vertical line). Photon free path increases faster with temperature decrease
than plasma radius. Corresponding 
maximum radius of equilibrium plasma is about 7 nm ($R/l_{e\gamma} \approx 3$) at $\Upsilon = 1$. 

The pair production or annihilation relaxation time Eq. (\ref{taugg}) is approximately $2\,10^{-2}$ fs at $T=2$ MeV, in agreement with condition (\ref{cheqtau}). Plasma drop with energy $0.5$ kJ is thermally and chemically equilibrated at $T > 2$ MeV and largest density with $\Upsilon = 1$. 

From figure~\ref{tauee1} we also see that the mean free path length of photon starts to increase exponentially at small temperatures $T < 0.5$ MeV.  
Therefore plasma size and energy at this temperature also increase exponentially. On the contrary the higher plasma temperature is the smaller photon free path length becomes. Higher temperature ($T>2$ MeV) opaque plasma can have smaller size ($R<7$ nm).  When $l$ (almost) satisfies Eq.(\ref{lee}), minimum radius of opaque plasma, allowed by opaqueness condition~(\ref{opcon}) is also $\propto 1/T^2$. If we focus laser pulse energy in this small volume, we obtain that the necessary energy is
\begin{equation}
E \propto T^4R^3 \propto 1/T^2.
\end{equation}  

%%%%%%%%%%%%%%%%%%%%%%%%%%%%%%%%%%%%%%
\subsection{Summary and Conclusion} \label{conc1}
%%%%%%%%%%%%%%%%%%%%%%%%%%%%%%%%%%%%%%

In this part we investigated physical conditions suitable  to create
opaque and, therefore, thermally and chemically equilibrated
$e^+,e^-,\gamma$-plasma
drop.  In order to address this question we evaluated Lorentz invariant
rates for the Compton scattering and pair production in thermally and chemically
equilibrated EP$^3$ plasma.  We then used these Lorentz invariant rates to
evaluate the corresponding mean free path length $l$ of particles.

Comparing $l$ to plasma drop size we showed that an opaque equilibrium
density plasma drop can be produced at energy $0.5$~kJ in the volume with largest
possible radius $R = 7$~nm. This volume corresponds to the smallest possible
 temperature $T = 2$~MeV. In order to reach  higher than 2~MeV temperature,
we need to increase energy of plasma (which is proportional to $T^4$)
or/and decrease plasma size. At higher temperature  opaque plasma can
be created at the total plasma energy  smaller than 0.5 MeV, since smaller
plasma drop size is in agreement with the opaqueness condition
Eq.\,(\ref{opcon}), as seen in figure \ref{tauee1}.

On the other hand in order to create opaque plasma with temperature
lower than 2 MeV, the necessary amount of  energy is larger than $0.5$~kJ.
This is so because the plasma size has to be large in order to satisfy
opaqueness condition Eq.\,(\ref{opcon}).

Our main result, perhaps unexpected at the first sight is illustrated in figure
\ref{tauee1}. For the  temperature $T>2$ MeV equilibrium plasma production with a
relatively small energy pulse (compared to lower temperature equilibrium plasma)  may be possible. However,
the challenge here is to focus the energy into the volume of size  $<10$~nm.

\section{Pion and muon production in relatvistic $e^+e^-\gamma$ plasma}\label{pimu}

\subsection{Introduction}
The elementary  properties of EP$^3$
have recently been reported, see~\cite{Thoma:2008my}, where  typical properties   are
explicitly  presented for $T=10$ MeV.
One of the challenges facing  a study of  EP$^3$ will be the understanding
of the fundamental mechanisms leading to its formation. We propose  here as a probe the production of
heavy particles with mass $m\gg  T$. Clearly, these processes occur during the history of the event
at the highest available temperature, and thus
information about the early stages of the plasma, and even pre-equilibrium state should
become accessible in this way.

We focus our attention on the strongly interacting  pions $\pi^\pm,\pi^0$ ($m_\pi c^2 \approx 140$ MeV),
and muons $\mu^\pm$($m_\mu c^2 \approx 106$ MeV), %the `heavy electron'
({\it in the following we use units in which
 $k=c=\hbar=1$ and thus  we omit these symbols from all equations.
Both, the particle mass, and  plasma  temperature, is thus given in the energy unit MeV.})
These very  heavy, compared to the electron  ($m_ec^2=0.511$ MeV), particles
are as noted natural `deep' diagnostic tools of the EP$^3$  drop. Of special interest
is  the neutral pion  $\pi^0$ which is, among all other heavy particles,
 most copiously produced for $T\ll m$.  The $\pi^0$    yield and spectrum  will
be therefore of great  interest in the study of the EP$^3$ properties.
 Conversely,   the study of  the in-medium  pion mass 
splitting  $\Delta m=m_{\pi^\pm}-m_{\pi^0}=4.594 $\,MeV    at a temperature 
$T\gtrsim \Delta m$    will contribute to the better
understanding of   this  relatively large mass splitting between
 $\pi^0$and $\pi^\pm$, $\Delta m/\overline m= 3.34\% $, believed to originate
in the isospin  symmetry breaking electromagnetic radiative corrections.

However, given its very short natural lifespan:
$$\pi^0\to \gamma+\gamma, \quad \tau_{\pi^0}^0=(8.4\pm0.6)10^{-17} {\rm s}.$$
$\pi^0$ is also the  particle most difficult to experimentally study among those we consider:
its decay products reach the detection system nearly at the same time as the electromagnetic
energy pulse of the decaying  plasma fireball, which is likely to `blind' the detectors.

This plasma drop we consider is a thousand times hotter than the center of the sun.
This implies presence of the  corresponding
high particle density $n$, energy density $\epsilon$ and pressure $P$. These quantities
 in the plasma can be  evaluated  using the relativistic
expressions:
\begin{eqnarray}
n_i &=&\int g_i f_i(p)d^3p,\\[0.3cm]
\epsilon &=& \int \sum_i g_iE_if_i(p)d^3p, \quad E_i=\sqrt{m_i^2+\vec p^{\,2}} \\[0.3cm]
P &=& \frac{1}{3}\int  \sum_i g_i\left(E_i-\frac{m_i^2}{E_i}\right)f_i(p)dp^3, \label{Ptherm}
\end{eqnarray}
where subscript $i\in \gamma$, $e^-,   e^+$, $\pi^0, \pi^+, \pi^-$, $\mu^-, \mu^+$,
$f_i(p)$ is the  momentum distribution of the particle $i$ and  and $g_i$ its degeneracy,
for $i=e^-,e^+, \gamma, \mu^-, \mu^+$ we have $g_i=2$, and $g_i=1$ for $\pi^0, \pi^- \pi^+$.
For a QED plasma which lives long enough so that electrons, positrons
are in thermal and chemical equilibrium with photons, ignoring small
QED interaction effects,  we use Fermi and
Bose momentum distribution, respectively, Eq.(\ref{fstat}).

It is convenient to parametrize the electron, positron  and photon
$ e^-e^+\gamma $ plasma properties   in terms of the
properties of the Stephan-Boltzmann law for  massless particles
(photons), presenting the physical properties in terms of the
effective degeneracy $g(T)$ comprising the count of all particles present at a given temperature $T$:
\begin{equation}\label{SB}
\frac{\cal E}{V}= \epsilon   = g(T)\sigma T^4,\qquad 3P = g^\prime(T) \sigma T^4,\qquad \sigma = \frac{\pi^2}{30}.
\end{equation}
we only have in this case truly massless
For temperatures $T\ll m_e$ photons and $g(T)\simeq g^\prime(T) \simeq 2_\gamma$.
Once temperature approaches and increases beyond $m_e$
we find  $g\simeq g^\prime(T) \simeq 2_\gamma+(7/8)(2_{e^-}+2_{e^+})=5.5$ degrees of freedom.
In principle these particles acquire additional in medium mass which reduces the degree of
freedom count, but this effect is compensated
by collective `plasmon' modes, thus we proceed with naive counting of nearly
free EP$^3$ components. The factor 7/8 expresses the difference in the evaluation of  Eq.\,(\ref{Ptherm})
for the  momentum distribution of  Fermions and Bosons Eq.\,(\ref{fstat}). Bosons
 provides the reference  point at low $T$, where only massless photons are
present. In passing, we note
that in the early Universe, there  would  be further present the neutrino degrees of freedom, not
considered here for  the laboratory experiments, considering their weak coupling to matter.

\begin{figure}
\centering\hspace*{0.5cm}
\includegraphics[width=8.6cm,height=10cm]{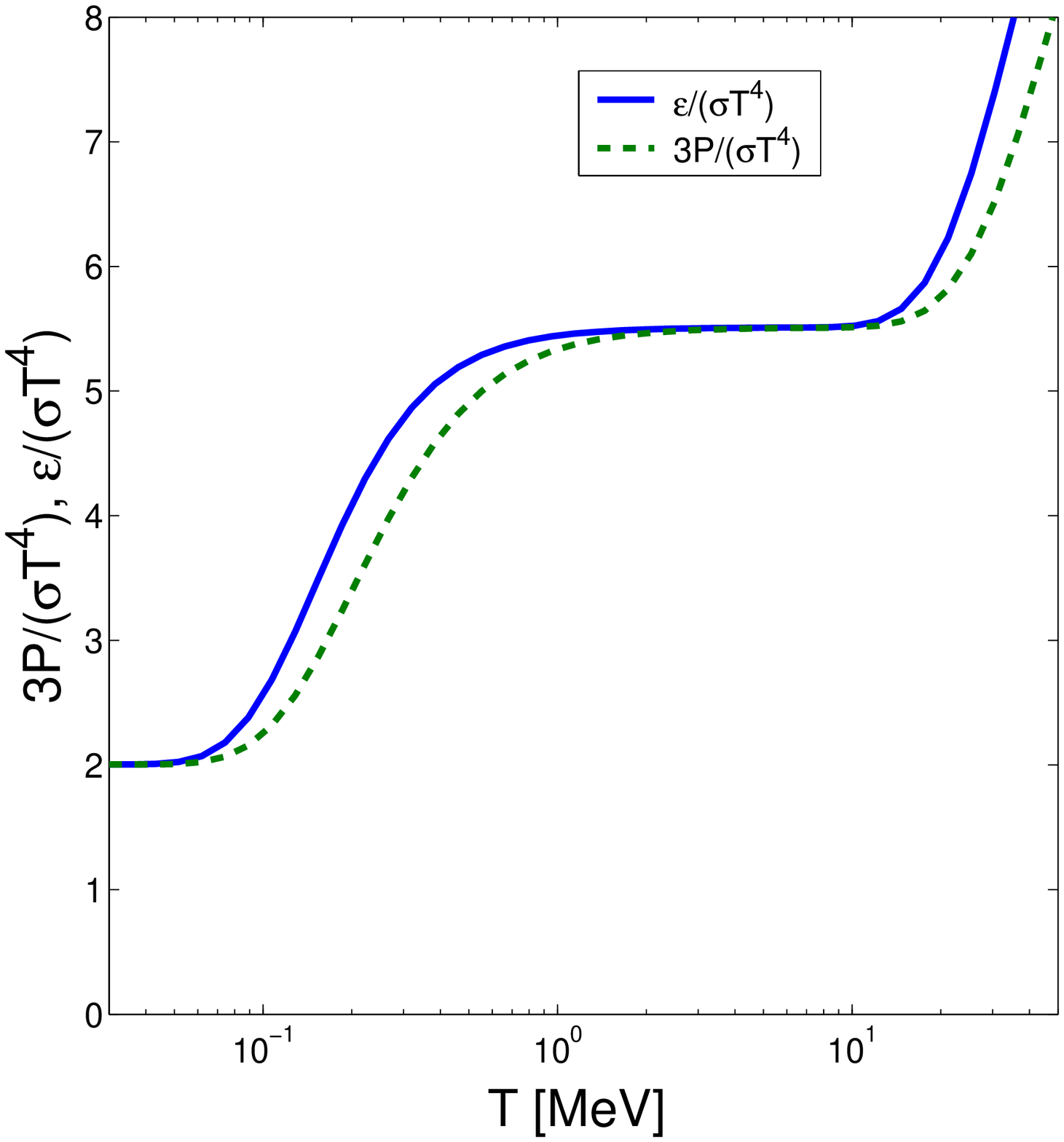}\hspace*{-0.5cm}
\includegraphics[width=8.6cm,height=10cm]{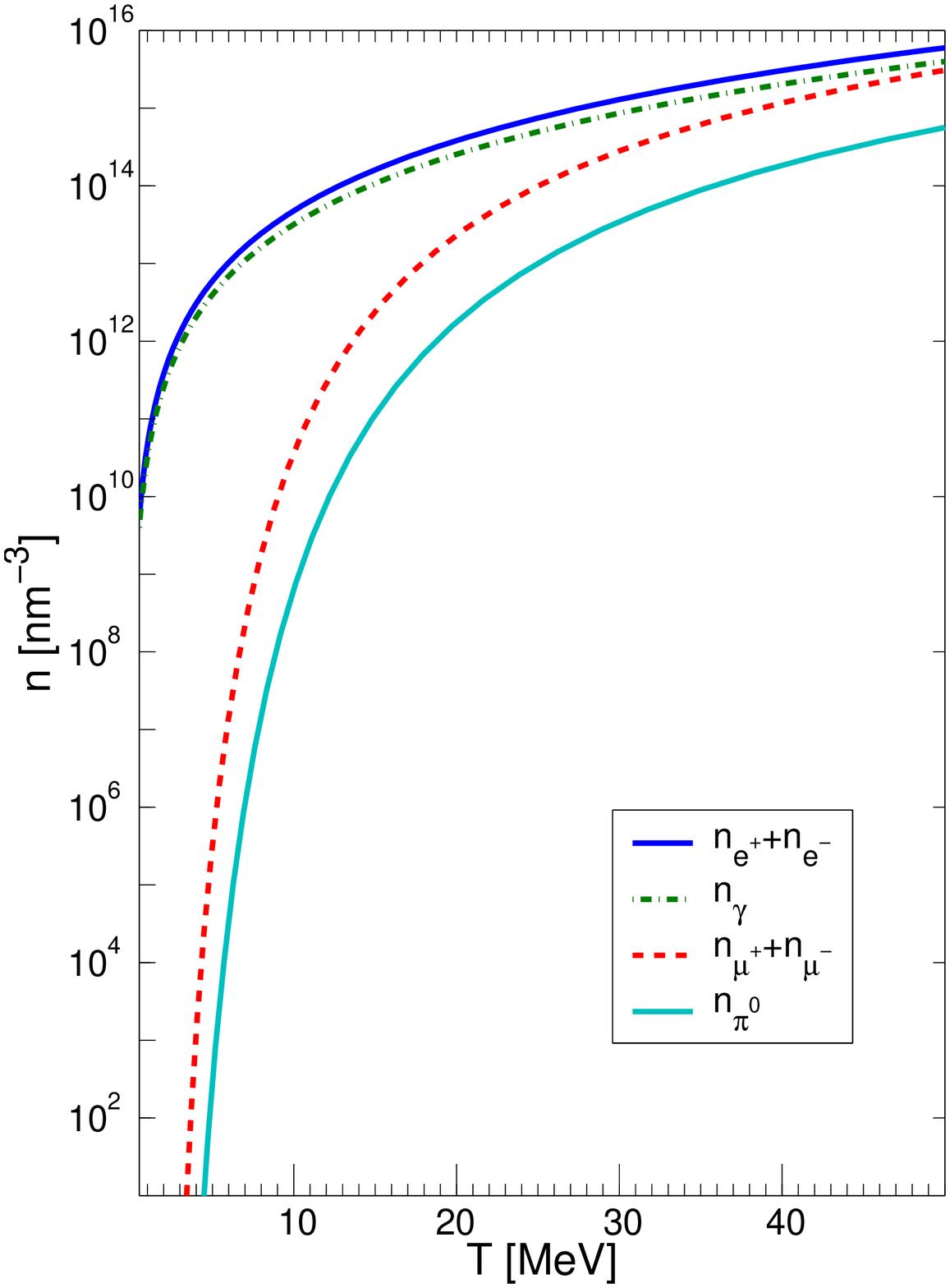}
\caption{\small{On left: the ratios $g\equiv \epsilon/\sigma T^4$ and $g'\equiv 3P/\sigma
T^4$  as a function of temperature $T$; on right: the equilibrium densities of electrons (blue, solid
line), photons (green, dash-dot line), muons (red, dashed line),
pions (blue dotted line) as functions of temperature $T$.}}
\label{energ}\label{nmupi}
\end{figure}

In figure \ref{energ} we present both $g(T)$ and  $g^\prime(T)$, as a function of temperature $T$
 in form of the energy density $\epsilon$ normalized by $\sigma T^4$, and, respectively,
the pressure $P$, normalized by $\sigma T^4/3$ .
The  $g(T)$   jumps more rapidly compared to  $g^\prime(T)$,
 between the limiting case of a black body photon
gas at  $T< 0.5$  MeV $(g=2)$ and the case $g=5.5 $ for $\gamma$, $e^-,  e^+$, since the energy density 
also contains the rest mass energy content of all particles present.
The rise of the  ratio at $T>15$ MeV indicates  the contribution of the excitation of  muons and  pions in
equilibrated plasma. We note that the plasma  produced   pions  (and muons) are
in general  not in chemical equilibrium. The
distribution functions which maximize entropy content at given particle number and
energy content are \cite{LLStat}:
\begin{equation}
f_{\pi} = \frac{1}{\Upsilon^{-1}_{\pi^0(\pi^{\pm})}e^{u\cdot p_{\pi}/T}-1},\quad
f_{\mu} = \frac{1}{\Upsilon^{-1}_{\mu}e^{u\cdot p_{\mu}/T} +1}, \label{dfpimu}
\end{equation}
where $\Upsilon_{\pi^0(\pi^{\pm})}$ and $\Upsilon_{\mu}$ are particles fugacities.
The chemical equilibrium corresponds to $\Upsilon_{\pi^0(\pi^{\pm})}=\Upsilon_{\mu}=1$ used in
figure \ref{nmupi} on right, since this is the maximum density that can be reached in the buildup of
these particles, for a given temperature.  We occasionally refer to $f_{\pi}\to f_-$ as the boson distribution
function  and to $f_\mu\to f_+$ as the Fermi distribution function. For $\Upsilon_i \to 0$ the quantum distributions
shown in Eq.\,(\ref{dfpimu}) turn into the classical Boltzmann distributions, with abundance prefactor $\Upsilon_i$.

In the case of interest here, when $T<m$, we also consider consider
the Boltzmann limit of the quantum distributions Eq.(\ref{dfpimu}),
that is to drop the `one' in the denominator, Eq.(\ref{fbol}) and taking the non-relativistic
limit Eq. (\ref{dist1}) we have:
\begin{equation}\label{bolpio}
\frac{N_{\pi}}{V}\equiv n_{\pi} = \Upsilon_{\pi}\frac{1}{2\pi^2}Tm_\pi^2 K_2(m_{\pi}/T)
    \to \Upsilon_{\pi}  \left( \frac{m_{\pi}T}{2\pi}\right)^{3/2}e^{-m_{\pi}/T}+\ldots ,
\end{equation}

%%%%%%%%%%%%%%%%%%%%%%%%%%%%%%
%%%%%%%%%%%%%%%%%%%%%%%%%%%

The particle densities are shown on right in figure \ref{nmupi}. The top solid line
is the sum of $n_{e^+}+n_{e^-}$, which is marginally bigger than the
photon density (dashed, blue) which follows below.  We also include in the figure the sum density of muons
$n_{\mu^+}+n_{\mu^-}$ (red, dashed), and the density of  the neutral pion $\pi^0$ (bottom solid  line),
both of which  appear comparatively very small in the temperature range of interest.  However, in magnitude they rival
the normal atomic density ($\simeq 10^2/{\rm nm}^3$)already at $T=4$ MeV, and 5 MeV, respectively.
This high particle density in the chemically equilibrated plasma
  explains the relatively large collision and reaction
rates we obtain in this work. In turn, this opens the question how such dense, chemically equilibrated EP$^3$  state
can be formed -- we observe that colliding two ultra intense
circularly polarized and focused laser beams on a heavy thin metal foil(s)  is the current line of approach.
Initial simulations were performed~\cite{Shen:2001}. Many strategies can be
envisaged aiming to deposit the laser pulse energy in the smallest possible spatial and temporal volume and
this interesting and challenging topic will without doubt keep us and others busy in years to come.

As it turns out, a small drop of  EP$^3$ plasma with a size scale of 1nm is, given the high particle density,
 opaque. The  mean free paths $l_i$ of  particles `i' are relatively short.
Where the reference energy values (31.1 and 27.5 MeV) correspond to the mean particle energy at $T=10$ MeV.
Photons are subject to Compton scattering, and electrons and positrons to charged particle scattering. In fact these
values of $l_i$ are likely to be upper limits, since Bremsstrahlung type processes are believed to further increase opaqueness
of the plasma~\cite{ThomaPriv}.
In our considerations  plasma  particles of energy above 70 MeV are of  interest,
since these are responsible for the production
of heavy particles. We see that the mean free path of such
particles has also nm scale magnitude.

We note that a EP$^3$ drop of radius 2nm at $T=10$ MeV
contains 13 kJ energy.  This is the expected energy content of a light pulse at ELI (European Light Infrastructure, in development)
with a pulse length of about $\Delta t=10^{-14}s$. For comparison,  the maximum energy
available in particle accelerators for at least 20, if not more, years
will be in head on Pb--Pb central collisions at LHC (Large Hadron Collider) at CERN, in its LHC-ion collider mode,
where per nucleon energy of about 3 TeV is reached. Thus the total energy available is 200 $\mu$J, of which about
10\%--20\% becomes thermalized. Thus ELI will have already an overall energy advantage of $10^9$, while
in the LHI-ion case the great advantage are a) the natural localization
of the energy at the length scale of $10^{-5}$ nm, given that the energy is
contained in colliding nuclei, and b) the high repetition rate of collisions.

As a purely academic exercise, we note that should one find a way
to `focus' the energy in ELI to nuclear dimensions, and scaling the energy density with $T^4$ up from
what is expected to be seen at CERN-LHC-ion ($T<1$GeV), we  exceed $T=150$ GeV, the presumed electro-weak
phase boundary. Such consideration lead the authors of Refs.~\cite{TajMou,TajMouBoul}  to suggest that
the electro-weak transition may be achieved at some future time using ultra-short laser pulses.

 Returning to present day physics, we are assuming here  that   $T$
near and in  MeV range  is achievable in foreseeable future, and that much higher values are obtainable  in
presence of pulses  with  $\Delta  t<10^{-18}s$, $c\Delta t< 0.3$nm. Hence
we consider production processes for  $\pi^0,\pi^\pm,\mu^\pm$ for $T<50$ MeV.
We   study here all two body reactions in EP$^3$ which lead to formation of the particles of interest,
excluding solely $e\gamma\to e\pi^0$, and the related $e^-e^+\to \gamma \pi^0$. The presence of a
significant (1.2\%) fraction of  $\pi^0\to e^+e^-\gamma$ decays
implies that these related two body processes could be  important in our considerations. 
However, these    reactions involve  the $\pi_0$ off-mass shell   couping to two photons, which
needs to be better understood before we can consider these reactions in our context. 

We also do not consider here the inverse three body reactions 
$ e^+e^-\gamma\to \pi^0$,  since  there is no exponential gain in using
$n>2$ particles to overcome an energy threshold, here $m_{\pi^0}$. 
The independent probability of  finding $n$ particles with energy 
$m_{\pi^0}/n$ each is the same for any value of $n$: 
\begin{equation}
P_1P_2....P_n \propto \left( e^{-m_{\pi^0}/nT}\right)^n=e^{-\frac{m_{\pi^0}}{T}}.
\end{equation}
This resolves the   argument  that more particles could  overcome 
more easily the reaction barrier. $n$-body reactions with $n>2$ 
are in fact suppressed in  EP$^3$  by the weakness of the 
electromagnetic (EM)  interaction, since adding an EM-interacting 
particle to the reactions process  requires  an EM-vertex with 
$\alpha=1/137$. Thus microscopic reactions in  EP$^3$
 involving $n>2$ are suppressed by a factor 100 
for each additional EM particle involved in the reaction.  
This does not mean that a collective/coherent  process of heavy 
particle production by many particles is similarly suppressed: 
for example fast time varying electromagnetic fields provide through 
$\vec E\cdot \vec B$ a collective source of $\pi^0$. We defer further 
study of this production mechanism which requires 
 multi MeV$^{-1}$ range oscillation  to be present in EP$^3$.

In the following section, we introduce the master equation governing
the production of pions and muons in plasma and formulate
the invariant rates in terms of know physical reactions. In section \ref{resnum}
we obtain the  numerical results for particles production rates and
reactions relaxation times which we present as figures.
In section \ref{concl} we  discuss these results further and consider
their implications.

%%%%%%%%%%%%%%%%%%%%%%%%%%%%%%%
\section{Particles production} \label{master}
\subsection{$\pi^0$ production}
%%%%%%%%%%%%%%%%%%%%%%%%%%%%%
$\pi^0$ in the QED plasma is  produced
predominantly  in the thermal two photon fusion Eq~\ref{ggpi0}, see chapter \ref{pi0pr}.
Much less probable is the production of $\pi_0$ in the reaction:
\begin{equation}
e^-+e^+ \rightarrow \pi^0. \label{pi0ee}
\end{equation}
These formation
processes are the inverse of the decay process of $\pi_0$.
The smallness of the electro-formation of $\pi_0$ is characterized by
the small  branching ratio in $\pi_0$ decay
$B=\Gamma_{ee}/\Gamma_{\gamma\gamma}=6.2\pm 0.5 10^{-8}$.
Other  decay processes involve more than two particles.  
$\pi^0$ can also be formed by charged pions in charge exchange
reactions. However,  in EP$^3$  in the domain of $T$ of interest
we find that at first the neutral pions will be produced. These
in turn produce charged pions. Therefore we introduce the pion charge exchange
process in the context  of charged pion formation  in  the subsection \ref{pichprod}.

Omitting all sub-dominant   processes, the resulting master equation for pion  number evolution is:
\begin{equation}
 \frac{1}{V}\frac{dN_{\pi^0}}{dt}= \frac{d^4W_{\gamma \gamma \rightarrow\pi^0}}{dVdt} -
  -\frac{d^4W_{\pi^0 \rightarrow \gamma \gamma}}{dVdt},
\label{piev}
\end{equation}
where $N_{\pi^0}$ is total number of $\pi^0$, $V$ is volume
of the system,${d^4W_{\gamma \gamma \rightarrow\pi^0}}/{dVdt}$ 
is the (Lorentz) invariant $\pi^0$ production rate per unit time and volume in photon fusion, and
$d^4W_{\pi^0 \rightarrow \gamma \gamma}/dVdt$ is the  invariant
$\pi^0$ decay rate per unit volume and time. The rates for $\pi^0$ decay and production can be calculated using 
Eq.(\ref{pd}) and (\ref{pp}).

We assume that in the laboratory frame the momentum
distribution of produced $\pi^0$ are characterized
by the ambient temperature.
Eq.\,(\ref{bolpio})   defines the relation of fugacity $\Upsilon_{\pi} $ to the
yield. This equation allows now to   study the production dynamics as if
we were dealing with a  $\pi^0$ in a thermal bath, and to exploit
the detailed balance between decay and production process in order
to estimate the rate of  $\pi^0$ production. This theoretical  consideration
should not be understood as assumption of equilibration of $\pi^0$,
which could upon production escape from the small plasma
drop.

Using the detailed balance relation and $R_{\gamma\gamma \rightarrow \pi^0 }$ definition~(\ref{db}), 
Eq.(\ref{piev}) can be written in the form:
\begin{equation}
\frac{1}{V}\frac{dN_{\pi^0}}{dt} = (\Upsilon_{\gamma}^2-{\Upsilon_{\pi^0}})R_{\gamma\gamma \rightarrow \pi^0 },
\label{pieqdyn}
\end{equation}
For $\Upsilon_{\pi_0} \to \Upsilon^2_{\gamma}=1$  we reach
chemical equilibrium, the time variation of density due to production and decay vanishes.

We introduce  the
pion equilibration  (relaxation) time constant by:
\begin{equation}\label{taupi0}
\tau_{\pi^0}  =  \frac{{dn_{\pi^0}}/{d\Upsilon_{\pi^0}}}{ R_{\gamma\gamma \rightarrow \pi^0} }  .
\end{equation}
Note that when the volume does not change in time on scale of $ \tau_{\pi^0} $ (absence of expansion
dilution) and thus $T$ is constant, the left hand side of Eq.(\ref{pieqdyn}) becomes
$dn_{\pi^0}/dt$. Given the relaxation time definition
Eq.(\ref{taupi0}) the time evolution  for of the pion fugacity for a
system  at  fixed time independent temperature   satisfies:
\begin{equation}
\label{pidynfug}
\tau_{\pi^0} \frac{d \Upsilon_{\pi^0}}{dt}=\Upsilon^2_{\gamma}-\Upsilon_{\pi^0},
\end{equation}
which has for $\Upsilon_{\pi^0}(t=0)=0$ the analytical solution $\Upsilon_{\pi^0}
=\Upsilon^2_{\gamma}\left(1-e^{-t/\tau_{\pi^0}}\right)$, justifying the proposed  definition of
the relaxation constant.

The relaxation time $\tau_{pi^0}$ is calculted in section \ref{taupi0} and shown in figure \ref{taupi0app}.
$\tau_{pi^0} \approx \tau^0_{\pi^0}$ for temperatures considered, because the relativistic time dilution effect
cancels with medium effect.

The $\pi^0$ production rate is thus related to the decay rate $1/\tau_{\pi^0}^0$ by the simple formula
\begin{equation}
R_{ \pi^0} \simeq
    \frac{{dn_{\pi^0}}/{d\Upsilon_{\pi^0}}}
{\tau_{\pi^0}^0 }\simeq \left(\frac{m_{\pi}T}{2\pi}\right)^{3/2}\frac{e^{-m_{\pi}/T} }{ \tau_{\pi^0}^0},
\label{taupi01}
\end{equation}
where in the last expression we have used Eq.\,(\ref{bolpio}) in the limit $m>>T$. It is important for
the reader to remember that   derivation of  Eq\,(\ref{taupi01}) is based on detailed balance in thermally
equilibrated plasma, and does not require chemical equilibrium to be established.

%%%%%%%%%%%%%%%%%%%%%%%%%%%%%%
\begin{figure}
\centering
\includegraphics[width=8.6cm,height=8.5cm]{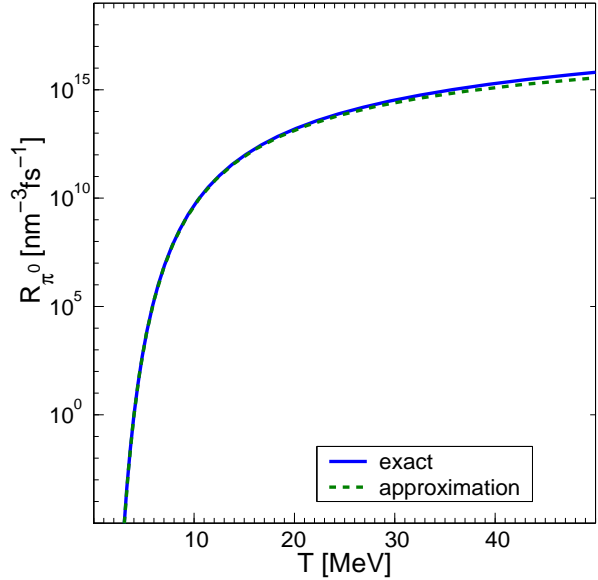}
\caption{\small{The $\pi_0$ production rate (blue, solid line)
and approximate rate from Eq.(\ref{taupi01}) (green dashed line)
as functions of temperature $T$.}} \label{Rpi0}
\end{figure}
%%%%%%%%%%%%%%%%%%%%%%%%%%

This exact result (blue, solid line) is compared to the approximate result  Eq.(\ref{taupi01}) (green, dashed line)  in figure  \ref{Rpi0}.
We note that it is hard to discern a difference on logarithmic scale, especially so for small temperatures where  the only (small) effect is
the relativistic time dilation. This implies that it is appropriate to use the simple intuitive result  Eq.(\ref{taupi01}) in the study of $\pi^0$ production.

Before closing this section we note that we can use exactly the same method to extract from the partial
width of the $\pi_0\to e^+e^-$ the reaction rate for the inverse process, which will be discussed below. All arguments carry
through in identical and exact fashion  replacing where appropriate the Bose by Fermi distributions and using Eq.\ref{FBrel}.

%%%%%%%%%%%%%%%%%%%%%%%%%%%%
%%%%%%%%%%%%%%%%%%%%%%%%%%%%%
\subsection{Muon  production}\label{muprod}
%%%%%%%%%%%%%%%%%%%%
In the plasma under consideration, muons can be directly  produced in photons or $e^+e^-$ fusions, reactions (\ref{ggmu}) and (\ref{eemu}).
For these reactions the master evolution equation developed for the study of thermal
strangeness in heavy ion collisions applies
~\cite{Biro:1981zi,Rafelski:1982pu,Matsui:1985eu,Koch:1986ud} (compared to these references our 
definition is changed  $R_{12 \rightarrow 34}\rightarrow 1/(\Upsilon_1\Upsilon_2)
R_{12 \rightarrow 34}$ )
\begin{equation}
\frac{1}{V}\frac{dN_{\mu}}{dt} = 
(\Upsilon_{\gamma}^2 -\Upsilon_{\mu}^2){R_{\gamma\gamma \rightarrow \mu^+\mu^-}} +
(\Upsilon_e^2 -\Upsilon_{\mu}^2)R_{e^+e^- \rightarrow \mu^+\mu^-}. \label{muev}
\end{equation}
Like before for $\pi^0$ we consider the master equation in order to find appropriate definition of the
relaxation time constant for $\mu^\pm$ production. In no way should this be understood to imply
that muons are retained in the small plasma drop.
The $\mu$ production relaxation time is defined by:
\begin{equation}\label{taumu}
\tau_{\mu} =
\frac{1}{a}\frac{dn_{\mu}/d\Upsilon_{\mu}}{\left(R_{\gamma\gamma
\rightarrow \mu^+\mu^-} + R_{e^+e^- \rightarrow \mu^+\mu^-}\right)},
\end{equation}
where a suitable choice is $a=1,2$ for $t=0,\infty$, respectively (see below).
The form of Eq.\,(\ref{taumu})  assures that, omitting the volume expansion, i.e. the dilution
effect, in chemically equilibrated EP$^3$ the evolution of the muon fugacity obeys the equation
\begin{equation}
a\tau_{\mu}  \frac{d\Upsilon_{\mu}}{dt}=1-\Upsilon_{\mu}^2,
\end{equation}
which  has $\Upsilon_\mu(t=0)=0$ the simple analytical solution~\cite{Rafelski:1982pu}:
  \begin{equation}
\Upsilon_{\mu}=\tanh t/a\tau_\mu.
\end{equation}
For $t\to \infty$, near to chemical equilibrium,  $\Upsilon_{\mu}\to 1-e^{-2t/a\tau_\mu}$, while
  for $t\to 0$, at the onset of particle production with small $\Upsilon_{\mu}$ we have
$\Upsilon_{\mu} ={t/(a\tau_\mu)}$ . Hence,  near to chemical equilibrium  it
is appropriate to use $a=2$ in definition of relaxation time Eq.(\ref{taumu}) ,
while at the onset of particle production, more
applicable to this work a more physical choice would be $a=1$. However,  following the convention, in
the results presented below the value $a=2$ is used.

For invariant muon production rates we use, Eq.\ref{pp4}, with photons (bosons) or $e^+e^-$ (fermions) in initial state.
It is interesting to note that despite
including of quantum effects (Bose stimulated emission and Fermi blocking), using rates as defined,
we don't change the master population equation form. Only modification is slight fugacity dependence 
of rates presented in Eq.(\ref{pp}).

The $\sum |M_{e^+e^-\rightarrow \mu^+\mu^-}|^2$ differs from often considered heavy quark production
$\sum |M_{q\bar{q}\rightarrow c\bar{c}}|^2$~\cite{Combridge:1978kx, Gluck:1977zm}
($m_c>>m_q$) by color factor $2/9$, and the coupling
$\alpha_s\to \alpha $ of QCD has to be changed to   $\alpha=1/137$ of QED. Then we obtain, based on above references:
\begin{equation}
\sum |M_{e^+e^-\rightarrow \mu^+\mu^-}|^2=g_e^28\pi^2\alpha^2\frac{(m^2-t)^2+(m^2-u)^2+2m^2s}{s^2}, \label{m2emu}
\end{equation}
where $m=106$ MeV is the muon mass, electron and positron degeneracy $g_e=2$, and  $s$, $t$, $u$ are the usual
Mandelstam variables: $s=(p_{1}+p_{2})^2$, $t=(p_{3}-p_{1})^2$, $u=(p_{3}-p_{2})^2$, $s+t+u=2m^2$.
For the total averaged over initial states $|M|^2$ for photon fusion we have
\begin{equation}
|M_{\gamma\gamma \rightarrow \mu^+\mu^-}|^2
 =g_\gamma^2 8 \pi^2 \alpha^2 \left(-4\left(\frac{m^2}{m^2-t}
 +\frac{m^2}{m^2-u}\right)^2+4\left(\frac{m^2}{m^2-t}+\frac{m^2}{m^2-u}\right)
+\frac{m^2-u}{m^2-t} + \frac{m^2-t}{m^2-u}\right), \label{m2gmu}
\end{equation}
where degeneracy $g_{\gamma}=2$.
Near threshold   $s \approx 4m^2$, with  $t, u  \approx - m^2$ we find
\begin{equation}
|M_{\gamma\gamma \rightarrow \mu^+\mu^-}|^2 = 64 \pi^2 \alpha^2,
\qquad
|M_{e^+e^-\rightarrow \mu^+\mu^-}|^2 = 32 \pi^2 \alpha^2.
\end{equation}
The $e^+e^-\to \mu^+\mu^- $ reaction  involves a single photon, and
thus  it is more constrained (by factor 2) compared to the photon
fusion, which is governed by two Compton type Feynman diagrams.
However, in the rate we compute below, the indistinguishability of
the two photons introduces an additional factor $1/2$, so that both
reactions differ only by the difference in the quantum Bose and
Fermi distributions.

Integrals in Eq.(\ref{pp4}) can be evaluated in spherical coordinates. We define:
\begin{equation}
q=p_1+p_2;\,\,\,\,p=\frac{1}{2}(p_1-p_2);\,\,\,\,q^{\prime}=p_3+p_4;\,\,\,\,p^{\prime}=\frac{1}{2}(p_3-p_4);
\end{equation}
z-axis is chosen in the direction of
$\overrightarrow{q}=\overrightarrow{p_1}+\overrightarrow{p_2}$:
$$q_{\mu}=(q_0,0,0,0),\,\,\,\,p_{\mu}=(p_0, p\sin\theta,0, p\cos\theta),\,\,\,\,
p_{\mu}^{\prime}=(p^{\prime}_0, p^{\prime}\sin\phi\sin\chi, p^{\prime}\sin\phi\cos\chi, p^{\prime}\cos\phi).$$
Now we obtain~\cite{Matsui:1985eu}:
\begin{eqnarray}
&& {R_{e^+ e^-(\gamma\gamma) \rightarrow \mu^+\mu^-}}= \frac{1}{1+I}\frac{(4\pi)(2\pi)}{(2\pi)^4 16} \int_{2m_{\mu}}^{\infty}
dq_0 \int_0^{\sqrt{s-q_0^2}}dq\int_{-\frac{q}{2}}^{\frac{q}{2}}dp_0\int_{-\frac{q^*}{2}}^{\frac{q^*}{2}}dp{\prime}_0
\int_0^{\infty}dp \int_0^{\infty}dp\prime
\nonumber\\[0.4cm]
&&\times\int^{1}_{-1}d(\cos{\theta})\int^{1}_{-1}d(\cos{\phi})\int_0^{2\pi}d{\chi}\delta\left(p-\left(p_0^2+\frac{s}{4}\right)^{1/2}\right)
 \delta\left(p{\prime}-\left(p{\prime}_0^2-{m_{\mu}^2}+\frac{s}{4}\right)^{1/2}\right)
\nonumber\\[0.4cm]
&&\times\delta \left(\cos{\theta}-\frac{q_0p_0}{qp}\right)\delta\left(\cos{\phi}-\frac{q{\prime}_0p{\prime}_0}{qp}\right)\sum|M_{e+e-(\gamma\gamma)\rightarrow \mu\mu}|^2
\Upsilon_{\mu}^{-2}f_{\mu}\left(\frac{q_0}{2}+p_0\right)f_{\mu}\left(\frac{q_0}{2}-p_0\right)
\nonumber\\&&\times
\Upsilon_{e(\gamma)}^{-2}f_{e(\gamma)}\left(\frac{q_0}{2}+p^{\prime}_0\right)f_{e(\gamma)}
\left(\frac{q_0}{2}-p^{\prime}_0\right)\exp{(q_0/T)}, \label{ratemu}
\end{eqnarray}
where
$q^*={q}\sqrt{1-\frac{m_{\mu}^2}{s}}$. The integration over $p$,
$p^{\prime}$, $\cos{\theta}$, $\cos{\phi}$ can be done analytically
considering the delta-functions. The other integrals can be evaluated
numerically. For the case of indistinguishable  colliding particles (two photons)   there is additional factor $1/2$
implemented by the value $I=1$, while for distinguishable colliding particles (here electron and positron) $I=0$.

%%%%%%%%%%%%%%%%%%%%%%%
%%%%%%%%%%%%%%%%%%%%%%%
\subsection{$\pi^{\pm}$ production}\label{pichprod}
%%%%%%%%%%%%%%%%%%%%%%%
$\pi^{\pm}$ can be produced in $\pi_0\pi_0$ charge exchange scattering~(\ref{pppp}) and photons or $e^+e^-$ fusion,
reactions~(\ref{ffpp}) and (\ref{eepp})

We find  that for $\pi^{\pm}$ production, the last two processes are much
slower compared to the   first,  in case that $\pi_0$ density is near chemical equilibrium upto temperatures
approximately 30 MeV.
As we mentioned before rate of two $\pi^0$ production in photons fusion, Eq.\ref{ggpi0pi0}, is much smaller than rate of one $\pi^0$ production
at considered temperatures.

The time evolution equations for the number
of $\pi^{\pm}$ are similar to Eq. (\ref{muev}):
\begin{eqnarray}
\frac{1}{V}\frac{dN_{\pi^{\pm}}}{dt} =
({\Upsilon_{\pi^0}^2}-{\Upsilon_{\pi^{\pm}}^2}){R_{\pi^0\pi^0\leftrightarrow \pi^+\pi^-}} 
+ (\Upsilon^2_{\gamma} - {\Upsilon_{\pi^{\pm}}^2}){R_{\gamma\gamma \leftrightarrow\pi^+\pi^-}}
+(\Upsilon_e^2 - {\Upsilon_{\pi^{\pm}}^2}){R_{e^+e^- \leftrightarrow\pi^+\pi^-}}. \label{pisc}
\end{eqnarray}

For the respective three cross sections we use, all results valid in the common range  $s\le 1 $ GeV$^2$ we consider here:
\begin{itemize}
\item
The  cross section for charge exchange $\pi^0$scattering reaction Eq.(\ref{pppp})
 have been considered in depth recently~\cite{Kaminski:2006qe}:
\begin{equation}
\sigma = \frac{16\pi}{9}\sqrt\frac{s-4M^2_{\pi^{\pm}}}{s-4M^2_{\pi^{0}}}
(a^{(0)}_0-a^{(2)}_0)^2; \label{pipipipi}
\end{equation}
where $a^{(0)}_0 - a^{(2)}_0 = 0.27/M_{\pi^{\pm}}$
This is the dominant process for charge pion production, subject to  presence of $\pi^0$.
\item
For process Eq.(\ref{ffpp}),  the cross section of $\pi^{\pm}$ production in photon fusion we use~\cite{Terazawa:1994at}:
\begin{equation}
\sigma_{\gamma\gamma \rightarrow \pi^+\pi^-} = \frac{2\pi \alpha^2}{s}\left(1-\frac{4m_{\pi}^2}{s}\right)^{1/2}
\left(\frac{m_V^4}{(1/2s + m_V^2)(1/4s+m_V^2)}\right), \label {ggpipi}
\end{equation}
where $m_V=1400.0$ MeV. As we will see from
numerical calculations given the cross sections for
$\gamma\gamma \rightarrow \pi^+\pi^-$ resulting production rates will be smaller than the charge exchange
$\pi^0\pi^0 \rightarrow \pi^+\pi^-$ reaction.
\item
For process Eq.(\ref{eepp}), the cross section of $\pi^{\pm}$ production in electron - positron fusion we use~\cite{Gounaris:1968mw}:
\begin{equation}
\sigma_{e^+e^-\rightarrow \pi^+\pi^-} = \frac{\pi \alpha^2}{3}\frac{(s-4m^2_{\pi})^{3/2}}{s^{5/2}}\left|F(s)\right|^2. \label{eepipi}
\end{equation}
The form factor $F(s)$ can be written in the form:
\begin{equation}
F(s) = \frac{m_{\rho}^2+m_{\rho} \Gamma_{\rho}d}{m_{\rho}^2-s+\Gamma_{\rho}(m_{\rho}^2/k_{\rho}^3)[k^2(h(s)-
h(m^2_{\rho}))+k_{\rho}^2h^{\prime}(m^2_{\rho})(m_{\rho}^2-s)]-im_{\rho}(k/k_{\rho})^3\Gamma_{\rho}(m_{\rho}/\sqrt{s})};
\end{equation}
where $h^\prime(s)=dh/ds$ and
\begin{equation*}
k=\left(\frac{1}{4}s-m_{\pi}^2\right)^{1/2};\quad
k_{\rho}=\left(\frac{1}{4}m_{\rho}^2-m_{\pi}^2\right)^{1/2};\quad
h(s) = \frac{2}{\pi}\frac{k}{\sqrt{s}}\ln\left(\frac{\sqrt{s}+2k}{2m_{\pi}}\right);
\end{equation*}
$m_{\rho}=775$ MeV, $\Gamma_{\rho}=130$ MeV, $d=0.48$. Given this cross section we also
find that the rate of charged pion production is  small when compared to $\pi_0$-charge exchange
scattering.
\item
For reaction (\ref{ggpi0pi0}) we have \cite{Mennessier:2007wk}:
\begin{equation}
\sigma(\gamma\gamma \rightarrow \pi^0\pi^0) =
\left(\frac{\alpha^2\sqrt{s-4m_{\pi}^2}}{8\pi^2\sqrt{s}}\right)\left[1+
\frac{m_{\pi}^2}{s}f_s\right]\sigma(\pi^+\pi^- \rightarrow \pi^0
\pi^0), \label{siggpi0}
\end{equation}
where
\begin{equation}
f_s =  2(\ln^2(z_+/z_-)-\pi^2)+\frac{m_{\pi}^2}{s}(\ln^2(z_+/z_-)+\pi^2)^2,
\end{equation}
and $z_{\pm} = (1/2)(1\pm \sqrt{s-4m_{\pi}^2})$.
\end{itemize}

\begin{figure}
\centering
\includegraphics[width=10.6cm,height=10cm]{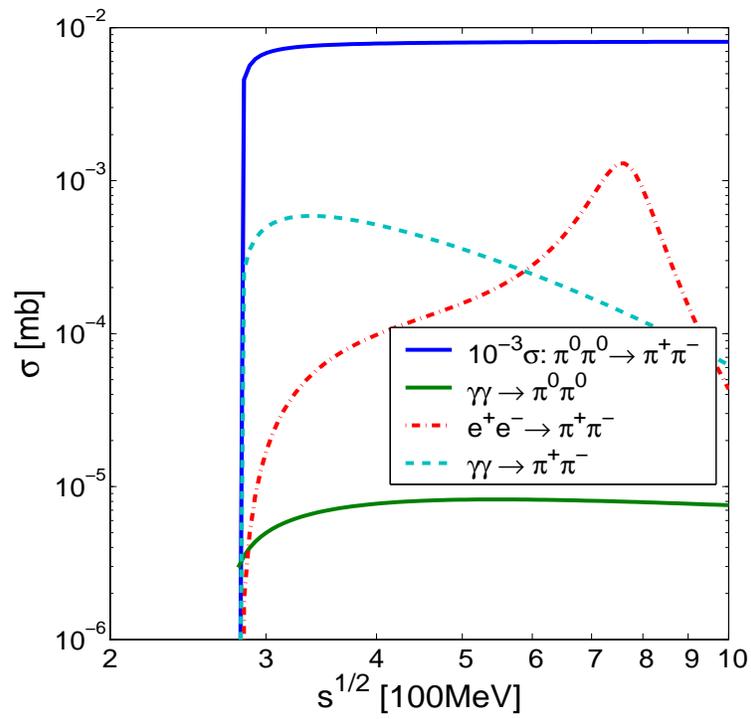}
\caption{\small{The cross section  $\sigma$ for
 pion  pair production, and pion charge exchange (solid top line),
as functions of $\sqrt{s}\le 1$ GeV$^2$.}}
\label{sigma}
\end{figure}

The cross sections for $\pi^+\pi^-$ pair  production, evaluated
using Eqs.(\ref{pipipipi}), (\ref{ggpipi}) and (\ref{eepipi}) are
presented in figure \ref{sigma} as functions of reaction energy
$\sqrt{s}$ . Top solid line (blue) is for charged pions production
in $\pi^0$ scattering Eq.(\ref{pppp}), the magnitude of this cross
section being very large we reduce it in presentation by factor
1000; the dashed line is for $\pi^+\pi^-$ production in photon
fusion Eq.(\ref{ffpp}); dash-doted line is for electron positron
fusion Eq.(\ref{eepp}). The bottom solid line (green) is for  photon
fusion into two neutral pions, Eq.(\ref{siggpi0}). The prediction
for $\sigma_{\gamma\gamma \rightarrow \pi^+\pi^-}$ is about 480 nb
(data 420 nb) at the peak near threshold \cite{Mennessier:2007wk},
which is in agreement with calculations presented here. The
reaction $\sigma_{\gamma\gamma \rightarrow \pi^0\pi^0}$(Eq.(\ref{ggpi0pi0}))
is much smaller than others and we do not consider this reaction further. We note
that some of these results are currently under intense theoretical
discussion as they relate to chiral symmetry. For our purposes the
level of precision of here presented reaction cross sections
is quite adequate.

%%%%%%%%%%%%%%%%%%%%%%%%%%%%%%
\section{Numerical results}\label{resnum}
%%%%%%%%%%%%%%%%%%%%%%%%%%%%%%

\subsection{Particle production relaxation times}
\begin{figure}
\centering
\includegraphics[width=10.6cm,height=14cm]{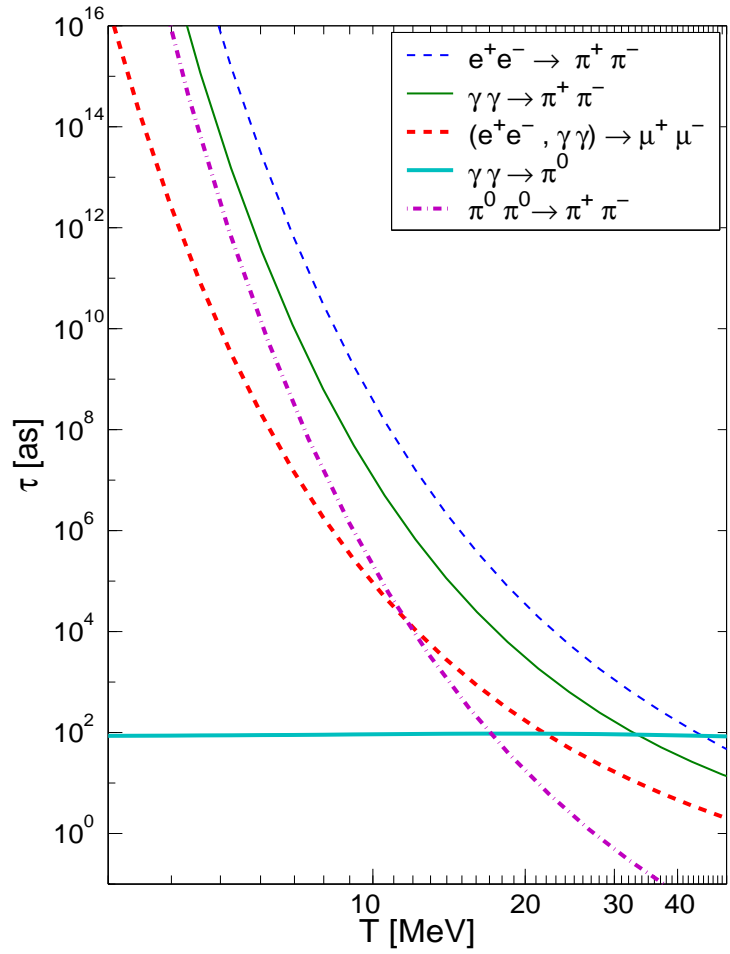}
\caption{\small{The relaxation time $\tau$ for  the different channels
of pion  and muon  production (see box), as functions of plasma
temperature $T$. }}
\label{taumupi}
\end{figure}

%%%%%%%%%%%%%%%%%%%%%%%%%%%%%%
%%%%%%%%%%%%%%%%%%%%%%%%%%%%%%
In figure \ref{taumupi} we show relaxation time $\tau$ for the different processes  considered as function of temperature $T\in [3,50]$ MeV.
Because of the  large difference in production rates which can be compensated by different densities of particles present
(magnitudes of fugacities) we introduce partial relaxation time for each of the three
 reactions $\pi^0\pi^0 \rightarrow \pi^+\pi^-$, $\gamma\gamma
\rightarrow \pi^+\pi^-$ and $e^++e^- \rightarrow \pi^+\pi^-$:
\begin{equation}\label{tausc}%\label{taugepi}
\tau_{\pi^0\pi^0\leftrightarrow\pi^{+}\pi^{-}} = \frac{\Upsilon_{\pi^0}^2}{2}\frac{{dn_{\pi^{\pm}}}/
{d\Upsilon_{\pi^{\pm}}}}{{R_{\pi^0\pi^0\leftrightarrow \pi^+\pi^-}}};\quad
\tau_{\gamma\gamma\leftrightarrow \pi^+\pi^-} =
    \frac{1}{2}\frac{{dn_{\pi^{\pm}}}/{d\Upsilon_{\pi^{\pm}}}}{R_{\gamma\gamma\leftrightarrow \pi^+\pi^-}}; \quad
\tau_{e^+e^-\leftrightarrow \pi^+\pi^-} =
    \frac{1}{2}\frac{{dn_{\pi^{\pm}}}/{d\Upsilon_{\pi^{\pm}}}}{R_{e^+e^-\leftrightarrow \pi^+\pi^-}};
\end{equation}
When $T\ll m$, we can use the Boltzmann approximation to the particle distribution functions. Since in this limit
the density is proportional to  $\Upsilon$ the  relaxation times  doesn't
depend on   $\Upsilon$. Moreover, even for $T\to 50$ MeV, we have for muons $e^{-m/T}\simeq 1/3$, thus quantum
correlations in phase space remain small, and the Boltzmann limit can be employed.
To account for small deviation from Boltzmann limit  arising towards the upper limit
of the temperature range we consider,  that is at $T \simeq 50$ MeV,
we used the exact equations with  $\Upsilon_i = 1$  to calculate $\tau$ for each case.
In addition to these three cases Eq.(\ref{tausc})  we show in figure \ref{taumupi}
the muon production relaxation time Eq.(\ref{taumu} ),
the two photon fusion into $\pi^0$ relaxation time Eq.(\ref{taupi0}), a nearly horizontal line (turquoise, bottom),
which is slightly greater than the free space $\pi^0$ decay rate. Finally, the thin dash-dot line at about $10^8$ times
greater value of time is the electron-positron fusion into $\pi^0$, Eq.(\ref{pi0ee}).

%%%%%%%%%%%%%%%%%%%%%%%%%%%%%%
%%%%%%%%%%%%%%%%%%%%%%%%%%%

%%%%%%%%%%%%%%%%%%%%%%%%%%%%%%
%%%%%%%%%%%%%%%%%%%%%%%%%%%
\subsection{Rates of pion and muon formation}
%%%%%%%%%%%%%%%%%%%%%%%%%%%%%%

%%%%%%%%%%%%%%%%%%%%%%%%%%%%%%
\begin{figure}
\centering \hspace*{0.99cm}
\includegraphics[width=8.6cm,height=12.5cm]{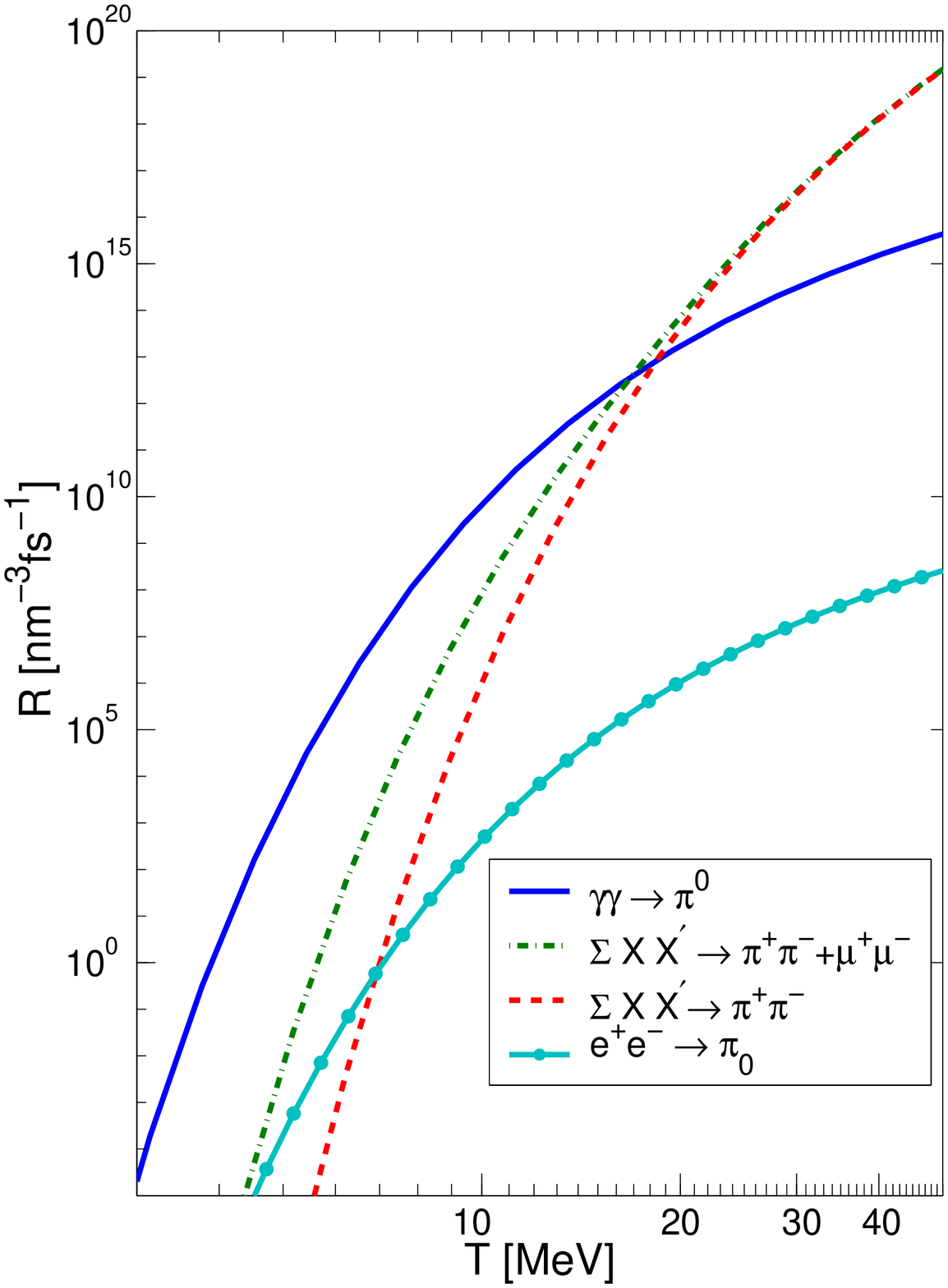}\hspace*{-0.69cm}
\includegraphics[width=8.6cm,height=12.5cm]{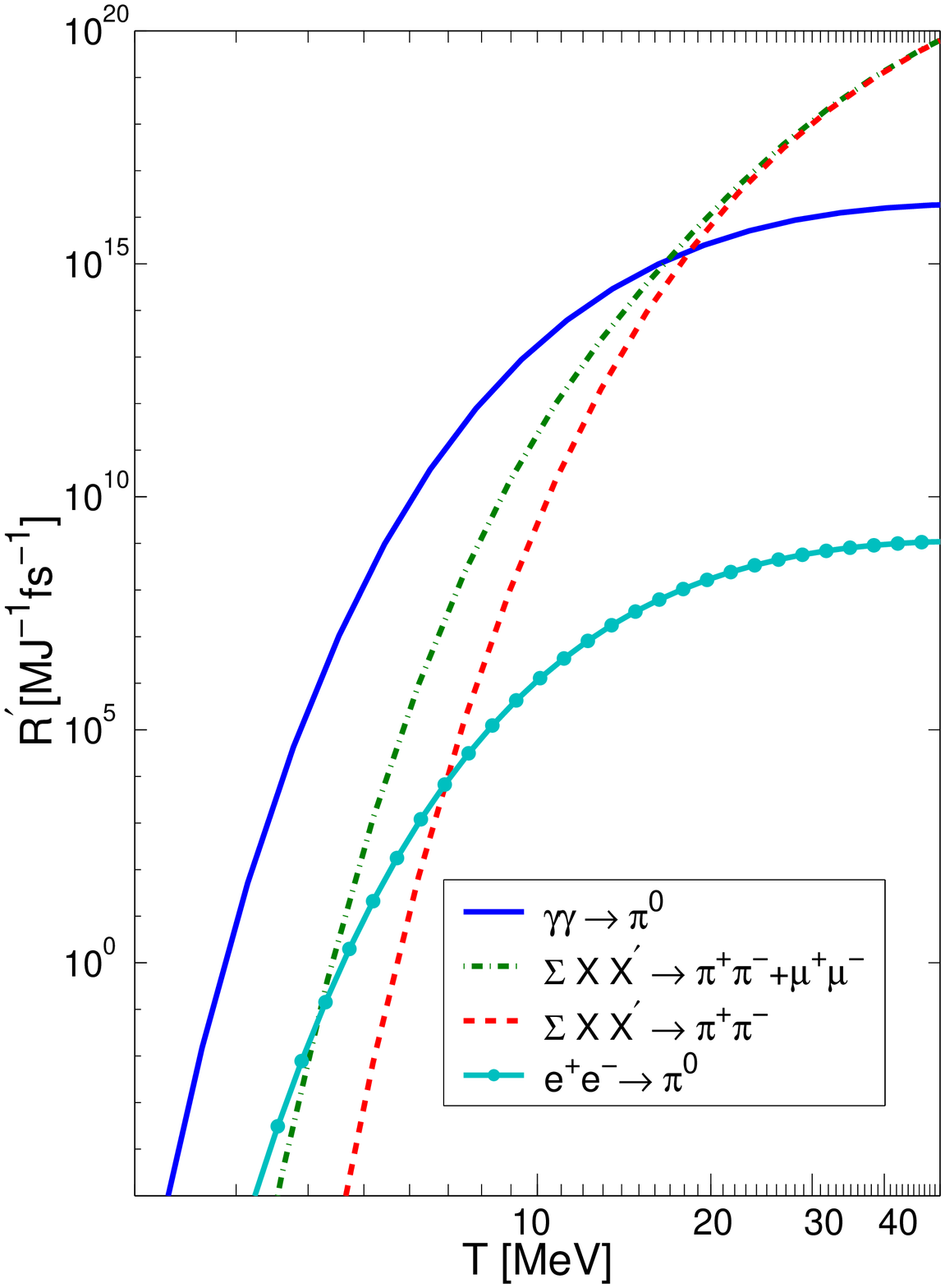}
\caption{\small{On left, the   invariant pion production rates in units of nm$^{-3}$fs$^{-1}$,
as a function of temperature $T$. On  right the production rate $R^\prime$ per  Joule energy content
in the fireball,  in units of MJ$^{-1}$fs$^{-1}$, in both cases for reactions shown in the box.}} \label{pions}
\end{figure}
%%%%%%%%%%%%%%%%%%%%%%%%%%%

In figure \ref{pions} we show on left as a solid (blue) line as a function of fireball temperature
 the rate per unit volume  and time for the process  $\gamma+\gamma\to \pi^0$, the dominant
mechanism of pion production. The other solid line with dots corresponds to $e^++e^-\to \pi^0$ reaction
which in essence remains, in comparison, insignificant. Its importance follows from the fact that it provides
the second most dominant path to $\pi_0$ formation at lowest temperatures considered, and it
operates even if and when photons are not confined to remain in the plasma drop.

We  improve   the rate presentation on the right hand side in figure \ref{pions}: considering that  the formation of a
plasma state involves an experimentally given fireball energy content $\cal E$ in Joules,
we use Eq.(\ref{SB})  to eliminate the volume $V$  at each temperature $T$:
\begin{equation}
R^{\prime}_{\pi^0}\equiv \frac {d^2 W^{\prime}_{\gamma\gamma\to \pi^0}}{dt d {\cal E} }
                      =\frac{1}{g\sigma T^4}\frac {d^4W_{\gamma\gamma\to \pi^0}}{  d V dt} = \frac{1}{g\sigma T^4}  R_{\pi^0}
\end{equation}
For chemical nonequilibrium, replace $\sigma \to \Upsilon^2_\gamma\sigma(\Upsilon)$.
Considering the (good) approximate Eq.(\ref{taupi01}) we  obtain:
\begin{equation}
R^{\prime}_{\pi^0} \simeq
      \left(\frac{m_{\pi} }{2\pi T}\right)^{3/2}\frac{e^{-m_{\pi}/T} }{ g\sigma T \tau_{\pi^0}^0 } .
\label{taupi02}
\end{equation}
We use units such that $\hbar=c=k=1$ and thus $R^{\prime}$  is a
dimensionless expression. Recalling the value of these constants, the  units we used for $R^{\prime}$ derive from
MeV s=1.603\,10$^{-4}$ MJ fs.  %1 MeV=1.603 10^{-13} Joule

The other lines in figure \ref{pions}  address the sum of  formation rates of charged pion pairs (dashed, red) by all reactions
considered in this work, $\pi^{0}+\pi^{0}\to \pi^{+}+\pi^{-}$,  $\gamma+\gamma \to \pi^{+}+\pi^{-}$, $e^++e^-\to \pi^{+}+\pi^{-}$.
We also present the sum of all reactions leading to either a charged pion pair, or muon pair (dot-dashed, green) lines,
adding in  $\gamma+\gamma \to \mu^{+}+\mu^{-}, e^++e^-\to \mu^{+}+\mu^{-}$.
The rationale for this presentation is that we  do not care how a heavy particle is produced, as long
as it can be observed. The dashed (red) line assumes that we specifically look for charged pions, and
dot-dashed (green) line that we wait till charged pions decays, being  interested in  the total final muon yield.
 The $\pi^0$ production
rate (blue, solid line) is calculated using Eq.(\ref{pi0pr}) and
yields on the logarithmic scale nearly indistinguishable result from
the approximation Eq.(\ref{taupi01}). For $\pi^\pm$ production we
refer to section \ref{pichprod} and for $\mu^\pm$ production we
refer to \ref{muprod}.

In table \ref{VTN} we show the values of key reaction rates $R$ and relaxation times $\tau$ at $T=5$ and $15$ MeV.
We note the extraordinarily fast rise of the rates with temperature, in some instances bridging 15 -- 20 orders in magnitude
when results for $T=5$ and $15$ MeV are compared.
%%%%%%%%%%%%%%%%%%%%%%%%%%%%%%%%%%%%%%%%%%%%%%%%%%%%%%TAble II
\begin{table}
\caption{Values of  rates, relaxation times for all reactions at $T=5$ MeV and  $T=15$ MeV}
 \label{VTN}
\begin{tabular}{|c|c|c|c|c|c|}
  \hline
             &   $T=5$ MeV  & $T=5$ MeV &$T=15$ MeV & $T=15$ MeV\\
 reaction & $\tau$ [as] & $R$ $[\rm{nm^{-3}fs^{-1}}]$& $\tau$ [as] & $R$ $[\rm{nm^{-3}fs^{-1}}]$   \\
  \hline
$\gamma\gamma \leftrightarrow \pi_0$ & $8.82\,10^2$ & $3.3\,10^3$ & $9.5\,10^2$ & $1.2\,10^{12}$\\
$e^+e^- \leftrightarrow \mu^+\mu^-$ & $1.2\,10^{10}$ &  $3.2\,10^{-3}$ & $1.9\,10^3$ & $1.5\,10^{11}$ \\
$\gamma\gamma \leftrightarrow \mu^+\mu^-$ & $1.0\,10^{10}$ & $3.7\,10^{-3}$ &$1.3\,10^3$& $2.1\,10^{11}$ \\
$\pi^0\pi^0 \leftrightarrow \pi^+\pi^-$& $2.9\,10^{12}$ & $2.1\,10^{-8}$& $4.6\,10^2$ & $9.5\,10^{10}$  \\
$\gamma\gamma \leftrightarrow \pi^+\pi^-$& $6.4 \, 10^{13}$ & $9.7\,10^{-10}$ & $5.1\,10^4$ & $8.7\,10^8$   \\
$e^+e^- \leftrightarrow \pi^+\pi^-$& $7.8\,10^{15}$ & $7.9\,10^{-12}$ & $9.5\,10^5$ & $4.6\,10^{7}$\\
\hline
\end{tabular}
\end{table}
%%%%%%%%%%%%%%%%%%%%%%%%%%%%%%%%%%%%%%%%%%%%%%%%%%%%%%%%%%%%%%%%

\begin{figure}
\centering\hspace*{0.99cm}
\includegraphics[width=8.6cm,height=11.5cm]{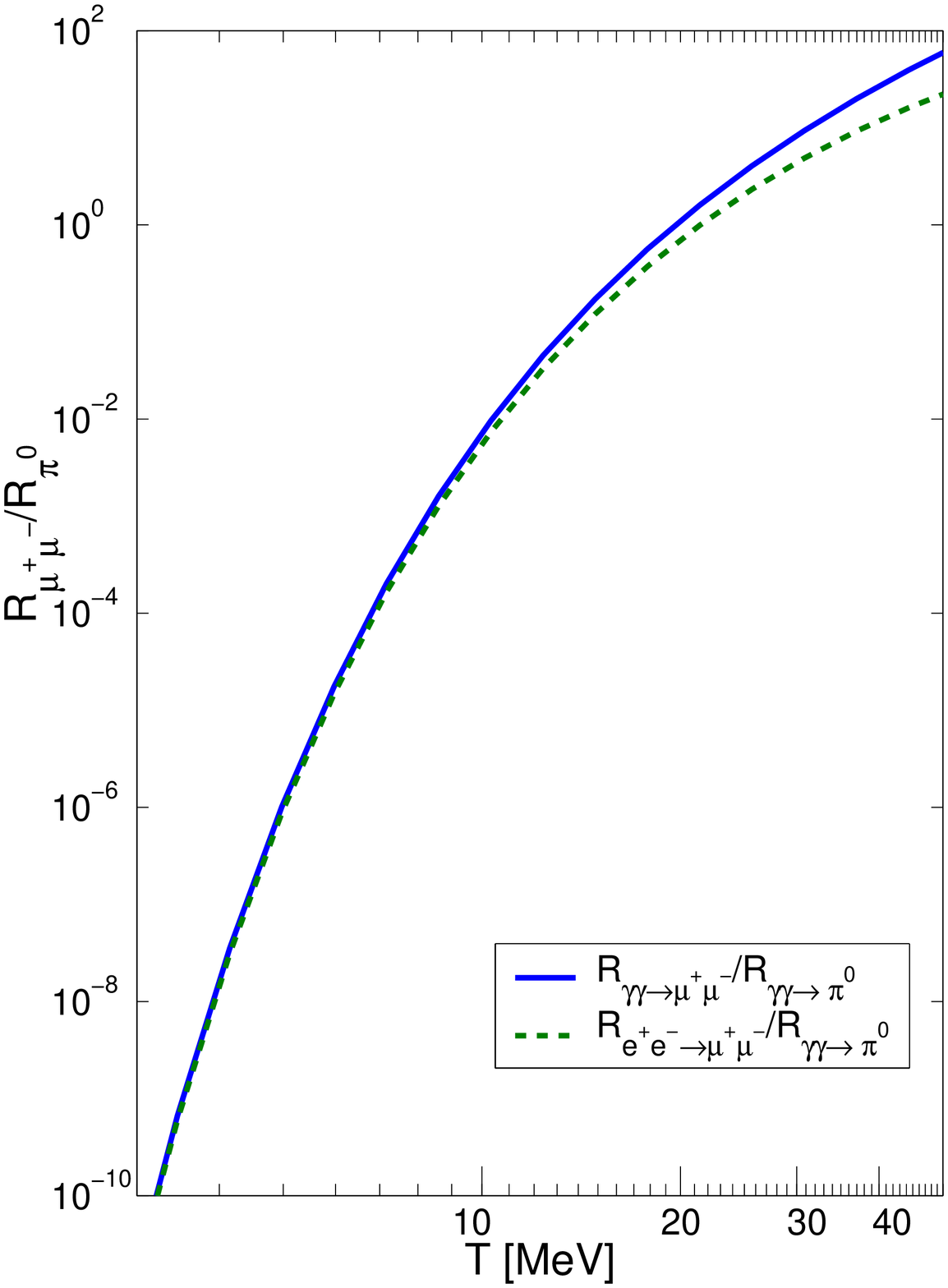}\hspace*{-0.7cm}
\includegraphics[width=8.6cm,height=11.5cm]{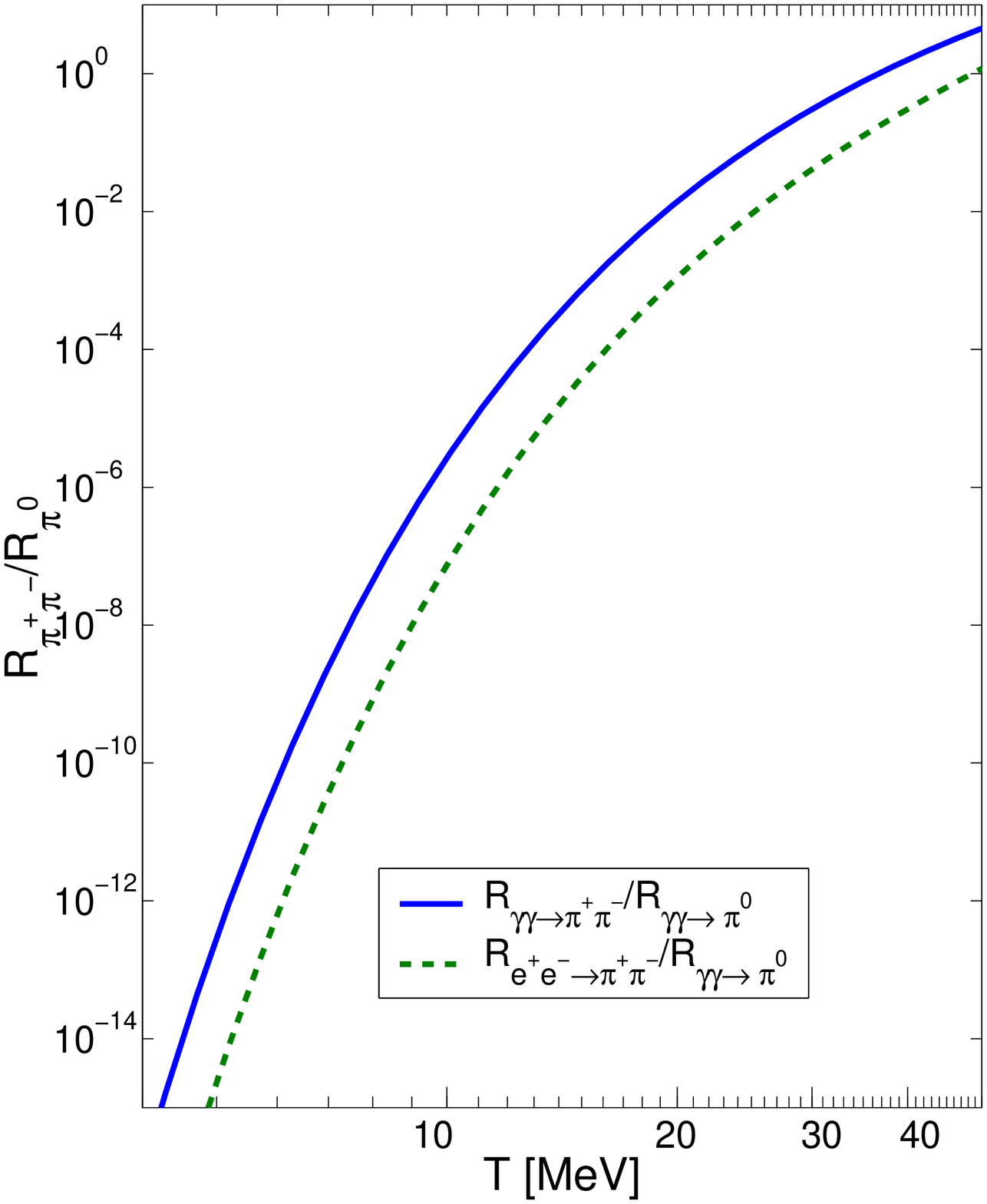}
\caption{\small{On left: Muon and on right charged pion production rates in electromagnetic processes
normalized by  $\pi^0$ production rate. Solid line (blue) for  $\gamma\gamma$ , dashed line  (green) for
 $e^+e^- $ induced process.}}
\label{mupir}
\end{figure}

In order to understand the individual contributions to the  different reactions
entering the sum of rates presented above, we show  as a function of temperature
in the figure \ref{mupir}   the relative strength of muon pair (left)
and charge pion (on right) electromagnetic ($\gamma+\gamma, e^++e^-$)
 production,  using as the reference the
 $\gamma+\gamma \rightarrow \pi^0$ reaction.
The $\mu^{\pm}$ production rates are calculated using
Eq.(\ref{ratemu}) with $|M|^2$ from Eq.(\ref{m2emu}) and
 Eq.(\ref{m2gmu}) respectively. This ratio is smaller than unity for $T \ll 20$ MeV.
For larger $T$, the  muon direct production rate becomes
larger than $\pi^0$ production rate. Charged pions (on right in figure  \ref{mupir})
can be produced in direct reaction at a rate larger than neutral pions
only for $T>35$ MeV. The photon channel dominates.

%%%%%%%%%%%%%%%%%%%%%%%%%%%%%%
%%%%%%%%%%%%%%%%%%%%%%%%%%%

\section{Discussion and Conclusions}\label{concl}
%%%%%%%%%%%%%%%%%%%%%%%%%%%%%%%%%%%%%%%%%%

We found that the production of $\pi^0$ is the dominant coupling of electromagnetic radiation
to heavy (hadronic) particles with $m \gg T$,  and as we have here demonstrated
 that noticeable particle yields can be expected already
at modest temperatures $T\in [3,10]$ MeV.
In present day  environment of 0.1 --1 J plasma  lasting a few fs, our results suggest that we can expect
integrated over space-time evolution of the EP$^3$ fireball a  $\pi^0$  yield at the limit of detectability.
For $T \to 15$ MeV the $\pi^0$ production rate  remains dominant and indeed
very large, reaching the production rate $R^{\prime}\simeq 10^{15}$[MJ$^{-1}$fs$^{-1}]$.
Charge exchange reactions convert some of the neutral pions into charged pions which are
more easy to detect.

In this situation it is realistic to consider the possibility of forming a chemically equilibrated fireball
with $\pi^0$, $\pi^{\pm}$, $\mu^{\pm}$ in chemical abundance equilibrium. The heavy particles are produced
in early stages when temperature reached is highest. Their abundance in the fireball  follows the fireball
expansion and cooling till their
freeze-out, that is decoupling of population equation production rates. The particle yields are than
given by the freeze-out conditions, specifically the chemical freeze-out temperature $T_f$ and
volume $V_f$, rather than the integral over the rate of production. In this situation the heavy
particle yields become diagnostic tools of the freeze-out conditions, with the mechanisms of
their formation being less accessible. However, one can avoid this condition by appropriate
staging of fireball properties.

The present study has not covered, especially for low temperature range all the possible
mechanisms, and we addressed some of these issues in the introduction. Here we note further
that the production of heavy particles
requires energies of the magnitude $m/2$ and thus   is due to collisions involving   the  (relatively speaking)
far tails of a thermal particle distribution. If these tails  fall off as a power law, instead of the
Boltzmann exponential  decay~\cite{Biro:2004qg},  a  much greater yield of heavy particles
 could ensue. There could further be present a   collective amplification
to the production process e.g. by residual matter flows, capable to
enhance the low temperature yields, or by collective plasma oscillations and inhomogeneities.

These are just some examples of many
reasons to hope and  expect a   greater particle yield than we
computed here in microscopic and controllable two particle reaction
approach. This consideration, and our encouraging
`conventional' results suggest that
 the study of  $\pi^0$ formation in  QED plasma  is of considerable  intrinsic interest. Our results
provide a lower limit for rate of particle production and when folded with models of EP$^3$ fireball
formation and evolution, final yield.

It is of some interest to note that  the study of pions in QED plasma allows exploration of pion properties
in electromagnetic medium. Specifically,    recall
 that  1.2\% fraction of  $\pi^0\to e^+e^-\gamma$ decays, which
implies that the associated processes such as  $e^++e^-\to \gamma +\pi_0$ are  important. We cannot
evaluate this process at present as it involves significant challenges in understanding of $\pi_0$
off-mass shell `anomalous' couping to two photons.

The experimental environment we considered here should allow
a detailed study of the properties of   pions (and also muons) in a thermal background.
There is considerable fundamental
interest in the study of   pion properties and specifically
pion mass splitting  in  QED plasma at temperature $T\gtrsim \Delta m$ and in presence of
electromagnetic fields. We already have shown that due to quantum statistics effects, the effective
in medium decay width of  $\pi^0$  differs  from the free  space value, see figure \ref{taupi0app}.
In addition,   modification of mass and  decay width due to ambient medium influence on
the pion internal structure is to be expected.
 Further we hope that the study of pions in the  EP$^3$ fireball  will contribute to the better
understanding of   the relatively large difference in mass between
 $\pi^0$and $\pi^\pm$. The  relatively large size of the PE$^3$ environment should 
make  such changes, albeit small, measurable.

The experimental study of  $\pi^0$ in QED plasma environment is not
an easy task. Normally, one would think that the study of the
  $\pi^0$ decay into two  67.5 MeV $\gamma$ (+ thermal Doppler shift motion)
produces a  characteristic signature. However, the   $\pi^0$ decay is
in time and also in location overlapping with the plasma formation and disintegration. The debris of the plasma,
 reaches any detection system  at practically the same time instance as does the 67.5 MeV $\gamma$. The large
amount of available radiation will disable  the detectors. On the other hand we realize that
the hard thermal component of the plasma, which leads to the production of   $\pi^0$ in the early fireball stage, is
most attenuated by plasma dynamical expansion.  Thus it seems possible to plan for  the  detection of
 $\pi^0$ e.g. in a heavily shielded detection system.

The decay time of charged pions being 26 ns, and that of charged muons being 2.2 $\mu$s
it is possible to separate in time the plasma debris from the decay signal of these particles.
Clearly, these heavy charged particles can be detected with much greater ease, also considering that
the decay product of interest is charged.
For this reason, we also have in depth considered all channels of production of charged pions
and muons. Noting that practically all charged pions turn into muons, we  have also compared
the production rates of $\pi^0$ with all heavy particles, see dot-dashed (green) line in figure \ref{mupir}.
This comparison suggests that for plasmas at a temperature reaching $T>10$ MeV the production
of final state muons will most probably be by far easier to detect. On the other hand for $T<5$ MeV
it would seem that the yield difference in favor of  $\pi^0$ outweighs the detection system/efficiency loss
considerations. Future work addressing non-conventional processes will show at how low  $T$
we can still expect observable heavy particle yields.

An   effort to detect $\pi^0$ directly is justified since   we can learn  about the
properties of the plasma (lifespan, volume and temperature in early stages) e.g.
from a comparative study of  the $\pi^0$ and  $\pi^\pm$ production.
We have found that  at about $T>16$ MeV, the pion charge exchange  $\pi^0\pi^0\to \pi^+\pi^-$
reaction for chemically equilibrated $\pi^0$ yield is  faster than the natural $\pi^0$ decay,
 and the chemical equilibration time constant,  see the dot-dashed line  in
figure \ref{taumupi}. Thus beyond this temperature the yield of charged pions can be expected to be
in/near chemical equilibrium for a plasma which lives at, or above this temperature, for longer than 100 as.

In such an environment the yield of $\pi^0$ is expected to be near chemical equilibrium, since the
decay rate is compensated by the production rate, and, within 100 as, the chemical
equilibrium yield is attained. Moreover,
the thermal speed of produced $\pi$ can be  obtained from the nonrelativistic
relation $\frac 12 m \langle v^2\rangle =\frac 3 2 T$,  thus $\overline v \propto  \sqrt T $ and, for $T=10 $ MeV,
$\overline v\simeq 0.5 $c. This is
nearly equal to the sound velocity of EP$^3$,  $v_s\simeq c/\sqrt{3}=0.58c$. Thus the heavy $\pi^0$ particles
can be seen as co-moving with the expanding/exploding  EP$^3$, which completes
the argument to justify their transient chemical
equilibrium yield in this condition.

The global production yield of neutral and charged pions
should  thus  allow the study of volume and temperature history of the QED  plasma.
More specifically, since with decreasing temperature, for $T<16$ MeV, there is a rapid  increase of the relaxation time for
the charge exchange  process, there is a rather rapid drop
of the  charged pion yield below chemical equilibrium --- we note that charge exchange   equilibration time at $T=10$ MeV is
a factor $10^5$ longer. We note that the study of two pion correlations provides an independent measure of
the source properties (HBT measurement).

The relaxation time of electromagnetic production of muon pairs wins over $\pi^0$
relaxation time for $T>22$ MeV, see dashed line, red, in figure \ref{taumupi},
the  direct  electromagnetic processes
of charged pion production (thin green, solid line for $ \gamma\gamma \rightarrow
\pi^+\pi^-$ and dashed, blue for $e^+e^- \rightarrow \pi^+\pi^-$) remain  sub-dominant.
Thus for $T>22$ MeV we expect, following the same chain of arguments for muons as above for
charged pions, a near chemical equilibrium yield. If the study of all these
$\pi^0,\pi^\pm,\mu^\pm$ yields,  their spectra and even pion correlations were possible, considerable insight into
 $ e^-,   e^+, \gamma $ plasma (EP$^3$) plasma formation and dynamics  at $T<25$ MeV can be achieved.

%%%%%%%%%%%%%%%%%%%%%%%%%%%%%%%

\chapter{PION AND MUON IN EARLY UNIVERSE}\label{earlyun}

In this chapter we begin to apply methods considere in previous chapters to early Universe. 
These all reactions of muon and pions production, considered in $e^+e^-\gamma$ plasma take place in early universe.
Here we show that $\pi^0$ is in chemical equilibrium with photons at all temperatures of interest. 

In expanding universe in metric \cite{Kolb:1988aj}
\begin{eqnarray}
ds^2&=&dt^2-R^2(t)\left(\frac{dr^2}{1-kr^2}+r^2d{\theta}^2+r^2sin^2{\theta}d{\phi}^2\right),\\
g&=-&\frac{R^6r^4sin^2{\theta}}{1-kr^2}
\end{eqnarray}
The Eq. (\ref{popeq}), which describes $\pi_0$ evolution, has dilution term:
\begin{equation}
\frac{d}{dt}n_{\pi}+3Hn_{\pi} =(1-\Upsilon_{\pi})A,
\label{npi1}
\end{equation}
where A is defined by Eq.(\ref{A})
\begin{equation}
\frac{dn_{\pi}}{dt}=\frac{dn_{\pi}}{d\Upsilon_{\pi}}{\dot{\Upsilon_{\pi}}}+\frac{dn_{\pi}}{dT} {\dot{T}}.
\end{equation}
Then dividing both sides of equation (\ref{npi1}) by $dn_{\pi}/d{\lambda}$ and using Eq.
(\ref{tau}) we obtain
\begin{equation}
\dot \Upsilon_{\pi} -\frac{1}{\tau_T}\Upsilon_{\pi}+3H \frac{n_{\pi}/\Upsilon_{\pi}}{dn_{\pi}/d{\Upsilon_{\pi}}}\Upsilon_{\pi} = (1 - \Upsilon_{\pi})\frac{1}{\tau},
\label{npi2}
\end{equation}
where
\begin{equation}
H=\frac{\dot R}{R},
\end{equation}
and
\begin{equation}
\frac{1}{\tau_T}=-\frac{dn_{\pi}/dT}{dn_{\pi}/d{\Upsilon_{\pi}}}\frac{\dot T}{\Upsilon_{\pi}}.
\end{equation}
We put '-' sign to this equation to have $\tau_T>0$. 

The temperature can be defined from entropy conservation for radiation dominated epoch we have
\begin{equation}
\frac{\dot T}{T}=-\frac{\dot R}{R}. \label{Tch}
\end{equation}

Now we estimate how large is time scale $1/{\tau_T}$.
For $m_{\pi}/T<<1$ the $n_{\pi} \propto T^3$ and using (\ref{Tch})
we have in chemical equilibrium ($\Upsilon_{\pi^0}=1$):
\begin{equation}
\frac{1}{\tau_T} = 3H.
\end{equation}

Therefore for ultrarelativistic particles dilution rate compensate the rate of density decrease with temperature change in expanding universe. So density of pions may stay as in chemical equlibrium for all temperatures while pions are ultrarelativistic. Dilution doesn't have direct effect on solution. However,  when interaction rate  becomes small compare to the expansion rate, if pions aren't equilibrium with photons for some reasons it takes large time compared to universe age at that moment to get to equilibrium density. So they become decoupled from radiation. This decoupling takes place when
\begin{equation}
3H>\frac{1}{\tau}. \label{frcon1}
\end{equation}

The value of $\dot R/R$ can be find from Friedmann equation \cite{Kolb:1988aj}
\begin{equation}
\frac{\dot R^2}{R^2} + \frac{k}{R^2} = \frac{8\pi G}{3}\rho. \label{FE}
\end{equation}
For radiation dominated epoch, if $k=1$ we have
\begin{equation}
\rho=\frac{\pi^2 g^* T^4}{30},
\end{equation}
where $g^*$ is number of degrees of freedom. Then
\begin{equation}
H=1.66 \sqrt{g^*}\frac{T^2}{m_{pl}} \label{H}
\end{equation}
If we assume $\tau \approx \tau_0 = 8.4\,10^{-17}$ s we have from
(\ref{frcon1}) condition for freeze-out temperature
\begin{equation}
T \approx 10^{5} \textrm{GeV}
\end{equation}
For $T<10^5$ GeV the expansion rate is small compare to $\pi^0$ production
and decay rates.

If $m/T > 1$ for Boltzmann distribution
\begin{equation}
n_{\pi}= \frac{1}{2\pi^2}
\lambda\sqrt\frac{{\pi}m_{\pi}^3T^3}{2}{\exp}(-m_{\pi}/T)
\end{equation}
we have
\begin{equation}
\frac{1}{\tau_T} \approx \frac{m_{\pi}}{T}\,H.
\end{equation}

For small $T<<m_{\pi}$ we have $1/{\tau_T} >> H$. If $1/{\tau_T}$
exceeds decay rate the pions may lose chemical equilibrium for small
T when universe is matter dominated. For matter dominated universe
\begin{equation}
\rho = \frac{3}{8\pi}m_{pl}^2H^2_0\frac{R_0^{3}}{R^{3}} = \frac{3}{8\pi}m_{pl}^2H^2_0\frac{T^3}{T^3_0}.
\end{equation}

Here we used equation (\ref{Tch}). From Eq.(\ref{FE}) we have
\begin{eqnarray}
&&H = H_0 \sqrt{\Omega_0}\left(\frac{T}{T_0}\right)^{3/2} \\
&&\frac{1}{\tau_T} =  H_0 \sqrt{\Omega_0} \frac{m_{\pi}T^{1/2}}{T_0^{3/2}},
\end{eqnarray}
where
\begin{equation}
\Omega_0=\frac{\rho_0}{3/8\pi G\,H_0^2} \approx 1.
\end{equation}
and $T_0 \approx 10^{-13} $ GeV. ${1}/{\tau_T}$ is decreasing as
$\sqrt T$ and it can't reach value of pions decay width
$\Gamma_0$,
\begin{equation}
\Gamma_0\tau_T = \frac{\Gamma_0T_0^{3/2}}{H_0 \sqrt{\Omega_0}m_{\pi}T^{1/2}} \approx \frac{10^{16}}{T_{\rm{GeV}}^{1/2}}>>1.
\end{equation}

Therefore for considered temperature range the $\pi^0$ is in chemical equilibrium with photons because of their fast decay rate. This is not always that decay is so fast to exceed universe expansion rate. For example the decay $n \rightarrow p + e^- + \nu_e$ is much slower $\tau = 885.7$ s. The dilution rate exceeds neutron decay rate at $T > 0.1$ MeV.

The relaxation time for $\mu^{\pm}$ and $\pi^{\pm}$ in reactions (\ref{ggmu}) and (\ref{pppp}) respectively 
become many orders of magnitude larger than $\tau_{\pi^0}$ at temperatures about few MeV, where these reaction have to freeze out. Therefore these particles are in chemical equilibrium and their densities are also relatively high (about nucleons density or higher) up to temperatures of few MeV.  

This process is important to understand how the hadronic component diminish with
the expansion of the Universe and the possible effects of hadronic relics in
the cosmic blackbody radiation spectrum, such as its fluctuations and
correlations.

\chapter{SUMMARY AND CONCLUSIONS} \label{sumconc}

In the first part of dissertation we studied heavy particles production at hadronization, resonance evolution in thermal hadronc gas after hadronization. 
In the second part $e^+e^-\gamma$ plasma equilibrium conditions were considered and pion and muon production in this type of plasma.

\section{Summary of heavy flavor production}

In chapter~\ref{hfhad} I considered heavy flavor (charm, bottom) hadrons production within statistical hadronization model. The new feature
compared to the others studies is that we assume entropy and strangeness conservation during hadronization. 

While I compare the yields to the expectations based on
chemical equilibrium yields of light and strange quark pairs,
I present results based on the hypothesis that
the QGP entropy and QGP flavor yields determine
the values of phase space occupancy $\gamma^\mathrm{H}_i$ $i=q,s,c,b$,
which are of direct interest in study of the heavy hadron yields.

For highest energy heavy ion
collisions the range of  values discussed  in literature is
$1\le \gamma^{\mathrm{H}}_q\le 1.65$ and
$0.7\le \gamma^{\mathrm{H}}_s/\gamma^{\mathrm{H}}_q\le 1.5$. However
$\gamma^{\mathrm{H}}_c$ and $\gamma^{\mathrm{H}}_b$
values which are much larger than unity arise. This is
due to the need to describe the large primary parton based
production, and considering that the   chemical
equilibrium yields   are suppressed by the factor $\exp(-m/T)$.

Our work is based on the grand canonical treatment of phase space.
 This approach is valid for charm hadron production at LHC,  since
the  canonical corrections, as we have discussed, are
 not material.  On the other hand, even at LHC 
the much smaller  yields of  bottom heavy hadrons are
subject to canonical suppression. The value of the parameter $\gamma_b^{H}$
obtained at a   fixed bottom yield  $N_b$, using either  the canonical, or the grand canonical methods,
are different, see e.g.  Eq.\,(15) in \cite{Rafelski:2001bu}.  Namely, to obtain  a given yield $N_b$ 
in canonical approach, a greater value of $\gamma_b^{H}$ is needed 
in order to compensate the canonical suppression effect. 
However, for any individual single-$b$ hadron, 
the  relative yields, i.g. $B/B_s$ do  not depend on $\gamma_b^{H}$
and thus such ratios are not influenced by  canonical phase spase effect.  Moreover, as long as 
the yield of single-$b$ hadrons dominates the total  bottom yield: 
$N_b\simeq B+B_s+\Lambda_b+\ldots$, also the $N_b$ scaled yields of
hadrons comprising one $b$-quark i.e. ratios such as $B/N_b$, $B_s/N_b$, $B_c/N_b$, etc,  
are not sensitive to the value of   $\gamma_b^{H}$ and  can be obtained
within either the canonical, or grand canonical method. 
 On the other hand for $b\bar{b}$ mesons and multi-$b$ baryons the
canonical effects should be considered. Study of the
yields of these particles is thus postponed.

I have addressed here how the yields of heavy hadrons are influenced by
$\gamma^{\mathrm{H}}_s/\gamma^{\mathrm{H}}_q\ne 1$ and $\gamma_q \ne 1$. The actual values
of $\gamma^{\mathrm{H}}_s/\gamma^{\mathrm{H}}_q$ we use are related to
the strangeness per entropy yield $s/S$ established in the QGP phase.
Because the final value $s/S$ is established well before hadronization,
and the properties of the hadron phase space are well understood,
the resulting $\gamma^{\mathrm{H}}_s/\gamma^{\mathrm{H}}_q$ are well
defined and turn out to be quite different from unity in the range of
temperatures in which we expect particle freeze-out to occur.
We consider in some detail the effect of QGP hadronization on
the values of $\gamma^{\mathrm{H}}_s$ and $\gamma^{\mathrm{H}}_q$.

One of first results I present (figure~\ref{JpsiD}) allows a test of the
statistical hadronization model for heavy flavor:
I show that the yield ratio
$c\bar c$ $s\bar s$/($c\bar s$ $\bar c s$) is nearly independent
of temperature and it is also nearly constant when the $\phi$ is
allowed to freeze-out later (figure~\ref{JpsiD2T}), provided that the condition of
production is at the same value of strangeness per entropy $s/S$.

I studied in depth how the (relative) yields of strange and non-strange
charmed mesons vary with strangeness content. For a chemically
equilibrated QGP source, there is considerable shift of the yield
from non-strange $D$ to the strange $D_s$
 for $s/S=0.04$ expected at LHC.
The expected fractional yield $D_s/N_c \simeq {B_s}/N_b\simeq 0.2$
when one assumes $\gamma^{\mathrm{H}}_s=\gamma^{\mathrm{H}}_q=1$,
 the expected
enhancement of the strange heavy mesons is at the level of 30\%
when $s/S=0.04$, and greater when greater strangeness yield is
available.

As the result we find a relative suppression of the
multi-heavy hadrons, except when they contain strangeness. This suppression depends on both factors $\gamma_s$ and $gamma_q$. When phase space occupancy of
light and strange quark is relatively high the probability for charm quarks to make hadrons with strange quarks increases and probability to find the second charm quark
among light and strange quarks decreases. Therefore the $c\bar c$ yield suppression increases when $\gamma_s/\gamma_q$ ratio increases for constant $\gamma_q$. This result
is qualitatively in agreement with experimental results obtained for SPS energies~~\cite{Becattini:2005yj}.

On the other hand, the yield of $c\bar{c}/N_c^2 \simeq 2 10^{-3}$ is found to be
almost independent on  hadronization 
temperature when entropy at hadronization is conserved. That is because for larger $T$ $\gamma_q$ decreases. The suppression effect 
decreasees, compared to SHM and
become even negative for $T>200$ MeV, resulting to the $c \bar c$ yield almost independent on temperature. 
We don't know exactly equation of state in QGP and so the value of
 $\gamma_q$ which is needed to conserve the entropy may be different.
 If $\gamma_q$ is larger for higher temperatures, suppression of
$c{\bar{c}}$ is larger for a fixed $s/S$.
The same result is found for $B_c \approx 5-6\,10^{-4} N_cN_b $,
that  yield remains considerably larger (by a factor 10 --- 100) compared to the scaled
yield in single nucleon nucleon collisions.

I have shown that the study of heavy flavor hadrons will provide
important information about the nature and properties of the QGP
hadronization. The yield of Bc($b\bar c$) mesons remains enhanced
while the hidden charm $c\bar c$ states encounter another suppression
mechanism, compensating for the greatly enhanced production due to
large charm yield at LHC.
The results are published in~\cite{Kuznetsova:2006bh}

\section{Summary on Chemical Equilibration Involving Decaying Particles at Finite Temperature}

In chapter~\ref{dpeq},I examined in detail the kinetic master equation for the
process involving formation of an unstable particle through the reaction Eq.(%
\ref{123}) in a relativistically covariant fashion. Assuming that
all the particles in the process are in thermal equilibrium, we
calculate the thermal averaged decay and formation rate of the
unstable particle based on the BUU equation. Using the time reversal
symmetry, we show that the time evolution of the density of the
unstable particle as Eq.(\ref{fe}).
Therefore in chemical equilibrium particles fugacities are connected by Eq.(%
\ref{equilcon}) as expected. We have explicit the thermal decay rate
of unstable particle, obtaining Eq.(\ref{Decay1-final}), which is
our principal result.

Using the formalism developed above, I examined the general properties of
the thermal particle decay/production rate. We see that for $T\ll m_{i}$
where the Boltzmann limit can be applied, the decay width is reduced to $%
\Upsilon _{1}/\tau _{0}$ and production width is $\Upsilon _{2}\Upsilon
_{3}/\tau _{0}$. For larger values of $T$ but $\Upsilon _{i}\ll 1$ so that
the Boltzmann approximation is valid, then decay width and production width
tend simply to $\Upsilon _{1}/\tau $ and $\Upsilon _{2}\Upsilon _{3}/\tau $,
respectively, where $\tau $ is essentially proportional to average Lorentz
factor and doesn't depend on $\Upsilon _{i}$. When some of $m_{i}/T$ and $%
\Upsilon _{i}$ are about unity or larger we see dependence of $\tau $ on $%
\Upsilon _{i}$.

I applied our formalism to $3$ examples, $\rho \leftrightarrow \pi +\pi $,
$\Sigma(1385) \leftrightarrow$ and $\pi ^{0}\leftrightarrow \gamma +\gamma .$ 
The first and second processes can take
place both in a hot hadronic gas created by the heavy ion collisions and in
the expanding early Universe. In particular for the heavy ion reaction case,
our analysis, coupled to the hydrodynamical expansion of the system will
furnish additional information of the dynamics of the system. We will study baryon resonances
evolution in heavy ions collisions in next chapter.
The relaxation time for $\pi^0$ decay remains close (within 50\%)  to relaxation
time in vacuum for large temperature range. In chapter~\ref{eeg} we will apply this for $\pi^0$
evolution in $e^+e^-\gamma$ plasma, created by the intensive laser pulse and in early universe. 
This part is going to be publised in~\cite{KuznKodRafl:2008}.
%%%%%%%%%%%%%%%%%%%%%%%%%%%%%%%%%%%%%%%%%%%%%%%%%%%%%%%%%%%%%%%%%%%%%%%%%%

\section{Summary of resonance production in heavy ions collisions}

In chapter~\ref{respr} I apply equations derived in chapter~\ref{dpeq} to baryon resonance densities evolution in thermal hadron gas
after QGP hadronization. The goal is to explain ratios $\Sigma(1385)/\Lambda^0$ and $\Lambda(1520)/\Lambda^0$ reported by RHIC experiment
and also predict $\Delta(1232)/N$ ratio. 

The resonant hadron states, considering their very  large decay and
reaction  rates, can interact beyond the chemical and thermal
freeze-out of stable particles. Thus the observed yield of
resonances is fixed by the physical conditions prevailing at a later
breakup of the fireball matter rather than the production of
non-resonantly interacting hadrons. Moreover,  resonances, observed
in  terms of the invariant mass signature,  are only visible when
emerging from a more dilute hadron system  given the ample potential
for rescattering of decay products. The combination of experimental
invariant mass method with a large resonant scattering makes the
here presented  population study of resonance kinetic  freeze-out
necessary. The evolution effects we find are greatly amplified at
low hadronization temperatures where greatest degree of initial
chemical equilibrium is present.

Our study quantifies the expectation that in a dense hadron  medium
narrow resonances are ``quenched''\cite{Rafelski:2001hp} that is,
effectively mixed with other states, and thus their observed
population is reduced. Since we follow here the particle density,
the effect we study is due to incoherent population  mixing of
$\Lambda(1520)$, in particular with $\Sigma^*$. This effect is
possible for particle densities out of chemical non-equilibrium.
However, this mixing can occur also at the amplitude (quantum
coherent) level. As the result the yield suppression effect  could
further increase, in some situations further improving the agreement
with experiment.

Our results show that the observable ratio $\Lambda(1520)_{\rm
ob}/\Lambda_{\rm tot}$ can be suppressed by two effects. First
$\Lambda(1520)$ yield is suppressed   due to excitation of heavy
$\Sigma^*$s in the resonance scattering process. Moreover, the final
$\Lambda(1520)_{\rm ob}$ yield is suppressed, because $\Sigma^*$s,
which decay to $\Lambda(1520)$, are suppressed at the end of the
kinetic phase evolution by their (asymmetric) decays to lower mass
hadrons, especially when dead channels are present (see
figure~\ref{Sigm1775r}). As a result, fewer  of these hadrons can
decay to $\Lambda(1520)_{\rm ob}$ during the following free
expansion. A contrary mechanism operates for the resonances such as
$\Sigma(1385), \Delta(1230)$. These resonances can be so strongly
enhanced, that in essence most final states strange and non-strange
baryons come from a resonance decay.

We note that despite a   scenario dependent resonance formation or suppression,
the stable particle yields used in study of chemical freeze-out remain unchanged, since
all resonances ultimately decay into the lowest ``stable'' hadron.   Therefore after a
description e.g. within a statistical hadronization model  of the yields
of stable hadrons, the understanding of resonance yields is a second, and  separate task
which helps to establish the consistency of our physical understanding of the hadron
production process.

We conclude noting the key result of this study, that   we can now understand the opposite behavior
of $\Lambda(1520)$ (suppression in high centrality reactions) and $\Sigma^(1385) $ (enhancement,
 and similarly $\Delta(1230)$) by considering their rescattering in matter. In order to explain both,
the behavior of the $\Lambda(1520)_{\rm ob}/\Lambda_{\rm tot}$
and $\Sigma(1385)/\Lambda_{\rm tot}$ ratios, one has to consider
$T=95-100$ MeV as the  favorite temperature of final kinetic
freeze-out  of hadron resonances, with   $T_0=140$ MeV being the
favored chemical freeze-out (hadronization, QGP break-up) temperature.
When there is little  matter  available to scatter, e.g. in peripheral  collisions,
  the average value of
$\Lambda(1520)_{\rm ob}/\Lambda_{\rm tot}$ ratio is higher,
approaching  the   expected chemical freeze-out  hadronization yield
for $T_0=140$ MeV. All these findings are in good agreement with
available experimental data.

This part is published in~\cite{Kuznetsova:2008zr} and~\cite{Kuznetsova:2008hb}.

\section{Results for relatvistic $e^+e^-\gamma$ plasma created by laser pulse}

In chapter~\ref{eeg} I consider $e^+e^-\gamma$ plasma. We study the freeze-out condition  for a relativistic $e^-e^+\gamma$ plasma,
when plasma density becomes so low that particles begin to stream freely. 

In order to address this question we evaluated Lorentz invariant
rates for the Compton scattering and pair production in thermally and chemically
equilibrated EP$^3$ plasma.  We then used these Lorentz invariant rates to
evaluate the corresponding mean free path length $l$ of particles.

Comparing $l$ to plasma drop size we showed that an opaque equilibrium
density plasma drop can be produced at energy $0.5$~kJ in the volume with largest
possible radius $R = 7$~nm. This volume corresponds to the smallest possible
 temperature $T = 2$~MeV. In order to reach  higher than 2~MeV temperature,
we need to increase energy of plasma (which is proportional to $T^4$)
or/and decrease plasma size. At higher temperature  opaque plasma can
be created at the total plasma energy  smaller than 0.5 MeV, since smaller
plasma drop size is in agreement with the opaqueness condition
Eq.\,(\ref{opcon}), as seen in figure \ref{tauee1}.

On the other hand in order to create opaque plasma with temperature
lower than 2 MeV, the necessary amount of  energy is larger than $0.5$~kJ.
This is so because the plasma size has to be large in order to satisfy
opaqueness condition Eq.\,(\ref{opcon}).

Our main result, perhaps unexpected at the first sight is illustrated in figure
\ref{tauee1}. For the  temperature $T>2$ MeV equilibrium plasma production with a
relatively small energy pulse (compared to lower temperature equilibrium plasma)  may be possible. However,
the challenge here is to focus the energy into the volume of size  $<10$~nm.

These results are in preparation~\cite{Inga2}. 

I also study
heavy particles (pion, muon) production in $e^{+}e^{-}\gamma$ plasma. 
We found that the production of $\pi^0$ is the dominant coupling of electromagnetic radiation
to heavy (hadronic) particles with $m \gg T$,  and as we have here demonstrated
 that noticeable particle yields can be expected already
at modest temperatures $T\in [3,10]$ MeV.
In present day  environment of 0.1 --1 J plasma  lasting a few fs, our results suggest that we can expect
integrated over space-time evolution of the EP$^3$ fireball a  $\pi^0$  yield at the limit of detectability.
For $T \to 15$ MeV the $\pi^0$ production rate  remains dominant and indeed
very large, reaching the production rate $R^{\prime}\simeq 10^{15}$[MJ$^{-1}$fs$^{-1}]$.
Charge exchange reactions convert some of the neutral pions into charged pions which are
more easy to detect.

In this situation it is realistic to consider the possibility of forming a chemically equilibrated fireball
with $\pi^0$, $\pi^{\pm}$, $\mu^{\pm}$ in chemical abundance equilibrium. The heavy particles are produced
in early stages when temperature reached is highest. Their abundance in the fireball  follows the fireball
expansion and cooling till their
freeze-out, that is decoupling of population equation production rates. The particle yields are than
given by the freeze-out conditions, specifically the chemical freeze-out temperature $T_f$ and
volume $V_f$, rather than the integral over the rate of production. In this situation the heavy
particle yields become diagnostic tools of the freeze-out conditions, with the mechanisms of
their formation being less accessible. However, one can avoid this condition by appropriate
staging of fireball properties.

The present study has not covered, especially for low temperature range all the possible
mechanisms, and we addressed some of these issues in the introduction. Here we note further
that the production of heavy particles
requires energies of the magnitude $m/2$ and thus   is due to collisions involving   the  (relatively speaking)
far tails of a thermal particle distribution. If these tails  fall off as a power law, instead of the
Boltzmann exponential  decay~\cite{Biro:2004qg},  a  much greater yield of heavy particles
 could ensue. There could further be present a   collective amplification
to the production process e.g. by residual matter flows, capable to
enhance the low temperature yields, or by collective plasma oscillations and inhomogeneities.

These are just some examples of many
reasons to hope and  expect a   greater particle yield than we
computed here in microscopic and controllable two particle reaction
approach. This consideration, and our encouraging
`conventional' results suggest that
 the study of  $\pi^0$ formation in  QED plasma  is of considerable  intrinsic interest. Our results
provide a lower limit for rate of particle production and when folded with models of EP$^3$ fireball
formation and evolution, final yield.

It is of some interest to note that  the study of pions in QED plasma allows exploration of pion properties
in electromagnetic medium. Specifically,    recall
 that  1.2\% fraction of  $\pi^0\to e^+e^-\gamma$ decays, which
implies that the associated processes such as  $e^++e^-\to \gamma +\pi_0$ are  important. We cannot
evaluate this process at present as it involves significant challenges in understanding of $\pi_0$
off-mass shell `anomalous' couping to two photons.

The experimental environment we considered here should allow
a detailed study of the properties of   pions (and also muons) in a thermal background.
There is considerable fundamental
interest in the study of   pion properties and specifically
pion mass splitting  in  QED plasma at temperature $T\gtrsim \Delta m$ and in presence of
electromagnetic fields. We already have shown that due to quantum statistics effects, the effective
in medium decay width of  $\pi^0$  differs  from the free  space value, see figure \ref{taupi0app}.
In addition,   modification of mass and  decay width due to ambient medium influence on
the pion internal structure is to be expected.
 Further we hope that the study of pions in the  EP$^3$ fireball  will contribute to the better
understanding of   the relatively large difference in mass between
 $\pi^0$and $\pi^\pm$. The  relatively large size of the PE$^3$ environment should 
make  such changes, albeit small, measurable.

The experimental study of  $\pi^0$ in QED plasma environment is not
an easy task. Normally, one would think that the study of the
  $\pi^0$ decay into two  67.5 MeV $\gamma$ (+ thermal Doppler shift motion)
produces a  characteristic signature. However, the   $\pi^0$ decay is
in time and also in location overlapping with the plasma formation and disintegration. The debris of the plasma,
 reaches any detection system  at practically the same time instance as does the 67.5 MeV $\gamma$. The large
amount of available radiation will disable  the detectors. On the other hand we realize that
the hard thermal component of the plasma, which leads to the production of   $\pi^0$ in the early fireball stage, is
most attenuated by plasma dynamical expansion.  Thus it seems possible to plan for  the  detection of
$\pi^0$ e.g. in a heavily shielded detection system.

The decay time of charged pions being 26 ns, and that of charged muons being 2.2 $\mu$s
it is possible to separate in time the plasma debris from the decay signal of these particles.
Clearly, these heavy charged particles can be detected with much greater ease, also considering that
the decay product of interest is charged.
For this reason, we also have in depth considered all channels of production of charged pions
and muons. Noting that practically all charged pions turn into muons, we  have also compared
the production rates of $\pi^0$ with all heavy particles, see dot-dashed (green) line in figure \ref{mupir}.
This comparison suggests that for plasmas at a temperature reaching $T>10$ MeV the production
of final state muons will most probably be by far easier to detect. On the other hand for $T<5$ MeV
it would seem that the yield difference in favor of  $\pi^0$ outweighs the detection system/efficiency loss
considerations. Future work addressing non-conventional processes will show at how low  $T$
we can still expect observable heavy particle yields.

An   effort to detect $\pi^0$ directly is justified since   we can learn  about the
properties of the plasma (lifespan, volume and temperature in early stages) e.g.
from a comparative study of  the $\pi^0$ and  $\pi^\pm$ production.
We have found that  at about $T>16$ MeV, the pion charge exchange  $\pi^0\pi^0\to \pi^+\pi^-$
reaction for chemically equilibrated $\pi^0$ yield is  faster than the natural $\pi^0$ decay,
 and the chemical equilibration time constant,  see the dot-dashed line  in
figure \ref{taumupi}. Thus beyond this temperature the yield of charged pions can be expected to be
in/near chemical equilibrium for a plasma which lives at, or above this temperature, for longer than 100 as.

In such an environment the yield of $\pi^0$ is expected to be near chemical equilibrium, since the
decay rate is compensated by the production rate, and, within 100 as, the chemical
equilibrium yield is attained. Moreover,
the thermal speed of produced $\pi$ can be  obtained from the nonrelativistic
relation $\frac 12 m \langle v^2\rangle =\frac 3 2 T$,  thus $\overline v \propto  \sqrt T $ and, for $T=10 $ MeV,
$\overline v\simeq 0.5 $c. This is
nearly equal to the sound velocity of EP$^3$,  $v_s\simeq c/\sqrt{3}=0.58c$. Thus the heavy $\pi^0$ particles
can be seen as co-moving with the expanding/exploding  EP$^3$, which completes
the argument to justify their transient chemical
equilibrium yield in this condition.

The global production yield of neutral and charged pions
should  thus  allow the study of volume and temperature history of the QED  plasma.
More specifically, since with decreasing temperature, for $T<16$ MeV, there is a rapid  increase of the relaxation time for
the charge exchange  process, there is a rather rapid drop
of the  charged pion yield below chemical equilibrium --- we note that charge exchange   equilibration time at $T=10$ MeV is
a factor $10^5$ longer. We note that the study of two pion correlations provides an independent measure of
the source properties (HBT measurement).

The relaxation time of electromagnetic production of muon pairs wins over $\pi^0$
relaxation time for $T>22$ MeV, see dashed line, red, in figure \ref{taumupi},
the  direct  electromagnetic processes
of charged pion production (thin green, solid line for $ \gamma\gamma \rightarrow
\pi^+\pi^-$ and dashed, blue for $e^+e^- \rightarrow \pi^+\pi^-$) remain  sub-dominant.
Thus for $T>22$ MeV we expect, following the same chain of arguments for muons as above for
charged pions, a near chemical equilibrium yield. If the study of all these
$\pi^0,\pi^\pm,\mu^\pm$ yields,  their spectra and even pion correlations were possible, considerable insight into
 $ e^-,   e^+, \gamma $ plasma (EP$^3$) plasma formation and dynamics  at $T<25$ MeV can be achieved.
This part is publised in~\cite{Kuznetsova:2008jt}

In chapter \ref{earlyun} we studied the pion equilibration in early universe. 
In early universe for temperature range of interest the $\pi^0$ are in chemical equilibrium with photons because of their fast decay rate. This is not always that decay is so fast to exceed universe expansion rate. For example the decay $n \rightarrow p + e^- + \nu_e$ is much slower $\tau = 885.7$ s. The dilution rate exceeds neutron decay rate at $T > 0.1$ MeV.

The relaxation time for $\mu^{\pm}$ and $\pi^{\pm}$ in reactions (\ref{ggmu}) and (\ref{pppp}) respectively 
become many orders of magnitude larger than $\tau_{\pi^0}$ at temperatures about few MeV, where these reaction have to freeze out. Therefore these particles are in chemical equilibrium and their density is also relatively high upto temperatures of few MeV.  

These processes are important to understand how the hadronic component diminish with
the expansion of the Universe and the possible effects of hadronic relics in
the cosmic blackbody radiation spectrum, such as its fluctuations and
correlations.
This part is going to be published in~\cite{KuznKodRafl:2008}.

%\setcounter{figure}{0}
%\setcounter{equation}{0}
%\setcounter{table}{0}
%\input{chapter8}
% Get ready for the appendices

% Create the ``References'' list
%\bibliographystyle{uabib}
%\bibliography{references}

\end{document}